\pdfoutput=1
\documentclass[usenatbib]{mn2e}
\usepackage{apjfonts}
\usepackage{amssymb}
\usepackage{amsmath}
\usepackage{ctable}
\usepackage{verbatim}
\usepackage{url}
\usepackage{fixltx2e} 
\usepackage[implicit=false,breaklinks,colorlinks,citecolor=blue]{hyperref}

\newcommand{\be}{\begin{equation}}
\newcommand{\ee}{\end{equation}}

\newcommand{\etal}{et al.}

\newcommand{\msun}{M_{\sun}}

\newcommand{\Alf}{{Alfv\'en}}

\newcommand{\ICsurl}{\href{http://www.tapir.caltech.edu/~phopkins/publicICs}{\url{http://www.tapir.caltech.edu/~phopkins/publicICs}}}
\newcommand{\FIREurl}{\href{http://fire.northwestern.edu}{\url{http://fire.northwestern.edu}}}
\newcommand{\gizmourl}{\href{http://www.tapir.caltech.edu/~phopkins/Site/GIZMO.html}{\url{http://www.tapir.caltech.edu/~phopkins/Site/GIZMO.html}}}
\newcommand{\movieurl}{\href{http://www.tapir.caltech.edu/~phopkins/Site/animations/}{\url{http://www.tapir.caltech.edu/~phopkins/Site/animations/}}}

\newcommand\plotonesize[2]
 {\centering \leavevmode \includegraphics[width={#2\columnwidth}]{#1}}
\newcommand{\plotsidesize}[2]
 {\centering \leavevmode \includegraphics[width={#2\textwidth}]{#1}}
\newcommand{\acknowledgments}{\begin{small}\section*{Acknowledgments}\end{small}}
\newcommand\altaffilmark[1]{$^{#1}$}
\newcommand\altaffiltext[1]{$^{#1}$}
\title[Cosmic Rays on FIRE]{But What About... Cosmic Rays, Magnetic Fields, Conduction, \&\ Viscosity in Galaxy Formation
\vspace{-0.5cm}}

\vspace{-0.2cm}
\author[Hopkins \etal]{
\parbox[t]{\textwidth}{ 
Philip F.~Hopkins\thanks{E-mail:phopkins@caltech.edu}\altaffilmark{1},
T.~K.\ Chan\altaffilmark{2},
Shea Garrison-Kimmel\altaffilmark{1},
Suoqing Ji\altaffilmark{1},
Kung-Yi Su\altaffilmark{1},
Cameron B.~Hummels\altaffilmark{1},
Du\v{s}an Kere\v{s}\altaffilmark{2}, 
Eliot Quataert\altaffilmark{3},
Claude-Andr{\'e} Faucher-Gigu{\`e}re\altaffilmark{4}
} 
\vspace*{6pt} \\
\altaffiltext{1}{TAPIR, Mailcode 350-17, California Institute of Technology, Pasadena, CA 91125, USA} \\
\altaffiltext{2}{Department of Physics, Center for Astrophysics and Space Sciences, University of California at San Diego, 9500 Gilman Drive, La Jolla, CA 92093} \\ 
\altaffiltext{3}{Department of Astronomy and Theoretical Astrophysics Center, University of California Berkeley, Berkeley, CA 94720} \\ 
\altaffiltext{4}{Department of Physics and Astronomy and CIERA, Northwestern University, 2145 Sheridan Road, Evanston, IL 60208, USA} 
\vspace{-0.5cm}
}

\date{Working Document\vspace{-0.6cm}}
\begin{document}
\maketitle
\label{firstpage}

\vspace{-0.2cm}
\begin{abstract}
We present and study a large suite of high-resolution cosmological zoom-in simulations, using the FIRE-2 treatment of mechanical and radiative feedback from massive stars, together with explicit treatment of magnetic fields, anisotropic conduction and viscosity (accounting for saturation and limitation by plasma instabilities at high-$\beta$), and cosmic rays (CRs) injected in supernovae shocks (including anisotropic diffusion, streaming, adiabatic, hadronic and Coulomb losses). We survey systems from ultra-faint dwarf ($M_{\ast}\sim 10^{4}\,M_{\sun}$, $M_{\rm halo}\sim 10^{9}\,M_{\sun}$) through Milky Way/Local Group (MW/LG) masses, systematically vary uncertain CR parameters (e.g.\ the diffusion coefficient $\kappa$ and streaming velocity), and study a broad ensemble of galaxy properties (masses, star formation [SF] histories, mass profiles, phase structure, morphologies, etc.). We confirm previous conclusions that magnetic fields, conduction, and viscosity on resolved ($\gtrsim 1\,$pc) scales have only small effects on bulk galaxy properties. CRs have relatively weak effects on all galaxy properties studied in dwarfs ($M_{\ast} \ll 10^{10}\,M_{\sun}$, $M_{\rm halo} \lesssim 10^{11}\,M_{\sun}$), or at high redshifts ($z\gtrsim 1-2$), for {\em any} physically-reasonable parameters. However at higher masses ($M_{\rm halo} \gtrsim 10^{11}\,M_{\sun}$) and $z\lesssim 1-2$, CRs can suppress SF and stellar masses by factors $\sim 2-4$, given reasonable injection efficiencies and relatively high effective diffusion coefficients $\kappa \gtrsim 3\times10^{29}\,{\rm cm^{2}\,s^{-1}}$. At lower $\kappa$, CRs take too long to escape dense star-forming gas and lose their energy to collisional hadronic losses, producing negligible effects on galaxies and violating empirical constraints from spallation and $\gamma$-ray emission. At much higher $\kappa$ CRs escape too efficiently to have appreciable effects even in the CGM. But around $\kappa\sim 3\times10^{29}\,{\rm cm^{2}\,s^{-1}}$, CRs escape the galaxy and build up a CR-pressure-dominated halo which maintains approximate virial equilibrium and supports relatively dense, cool ($T\ll 10^{6}$\,K) gas that would otherwise rain onto the galaxy. CR ``heating'' (from collisional and streaming losses) is never dominant.
\end{abstract}

\begin{keywords}
galaxies: formation --- galaxies: evolution --- galaxies: active --- 
stars: formation --- cosmology: theory\vspace{-0.5cm}
\end{keywords}

\section{Introduction}
\label{sec:intro}

\begin{footnotesize}
\ctable[
  caption={{\normalsize Zoom-in simulation volumes run to $z=0$ (see \citealt{hopkins:fire2.methods} for details). All units are physical.}\label{tbl:sims}},center,star
  ]{lcccccr}{
\tnote[ ]{Halo/stellar properties listed refer only to the original ``target'' halo around which the high-resolution volume is centered: these volumes can reach up to $\sim (1-10\,{\rm Mpc})^{3}$ comoving, so there are actually several hundred resolved galaxies in total. 
{\bf (1)} Simulation Name: Designation used throughout this paper. 
{\bf (2)} $M_{\rm halo}^{\rm vir}$: Virial mass \citep[following][]{bryan.norman:1998.mvir.definition} of the ``target'' halo at $z=0$.
{\bf (3)} $M_{\ast}^{\rm MHD+}$: Stellar mass of the central galaxy at $z=0$, in our non-CR, but otherwise full-physics (``MHD+'') run. 
{\bf (4)} $M_{\ast}^{\rm CR+}$: Stellar mass of the central galaxy at $z=0$, in our ``default'' (observationally-favored) CR+ ($\kappa=3e29$) run.
{\bf (5)} $m_{i,\,1000}$: Mass resolution: the baryonic (gas or star) particle/element mass, in units of $1000\,\msun$. The DM particle mass is always larger by the universal ratio, a factor $\approx 5$. 
{\bf (6)} $\langle \epsilon_{\rm gas} \rangle^{\rm sf}$: Spatial resolution: the gravitational force softening (Plummer-equivalent) at the mean density of star formation (gas softenings are adaptive and match the hydrodynamic resolution, so this varies), in the MHD+ run. Time resolution reaches $\sim 10-100\,$yr and density resolution $\sim 10^{3}-10^{4}\,{\rm cm^{-3}}$.
{\bf (7)} Additional notes.\vspace{-0.5cm}
}
}{
\hline\hline
Simulation & $M_{\rm halo}^{\rm vir}$ & $M_{\ast}^{\rm MHD+}$ & $M_{\ast}^{\rm CR+}$ & $m_{i,\,1000}$ & $\langle \epsilon_{\rm gas} \rangle^{\rm sf}$ & Notes \\
Name \, & $[\msun]$ &  $[\msun]$  &   $[\msun]$  &$[1000\,\msun]$ & $[{\rm pc}]$ & \, \\ 
\hline 
{\bf m09} & 2.4e9 &  2e4 & 3e4 & 0.25 & 0.7 & early-forming, ultra-faint field dwarf \\
{\bf m10v} & 8.3e9 & 2e5 & 3e5 & 0.25 & 0.7 & isolated dwarf in a late-forming halo \\
{\bf m10q} & 8.0e9 & 2e6 & 2e6 & 0.25 & 0.8 & isolated dwarf in an early-forming halo \\
{\bf m10y} & 1.4e10 & 1e7 & 1e7 & 0.25 & 0.7 & early-forming dwarf, with a large dark matter ``core'' \\
{\bf m10z} & 3.4e10 & 4e7 & 3e7 & 0.25 & 0.8 & ultra-diffuse dwarf galaxy, with companions \\
\hline
{\bf m11a} & 3.5e10 & 6e7 & 5e7 & 2.1 & 1.6 & classical dwarf spheroidal \\
{\bf m11b} & 4.3e10 & 8e7 & 8e7 & 2.1 & 1.6 & disky (rapidly-rotating) dwarf \\
{\bf m11i} & 6.8e10 & 6e8 & 2e8 & 7.0 & 1.8 & dwarf with late mergers \&\ accretion \\
{\bf m11e} & 1.4e11 & 1e9 & 7e8 & 7.0 & 2.0 & low surface-brightness dwarf \\
{\bf m11c} & 1.4e11 & 1e9 & 9e8 & 2.1 & 1.3 & late-forming, LMC-mass halo \\
\hline
{\bf m11q} & 1.5e11 & 1e9 & 1e9 & 0.88 & 1.0 & early-forming, large-core diffuse galaxy\\
{\bf m11v} & 3.2e11 & 2e9 & 1e9 & 7.0 & 2.4 & has a multiple-merger ongoing at $z\sim0$ \\
{\bf m11h} & 2.0e11 & 4e9 & 3e9 & 7.0 & 1.9 & early-forming, compact halo \\
{\bf m11d} & 3.3e11 & 4e9 & 2e9 & 7.0 & 2.1 & late-forming, ``fluffy'' halo and galaxy \\
{\bf m11f} & 5.2e11 & 3e10 & 1e10 & 12 & 2.6 & early-forming, intermediate-mass halo \\
\hline
{\bf m11g} & 6.6e11 & 5e10 & 1e10  & 12 & 2.9 & late-forming, intermediate-mass halo \\
{\bf m12z} & 8.7e11 & 2e10 & 8e9 & 4.0 & 1.8 & disk with little bulge, ongoing merger at $z\sim0$ \\
{\bf m12r} & 8.9e11 & 2e10 & 9e9 & 7.0 & 2.0 & late-forming, barred thick-disk \\
{\bf m12w} & 1.0e12 & 6e10 & 2e10 & 7.0 & 2.1 & forms a low surface-brightness / diffuse disk \\
\hline
{\bf m12i} & 1.2e12 & 7e10 & 3e10 & 7.0 & 2.0 & ``Latte'' halo, later-forming MW-mass halo, massive disk\\  
{\bf m12b} & 1.3e12 & 9e10 & 4e10 & 7.0 & 2.2 & early-forming, compact bulge+thin disk \\
{\bf m12c} & 1.3e12 & 6e10 & 2e10 & 7.0 & 1.9 & MW-mass halo with $z\sim1$ major merger(s) \\
{\bf m12m} & 1.5e12 & 1e11 & 3e10 & 7.0 & 2.3 & earlier-forming halo, features strong bar at late times \\ 
{\bf m12f} & 1.6e12 & 8e10 & 4e10 & 7.0 & 1.9 & MW-like disk, merges with LMC-like companion \\ 
\hline\hline
}
\end{footnotesize}

\begin{footnotesize}
\ctable[
  caption={{\normalsize Additional high-redshift, massive-halo simulations ($M_{\rm halo}^{\rm vir} \gtrsim 10^{12}\,M_{\sun}$). All units are physical.}\label{tbl:HiZ}},center,star
  ]{lcccccll}{
\tnote[ ]{Halo/stellar properties (as Table~\ref{tbl:sims}) of simulations which feature halos that reach $\sim 10^{12}\,M_{\sun}$ at various redshifts $z$ (i.e.\ are more massive at $z\sim0$). Lacking AGN feedback, we do not evolve these halos past masses $\sim 10^{12-13}\,M_{\sun}$. Columns show:
{\bf (1)} Simulation name. 
{\bf (2)} $z_{12}$: Redshift at which the halo virial mass reaches $\sim 10^{12}\,M_{\sun}$.
{\bf (3)} $z_{f}$: Lowest redshift to which the simulation is evolved.
{\bf (4)} $M_{\ast}^{\rm MHD+}(z_{f})$: Stellar mass (in $M_{\sun}$) of the primary galaxy at $z_{f}$, in the MHD+ run.
{\bf (5)} $M_{\ast}^{\rm CR+}(z_{f})$:  Stellar mass (in $M_{\sun}$) of the primary galaxy at $z_{f}$, in the CR+($\kappa=3e29$) run.
{\bf (6)} $m_{i,\,1000}$: Mass resolution in $1000\,M_{\sun}$.
{\bf (7)} Reference (paper in which this IC first appeared).
{\bf (8)} Additional notes.\vspace{-0.2cm}
}
}{
\hline\hline
Simulation & $z_{12}$ & $z_{f}$ & $M_{\ast}^{\rm MHD+}$ & $M_{\ast}^{\rm CR+}$ & $m_{i,\,1000}$ & Reference & Notes \\
\hline
{\bf m12z10} & 10 & 10 & 2e9 & 3e9 & 7 & \citet{ma:fire.reionization.epoch.galaxies.clumpiness.luminosities} & clumpy, multiply-merging, no defined center \\
{\bf m12z7} & 7 & 7 & 1e10 & 1e10 & 7 & \citet{ma:fire.reionization.epoch.galaxies.clumpiness.luminosities} & well-defined center, but still clumpy \&\ extended \\
{\bf m12z5} & 5 & 5 & 2e10 & 2e10 & 7 & \citet{ma:fire.reionization.epoch.galaxies.clumpiness.luminosities} & ordered structure emerging, bursty SFH \\
{\bf m12z4} & 4 & 2 & 3e11 & 3e11 & 32 & \citet{feldmann.2016:quiescent.massive.highz.galaxies.fire} & massive $z\sim4-5$ starburst forms extremely-dense bulge \\
{\bf m12z3} & 3 & 2.5 & 2e11 & 3e11 & 32 & \citet{daa:BHs.on.FIRE} & submillimeter galaxy, $z\sim 2-3$ starburst leaves dense bulge \\
{\bf m12z2} & 2 & 1 & 2e11 & 2e11 & 56 & \citet{faucher-giguere:2014.fire.neutral.hydrogen.absorption} & $z\sim1-2$ massive LBG-like galaxy \\
{\bf m12z1} & 1 & 0 & 2e11 & 1e11 & 56 & \citet{hopkins:2013.fire} & very little $z<1$ growth ({\bf m12q} in  \citet{hopkins:2013.fire}) \\
\hline\hline
}
\end{footnotesize}

\begin{footnotesize}
\ctable[
  caption={{\normalsize Default physics ``suites'' in this paper}\label{tbl:physics}},center,star
  ]{ll}{
\tnote[ ]{Description of the basic physics in the simulation sets described here. This notation is used throughout the paper.}
}{
\hline\hline
Name & Notes \\
\hline 
Hydro+ & ``Default'' FIRE-2 physics. Includes stellar feedback (SNe, radiation, mass-loss), cooling, gravity, but no MHD or CRs \\
MHD+ & Includes all ``Default'' FIRE-2 physics, plus ideal MHD, with anisotropic Spitzer-Braginskii conduction \&\ viscosity \\
CR+ & Includes all ``MHD+'' physics, with CR injection in SNe, CR streaming, diffusion (with coefficient $\kappa$ in cgs), hadronic and Coulomb losses\\
& - in this category we survey these properties, e.g.\ ``CR+($\kappa=3e28$)'' and ``CR+($\kappa=3e29$)'' vary $\kappa$ as labeled\\
\hline\hline
}
\end{footnotesize}

Galaxy and star formation are intrinsically multi-physics processes that involve a competition between gravity, collisionless dynamics, fluid dynamics and turbulence, radiation-matter coupling and chemistry, relativistic particles, magnetic fields, and more. Many of these processes enter most dramatically via ``feedback'' from massive stars, whereby the radiation, winds, and explosions from massive stars dramatically alter subsequent generations of star and galaxy formation. 

In recent years, there has been tremendous progress in modeling and understanding the effects of multi-phase gas in the interstellar and circum/inter-galactic medium (ISM, CGM, and IGM), radiative cooling, turbulence, and self-gravity, and how these processes couple to stellar feedback. For example, it is now well-established that without feedback, these processes lead to runaway fragmentation and gravitational collapse that turns dense gas into stars on a single gravitational free-fall time (on both giant molecular cloud [GMC] and galactic scales; \citealt{tasker:2011.photoion.heating.gmc.evol,hopkins:rad.pressure.sf.fb,dobbs:2011.why.gmcs.unbound,harper-clark:2011.gmc.sims}), and transforms most of the baryons in the Universe into stars \citep{katz:treesph,somerville99:sam,cole:durham.sam.initial,springel:lcdm.sfh,keres:fb.constraints.from.cosmo.sims}, both in stark contrast to observations. Moreover direct and indirect effects of feedback are ubiquitously observed in outflows and enrichment of the CGM and IGM \citep{martin99:outflow.vs.m,heckman:superwind.abs.kinematics,pettini:2003.igm.metal.evol,songaila:2005.igm.metal.evol,martin:2010.metal.enriched.regions,sato:2009.ulirg.outflows,weiner:z1.outflows,steidel:2010.outflow.kinematics}. Of course, feedback (even restricting just to feedback from stars) comes in a variety of forms, including radiative (ionization, photo-heating and radiation pressure), mechanical (thermal and kinetic energy associated with supernovae Types Ia \&\ II, stellar mass-loss, and protostellar jets), the injection of magnetic fields, and acceleration of cosmic rays. Numerical simulations have begun to directly resolve the relevant scales of some of these processes and therefore have begun to explicitly treat some of these feedback channels and their interactions on ISM and galactic or inter-galactic scales \citep[e.g.][]{tasker:2011.photoion.heating.gmc.evol,hopkins:rad.pressure.sf.fb,hopkins:fb.ism.prop,wise:2012.rad.pressure.effects,kannan:2013.early.fb.gives.good.highz.mgal.mhalo,agertz:2013.new.stellar.fb.model,roskar:2014.stellar.rad.fx.approx.model}. 

One example is the suite of simulations studied in the ``Feedback In Realistic Environments'' (FIRE)\footnote{\label{foot:movie}See the {\small FIRE} project website:\\
\FIREurl \\
For additional movies and images of FIRE simulations, see:\\
\movieurl} project \citep{hopkins:2013.fire}. 
These simulations have been used extensively in recent years to explore the interplay between a multi-phase ISM and CGM and both radiative \citep{hopkins:radiation.methods} and mechanical feedback \citep{hopkins:sne.methods} processes from stars. These processes alone, coupled to the physics of radiative cooling, gravity, and star formation, appear to explain a wide variety of observed phenomena in galaxies, including their abundances \citep{ma:2015.fire.mass.metallicity,escala:turbulent.metal.diffusion.fire}, star formation ``main sequence'' and fluctuations \citep{sparre.2015:bursty.star.formation.main.sequence.fire}, satellite mass functions \citep{wetzel.2016:latte,garrisonkimmel:local.group.fire.tbtf.missing.satellites}, color distributions \citep{feldmann.2016:quiescent.massive.highz.galaxies.fire}, stellar \citep{wheeler:dwarf.satellites,wheeler.2015:dwarfs.isolated.not.rotating} and gas-phase \citep{elbadry:fire.morph.momentum,elbadry:HI.obs.gal.kinematics} kinematics, radial gradients and internal thick/thin disk structure \citep{ma:radial.gradients,ma:2016.disk.structure,bonaca:gaia.structure.vs.fire}, stellar halos \citep{sanderson:stellar.halo.mass.vs.fire.comparison,elbadry:most.ancient.mw.halo.stars.kicked}, multi-phase fast outflows \citep{muratov:2015.fire.winds,muratov:2016.fire.metal.outflow.loading}, dark matter profiles \citep{onorbe:2015.fire.cores,chan:fire.dwarf.cusps}, and more. 

However, essentially all of the conclusions above were based on simulations that treated the gas in the hydrodynamic limit. The detailed plasma physics of the ISM and CGM/IGM is of course quite complex and still a subject of active research. But it is well-established that magnetic fields are ubiquitous and important for local particle transport, that anisotropic transport processes (e.g.\ conduction and viscosity) can become significant in hot, tenuous gas, and that the ISM and IGM contain a spectrum of relativistic charged particles (cosmic rays [CRs]). In a very broad sense at least in the local Solar-neighborhood ISM, magnetic field and CR energy densities are order-of-magnitude comparable to thermal and turbulent energy densities \citep{Ginz85,Boul90}, suggesting they may not be negligible dynamically. 

Magnetic fields have been studied and discussed in the galactic and extra-galactic context for decades \citep[for reviews, see][]{1996ARA&A..34..155B,2009ASTRA...5...43B}. They can, in principle, slow star formation in dense gas \citep{2005ApJ...629..849P,2007ApJ...663..183P,2009ApJ...696...96W,2012MNRAS.422.2152B,2013MNRAS.432..176P,2015ApJ...815...67K}, or alter fluid mixing instabilities \citep{1995ApJ...453..332J, 2015MNRAS.449....2M,2017MNRAS.470..114A} and the evolution of SNe remnants \citep{1996ApJ...465..800J,1996ApJ...472..245J,1999ApJ...511..774J,2000ApJ...534..915T, 2015ApJ...815...67K}, and of course determine the actual dynamics of anisotropic transport. Anisotropic thermal conduction and viscosity in hot gas have also been studied extensively in the past (albeit not quite as widely), and it has been widely suggested that both could be important for plasma heating and dynamics on galaxy cluster scales \citep{2005MNRAS.357..242R,2006MNRAS.371.1025S,2007PhR...443....1M,2009ApJ...699..348S,2010ApJ...720..652S, 2012MNRAS.422..704P, 2012ApJ...747...86C,2017MNRAS.470..114A} or, again, in SNe remnants (see references above), or for the mixing/survival of cool clouds in hot galactic outflows or the CGM \citep{2016ApJ...822...31B,2017MNRAS.470..114A}. However, it is not clear if these processes are particularly important for {\em galaxy} properties. In fact, most studies in the past have argued the effects of these physics in dwarfs and Milky Way (MW)-mass ($\sim L_{\ast}$) galaxies are relatively small -- perhaps not surprising since the magnetic dynamo in super-sonic turbulence appears to saturate with magnetic fields always sub-dominant to turbulence \citep[effectively, passively-amplified; see][]{federrath:supersonic.turb.dynamo,su:fire.feedback.alters.magnetic.amplification.morphology,squire.hopkins:turb.density.pdf,colbrook:passive.scalar.scalings,2017MNRAS.471.2674R,2018MNRAS.473.3454B,2018MNRAS.479.3343M} and conduction/viscosity depend strongly on gas temperature and are weaker in the cooler gas of sub-$L_{\ast}$ galaxy halos.\footnote{Throughout this manuscript, we use ``halo'' to refer to the extended circum-galactic gas, stars, and/or dark matter extending from outside the ``luminous'' galaxy to around the virial radius, i.e.\ from $\sim 10-30\,$kpc to $\sim 200-400\,$kpc. This is standard notation in many fields (e.g.\ cosmology, CGM/IGM, galaxy formation), but we emphasize that this is quite different from what is usually called ``halo'' gas in much of the Galactic CR and magnetic field literature, where ``halo'' is often used to refer to gas within a scale height $\lesssim 1-10\,$kpc of the disk midplane out to $\sim 10\,$kpc (i.e.\ the ``thick disk'' in standard galaxy/CGM notation; see \citealt{2015ASSL..407..483H,2017ARA&A..55..111H}).} In \citet{su:2016.weak.mhd.cond.visc.turbdiff.fx}, we attempted to study the effect of magnetic fields, anisotropic conduction, and viscosity in addition to the physics described above in FIRE simulations, and concluded the effects were minimal. However, the simulations in that paper were mostly non-cosmological (although it did include two cosmological cases), so might not capture all the important effects in the CGM. Moreover, a swathe of recent work in plasma physics has argued that conductivity and viscosity of dilute plasmas might be self-limiting under exactly the relevant conditions of the ISM and CGM \citep[see][and references therein]{kunz:firehose,riquelme:viscosity.limits,roberg:2018.whistler.turb.conduction.suppression,squire:2017.max.anisotropy.kinetic.mhd,komarov:whistler.instability.limiting.transport}, and these effects were not accounted for in previous studies (although they generally act to {\em weaken} the conductivity and viscosity). 

The situation with CRs is much less clear. In the MW (and, it is widely believed, most dwarf and $\sim L_{\ast}$ or star-forming galaxies), the CR pressure and energy density (and correspondingly, effects on both gas dynamics and heating/cooling rates of gas via hadronic or Coulomb collisions or excitation of \Alf\ waves in various plasma ``streaming instabilities''; \citealt{Mann94,Enss07,guo.oh:cosmic.rays}) are dominated by mildly-relativistic $\sim$\,GeV protons accelerated primarily in supernova remnants (with $\sim 10\%$ of the SNe ejecta energy ultimately in CRs; \citealt{Bell.cosmic.rays}). In more massive galaxies, which host supermassive black holes (BHs) and have little star formation, most of the CR production appears to be associated with AGN jets and ``bubbles.'' CRs and their influence on galaxy evolution have been a subject of interest in both analytic \citep{Ipav75,Brei91,Brei93,Zira96,Socr08,Ever08,Dorf12,Mao18} and numerical simulation \citep{jubelgas:2008.cosmic.ray.outflows,uhlig:2012.cosmic.ray.streaming.winds,Boot13,Sale14,Rusz17,Farb18} studies for decades -- with an explosion of work in recent years. This work has argued that CRs could, in principle, drive galactic outflows \citep{Simp16,Giri16,Pakm16,Wien17}, suppress star formation in low (or high) mass galaxies \citep{Hana13,Chen16,Jaco18}, provide additional pressure to ``thicken'' galactic gas disks \citep{Wien13,Sale14cos}, alter the phase structure of the CGM \citep{Sale16,Giri18,Buts18}, ``open up'' magnetic field lines or otherwise alter the galactic dynamo \citep{Park92,Hana09,Kulp11,Kulp15}, and more. 

However, a number of major uncertainties and limitations remain in this field. First, the actual CR transport processes, and their coupling to the gas, remain deeply uncertain (owing to the extremely complicated plasma processes involved) -- the physics that gives rise to some ``effective diffusivity'' and/or CR streaming is still debated \citep[see][]{Stro07,Zwei13,Gren15}, and there is no widely-accepted {\em a priori} model which predicts the relevant transport coefficients in the way of, say, Spitzer-Braginskii conductivity and viscosity. There are some empirical constraints from e.g.\ $\gamma$-ray emission in nearby galaxies or more detailed products (e.g.\ spallation) in the MW, but critically any inferred constraint on the ``effective diffusion coefficient'' or ``streaming speed'' of CRs is {\em strongly model-dependent} (as it depends on the density distribution the CRs propagate through, the magnetic field configuration, etc.). Really, one must forward-model these constraints in any galaxy model, to test whether the adopted CR transport assumptions are consistent with the observations. Second, almost all previous studies of CRs on galaxies either focused on (a) idealized ``patches'' of the ISM or CGM, ignoring the global dynamics of accretion, outflows, star formation, etc., or (b) galaxy simulations with (intentionally) highly-simplified models for the turbulent, multi-phase ISM, star formation, stellar feedback from supernovae, stellar winds, radiation, and more. But these details are critical for determining the balance of CR heating and cooling, how CRs will be trapped or escape galaxies, whether CRs will influence outflows or gas in the CGM ``lofted up''  by other processes, and whether CRs ultimately matter compared to the order-of-magnitude larger energy input in mechanical (thermal+kinetic) form in SNe. To give an extreme example: almost anything will have a large effect relative to a ``baseline'' model which includes weak or no stellar feedback. It is much less clear whether the inclusion of CR physics will ``matter'' once mechanical and radiative feedback from stars is already accounted for. 

Working towards this goal, \citet{chan:2018.cosmicray.fire.gammaray} performed and presented the first simulations combining the specific physics from the FIRE simulations, described above, with explicit CR injection and transport, accounting for advection and fully-anisotropic streaming and diffusion, as well as hadronic and Coulomb collisional and streaming (\Alf) losses. They systematically varied the transport coefficients and treatment, and compared with observational constraints, to argue that -- at least {\em given} this particular physics set and treatment of CR transport -- the observations required diffusivities $\gtrsim 10^{29}\,{\rm cm^{2}\,s^{-1}}$, and that within the allowed range of diffusivities, the effects on galaxy star formation rates and gas density distributions were modest. However, these simulations were restricted to non-cosmological, isolated galaxies, representative of just a couple of $z=0$ galaxy types (e.g.\ one dwarf, one MW-mass system). Predicting the consequences for galactic winds or the CGM and therefore long-term galaxy evolution (e.g. stellar masses, etc.) requires cosmological simulations. 

In this paper, we therefore introduce and explore a new, high-resolution suite of FIRE simulations, with $>150$ fully-cosmological simulations spanning halo masses from ultra-faint dwarfs through MW-mass systems at a range of redshifts (reaching $\sim$\,pc-scale resolution), and systematically exploring all of the physics above. Specifically, we compare our standard physics assuming hydrodynamics, to simulations with explicit MHD and anisotropic conduction and viscosity as in \citet{su:2016.weak.mhd.cond.visc.turbdiff.fx}, and simulations with all of the above plus explicit treatment of cosmic rays as in \citet{chan:2018.cosmicray.fire.gammaray}. We moreover systematically survey the treatment of CR transport physics and coefficients, and compare with observations where possible to constrain the allowed range of assumptions. Our intention here is to identify which physics might have an influence on {\em bulk galaxy} properties (e.g.\ SFRs, stellar masses, morphologies), and where uncertain parameters exist (e.g.\ CR diffusivities), what range of those parameters is allowed and how the effects (if any) on galaxies depend on them within the allowed range. We also limit our study to dwarf and $\sim L_{\ast}$ galaxies where it is widely believed that SNe dominate the CR injection. In companion papers (e.g.\ \citealt{su:turb.crs.quench}, Su et al., in prep.) we will study the complementary role of AGN injecting CRs in much more massive galaxies, and in other companion papers (e.g.\ \citealt{ji:fire.cr.cgm} and Chan et al., in prep) we will study the (potentially much larger) effects of CRs on the CGM around galaxies and the origin and properties of the weak, CR-driven outflows.

In \S~\ref{sec:methods} we review the numerical methods and describe the simulation suite. Before analyzing the simulations, \S~\ref{sec:toy} presents a simple analytic model for the effects and equilibrium distribution of CRs, given our assumptions in the simulations, which allows us to predict and estimate (with surprising accuracy) many of the scalings we will observe in the cosmological simulations. \S~\ref{sec:results} briefly presents the key results from the simulations, which we discuss and analyze in more detail -- attempting to break down the effects of different physics on different scales -- in \S~\ref{sec:discussion}. Finally, we summarize and conclude in \S~\ref{sec:conclusions}. 

\begin{figure*}
    \plotsidesize{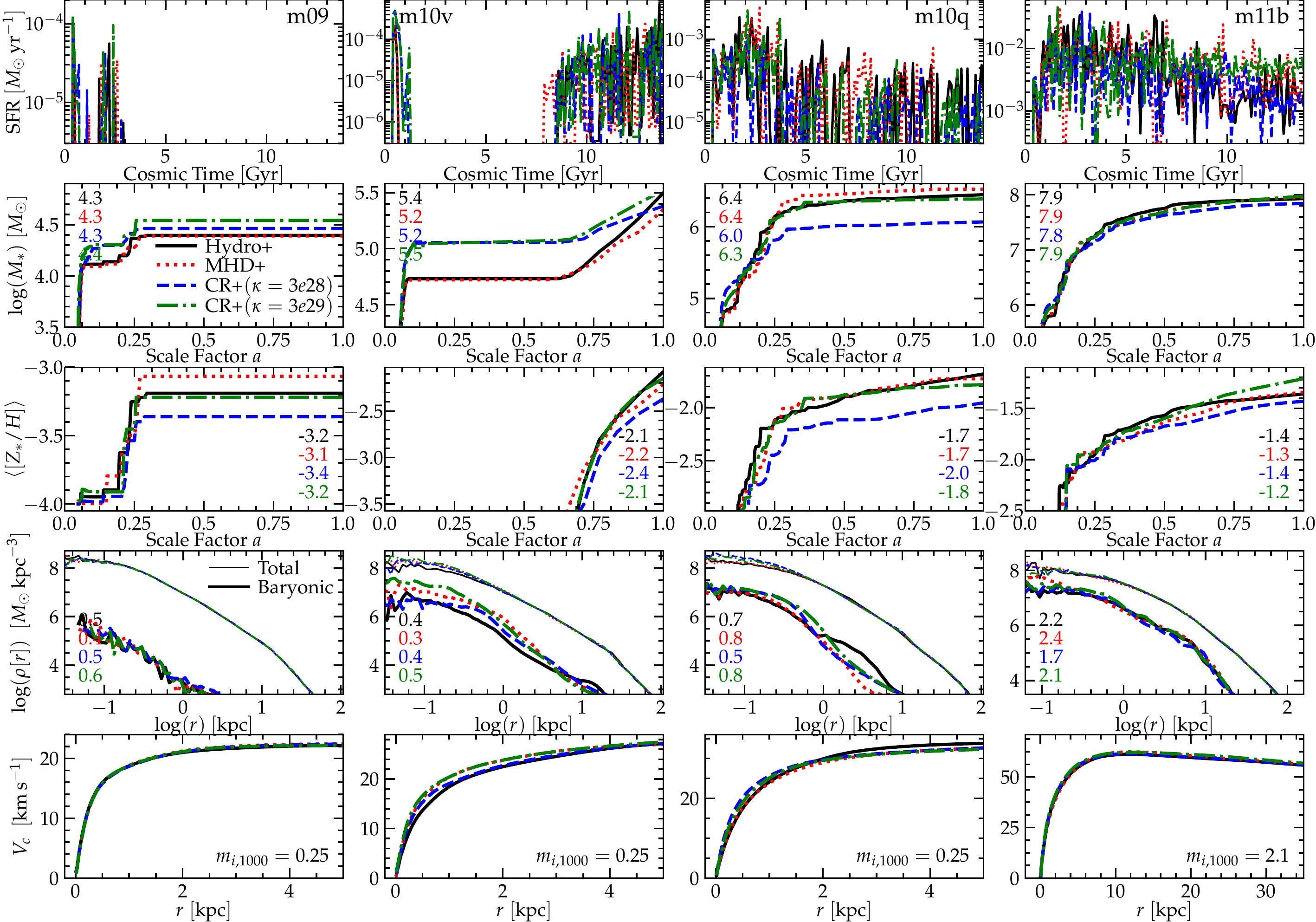}{0.99}
    \vspace{-0.25cm}
    \caption{Properties of our default physics ``suites.'' We compare different galaxies (columns, from Table~\ref{tbl:sims}), simulated with different physics ``suites'' (lines as labeled, from Table~\ref{tbl:physics}). Recall, all include the same cooling, stellar feedback, etc, but the MHD+ runs include MHD and fully-anisotropic conduction \&\ viscosity, and the CR+ runs include MHD, conduction, viscosity, CR injection, losses, streaming, and diffusion (with fixed CR diffusion constant $\kappa$ as labeled). 
    We compare:
    {\em Top:} Star formation history (averaged in $100\,$Myr intervals) of the primary ($z=0$) galaxy. 
     {\em Second:} Total stellar mass in box (dominated by primary) vs.\ scale factor ($a=1/(1+z)$). 
     The $log M_{\ast}/M_{\odot}$ value at $z=0$ for each run is shown as the number in the panel. 
     {\em Middle:} Stellar mass-weighted average metallicity vs.\ scale factor ($z=0$ value shown). 
     {\em Fourth:} Baryonic ({\em thick}) and total ({\em thin}) mass density profiles (averaged in spherical shells) as a function of radius around the primary galaxy at $z=0$. Number is the stellar effective ($1/2$-mass) radius at $z=0$ in kpc. 
     {\em Bottom:} Rotation curves (circular velocity $V_{c}$ versus radius) in the primary galaxy. 
     Value $m_{i,\,1000} = m_{i}/1000\,M_{\odot}$ of the mass resolution is shown. 
     The galaxies here are dwarfs, from lowest-to-highest mass. Though there are some effects (e.g.\ {\bf m10v} rises to higher stellar mass in its initial high-$z$ burst in the CR runs, and {\bf m10q} has a lower $M_{\ast}$ in one CR run), they are well within the range of stochastic run-to-run variations. There does not appear to be a large systematic effect of CRs or MHD or conduction/viscosity (perhaps a small suppression of $M_{\ast}$ with CRs and low diffusion coefficients, but see Fig.~\ref{fig:diffusioncoeff}). This Figure is continued in Figs.~\ref{fig:cr.demo.m11}-\ref{fig:cr.demo.m12}.
    \label{fig:cr.demo.m10}}
\end{figure*}
\begin{figure*}
    \plotsidesize{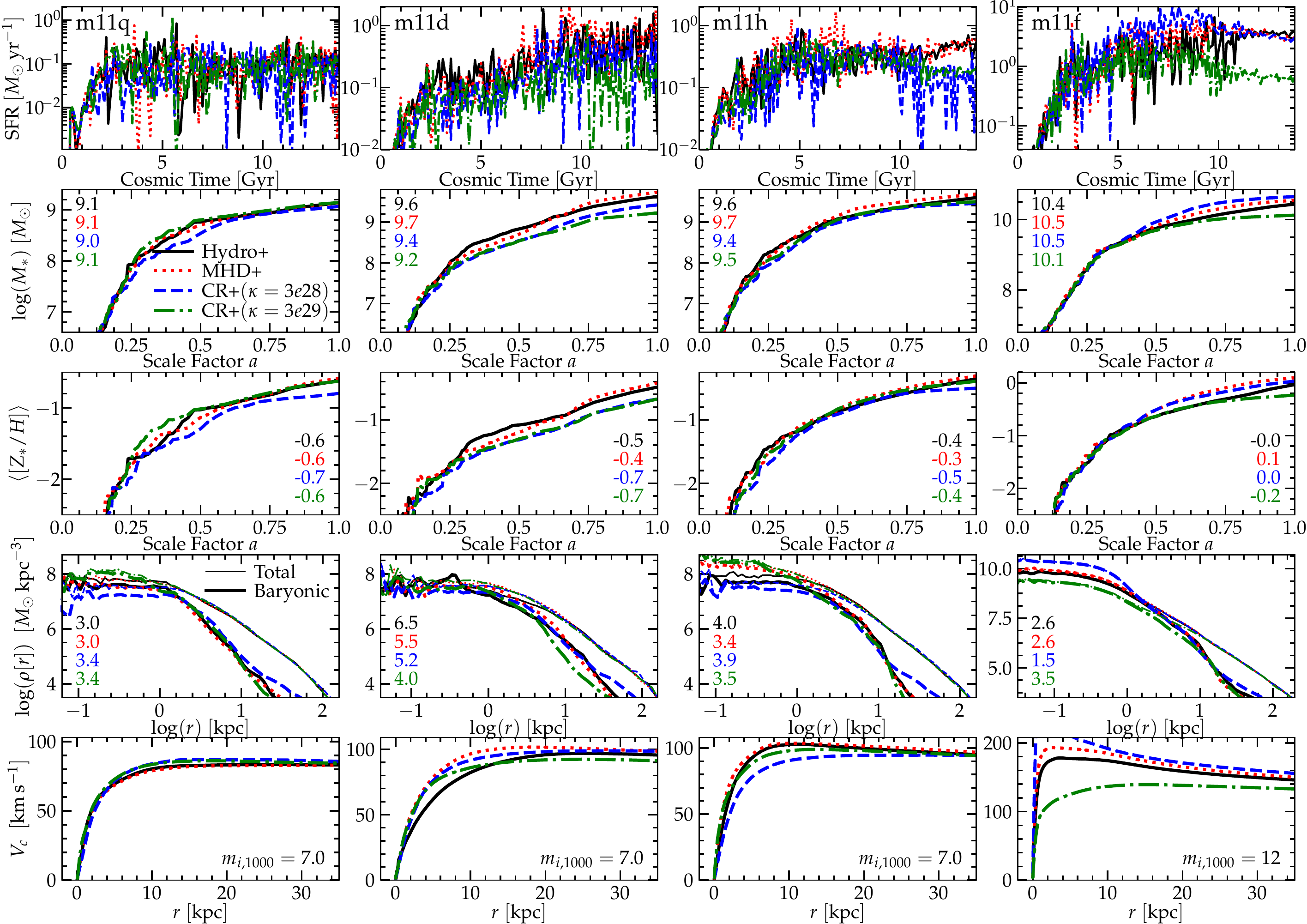}{0.99}
    \vspace{-0.25cm}
    \caption{Fig.~\ref{fig:cr.demo.m10}, continued to higher stellar and halo masses. There is no systematic discernable effect of MHD/conduction/viscosity. CRs may have a small systematic effect at low diffusion coefficient. At high $\kappa \sim 3\times10^{29}\,{\rm cm^{2}\,s^{-1}}$, they have a larger effect, most pronounced at late times ($z < 1$) in the SFRs $\gg 0.1\,M_{\odot}\,{\rm yr^{-1}}$. The effect gets larger at larger stellar masses.
    \label{fig:cr.demo.m11}}
\end{figure*}
\begin{figure*}
    \plotsidesize{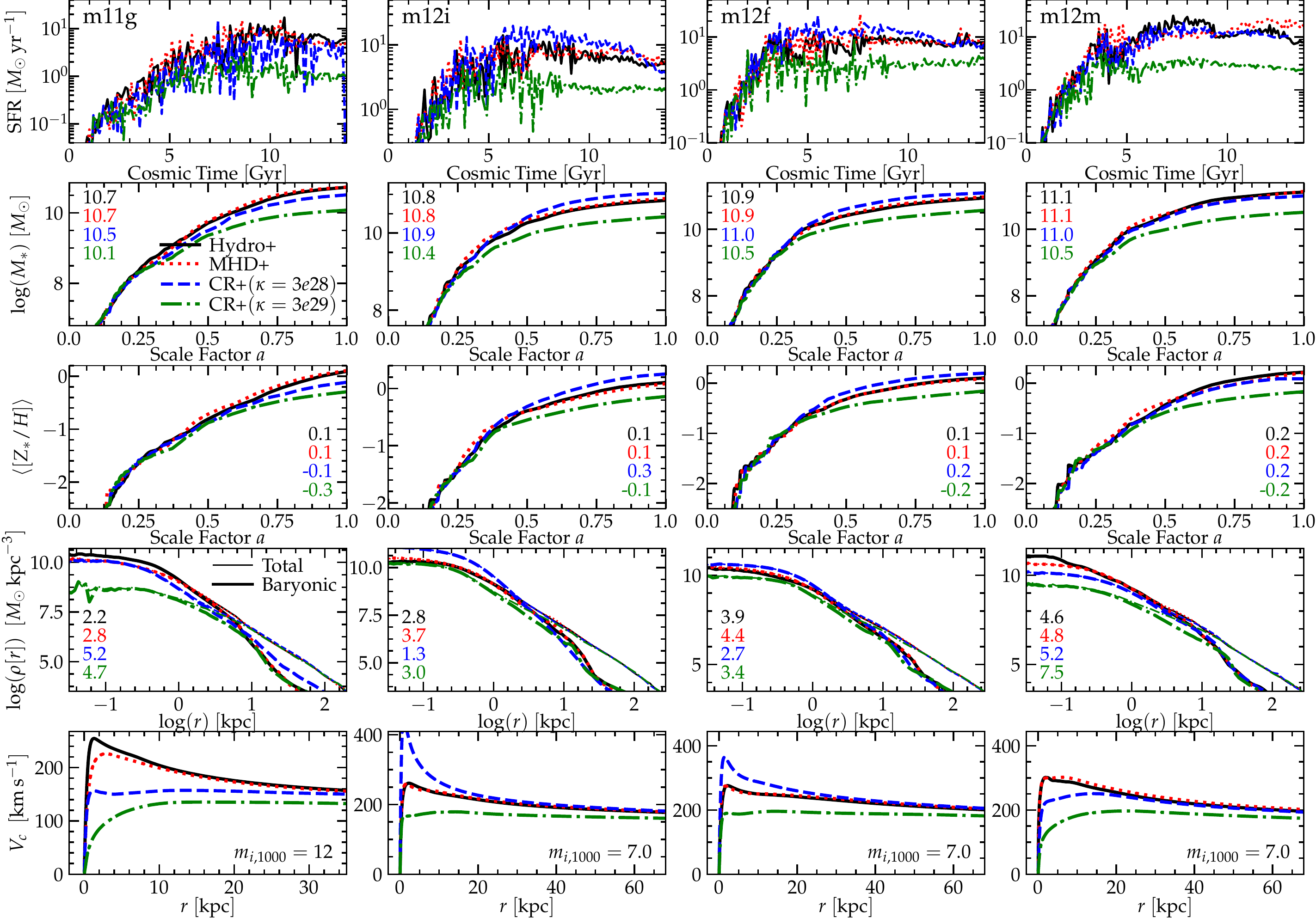}{0.99}
    \vspace{-0.25cm}
    \caption{Figs.~\ref{fig:cr.demo.m10}-\ref{fig:cr.demo.m11}, continued to Milky Way-mass ($M_{\rm halo}\sim 10^{12}\,M_{\odot}$) halos. Again MHD/conduction/viscosity have no discernable effect. CRs with lower diffusion coefficients $\lesssim 10^{29}\,{\rm cm^{2}\,s^{-1}}$ also produce no systematic effect. But CRs with higher diffusion coefficients produce substantial suppression of the SFRs and stellar masses above $M_{\ast} \gtrsim 10^{10}\,M_{\odot}$. This in turn strongly reduces the central "spike" in the rotation curves in these galaxies. The final ($z=0$) stellar masses and SFRs are suppressed by factors $\sim 3$.
    \label{fig:cr.demo.m12}}
\end{figure*}

\vspace{-0.5cm}
\section{Methods}
\label{sec:methods}

The simulations in this paper were run with the multi-physics code {\small GIZMO}\footnote{A public version of {\small GIZMO} is available at \gizmourl} \citep{hopkins:gizmo}, in its meshless finite-mass MFM mode. This is a mesh-free, finite-volume Lagrangian Godunov method which provides adaptive spatial resolution together with conservation of mass, energy, momentum, and angular momentum, and the ability to accurately capture shocks and fluid mixing instabilities (combining advantages of both grid-based and smoothed-particle hydrodynamics methods). We solve the equations of ideal magneto-hydrodynamics (MHD), as described and tested in detail in \citet{hopkins:mhd.gizmo,hopkins:cg.mhd.gizmo}, with fully-anisotropic Spitzer-Braginskii conduction and viscosity and other diffusion operators implemented as described in \citet{hopkins:gizmo.diffusion}. Gravity is solved for gas and collisionless (stars and dark matter) species with fully-adaptive Lagrangian force softening (so hydrodynamic and force resolutions are consistently matched). 

Our simulations are fully-cosmological ``zoom-in'' runs, with a high-resolution Lagrangian region identified surrounding a $z=0$ ``primary'' halo of interest in a large cosmological box \citep[see][]{onorbe:2013.zoom.methods}. Tables~\ref{tbl:sims}-\ref{tbl:HiZ} list the specific volumes run, and the properties of each ``primary'' halo.\footnote{For the MUSIC \citep{hahn:2011.music.code.paper} files necessary to generate all ICs here, see:\\
\ICsurl} We note that the high-resolution volumes reach as large as $\sim(10\,{\rm Mpc})^{3}$ in the largest runs, so there are many galaxies present (our set has hundreds of galaxies with $>100$ star particles each). Also, although we focus on $\sim 30$ zoom-in volumes, because we systematically vary the physics and CR parameters like the diffusion coefficient, our default simulation set includes well over 100 full-physics high-resolution simulations. However, to simplify our analysis and presentation, avoid ambiguities in galaxy matching and separating systematic differences between satellite and field galaxies, and to focus on the best-resolved galaxies possible, we focus only on the most massive ``primary'' galaxies in each box. We note though that a brief comparison indicates that our conclusions appear to apply to all galaxies in the box. Unless otherwise specified (e.g.\ Table~\ref{tbl:HiZ}), all are run to $z=0$.

\vspace{-0.5cm}
\subsection{Default FIRE-2 (``Hydro+'') Physics}

{\em All} our simulations here include the physics of cooling, star formation, and stellar feedback from the FIRE-2 version of the Feedback in Realistic Environments (FIRE) project, described in detail in \citet{hopkins:fire2.methods}, but briefly summarized here.

Gas cooling is followed from $T=10-10^{10}\,$K including free-free, Compton, metal-line, molecular, fine-structure and dust collisional, and we also follow gas heating from photo-electric and photo-ionization by both local sources and a uniform meta-galactic background including the effect of self-shielding.\footnote{As detailed in \citet{hopkins:fire2.methods} Appendix~B, in our runs that {\em do not} include explicit CR transport (``Hydro+'' and ``MHD+''), the cooling/ionization tables do assume a uniform MW-like CR background ($\sim 1\,{\rm eV\,cm^{-3}}$) for gas at densities $>0.01\,{\rm cm^{-3}}$. This is generally negligible for heating, but is important for e.g.\ the small ionized fraction in GMCs. In our runs with explicit CR transport, these terms are replaced with the explicitly-evolved CR background and collisional+streaming heating rates described below.} We explicitly follow 11 different abundances, including explicit treatment of turbulent diffusion of metals and passive scalars as in \citet{colbrook:passive.scalar.scalings,escala:turbulent.metal.diffusion.fire}. Gas is turned into stars using a sink-particle prescription: gas which is locally self-gravitating at the resolution scale following \citealt{hopkins:virial.sf}, self-shielding/molecular following \citealt{krumholz:2011.molecular.prescription}, Jeans unstable, {\em and} denser than $n_{\rm crit}>1000\,{\rm cm^{-3}}$ is converted into star particles on a free-fall time. Star particles are then treated as single-age stellar populations with all IMF-averaged feedback properties calculated from {\small STARBURST99} \citep{starburst99} assuming a \citet{kroupa:2001.imf.var} IMF. We then explicitly treat feedback from SNe (both Types Ia and II), stellar mass loss (O/B and AGB mass-loss), and radiation (photo-ionization and photo-electric heating and UV/optical/IR radiation pressure), with implementations at the resolution-scale described in \citet{hopkins:fire2.methods}, \citet{hopkins:sne.methods}, and \citet{hopkins:radiation.methods}.

The simulations labeled ``Hydro+'' in this paper include all of the physics above, but do not include magnetic fields, physical conduction or viscosity, or explicit treatment of cosmic rays.

\vspace{-0.5cm}
\subsection{Magnetic Fields, Conduction, \&\ Viscosity (``MHD+'')}
\label{sec:methods.mhd}

Our simulations labeled ``MHD+'' in this paper include all of the ``Hydro+'' physics (e.g.\ radiative cooling, star formation, stellar feedback), but add magnetic fields and physical, fully-anisotropic conduction and viscosity. These are described in \citet{su:2016.weak.mhd.cond.visc.turbdiff.fx} but we briefly summarize here. 

As noted above, for magnetic fields we solve the equations of ideal MHD, as described in \citet{hopkins:mhd.gizmo,hopkins:cg.mhd.gizmo}. We can optionally include ambipolar diffusion, the Hall effect, and Ohmic resistivity as in \citealt{hopkins:gizmo.diffusion}, but these are completely negligible at all resolved scales in our simulations here.\footnote{We have, in fact, run one {\bf m10q} and one {\bf m12i} run from $z=0.1$ to $z=0$ with the full non-ideal MHD terms enabled and the coefficients calculated following \citealt{zhu:2014.dust.power.spectra.mri.turbulent.disks}, to confirm that they are negligible corrections to the induction equation at the scales resolved here (usually by several orders of magnitude, except for ambipolar diffusion, which can be a $\sim 10\%$-level correction to the induction equation in the dense, neutral gas in GMCs, but this is far smaller than other physical uncertainties).}

For conduction and viscosity we include the physical scalings for fully-anisotropic Spitzer-Braginskii transport, with the conductive heat flux $\kappa_{\rm cond}\,\hat{\bf B}\,(\hat{\bf B}\cdot \nabla T)$, where
\begin{align}
\kappa_{\rm cond} &\equiv \frac{0.96\,k_{B}\,(k_{B}\,T)^{5/2}}{m_{e}^{1/2}\,e^{4}\,\ln{\Lambda_{c}}}\,\frac{f_{i}}{1 + (4.2+\beta/3)\,\ell_{e}/\ell_{T}}
\end{align}
\citep{spitzer:conductivity,braginskii:viscosity}, where $m_{e}$ and $e$ are the electron mass and charge, $k_{B}$ the Boltzmann constant, $f_{i}$ is the ionized fraction, $\ln{\Lambda_{c}}\sim37$ is a Coulomb logarithm and $\ell_{e}= 3^{3/2}\,(k_{B}\,T)^{2}/4\,n_{e}\,\pi^{1/2}\,e^{4}\,\ln{\Lambda_{c}}$ is the electron deflection length \citep{sarazin:coulomb.log}, $\ell_{T}\equiv T/|\hat{\bf B}\cdot \nabla T|$ is the parallel temperature gradient scale-length, and the plasma $\beta \equiv P_{\rm thermal}/P_{\rm magnetic}$ is the usual ratio of thermal-to-magnetic pressure. Note that the $\ell_{e}/\ell_{T}$ term ensures proper behavior in the saturated limit (and a smooth transition between un-saturated and saturated limits, e.g.\ \citealt{cowie:1977.evaporation}), while the $\beta$ term accounts for micro-scale plasma instabilities (e.g.\ the Whistler instability) limiting the heat flux in high-$\beta$ plasmas \citep[see][]{komarov:whistler.instability.limiting.transport}. For viscosity we modify the momentum and energy equations with the addition of the viscous stress tensor $\Pi \equiv -3\,\eta_{\rm visc}\,(\hat{\bf B}\otimes\hat{\bf B} - \mathbb{I}/3)\,(\hat{\bf B}\otimes\hat{\bf B} - \mathbb{I}/3) : (\nabla\otimes{\bf v})$ (where $\otimes$ is the outer product, $\mathbb{I}$ the identity matrix, and $:$ the double-dot-product), with
\begin{align}
\eta_{\rm visc} &\equiv \frac{0.406\,m_{i}^{1/2}\,(k_{B}\,T)^{5/2}}{(Z_{i}\,e)^{4}\,\ln{\Lambda_{c}}}\,\frac{f_{i}}{1+(4+\beta^{-1/2})\,\ell_{i}/\ell_{|v|}} \\ 
-2\,&P_{\rm magnetic} < {\eta_{\rm visc}\,(\hat{\bf B}\otimes\hat{\bf B}):(\nabla \otimes {\bf v})} < P_{\rm magnetic}
\end{align}
where like the above $m_{i}$, $\ell_{i}$, and $Z_{i}\,e$ are the mean ion mass, deflection-length, and charge, and $\ell_{|v|} \equiv |{\bf v}| / |\hat{\bf B}\otimes\hat{\bf B}:\nabla \otimes {\bf v}|$ is the parallel velocity gradient scale-length. The upper and lower allowed values of the anisotropic stress are limited (capped) according to the latter expression, to account again for plasma instabilities (the mirror and firehose at positive and negative anisotropy, respectively) limiting the flux \citep[see][]{squire:2017.max.braginskii.scalings,squire:2017.kinetic.mhd.alfven,squire:2017.max.anisotropy.kinetic.mhd}.

These numerical methods have been extensively tested and discussed in previous papers, to which we refer for details \citep[e.g.][]{hopkins:mhd.gizmo,hopkins:cg.mhd.gizmo,hopkins:gizmo.diffusion,hopkins.conroy.2015:metal.poor.star.abundances.dust,su:2016.weak.mhd.cond.visc.turbdiff.fx,su:fire.feedback.alters.magnetic.amplification.morphology,lee:dynamics.charged.dust.gmcs,colbrook:passive.scalar.scalings,seligman:2018.mhd.rdi.sims}. We note that, given the mass resolution and Lagrangian nature of the code, the \citet{field:length} length $\lambda_{\rm F}$ (approximately, the scale below which thermal conduction is faster than cooling) is resolved ($\Delta x<\lambda_{\rm F}$) in the simulations here in gas hotter than $T \gtrsim 2\times10^{5}\,{\rm K}\,(n/0.01\,{\rm cm^{-3}})^{0.4}$.

\vspace{-0.5cm}
\subsection{Cosmic Rays (``CR+'')}
\label{sec:methods:crplus}

Our simulations labeled ``CR+'' in this paper include all of the ``MHD+'' physics, and add our ``full physics'' treatment of CRs. The CR physics is described in detail in \citet{chan:2018.cosmicray.fire.gammaray} but we again summarize here: we include injection in SNe shocks, fully-anisotropic CR transport with streaming and advection/diffusion, CR losses (hadronic and Coulomb, adiabatic, streaming), and self-consistent CR-gas coupling. 

CRs are treated as an ultra-relativistic fluid\footnote{The fluid limit adopted here is motivated by the fact that $\sim$\,GeV CR gyro radii $r_{\rm L} \ll {\rm au}$ are always much smaller than any resolved spatial scales, and CR ``deflection lengths'' or scattering lengths $\sim \kappa/c \sim $\,pc are also always smaller than e.g.\ the CR pressure gradient scale lengths. However, we emphasize that fundamental questions remain about the validity of other assumptions (like the form of the CR distribution function) when the CR diffusivity is large (scattering rates are low), particularly when $|\kappa_{\|} \nabla_{\|} P_{\rm cr}| \gg v_{A}\,(e_{\rm cr}+P_{\rm cr})$, which occurs in our favored models at galacto-centric radii $\ll 30\,$kpc, as discussed below.} (adiabatic index $\gamma_{\rm cr}=4/3$) in the ``single bin'' approximation, which we can think of either as evolving only the CR energy density ($e_{\rm cr}$) at $\sim$\,GeV energies that dominate the CR pressure ($P_{\rm cr} \equiv (\gamma_{\rm cr}-1)\,e_{\rm cr}$), or equivalently assuming a universal CR energy spectral shape. CR pressure contributes to the total pressure and effective sound speed in the Riemann problem for the gas equations-of-motion according to the {\em local strong-coupling approximation}, i.e.\ $P = P_{\rm gas} + P_{\rm cr}$ and $c_{s,\,{\rm eff}}^{2} = \partial P/\partial \rho = c_{s}^{2} + \gamma_{\rm cr}\,P_{\rm cr}/\rho$. Integrating over the CR distribution function and energy spectrum, we evolve the CR energy density as \citet{mckenzie.voelk:1982.cr.equations}: 
\begin{align}
\label{eqn:cr.eom}
\frac{\partial e_{\rm cr}}{\partial t} + \nabla \cdot {\bf F}_{\rm cr} &= \langle {\bf v}_{\rm cr} \rangle \cdot \nabla P_{\rm cr} + S_{\rm cr} - \Gamma_{\rm cr} 
\end{align}
where $S_{\rm cr}$ and $\Gamma_{\rm cr}$ are source and sink terms; $\langle {\bf v}_{\rm cr} \rangle \equiv {\bf v}_{\rm gas} + {\bf v}_{\rm stream}$ is the bulk CR advection velocity, de-composed into the gas velocity ${\bf v}_{\rm gas}$ and the ``streaming velocity'' ${\bf v}_{\rm stream}$; and ${\bf F}_{\rm cr}$ is the lab-frame or total CR energy flux which can be de-composed into ${\bf F}_{\rm cr} \equiv \langle {\bf v}_{\rm cr} \rangle\,(e_{\rm cr} + P_{\rm cr}) + {\bf F}_{\rm di} \equiv {\bf v}_{\rm gas}\,(e_{\rm cr} + P_{\rm cr}) + \tilde{\bf F}_{\rm cr}$, where $\tilde{\bf F}_{\rm cr}$ is the flux in the fluid frame (with ${\bf F}_{\rm di}$ the ``diffusive'' flux). 

For $S_{\rm cr}$, we assume CR injection in SNe shocks, with a fixed fraction $\epsilon_{\rm cr}$ of the initial ejecta kinetic energy ($\approx 10^{51}\,{\rm erg}$) of every SNe going into CRs. In our default simulations, $\epsilon_{\rm cr}=0.1$ is adopted. This is coupled directly to gas in the immediate vicinity of each explosion, alongside the thermal and kinetic energy, mass, and metals, as described in \citet{hopkins:sne.methods}. For $\Gamma_{\rm cr}$, we follow \citet{guo.oh:cosmic.rays} and account for both hadronic/catastropic and Coulomb losses with $\Gamma_{\rm cr} = e_{\rm cr}\,n_{\rm n}\,(1 + 0.28\,x_{e}) / t_{0}\,n_{0}$, where $n_{\rm n}$ is the nucleon number density, $t_{0}\,n_{0} \equiv 1.72 \times10^{15}\,{\rm s\,cm^{-3}}$, and $x_{e}$ is the number of free electrons per nucleon. A fraction $\sim 1/6$ of the hadronic products and all Coulomb losses thermalize, and contribute a volumetric gas heating term $Q_{\rm gas} = e_{\rm cr}\,n_{\rm n}\,(0.17 + 0.28\,x_{e}) / t_{0}\,n_{0}$. 

As shown in \citet{chan:2018.cosmicray.fire.gammaray} the remaining terms in Eq.~\ref{eqn:cr.eom} can be decomposed (in Lagrangian form) into simple advection (automatically handled in our Lagrangian formulation) and adiabatic ``PdV work'' terms (solved in an exactly-conservative manner with our usual MFM solver), the $\tilde{\bf F}_{\rm cr}$ term (see below), and a ``streaming loss''  term\footnote{As discussed at length in \citet{chan:2018.cosmicray.fire.gammaray} and studied in the Appendices therein, we have considered simulations where the ``streaming losses'' scale either as ${\bf v}_{\rm stream}\cdot \nabla P_{\rm cr}$ or ${\bf v}_{A}\cdot \nabla P_{\rm cr}$. The latter (streaming losses limited to the \Alf\ speed, as they arise from damped \Alf\ waves) is our default choice, even if $v_{\rm stream} > v_{A}$. However we do not find this choice significantly alters any of our conclusions in this paper.} ${\bf v}_{A}\cdot \nabla P_{\rm cr}$ which is negative definite and represents energy loss via streaming instabilities that excite high-frequency \Alf\ waves (frequency of order the CR gyro frequency, well below our resolution; \citealt{wentzel:1968.mhd.wave.cr.coupling,kulsrud.1969:streaming.instability}) that damp and thermalize almost instantaneously, so the energy lost via this term is added to the gas thermal energy each timestep. The streaming velocity always points down the CR pressure gradient, projected along the magnetic field, so ${\bf v}_{\rm stream} = -v_{\rm stream}\,\hat{\bf B}\,(\hat{\bf B}\cdot \hat{\nabla} P_{\rm cr})$. 

It is widely argued that micro-scale instabilities regulate the streaming speed to of order the \Alf\ speed \citep{skilling:1971.cr.diffusion,kulsrud:plasma.astro.book,yan.lazarian.2008:cr.propagation.with.streaming,ensslin:2011.cr.transport.clusters}, although super-\Alf{ic} streaming can easily emerge in self-confinement models for CR transport \citep{wentzel:1968.mhd.wave.cr.coupling,holman:1979.cr.streaming.speed,achterberg:1981.cr.streaming,wiener:cr.supersonic.streaming.deriv,lazarian:2016.cr.wave.damping}; moreover in partially-neutral gas (ionized fraction $f_{\rm ion} <  1$) there is an ambiguity about whether the appropriate \Alf\ speed is the ideal-MHD \Alf\ speed $v_{A} \equiv |\bf B|/(4\pi\,\rho)^{1/2}$ or the (larger) ion \Alf\ speed $v_{A}^{\rm ion} \equiv |{\bf B}|/({4\pi\,\rho_{\rm ion}})^{1/2} \sim f_{\rm ion}^{-1/2}\,v_{A}$ \citep{1975MNRAS.172..557S,Zwei13,farber:decoupled.crs.in.neutral.gas}. So we simply adopt the ad-hoc $v_{\rm stream} \approx 3\,v_{A}$ as our default, although we vary this  widely below setting $v_{\rm stream}=v_{A}$, disabling streaming entirely, or allowing highly super-sonic/\Alf{ic} streaming with $v_{\rm stream} = 3\,(v_{A}^{2} + c_{s}^{2})^{1/2}$ (several times the fastest possible MHD wavespeed), and show (both here and in \citealt{chan:2018.cosmicray.fire.gammaray}) it has little effect on our conclusions. 

Following \citet{chan:2018.cosmicray.fire.gammaray} we treat $\tilde{\bf F}_{\rm cr}$ using a two-moment scheme (similar to other recent implementations by e.g.\ \citealt{jiang.oh:2018.cr.transport.m1.scheme} and \citealt{thomas.pfrommer.18:alfven.reg.cr.transport}), solving 
\begin{align}
\label{eqn:cr.flux} \frac{1}{\tilde{c}^{2}}\,\left[ \frac{\partial \tilde{\bf F}_{\rm cr}}{\partial t} + \nabla\cdot\left( {\bf v}_{\rm gas} \otimes \tilde{\bf F}_{\rm cr} \right) \right] + \nabla_{\|} P_{\rm cr} &= -\frac{(\gamma_{\rm cr}-1)}{\kappa_{\ast}}\,\tilde{\bf F}_{\rm cr}
\end{align}
where $\nabla_{\|} P_{\rm cr}  = (\hat{\bf B} \otimes \hat{\bf B}) \cdot (\nabla P_{\rm cr}) = (\gamma_{\rm cr}-1)\,\hat{\bf B}\,(\hat{\bf B} \cdot \nabla e_{\rm cr})$ is the parallel derivative of the CR pressure tensor, 
$\tilde{c}$ is the maximum allowed CR ``free streaming'' speed,\footnote{Following \citet{chan:2018.cosmicray.fire.gammaray} we note that $\tilde{c}$ is a nuisance parameter (the simulations evolve to identical solutions independent of $\tilde{c}$, so long as it is faster than other bulk flow speeds in the problem), so rather than adopt the microphysical $\tilde{c}=c$ (speed of light), we adopt $\tilde{c}=1000\,{\rm km\,s^{-1}}$ by default (but show below that our solutions are independent of $\tilde{c}$ over a wide range).} 
and $\kappa_{\ast} \equiv \kappa_{\|} + \gamma_{\rm cr}\,v_{\rm st}\,P_{\rm cr}/|\nabla_{\|}P_{\rm cr}|$ is the effective parallel diffusivity (we are implicitly taking the perpendicular $\kappa_{\bot}=0$).
We note this reduces to the simpler pure-anisotropic diffusion+streaming equation in steady state and/or on large spatial and time-scales ($\tilde{\bf F}_{\rm cr} \rightarrow -\kappa_{\ast} \nabla_{\|} e_{\rm cr} 
= \kappa_{\|} \nabla_{\|} e_{\rm cr} +  {\bf v}_{\rm stream}\,(e_{\rm cr}+P_{\rm cr})$), but unlike a pure-diffusion equation (where one {\em forces} $\tilde{\bf F}_{\rm cr}$ to always be exactly $-\kappa_{\ast}\nabla_{\|} e_{\rm cr})$)  it correctly handles the transition between streaming and diffusion and prevents un-physical super-luminal CR transport.\footnote{Super-luminal CR transport would occur in the ``pure-diffusion'' approximation for $\tilde{\bf F}_{\rm cr}$ wherever the resolution scale $\Delta x \lesssim \kappa/c \sim 3\,{\rm pc}\,(\kappa/3\times10^{29}\,{\rm cm^{2}\,s^{-1}})$. The simulations in this paper routinely reach this or better spatial resolution, so this distinction is important.} 

As discussed in \citet{chan:2018.cosmicray.fire.gammaray}, if the ``streaming loss'' term is limited or ``capped'' to scale with the \Alf\ speed ($\sim v_{A}\nabla P_{\rm cr}$; see above), and streaming is super-\Alf{ic} ($v_{\rm stream} \gg v_{A}$), then only the ``effective'' diffusivity $\kappa_{\ast}$ (which can arise from a combination of microphysical diffusion and/or streaming) -- as compared to $\kappa_{\|}$ or $v_{\rm stream}$ individually -- enters the large-scale dynamics. This effective $\kappa_{\ast}$, or equivalent CR transport speed $v_{\rm cr,\,eff} \sim \kappa_{\ast}\,|\nabla P_{\rm cr}|/P_{\rm cr}$, is what we actually constrain in our study here. 

There are many approximations in this description, and the effective ``diffusion coefficient'' $\kappa_{\ast}$ for CRs on these (energy, spatial, and time) scales remains both theoretically and observationally uncertain. Therefore, we treat $\kappa$ as a constant but vary it systematically in a parameter survey, with values motivated by the comparison with observational constraints in \citet{chan:2018.cosmicray.fire.gammaray}.

\begin{figure}
    \plotonesize{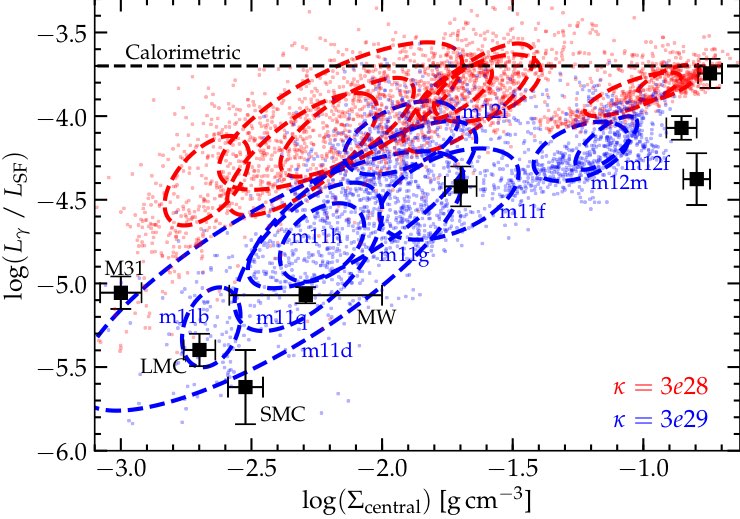}{0.99}
    \plotonesize{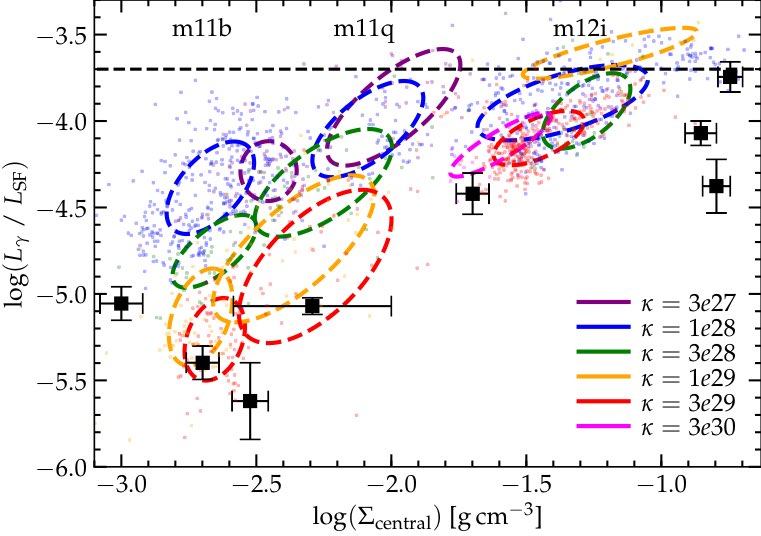}{0.99}
    \plotonesize{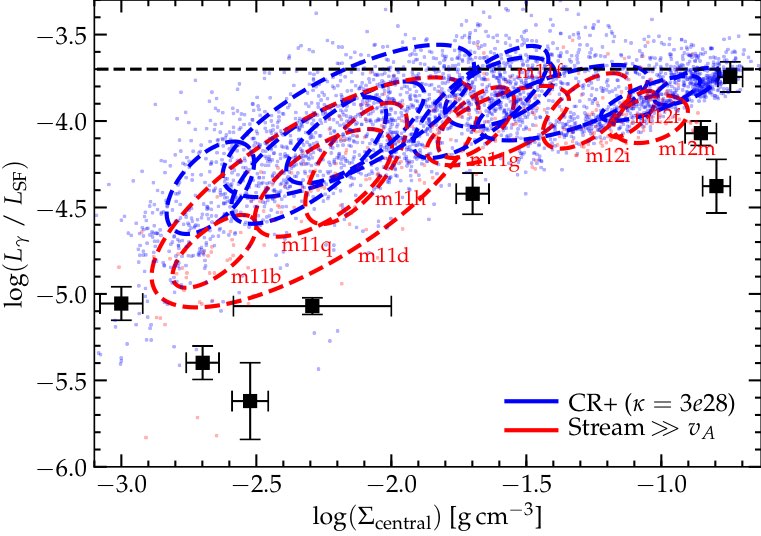}{0.99}
    \vspace{-0.25cm}
        \caption{{\em Top:} Predicted ratio of $\gamma$-ray luminosity from hadronic collisions of CRs ($L_{\gamma}$) to luminosity from star formation/massive stars ($L_{\rm SF}$), as a function of the central gas surface density of the galaxy. Points show each snapshot (every $\sim 20\,$Myr) at $z<1$, while dashed lines show the $1\,\sigma$ ellipsoid for each galaxy (labeled). We show the galaxies in Figs.~\ref{fig:cr.demo.m10}-\ref{fig:cr.demo.m12} (halos $\lesssim 10^{10}\,\msun$ continue the trend but fall off the plot to smaller $\Sigma_{\rm central}$). We compare the observed points from the MW, Local Group, and other Fermi detections (black squares with error bars) from \citet{lacki:2011.cosmic.ray.sub.calorimetric}. Horizontal dashed line is the steady-state, constant-SFR calorimetric limit. The $\kappa\sim 3\times10^{29}\,{\rm cm^{2}\,s^{-1}}$ runs appear to agree well with observations; lower $\kappa$ produces excessive $\gamma$-ray flux. 
        {\em Middle:} Same, for just three galaxies (each at distinct $\Sigma_{\rm central}$, as labeled) varying $\kappa$. Increasing $\kappa$ systematically lowers $L_{\gamma}$, on average. 
        {\em Bottom:} Same, comparing runs with lower $\kappa$ but default (\Alf{ic}) streaming ({\em blue}) or super-sonic/\Alf{ic} streaming ($v_{\rm stream} = 3\,(v_{A}^{2}+c_{s}^{2})^{1/2}$; {\em red}). Faster streaming reduces $L_{\gamma}$ (as CRs escape faster and thermalize energy in streaming losses which do not appear in $\gamma$-rays), but for the values here the effect is equivalent to a small (factor $<2$) increase in $\kappa$.
    \label{fig:Lgamma}}
\end{figure}

\begin{figure*}
\begin{centering}
\includegraphics[width={0.245\textwidth}]{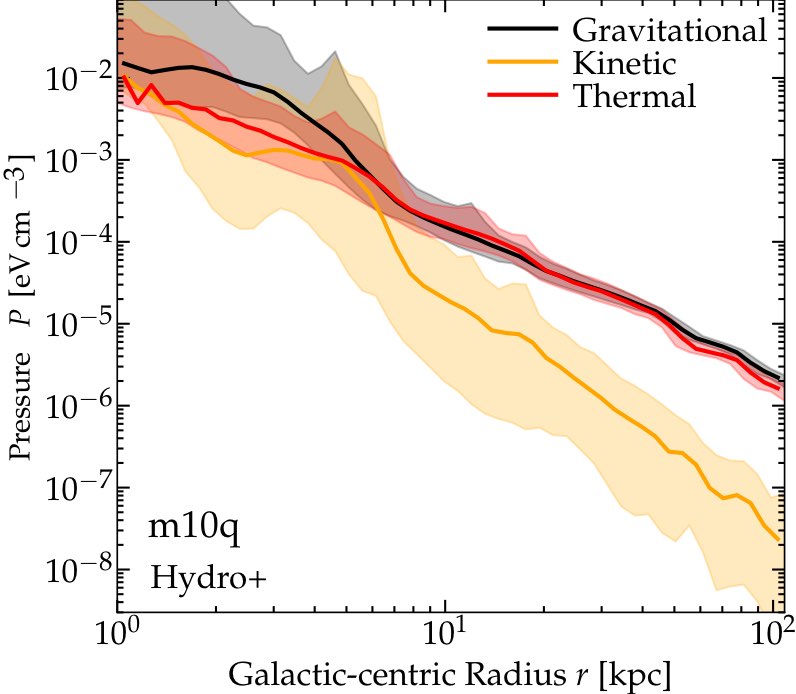}
\includegraphics[width={0.245\textwidth}]{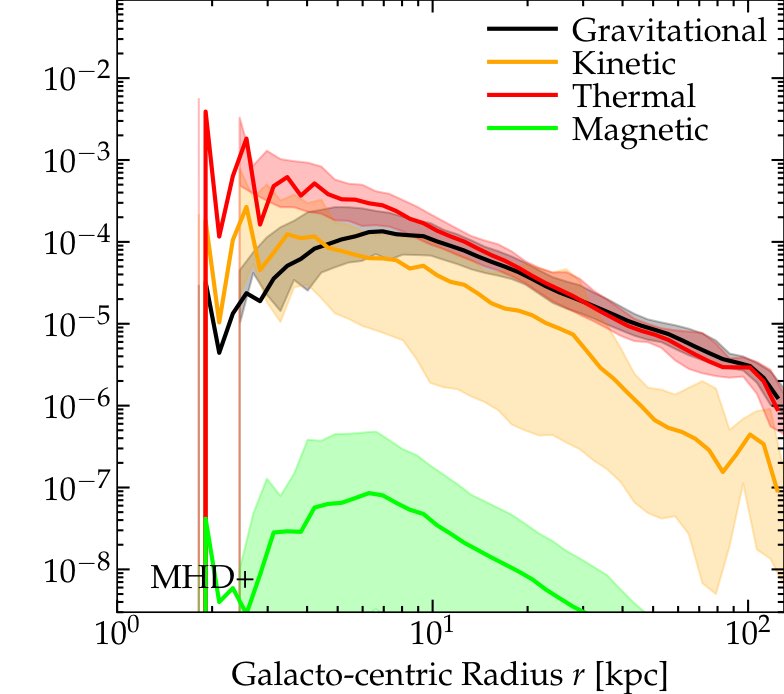}
\includegraphics[width={0.245\textwidth}]{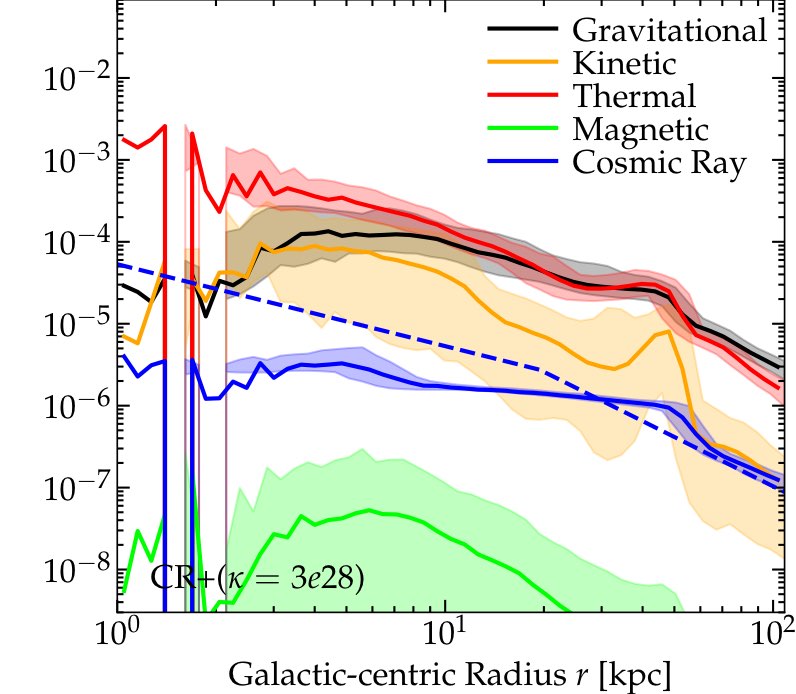}
\includegraphics[width={0.245\textwidth}]{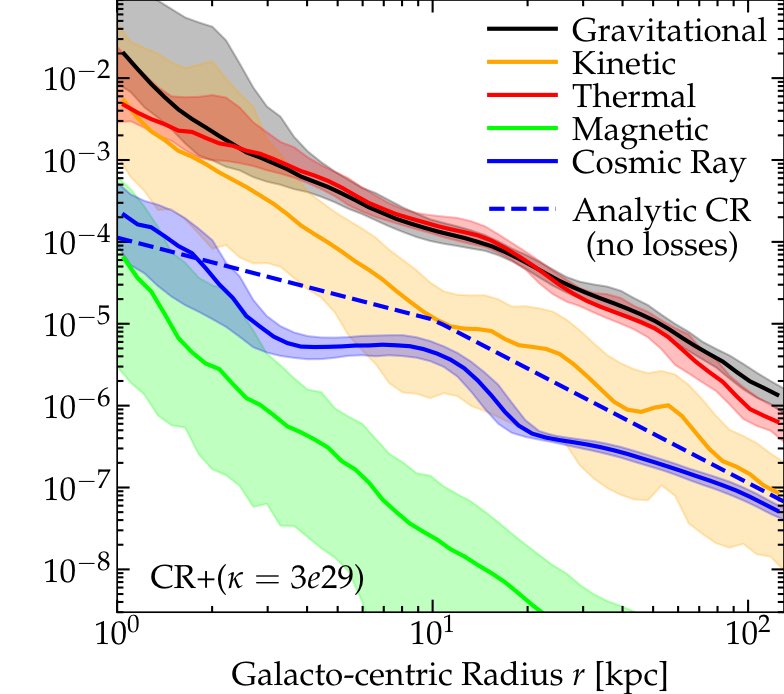} \\
\includegraphics[width={0.245\textwidth}]{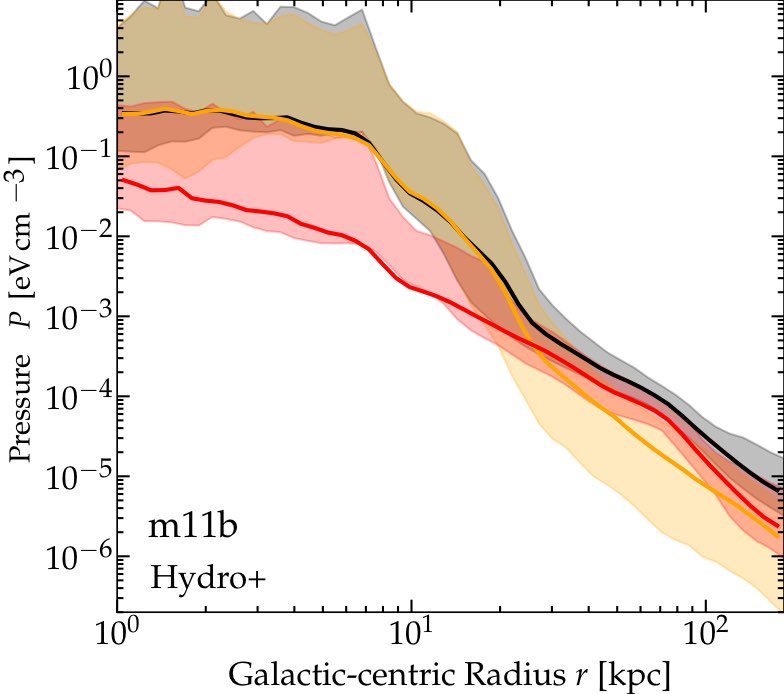}
\includegraphics[width={0.245\textwidth}]{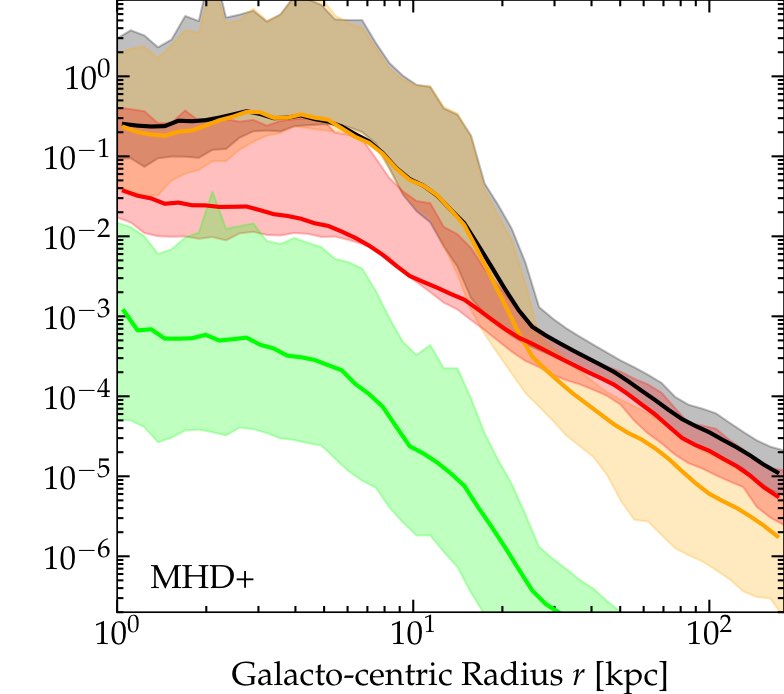}
\includegraphics[width={0.245\textwidth}]{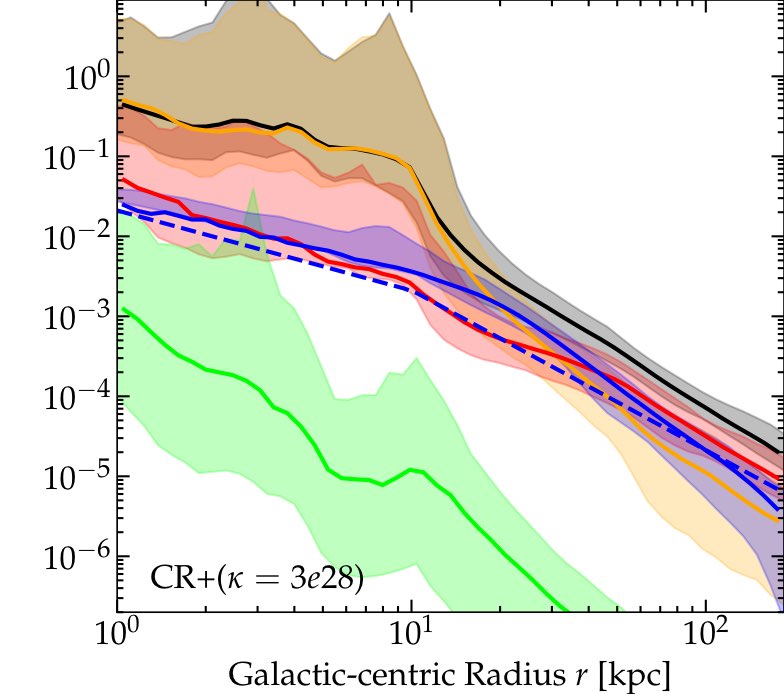}
\includegraphics[width={0.245\textwidth}]{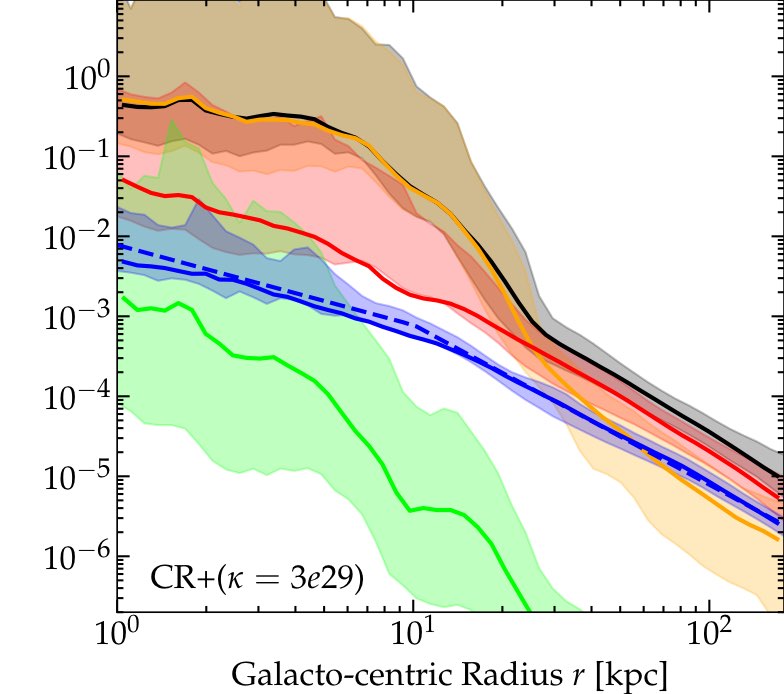} \\
\includegraphics[width={0.245\textwidth}]{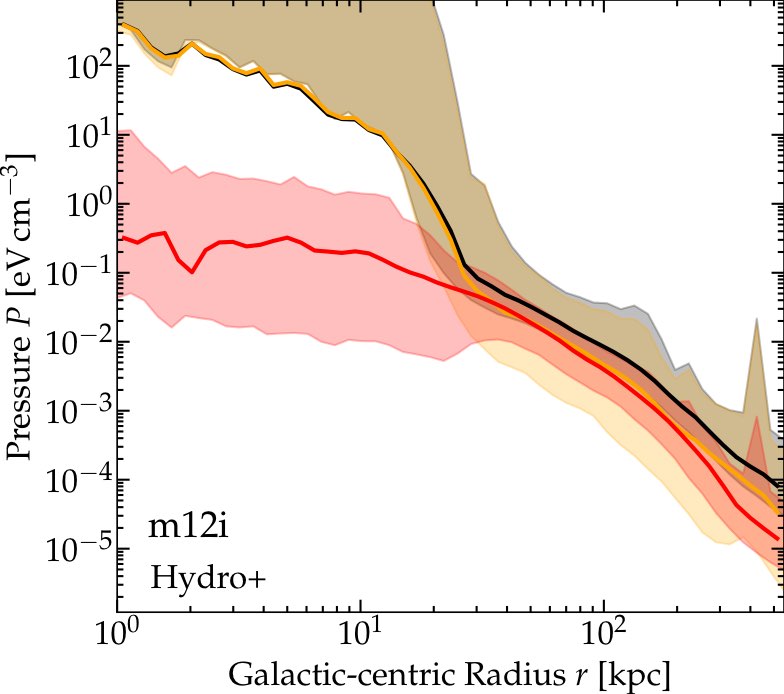}
\includegraphics[width={0.245\textwidth}]{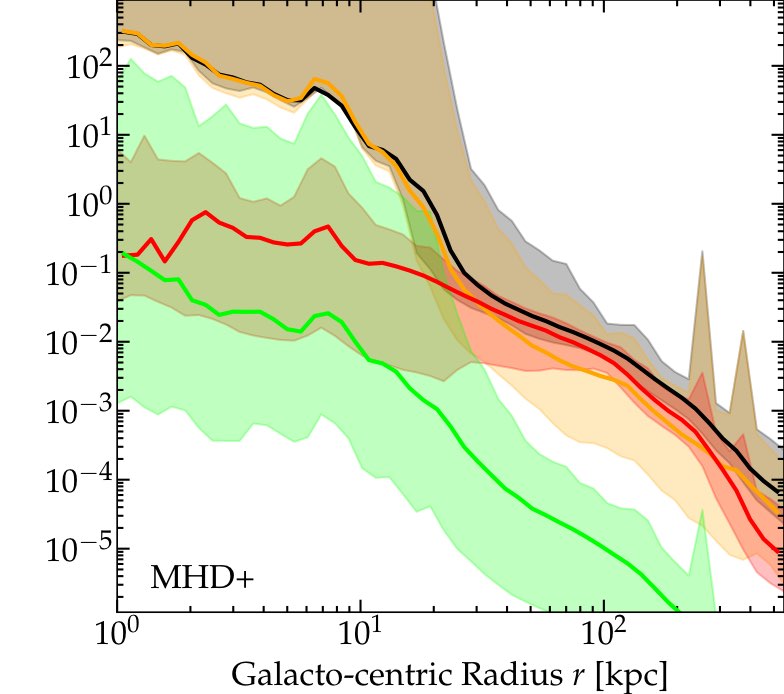}
\includegraphics[width={0.245\textwidth}]{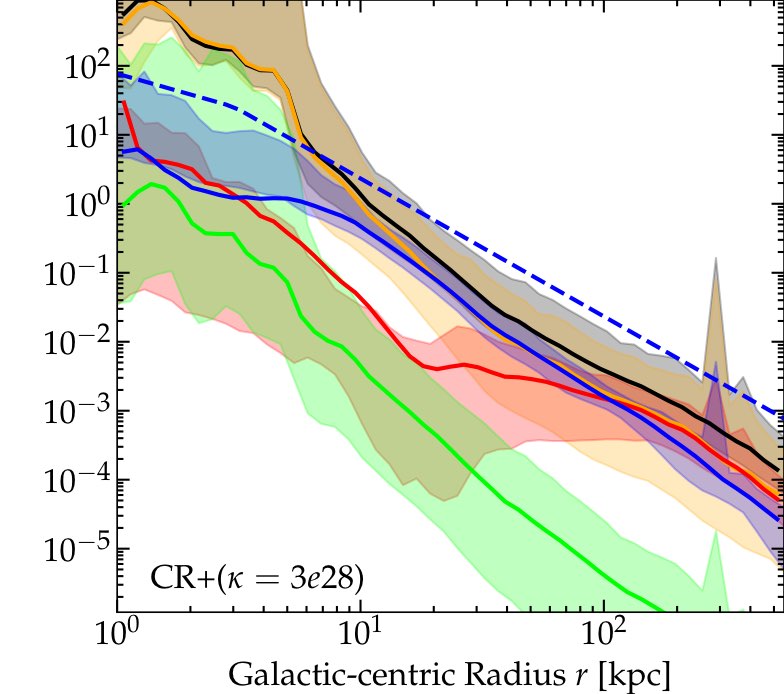}
\includegraphics[width={0.245\textwidth}]{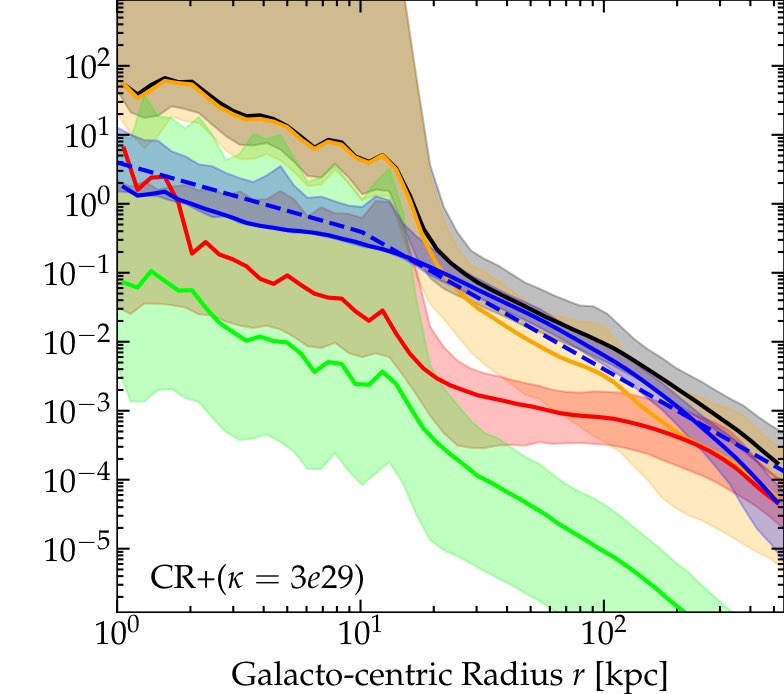} \\
    \end{centering}
    \vspace{-0.25cm}
        \caption{Gas pressure profiles for {\bf m10q} ({\em top}), {\bf m11b} ({\em middle}), {\bf m12i} ({\em bottom}), runs ``Hydro+'', ``MHD+'', ``CR+($\kappa=3e28$),'' ``CR+($\kappa=3e29$)'' (left-to-right) at $z=0$. 
        In each, solid lines show volume-averaged profiles (in spherical shells); shaded range shows the $5-95\%$ inclusion interval of all resolution elements (gas-mass-weighted sampling) at each radius around the galaxy. 
        We compare thermal ($n\,k_{B}\,T$), magnetic ($|{\bf B}|^{2}/8\pi$), CR ($(\gamma_{\rm cr}-1)\,e_{\rm cr}$), ``gravitational'' ($\equiv \rho\,V_{c}^{2}/2$), and ``kinetic'' ($\equiv \langle \rho\,|{\bf v}|^{2}/2$) pressures.  
        Magnetic pressure is sub-dominant to thermal ($\beta\gg1$) especially at large radii, as expected. 
        Both (and CRs) are well below gravity within the disk, where gas is primarily rotation-supported ($|{\bf v}|\sim V_{c}$) -- i.e.\ gas is primarily in a thin or turbulent structure inside $<10\,$kpc. 
        With CRs present, CR pressure can dominate and balance gravity at large $r$ (in massive galaxies), supporting denser and/or cooler gas which would otherwise accrete onto the galaxy. 
        Dashed line shows the analytic prediction from \S~\ref{sec:toy} for CR pressure with negligible collisional losses. This is an excellent approximation at high-$\kappa$ (the ``turnover'' at $\sim 30-100$\,kpc is where streaming begins to dominate transport). 
        At low-$\kappa$, the CR pressure is much lower, indicating that most CR energy has been lost. 
        Although CRs still balance gravity, the lower energy means gas has already cooled onto the galaxy (forming stars) so the remaining ``weight'' (magnitude of gravitational pressure to be supported) is much lower.
    \label{fig:profile.pressure}}
\end{figure*}

\begin{figure}
\begin{centering}
\includegraphics[width={0.49\columnwidth}]{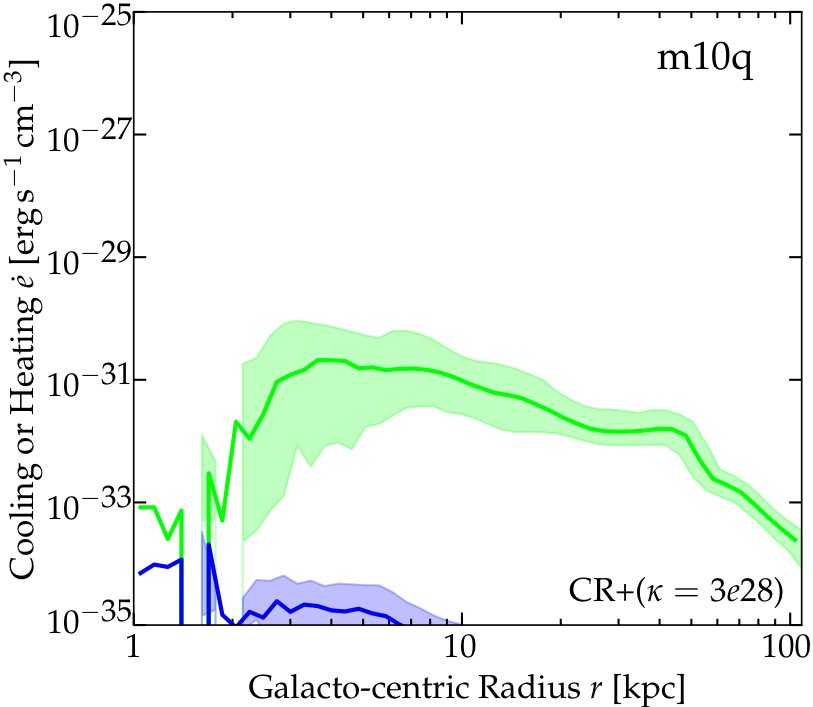}
\includegraphics[width={0.49\columnwidth}]{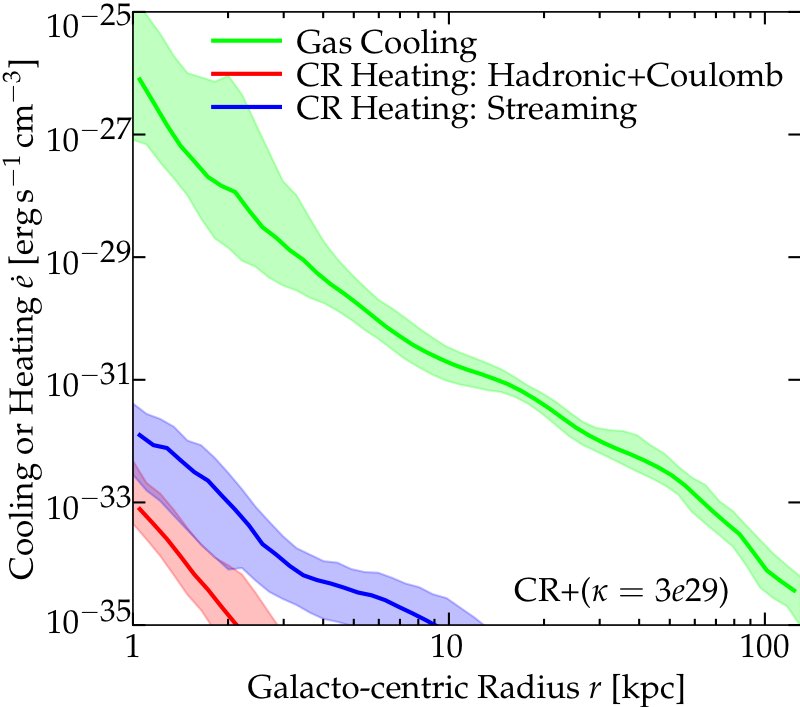} \\
\includegraphics[width={0.49\columnwidth}]{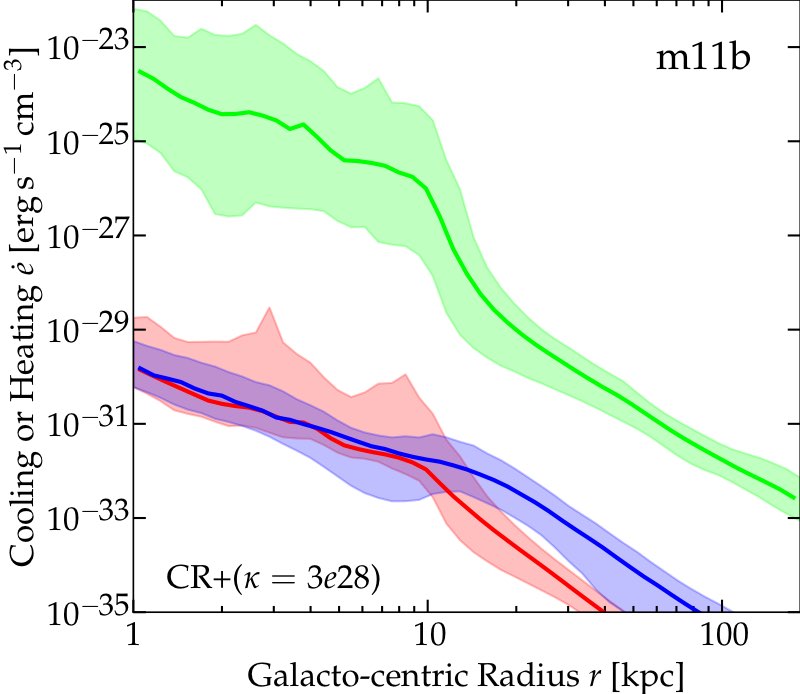}
\includegraphics[width={0.49\columnwidth}]{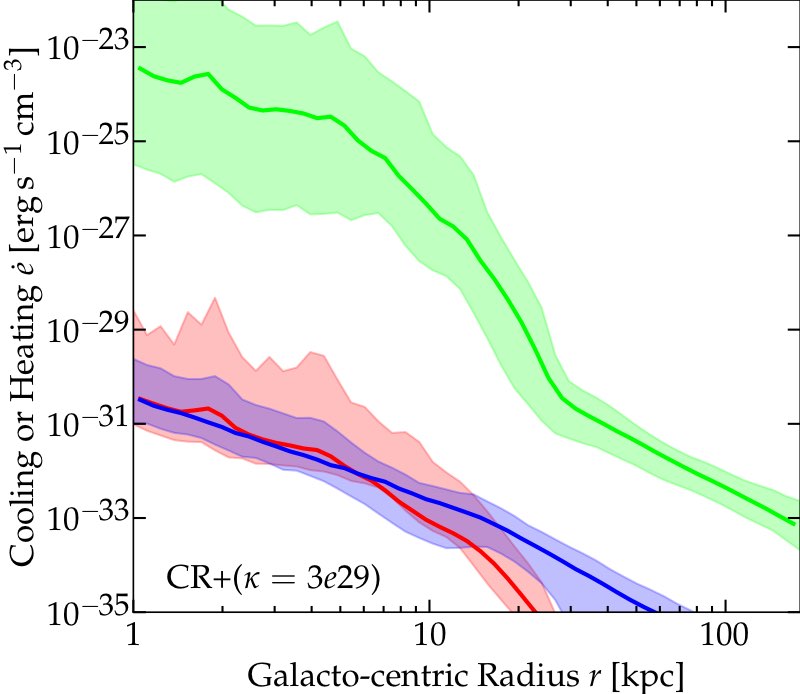} \\
\includegraphics[width={0.49\columnwidth}]{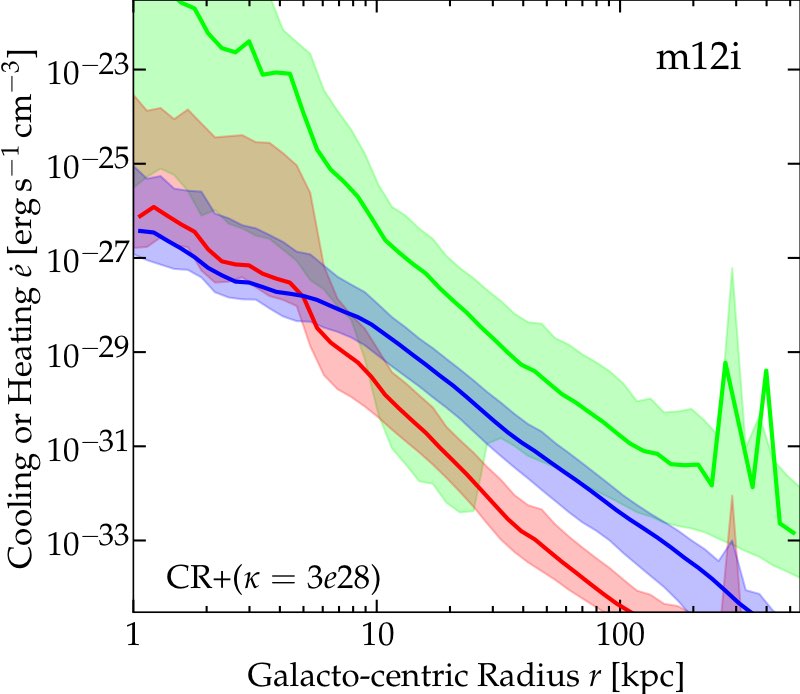}
\includegraphics[width={0.49\columnwidth}]{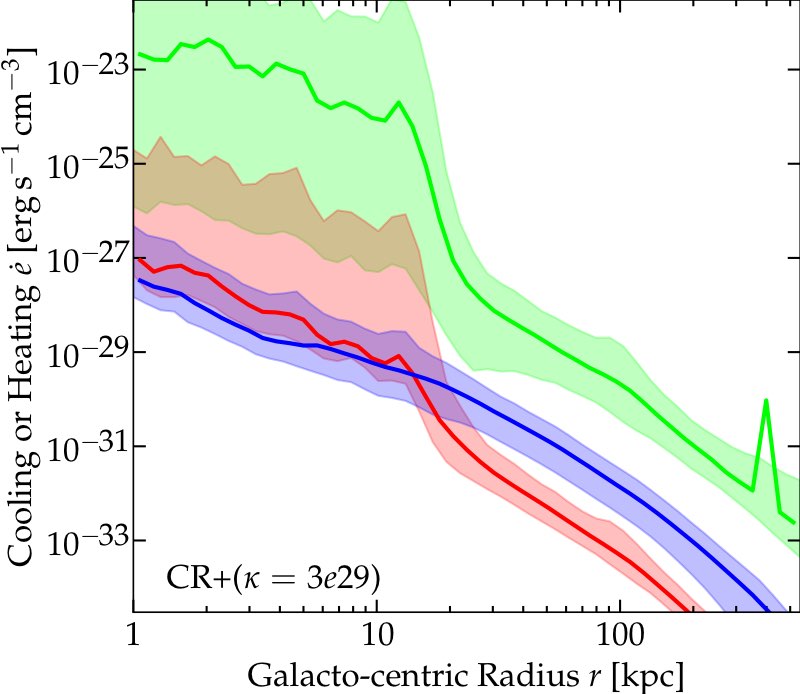} \\
    \end{centering}
    \vspace{-0.25cm}
        \caption{Profiles of gas heating/cooling rate around the central galaxy, at $z=0$, as Fig.~\ref{fig:profile.pressure}. We show the total gas cooling rate, vs.\ the heating rate from CRs via collisional (hadronic+Coulomb) and streaming losses. Hadronic losses ($\propto n_{\rm gas}$) dominate in dense gas (e.g.\ within the disks at $r\lesssim 10\,$kpc of progressively more massive galaxies), streaming dominates in more diffuse gas. Both are orders-of-magnitude below cooling rates (recall, these are dwarf and MW-mass halos, where even virial-temperature gas generally has $T \lesssim 10^{6}\,$K and so cools relatively rapidly). Streaming rates for the runs with super-\Alf{ic} streaming are larger but the difference is negligible compared to gas cooling.
    \label{fig:profile.heating}}
\end{figure}

\vspace{-0.5cm}
\section{A Theoretical Toy Model for CRs}
\label{sec:toy}

In this section, we develop a simple theoretical ``toy model'' for CRs, which provides considerable insight into the phenomena we see in the simulations. 

Assume a galaxy has a quasi-steady SFR $\dot{M}_{\ast}$, so the associated CR injection rate is $\dot{E}_{\rm cr} = \epsilon_{\rm cr}\,\epsilon_{\rm SNe}\,\dot{M}_{\ast}$, where $\epsilon_{\rm SNe} \sim (10^{51}\,{\rm erg}/70\,M_{\odot})$ is the time-and-IMF-averaged energetic yield per solar mass of star formation from SNe (for the IMF here including Type-II and prompt Ia events). Also assume that (since we are primarily interested in large scales) the CR injection is concentrated on relatively small scales, so it can be approximated as point-like, and assume -- for now -- that diffusion with constant {\em effective} coefficient $\tilde{\kappa} \sim  \langle |\hat{\bf B} \cdot \hat{\nabla}e_{\rm cr}|^{2} \rangle\,\kappa_{\ast} \sim \kappa_{\ast}/3$ dominates the transport (e.g.\ collisional and streaming losses are negligible; we will return to these below). This has the trivial equilibrium solution:\footnote{Eq.~\ref{eqn:cr.energy.profile} is still approximately valid (up to an order-unity constant) in most cases for a non-constant $\tilde{\kappa}$ or effective transport/streaming speed $v_{\rm cr}$ (replacing $\tilde{\kappa} \rightarrow v_{\rm cr}\,r$). This follows from the fact that the radial CR flux (in a spherical hydrostatic system with no losses) is ${\bf F}_{\rm cr} \rightarrow -\bar{\kappa}\,(\partial e_{\rm cr}/\partial r)\,\hat{r} \sim -\bar{\kappa}\,(e_{\rm cr}/r)\,\hat{r} \sim v_{\rm cr}\,e_{\rm cr}\,\hat{r}$.}
\begin{align}
e_{\rm cr} &= \frac{\dot{E}_{\rm cr}}{4\pi\,\tilde{\kappa}\,r} \label{eqn:cr.energy.profile}
\end{align}
This gives rise to the CR pressure gradient $\nabla P_{\rm cr} = -e_{\rm cr}\,\hat{r}/(3\,r)$.

Now also assume, for simplicity, the (diffuse) gas and DM are in an isothermal sphere (the detailed profile shape is not important) with gas fraction $f_{\rm gas}$ and circular velocity $V_{\rm c} \sim V_{\rm max}$, with some characteristic halo scale radius $R_{s} \sim R_{\rm vir}/c$. The ratio of the CR pressure gradient to the gravitational force at $R_{s}$ is ${|\nabla P_{\rm cr}|}/{|\rho\,\nabla\Phi|} \sim \dot{E}_{\rm cr}\,G\,R_{s}/(3\,\tilde{\kappa}\,f_{\rm gas}\,V_{\rm max}^{4})$ (this is identical up to an order-unity constant if we assume the gas is in e.g.\ an NFW profile or a Mestel disk). Now recall, $\dot{E}_{\rm cr} \propto \dot{M}_{\ast}$. For star-forming (sub-$L_{\ast}$) galaxies, $\dot{M}_{\ast} = \alpha\,M_{\ast}/t_{\rm Hubble}$ where $t_{\rm Hubble}(z)$ is the Hubble time at redshift $z$ and $\alpha\approx1-2$ is (in a time-and-sample averaged sense) very weakly dependent on galaxy mass \citep[see e.g.][]{2017MNRAS.464.2766M}. Using this and the canonical scaling relations\footnote{For Eq.~\ref{eqn:pressure.ratio}, we approximate the system around the radii of interest as an isothermal sphere with  $V_{c}\approx V_{\rm max}$, gas $\rho=f_{\rm gas}\,V_{c}^{2}/(4\pi\,G\,r^{2})$, $|\nabla\Phi | = G\,M_{\rm enc}/r^{2}  = V_{c}^{2}/r$, and $V_{\rm max} \approx  (2\,G\,M_{\rm vir}/R_{\rm vir})^{1/2}$ (a reasonable approximation for NFW-type halos with concentrations $\sim 1-20$), and $\dot{E}_{\rm cr} = \epsilon_{\rm cr}\,u_{\rm SNe}\,\dot{M}_{\ast} = \epsilon_{\rm cr}\,u_{\rm SNe}\,(\alpha\,M_{\ast}/t_{\rm Hubble}[z])$ (with $u_{\rm SNe}\approx 10^{51}\,{\rm erg}/70\,M_{\sun}$). For simplicity we adopt redshift scalings for a matter-dominated Universe: $M_{\rm  halo} = M_{\rm vir} \equiv (4\pi/3)\,\Delta_{c}\,\rho_{c}\,R_{\rm vir}^{3}$ with $\Delta_{c}\approx 180$ (and $\rho_{c}=3\,H_{0}^{2}\,(1+z)^{3}/8\pi\,G$  the critical density) and $t_{\rm Hubble}[z]\approx 14\,(1+z)^{-3/2}\,{\rm Gyr}$, and evaluate everything at a halo scale radius $r\sim R_{s} \sim 0.1\,R_{\rm vir}$, but the choice of cosmology or exact radius do not qualitatively alter our conclusions.} for halo $R_{s}$, $R_{\rm vir}$, and $M_{\rm halo}=M_{\rm vir}$, we can re-write this ratio as:
\begin{align}
\label{eqn:pressure.ratio}
\frac{|\nabla P_{\rm cr}|}{|\rho\,\nabla\Phi|} \sim \frac{\dot{E}_{\rm cr}\,G\,R_{s}}{3\,\tilde{\kappa}\,f_{\rm gas}\,V_{\rm max}^{4}} 
\sim \frac{0.5\,\alpha\,\epsilon_{{\rm cr},\,0.1}}{f_{{\rm gas},\,0.1}\,\tilde{\kappa}_{29}\,(1+z)^{3/2}}\, \left( \frac{M_{\ast}}{f_{b}\,M_{\rm halo}} \right)
\end{align}
where $\epsilon_{{\rm cr},\,0.1} \equiv \epsilon_{\rm cr}/0.1$, $f_{{\rm gas},\,0.1}\equiv f_{\rm gas}/0.1$, $\tilde{\kappa}_{29} \equiv \tilde{\kappa} / 10^{29}\,{\rm cm^{2}\,s^{-1}}$, and $f_{b}\equiv\Omega_{b}/\Omega_{m}$ is the universal baryon fraction, so ${M_{\ast}}/{f_{b}\,M_{\rm halo}}$ is the stellar mass relative to what would be obtained converting all baryons into stars.

This leads immediately to the prediction that CRs can have a large effect in relatively {\em massive} (intermediate, LMC-like through MW-mass; $M_{\rm halo} \sim 10^{11-12}\,M_{\sun}$) halos, at low-to-moderate redshifts ($z\lesssim 1-2$), while their effect is limited at high redshifts or in low-mass halos: $M_{\ast} / M_{\rm halo}$ is a strongly-increasing function of mass (roughly, $M_{\ast} \propto M_{\rm halo}^{2}$ at low masses). So for a MW mass halo, CR pressure is approximately able to balance gravitational forces for $\tilde{\kappa}_{29}\sim 1$, while for a true dwarf halo like {\bf m10q}, the CR pressure will be order-of-magnitude too small, because the SFR (hence the CR injection rate, which is proportional to the CR energy density/pressure/pressure gradient in steady state) is several orders of magnitude smaller in such a tiny dwarf. We also show below that in dwarfs, CRs escape efficiently from both the galaxy and CGM, while SNe cool less efficiently, all of which make CRs relatively less-dominant compared to mechanical stellar feedback.

Note also that our derivation above, at face value, would imply that a {\em lower} diffusion coefficient would give stronger effects of CR pressure (because the CRs are more ``bottled up'' so the steady-state pressure is larger). However, we neglected collisional and streaming losses: we show below that if $\tilde{\kappa}_{29} \ll  1$, these quickly dominate and prevent CRs from having any significant effects.

\vspace{-0.5cm}
\subsection{Collisional Losses at Low-$\kappa$ \&\ Observational Constraints}

We neglected collisional losses above. Assuming ionized gas (the difference is small, this just determines the contribution from Coulomb terms) these scale as $\dot{e}_{\rm loss} = e_{\rm cr}\,n/(t_{0}\,n_{0})$ with $t_{0}\,n_{0} \sim 1.7\times10^{15}\,{\rm s\,cm^{-3}}$ (\S~\ref{sec:methods:crplus}). If we take the same isothermal sphere model above, and calculate the steady-state volume-integrated CR loss rate $\dot{E}_{\rm loss}$, we obtain: 
\begin{align}
\label{eqn:hadronic.loss.analytic.1}  \frac{\dot{E}_{\rm loss}}{\dot{E}_{\rm cr}} \approx \frac{\ln{(r_{\rm max}/r_{\rm min})}}{4\pi\,\tilde{\kappa}\,m_{p}\,t_{0}\,n_{0}}\,\frac{f_{\rm gas}\,V_{\rm max}^{2}}{G} \sim \frac{0.15}{\tilde{\kappa}_{29}}\,\left( \frac{\Sigma_{\rm gas}\,R_{\rm gas}}{0.01\,{\rm g\,cm^{-2}\,kpc}}\right)
\end{align}
where in the latter equality we use $f_{\rm gas}\,V_{\rm max}^{2}/G = M_{\rm gas}/R_{\rm gas} \approx \pi\,\Sigma_{\rm gas}\,R_{\rm gas}$ (and $\ln{(r_{\rm max}/r_{\rm min})} \sim 5$). Using the observational facts that (sub-$L_{\ast}$, star-forming) galactic disks have approximately constant effective $\Sigma_{\rm gas} \sim 10-30\,M_{\odot}\,{\rm pc^{-2}}$ and $R_{\rm gas} \sim 2\,R_{\ast}\sim 10\,{\rm kpc}\,(M_{\ast}/10^{11}\,M_{\odot})^{1/3}$ \citep[e.g.][]{courteau:disk.scalings}, we can equivalently  write: 
\begin{align}
\label{eqn:hadronic.loss.analytic.2}  \frac{\dot{E}_{\rm loss}}{\dot{E}_{\rm cr}} &\sim \frac{1}{\tilde{\kappa}_{29}}\,\left( \frac{\Sigma_{\rm gas}}{10\,M_{\odot}\,{\rm pc^{-2}}} \right)\,\left( \frac{M_{\ast}}{10^{10}\,M_{\odot}} \right)^{1/3} 
\end{align}
Of course, $\dot{E}_{\rm loss}$ cannot exceed $\dot{E}_{\rm cr}$ in steady-state: this gives an upper bound to the expression above at the``calometric limit'' ($\dot{E}_{\rm loss}=\dot{E}_{\rm cr}$) at which point all input CR energy is lost to collisions.

The ratio $\dot{E}_{\rm loss}/\dot{E}_{\rm cr}$ is directly related to the observed ratio of GeV $\gamma$-ray flux or luminosity  ($F_{\gamma} \propto L_{\rm gamma} \propto \dot{E}_{\rm loss}$) to bolometric flux  from massive stars ($F_{\rm SF} \propto L_{\rm SF} \propto \dot{M}_{\ast} \propto \dot{E}_{\rm cr}$). Using the values given in \S~\ref{sec:results} below for conversion factors between collisional losses and $\gamma$-ray luminosity or flux, Eq.~\ref{eqn:hadronic.loss.analytic.1} becomes:
\begin{align}
\label{eqn:fgamma.fsf.analytic}  \frac{F_{\gamma}}{F_{\rm SF}} &\sim \frac{3\times10^{-5}}{\tilde{\kappa}_{29}}\,\left( \frac{\Sigma_{\rm gas}\,R_{\rm gas}}{0.01\,{\rm g\,cm^{-2}\,kpc}}\right) 
\end{align}
with the calorimetric limit ``capping'' this at $F_{\gamma}/F_{\rm  SF} \sim 2\times10^{-4}$.

Note that the geometry of the gas is not especially important here. If we assume instead the gas is in a thin, exponential disk with scale height $H/R \ll 1$, and effective radius $R_{e}$, but the CRs diffuse approximately spherically (as a random walk), then we obtain, a result which differs only by an order-unity constant from the isothermal-sphere scaling above even as $H/R \rightarrow 0$ (because even though the disk occupies a vanishingly small volume, for fixed surface density $\Sigma$, the three-dimensional density $n$ in the disk, hence the CR loss rate while within it, increases inversely with $H/R$). 

Three immediate consequences follow from this. First, essentially all observed galaxies with SFRs below $\sim 10\,{\rm M_{\odot}\,yr^{-1}}$ (including the MW, LMC, SMC, M31, and M33) have observed $F_{\gamma}/F_{\rm SF} \le 10^{-5}$ \citep{lacki:2011.cosmic.ray.sub.calorimetric}, so we predict that a diffusion coefficient $\tilde{\kappa}_{29} \gtrsim 1$ is {\em required} to match the observations. The same $\tilde{\kappa}$ is required to reproduce the canonical Milky Way constraints -- most recent studies agree that for a MW with a diffuse gaseous halo of scale length $\gtrsim 10\,$kpc (appropriate for the simulations here, since the galaxies have extended halos and the diffusivity is, by assumption, constant) an effective, {\em isotropically-averaged} diffusivity $\tilde{\kappa}_{29} \gtrsim 1$ is needed \citep{blasi:cr.propagation.constraints,vladimirov:cr.highegy.diff,gaggero:2015.cr.diffusion.coefficient,2016ApJ...819...54G,2016ApJ...824...16J,2016ApJ...831...18C,2016PhRvD..94l3019K,evoli:dragon2.cr.prop,2018AdSpR..62.2731A}.\footnote{For the MW, we note that the observations (e.g.\ secondary-to-primary ratios and the like) do not really constrain the ``diffusion coefficient'' or ``residence time'' (these are model-dependent inferences), but rather the effective column density or ``grammage'' $X_{s} \equiv \int_{\rm CR\,path}\,n_{\rm gas}\,d\ell_{\rm CR} = \int_{\rm CR\,path}\,n_{\rm gas}\,c\,dt$ integrated over the path of individual CRs from their source locations to the Earth, with $X_{s} \approx 3\times10^{24}\,{\rm cm^{-2}}$ measured. Repeating our calculation above for either an isothermal sphere or thin-disk gas distribution, it is straightforward to show that the grammage (integrated to infinity, as opposed to the solar circle) is {\em directly} related to the hadronic losses as $\dot{E}_{\rm loss}/\dot{E}_{\rm cr} = X_{s}^{\infty}/(2\,t_{0}\,n_{0}\,c) \sim 0.01\,(X_{s}^{\infty} / 3\times10^{24}\,{\rm cm^{-2}})$. Thus the constraints from matching the direct MW observations are essentially equivalent to matching the observationally ``inferred'' $F_{\gamma}/F_{\rm SF} \sim (0.3-1)\times10^{-5}$ in the MW.}

Second, we see why low diffusion coefficients cannot be invoked to increase the CR pressure (as noted above) -- if one lowers $\tilde{\kappa}_{29} \ll 1$, then not only will the model fail to reproduce the observations, but the collisional losses will quickly dominate. If $\dot{E}_{\rm loss} \gtrsim \dot{E}_{\rm cr}$, it means that all CR energy is rapidly lost in the ISM, so there is no steady-state, high-pressure CR halo (which requires they escape the galaxy in the first place). In small dwarfs with low densities (low $\Sigma_{\rm gas}\,R_{\rm gas}$) one may be able to make $\tilde{\kappa}$ slightly lower without losing all of the CR energy, but we show below this still leads to large violations of the observational constraints. 

Third, this means we do {\em not} expect CRs to (at least locally) have strong effects in extreme starburst-type galaxies, at either low or high redshift (where they are more common), owing to the very high $\Sigma_{\rm gas}$ observed in such systems, which should lead to efficient losses (see \citealt{chan:2018.cosmicray.fire.gammaray} for more discussion).

\vspace{-0.5cm}
\subsection{Streaming \&\ The Critical Radius}
\label{sec:streaming_toy}

Now consider the streaming terms explicitly. If the streaming velocity is $\sim v_{A} \sim \beta^{-1/2}\,c_{s}$, and $\beta$ is approximately constant (as one might expect from e.g.\ a transsonic turbulent dynamo), then in an isothermal halo $v_{A}$ is also constant, so the radial streaming ``flux'' is $\sim v_{A}\,e_{\rm cr}$. The diffusive flux is $\sim \tilde{\kappa}\,\nabla e_{\rm cr} \sim \tilde{\kappa}\,e_{\rm cr}/r$, so the streaming dominates at $r \gtrsim r_{\rm stream} \sim \tilde{\kappa}/v_{A}$ (note we could have derived this instead in terms of the transport {\em timescales} to reach a given radius, and would reach the identical conclusion). This gives: 
\begin{align}
\label{eqn:rstream}
r_{\rm stream} \sim 30\,{\rm kpc}\,\frac{\tilde{\kappa}_{29}}{v_{A,\,10}} \sim 30\,{\rm kpc}\,\frac{\beta_{100}^{1/2}\,\tilde{\kappa}_{29}}{M_{{\rm vir},\,12}^{1/3}\,(1+z)^{1/2}}
\end{align}
where $v_{A,\,10} \equiv v_{A}/10\,{\rm km\,s^{-1}}$, $\beta_{100}\equiv \beta/100$, and $M_{{\rm vir},\,12}\equiv M_{\rm halo} / 10^{12}\,M_{\odot}$. 

Beyond this radius, if we continue to neglect losses, the solution would become that for a steady-state wind with constant velocity, i.e.\ 
$e_{\rm cr}(r > r_{\rm stream}) \sim  {\dot{E}_{\rm cr}}/({4\pi\,v_{A}\,r^{2}})$, so it falls more steeply compared to the diffusion-dominated case  ($e_{\rm cr} \propto r^{-1}$). This has several important consequences. First (1) if trans-\Alf{ic} streaming dominates transport, then the streaming {\em transport timescale} $\sim r/v_{\rm stream}$ is always of the same order as the streaming {\em loss timescale} $\sim e_{\rm cr}/\dot{e}^{\rm stream}_{\rm loss} \sim e_{\rm cr} / |{\bf v}_{\rm stream}\cdot \nabla P_{\rm cr} | \sim (3/2)\,r/v_{\rm stream}$. So the CRs will, by definition, lose energy to \Alf{ic} and ultimately thermal energy as they stream at the same rate they propagate out, at this radius. This means the energy density must eventually decay, further accelerating streaming losses. So (2) a non-negligible fraction of the CR energy is thermalized within a factor of a few of this radius. And (3) since the CR pressure drops more rapidly, the CRs eventually provide small pressure support at $r \gg r_{\rm stream}$, {\em even if} they dominate over gravitational pressure at $r \lesssim r_{\rm stream}$ according to our arguments above.

Thus, in general, the dominant role of CRs is predicted to be confined to an ``inner CGM'' CR ``halo'' at $\lesssim 30-100\,{\rm kpc}$. For fixed $\beta$ and $\tilde{\kappa}$, lower $v_{A}$ means that this radius extends further in smaller halos, but (as we argued above), the CR injection and energy density also declines rapidly in smaller halos, so this simply means that the (relatively small) CR energy density more efficiently escapes, rather than being thermalized, in smaller halos. In massive halos, $r_{\rm stream}$ becomes comparable to the halo scale-radius.

Streaming losses still occur in the inner parts of the halo, even if streaming does not dominate the transport. Integrating the streaming loss rate $={\bf v}_{\rm stream}\cdot\nabla P_{\rm cr}$ within a maximum radius $r$, we obtain:
\begin{align}
\frac{\dot{E}^{\rm stream}_{\rm loss}}{\dot{E}_{\rm cr}}(r \ll  r_{\rm stream}) = \frac{v_{A}\,r}{3\,\tilde{\kappa}} = \frac{1}{3}\,\left( \frac{r}{r_{\rm stream}} \right) 
\end{align}
so this is always a small (but not negligible) fraction of $\dot{E}_{\rm cr}$ within $r < r_{\rm stream}$. However, from the scaling of $r_{\rm stream}$, we see that streaming imposes {\em yet another} reason why CRs will not have a large effect at very low diffusion coefficients. If $\tilde{\kappa}$ is too low, then even if the gas density is so low that collisional losses are negligible, the CRs will lose much of their energy to \Alf{ic} damping (``streaming losses'') at very small $r$. Effectively, if $r_{\rm stream}$ is smaller than the gas disk effective radius, it means that the CR energy is thermalized before it can escape the star-forming disk. 

Note that if we make $v_{\rm stream}$ much larger {\em and} allow the ``streaming loss'' term to increase with $\sim v_{\rm stream} \nabla P_{\rm cr}$ (instead of limiting it at $\sim v_{A}\nabla P_{\rm cr}$), then although streaming moves CRs {\em faster}, it also means streaming losses occur much faster and closer to the galaxy, where they can be radiated away more efficiently. In Eq.~\ref{eqn:rstream}, for example, if we set $v_{\rm stream}^{2} = v_{A}^{2} + c_{s}^{2}$, then for $\beta \gg 1$ the expression becomes $r_{\rm stream} \sim 3\,{\rm kpc}\,\tilde{\kappa}_{29}\,M_{{\rm vir},\,12}^{-1/3}\,(1+z)^{-1/2}$. Thus, especially if combined with a lower diffusion coefficient, this particular super-\Alf{ic} streaming model means that streaming losses would thermalize most of the CR energy within the galaxy and ISM, where it would be efficiently radiated away.

Note that because $v_{A}$ and $r_{\rm stream}$ depend directly on $|{\bf B}|$ or $\beta$, if magnetic fields are stronger (weaker) it shifts our predictions for CR transport accordingly. For CRs, this is effectively degenerate with how we treat the scaling of $v_{\rm stream}$ with $v_{A}$: our default model ($v_{\rm st} \approx 3\,v_{A}$) is akin to a model with $v_{\rm st} \approx v_{A}$ but $10$ times lower $\beta$. Our experiments with different $v_{\rm st}$ therefore give some insight into how our predictions depend on the ultimate strength of magnetic fields.

\vspace{-0.5cm}
\subsection{What about CR Heating?}

We have argued for the importance of the CR pressure/adiabatic terms above, and discussed CR losses. But can the CRs also be important as a thermal heating mechanism? 

If we assume $100\%$ of the CR energy is thermalized (of course an upper limit, since some CRs escape, some energy is lost doing adiabatic work, and for hadronic losses $5/6$ of the energy goes into products like $\gamma$-rays which escape rather than thermalize), we can compare this to the cooling luminosity $L_{\rm cool}^{\rm tot} = \int\,\Lambda\,n^{2}\,d^{3}{\bf x}$, where $\Lambda \sim \Lambda_{22}\,10^{-22}\,{\rm erg\,s^{-1}\,cm^{3}}$ is roughly constant over the temperature range of interest. Using this and the various scalings above (for fully-ionized gas), we obtain:
\begin{align}
\label{eqn:heating.all} \frac{\dot{E}_{\rm cr}}{L_{\rm cool}^{\rm tot}} \sim \frac{0.16\,\alpha\,\epsilon_{{\rm cr},\,0.1}\,(1+z)^{3/2}}{\Lambda_{22}\,(f_{\rm gas}/f_{b})}\, \left( \frac{M_{\ast}}{f_{b}\,M_{\rm halo}} \right) \, \left( \frac{10^{-4}\,{\rm cm^{-3}}}{\langle n \rangle_{\rm cool}} \right)
\end{align}
where $\langle n \rangle_{\rm cool}$ is the {\em cooling-luminosity-weighted} density (i.e.\ density where most of the cooling occurs).
We immediately see, again, that CR heating cannot have a large effect in dwarf galaxies, owing to their very low $M_{\ast}/M_{\rm halo}$ (and correspondingly low SFRs and CR energy production), even if the CRs couple at very low CGM densities. In massive halos, at ISM densities  ($\langle n \rangle_{\rm cool}\sim 1$), we see that there is no possible way CR heating can compete with radiative cooling. Moreover, the CR energy injection rate is an order-of-magnitude smaller than that from mechanical energy in SNe shocks.\footnote{CR heating can be non-negligible in cold gas with $T \ll 10^{4}\,$K, where $\Lambda$ is much smaller, and the gas is strongly self-shielded so photo-electric heating is negligible, provided the local CR energy density is high. However this has a negligible effect on the dynamics of the cold gas (it is mostly important for accurate ionization fraction calculations), or on the total cooling budget of the ISM which is dominated by warmer gas cooling down to these low temperatures.} 

In the very lowest-density CGM near the virial radius (around the mean gas density of the halo, $\langle n \rangle_{\rm cool} \sim 10^{-4}\,(1+z)^{3}\,{\rm cm^{-3}}$), Eq.~\ref{eqn:heating.all} suggests CR heating could become significant, but that assumes {\em all} the CR energy couples in the least-dense gas just inside $R_{\rm vir}$ (and ignores gas outside/inside). More accurately, if we assume streaming losses dominate (with the equilibrium $e_{\rm cr}$ for streaming at large $r > r_{\rm stream}$, defined above),\footnote{In Eq.~\ref{eqn:edot.stream}, because we assume streaming dominates the transport ($r>r_{\rm stream}$), the steady-state CR energy  $e_{\rm cr} \propto \dot{E}_{\rm cr}/(v_{\rm stream}\,r^{2})$ while the streaming  losses scale as $\sim v_{A}\,\nabla P_{\rm cr} \propto (v_{A}/v_{\rm stream})\, \dot{E}_{\rm cr} / r^{3}$. For $v_{\rm stream} \sim v_{A}$ (assumed in Eq.~\ref{eqn:edot.stream}), the dependence on $v_{A}$ vanishes, while for super-\Alf{ic} streaming, the heating rate from streaming losses is reduced by a factor $\sim v_{A}/v_{\rm stream}$. Likewise, for streaming inside of $r <  r_{\rm stream}$ where diffusion dominates transport,  the heating rate is reduced by a factor $\sim v_{A}\,r / \kappa \sim r / r_{\rm stream} < 1$.} and that these are instantly thermalized, and that the gas is in an isothermal sphere, we can then calculate the ratio of the local thermal heating rate from CRs to the cooling rate:
\begin{align}
\label{eqn:edot.stream} \frac{\dot{e}^{\rm stream}_{\rm heat}}{\dot{e}_{\rm cool}} \sim \frac{0.02\alpha\,\epsilon_{{\rm cr},\,0.1}}{\Lambda_{22}\,(f_{\rm gas}/f_{b})^{2}\,(1+z)^{3/2}} \left( \frac{M_{\ast}}{f_{b}\,M_{\rm halo}} \right)\,\left(\frac{r}{R_{\rm vir}} \right)
\end{align}
So the CR thermal heating is unlikely to be relevant at any radius (at least for the  halos of interest here).

\vspace{-0.5cm}
\subsection{(Lack of) Effects Interior to Star-Forming Galactic Disks}

The model above immediately implies that CRs have very weak effects {\em within} the star-forming galaxy disk at any mass scale. In order to avoid losing all the energy to collisional losses, it is required that the CR diffusion time ($\sim 0.03\,{\rm Myr}\,(L/100\,{\rm pc})^{2}\,\tilde{\kappa}_{29}$ for diffusion on scale $L$) is much faster than dynamical times in the disk. This essentially means CRs ``diffuse out'' of any locally dynamically interesting region of the disk (e.g.\ a GMC or strong shock) well before they can do interesting adiabatic work on that region. As a result, the CRs form (as assumed here) a quasi-spherical profile. 

For a MW-like galaxy, our Eq.~\ref{eqn:cr.energy.profile} predicts a CR energy density at the solar circle of $e_{\rm cr} \sim 1\,{\rm eV\,cm^{-3}}\,(\dot{M}_{\ast}/{\rm M_{\odot}\,yr^{-1}})\,\epsilon_{{\rm cr},\,0.1}\,(r/8\,{\rm kpc})^{-1}\,\tilde{\kappa}_{29}^{-1}$, more or less exactly the canonical value, and in order-of-magnitude equipartition with other disk-midplane energy densities. However, because of rapid diffusion, the CR pressure {\em gradients} are necessarily weak in the disk. If we assume a vertically-exponential gas disk balancing gravity in vertical hydrostatic equilibrium (disk midplane pressure $P_{\rm mid} \approx \pi\,G\,\Sigma_{\rm gas}^{2}$), then the ratio of the vertical pressure gradients $|\partial P_{\rm cr}/\partial z | / |\partial  P_{\rm mid}^{\rm other}/\partial z|$ (where $\partial P_{\rm mid}^{\rm other}/\partial z$ is the gradient of thermal/magnetic/turbulent pressure in vertical hydrostatic equilibrium) scales as $\sim (P_{\rm cr}/P_{\rm mid}^{\rm other})\,(H/R)^{2}$ -- i.e.\ the CR pressure {\em forces} (or gradients) are sub-dominant by a factor of $\sim (R/H)^{2} \gg 100$ in the MW midplane. The difference is even more dramatic if we consider still smaller-scale sub-structure (e.g.\ turbulent substructure in the ISM or clumps/cores in GMCs, where the relevant turbulent or magnetic/thermal support terms have structure on sub-pc scales).

\vspace{-0.5cm}
\subsection{Summary: The ``Sweet Spot'' for CRs}

This toy model illustrates that, although for realistic (or observationally  allowed) parameters we do not expect CRs to be dynamically dominant in the evolution of small dwarf galaxies, there is a potential ``sweet spot'' of galaxy parameter space in which CRs (from SNe) might be quite important for intermediate (LMC) through massive (MW-mass) galaxies ($M_{\rm halo} \sim 10^{11-12}\,M_{\sun}$) at low-to-intermediate redshifts ($z\lesssim 1-2$), via the creation of an extended CR halo in the inner CGM with pressure sufficient to maintain virial equilibrium and therefore support gas which might otherwise accrete onto the galaxy.

However even this requires some ``sweet spot'' in the {\em cosmic ray transport} parameter space. If $\tilde{\kappa}$ is too low, the CRs are trapped and collisional+streaming losses dissipate all their energy rapidly (in contradiction to all present observational constraints for Local Group galaxies). If $\tilde{\kappa}$ is too high (which may be observationally allowed), or is not constant but rises very rapidly outside of the galaxy (certainly allowed observationally), CRs will simply ``free stream'' out of the CGM without building up a significant pressure gradient or thermalizing their energy -- although this may require extremely high $\tilde{\kappa}$. 
If the CR injection fraction $\epsilon_{\rm cr} \ll 0.1$, there is simply not enough energy in CRs to have an effect at any mass scale, while if it is too large ($\epsilon_{\rm cr}\sim 1$) it would violate direct observational constraints. 

That is not to say this ``sweet spot'' requires very special fine-tuning. In fact, the characteristic parameters in the ``sweet spot'': $\tilde{\kappa} \sim 10^{29} - 10^{30}\,{\rm cm^{2}\,s^{-1}}$, $v_{\rm stream} \sim v_{A}$, $\epsilon_{\rm cr} \sim 0.1$ are both theoretically predicted (or at least plausible) and observationally allowed. These parameters are also in good agreement with the "sweet spot" values identified in idealized galaxy simulations in \citet{chan:2018.cosmicray.fire.gammaray}. And we stress that each of these can be varied by a factor of several (as we have done) without radically altering our conclusions. However, caution is needed, as these remain deeply uncertain.

\begin{figure}
\begin{centering}
\includegraphics[width={0.95\columnwidth}]{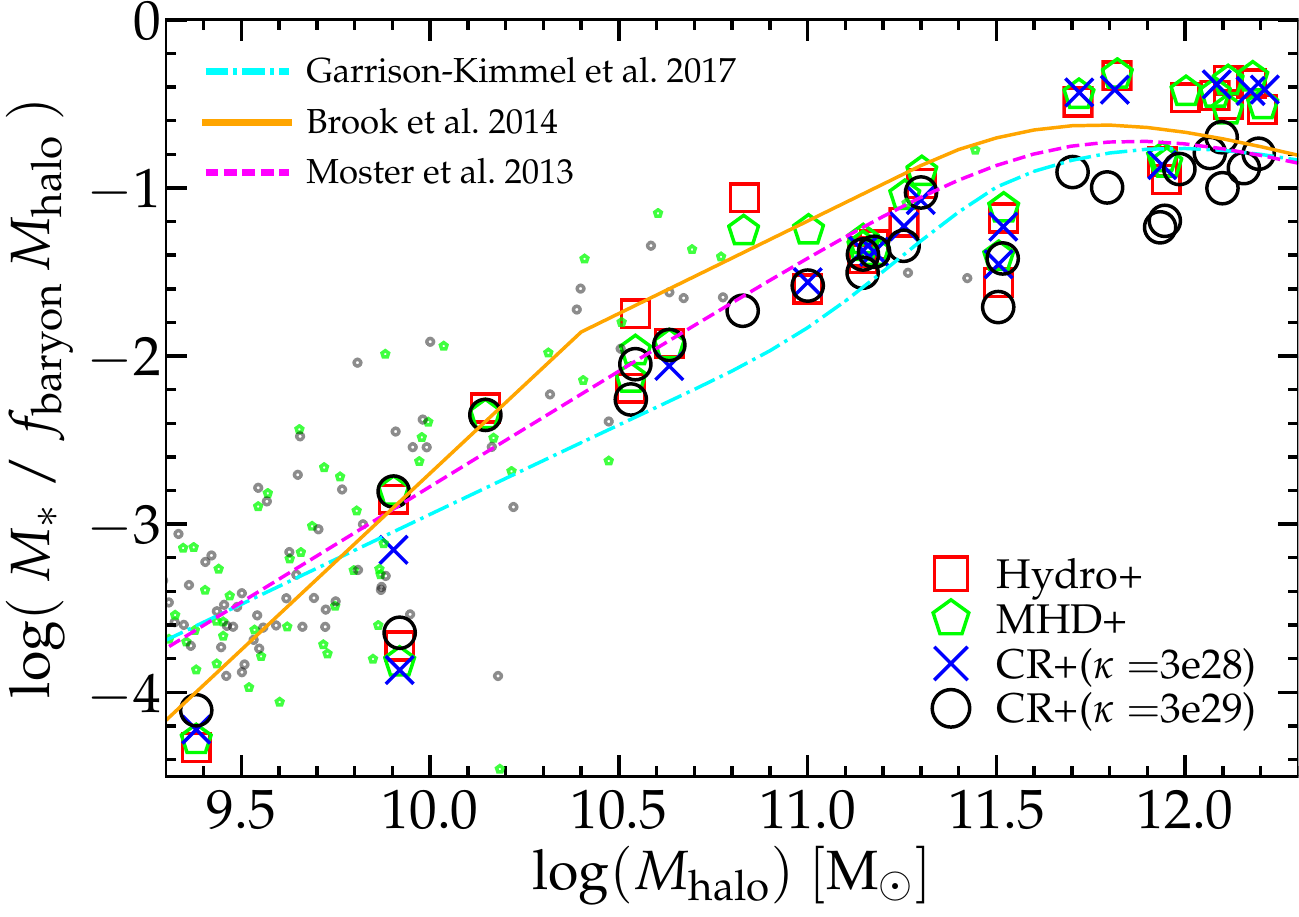}
\includegraphics[width={0.95\columnwidth}]{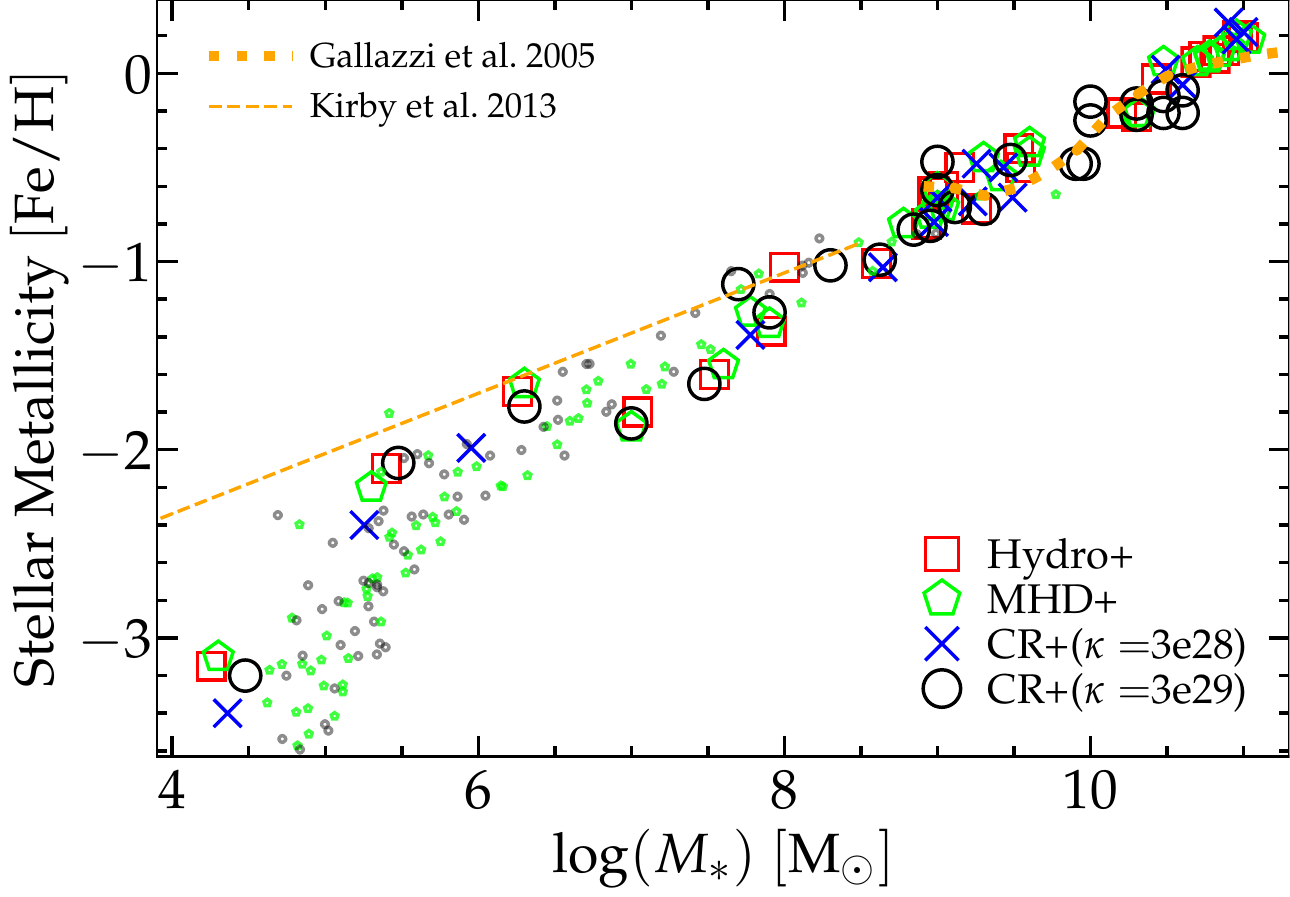}
    \end{centering}
    \vspace{-0.25cm}
        \caption{{\em Top:} Stellar mass (normalized to  the universal baryon fraction) vs.\ halo mass relation of the simulated galaxies (Table~\ref{tbl:sims}) at $z=0$. Lines compare observational estimates of the median relation from abundance-matching to either isolated massive galaxies \citep[][extrapolated to low  masses]{moster:2013.abundance.matching.sfhs} or Local Group dwarfs  \citep{brook:2014.mgal.mhalo.local.group,sgk:2016.mgal.mhalo.lowmass.scatter}.
        {\em Bottom:} Stellar mass vs.\ metallicity of the same  galaxies at $z=0$. Lines show observational estimates for Local Group satellites \citep{kirby:2013.mass.metallicity} and isolated massive galaxies \citep{gallazzi:ssps}. 
        Definitions of mass and metallicity match Figs.~\ref{fig:cr.demo.m10}-\ref{fig:cr.demo.m12}.
        Large symbols show each ``primary'' galaxy.  
        To illustrate that the results are robust across the larger set of halos in our simulated volumes, as well as resolution,  small points show $\sim 50$ randomly-selected non-satellite halos  (distance $>500\,$kpc from  the ``primary'') from the ``MHD+'' and ``CR+($\kappa=3e29$)'' runs of boxes {\bf m12i,f,m} (resolution $m_{\rm i,\,1000}=7$, much poorer than our ``primary'' small-dwarf galaxies). 
        The ``CR+'' runs with higher-$\kappa$ exhibit systematically lower stellar masses at $M_{\rm halo} \gtrsim 10^{10.5-11}\,M_{\odot}$'' (other suites do not differ significantly). However, the galaxies move {\em along}, not off of, a tight mass-metallicity relation.
    \label{fig:mgal.mhalo}}
\end{figure}

\vspace{-0.5cm}
\section{Results}
\label{sec:results}

We now present the results of our simulations. Across all the parameter surveys described below, we study $>150$ high-resolution full-physics cosmological zoom-in simulations run to $z=0$. We discuss the implications of these results, and use them to test our simple model expectations, in \S~\ref{sec:discussion} below.

\subsection{Case Studies}
\label{sec:results:overview}

Figs.~\ref{fig:cr.demo.m10}, \ref{fig:cr.demo.m11}, \&\ \ref{fig:cr.demo.m12}, summarize a number of basic properties of the simulated galaxies: the ``archeological'' star formation history (distribution of star formation times); galaxy stellar mass and metallicity versus redshift; dark matter and baryonic mass profiles, stellar effective radii, and circular velocity curves at $z=0$. A number of other properties can be directly inferred from this (or are trivially related to information here), including e.g.: the ``burstiness'' of star formation, the distribution of stellar ages, the evolution of the mass-metallicity relation and stellar mass-halo mass relation, the existence of ``cusps'' or ``cores'' in the dark matter profile, baryon content of the halo, baryonic mass concentration, etc. We explicitly note the $z=0$ stellar mass, stellar effective radius, and metallicity. These figures intentionally parallel our previous comparison of feedback physics (adding or removing SNe, radiation, etc.), numerical methods and resolution in \citet{hopkins:fire2.methods} -- additional details of how each quantity is measured are given in that paper.

We show this for each of the ``primary'' (most well-resolved) galaxies in the high-resolution region of a number of representative simulations from Table~\ref{tbl:sims}. For each, we compare the different physics variations in Table~\ref{tbl:physics}: {\bf (1)} our default FIRE-2 physics ``Hydro+'' run, {\bf (2)} the FIRE-2 + MHD + anisotropic conduction and viscosity ``MHD+'' run, {\bf (3)} the FIRE-2 + MHD + anisotropic conduction and viscosity + full CR physics ``CR+'' run with a ``lower'' parallel diffusion coefficient $\kappa \equiv 3\times10^{28}\,{\rm cm^{2}\,s^{-1}}$, and {\bf (4)} a ``CR+'' run with higher parallel diffusion coefficient $\kappa \equiv 3\times10^{29}\,{\rm cm^{2}\,s^{-1}}$.

Simulations from Table~\ref{tbl:sims} not shown explicitly in Figs.~\ref{fig:cr.demo.m10}-\ref{fig:cr.demo.m12}, as well as the (many) less-massive galaxies in each box, are omitted for brevity, but exhibit very similar behavior to those shown at similar masses. Likewise, Appendix~\ref{sec:additional.tests} shows additional simulations with varied resolution: the qualitative conclusions are similar to the survey in Figs.~\ref{fig:cr.demo.m10}-\ref{fig:cr.demo.m12}.

Qualitatively, we see that ``MHD+'' and ``Hydro+'' runs are very  similar in all respects and ``CR+($\kappa=3e28$)'' runs generally differ by only a small amount (with a couple exceptions). However ``CR+($\kappa=3e29$)'' runs produce suppressed SFRs and stellar masses in massive ($M_{\rm halo} \gtrsim 10^{11}\,M_{\sun}$) systems at low redshifts ($z\lesssim 1-2$).

\subsection{$\gamma$-Ray Emission}
\label{sec:results:Lgamma}

Fig.~\ref{fig:Lgamma} compares the $\gamma$-ray emission  predicted in  our simulations to observational constraints. With the exception of the MW, where more detailed constraints from spallation and other measurements exist (see \S~\ref{sec:toy}), $\gamma$-ray emission represents one of the most direct observational constraints available on the CR energy density in nearby galaxies. This was studied in detail in \citet{chan:2018.cosmicray.fire.gammaray}, in non-cosmological simulations, so we extend that comparison here with the addition of our cosmological runs. Many of the $\gamma$-ray observations (and the equivalent constraint for the MW, essentially the measured grammage or ``residence time'') are collected in \citet{lacki:2011.cosmic.ray.sub.calorimetric} (note that more recent studies, e.g.\ \citealt{tang:2014.ngc.2146.proton.calorimeter,griffin:2016.arp220.detection.gammarays,fu:2017.m33.revised.cr.upper.limit,wjac:2017.4945.gamma.rays,wang:2018.starbursts.are.proton.calorimeters,lopez:2018.smc.below.calorimetric.crs}, find consistent results). We compare directly to these constraints in Fig.~\ref{fig:Lgamma}, mimicking the observations as best as possible (see \citealt{chan:2018.cosmicray.fire.gammaray} for additional details). Specifically, we compute the predicted luminosity $L_{\gamma}$ in $\sim$\,GeV $\gamma$-rays produced by hadronic CR collisions. To do so, we follow \citet{guo.oh:cosmic.rays,chan:2018.cosmicray.fire.gammaray} and take the exact hadronic loss rate computed in-code, assume $5/6$ of the losses go to pions with a branching ratio of $1/3$ to $\pi^{0}$, which decay to $\gamma$-rays with a spectrum that gives $\sim 70\%$ of the energy at $>1\,$GeV, and integrate this inside a $\sim 5\,$kpc aperture for dwarfs ($M_{\rm halo} < 3\times10^{11}\,M_{\sun}$) or $\sim 10\,$kpc aperture for more massive galaxies (similar to the effective areas used for observations, although this has a relatively small effect). We also compute the central\footnote{``Central'' radius is defined here following \citet{lacki:2011.cosmic.ray.sub.calorimetric} as a projected radius $\sim 2\,$kpc for dwarfs, and  $\sim 4\,$kpc for $L_{\ast}$ galaxies. However using a constant $\sim1\,$kpc or half the  effective radius gives qualitatively similar results.} sightline-averaged gas surface density $\Sigma_{\rm central}$, and the luminosity from young stars $L_{\rm SF}$ (computed with {\small STARBURST99} convolving all star particles $<100\,$Myr old with their appropriate ages and metallicities). This defines the ratio $L_{\gamma}/L_{\rm SF}$, versus $\Sigma_{\rm central}$, as measured in \citet{lacki:2011.cosmic.ray.sub.calorimetric}. 

If {\em all} of the CR energy injected by SNe were lost collisionally, in steady-state with a time-constant SFR, this would produce a steady-state $L_{\gamma}/L_{\rm SF} \sim 2\times10^{-4}$, which we label as the ``calorimetric limit.'' Of course, galaxies can violate this in transient events (or by a small systematic amount if e.g.\ star formation is non-constant in time). The ratio of $L_{\gamma}/L_{\rm SF}$ to calorimetric gives, approximately, the fraction of CR energy lost collisionally. We compare this for our ``default'' suite from Figs.~\ref{fig:cr.demo.m10}-\ref{fig:cr.demo.m12}, as well as a detailed surveys varying $\kappa$ (discussed below) and the streaming speed. For the streaming-speed study, we specifically re-run simulations  {\bf m10q}, {\bf m11b}, {\bf m11q}, {\bf m11d}, {\bf m11h},  {\bf m11f}, {\bf m11g}, {\bf m12f}, {\bf m12i}, and {\bf m12m}, all with $\kappa=3\times10^{28}\,{\rm cm^{2}\,s^{-1}}$, either with our default streaming speed,  or  with an arbitrarily much larger speed ($=3\,(c_{s}^{2}+v_{A}^{2})^{1/2}$, chosen {\em ad hoc} to be faster by a factor of  a few  than the largest MHD wavespeed). 

We see that increasing $\kappa$ leads to more-efficient CR escape from the dense galactic gas, lowering $L_{\gamma}$ (as expected). Increased streaming speed produces a similar but much weaker effect, for the values we consider. The observations appear to strongly rule out $\kappa  \ll 10^{29}\,{\rm  cm^{2}\,s^{-1}}$: reproducing them requires $\kappa \sim 3-30\times10^{29}\,{\rm cm^{2}\,s^{-1}}$. In this regime, galaxies with dense nuclei (e.g.\ starbursts and bulge centers) are approximate proton calorimeters, but less-extreme systems (essentially all dwarfs and much of the volume of $\sim L_{\ast}$ galaxies) see the large majority ($>90\%$) of the CR energy escape the ISM without producing $\gamma$-rays.

\begin{figure}
\begin{centering}
\includegraphics[width={0.95\columnwidth}]{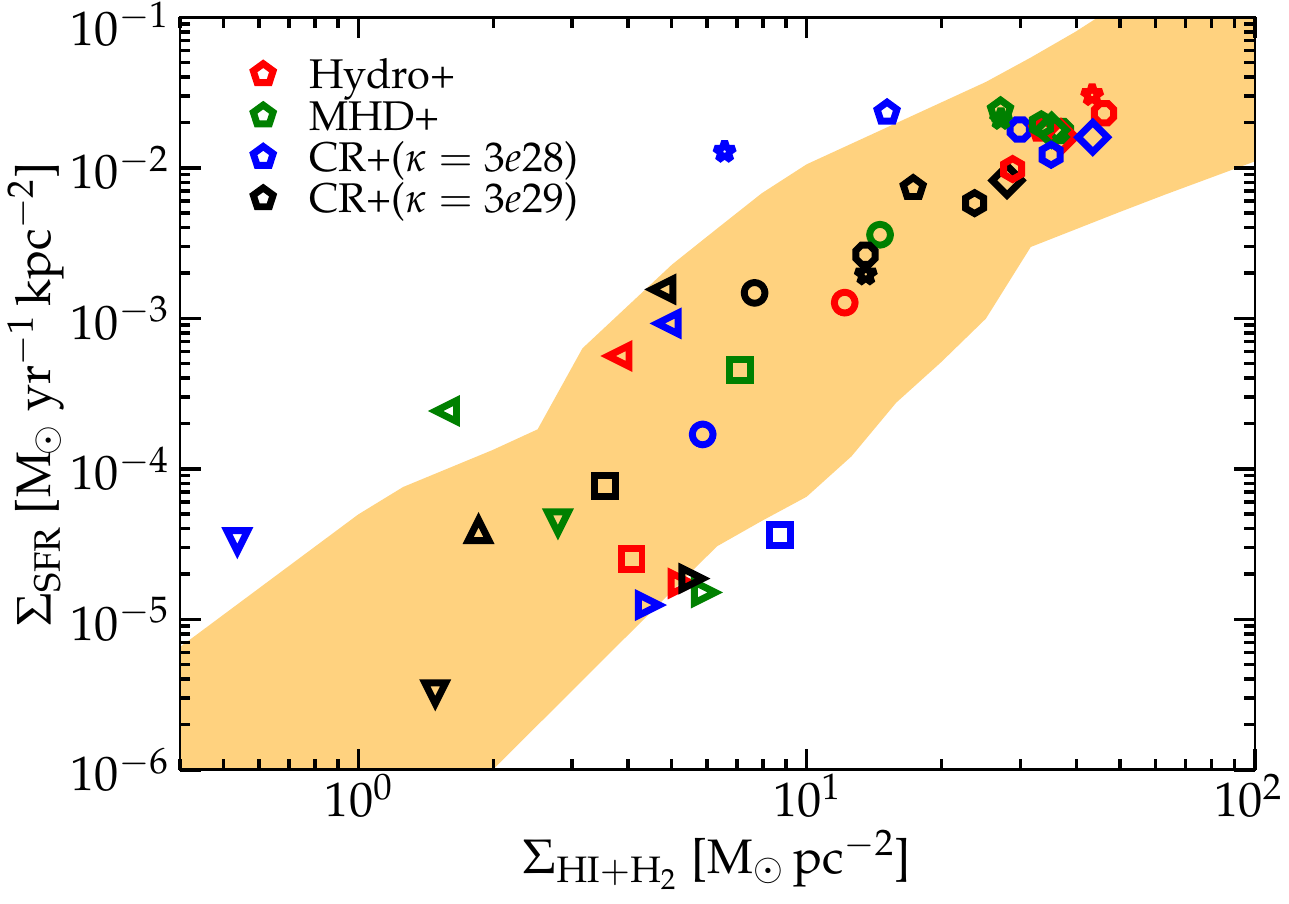}
    \end{centering}
    \vspace{-0.25cm}
        \caption{Location of the suite of runs from Figs.~\ref{fig:cr.demo.m10}-\ref{fig:cr.demo.m12} on the Schmidt-Kennicutt relation at $z=0$. For each zoom-in region (different symbols), we plot the $z=0$ value only of the neutral gas surface density ($\Sigma_{\rm HI+H_{2}}$) and star formation (averaged over the last $<10\,$Myr) surface density ($\Sigma_{\rm SFR}$), averaged over $\sim 10$ different random line-of-sight projections, within a circular aperture containing $90\%$ of the $V$-band luminosity. 
         We compare (colors, as labeled) different physics sets as Figs.~\ref{fig:cr.demo.m10}-\ref{fig:cr.demo.m12}. For reference we show (shaded contour) the $5-95\%$ inclusion interval at 
         each $\Sigma_{\rm HI+H_{2}}$ of observed galaxies compiled from \citet{kennicutt:m51.resolved.sfr,bigiel:2008.mol.kennicutt.on.sub.kpc.scales,genzel:2010.ks.law}. We note that the scatter in time for any individual simulation is large, comparable to the scatter observed (see \citet{orr:ks.law} for a detailed study), so the deviations between individual runs are all consistent within this scatter. The robust conclusion is that there is no systematic trend towards lower/higher ``star formation efficiency'' (normalization of the relation here) with different physics studied: to the extent that some physics produce higher/lower SFRs, galaxies move {\em along} the relation rather than off of it (e.g.\ {\bf m12i}, in pentagons, which shows lower SFR in the ``CR+($\kappa=3e29$)'' run).
    \label{fig:kslaw}}
\end{figure}

\begin{figure}
\begin{centering}
\includegraphics[width={0.48\textwidth}]{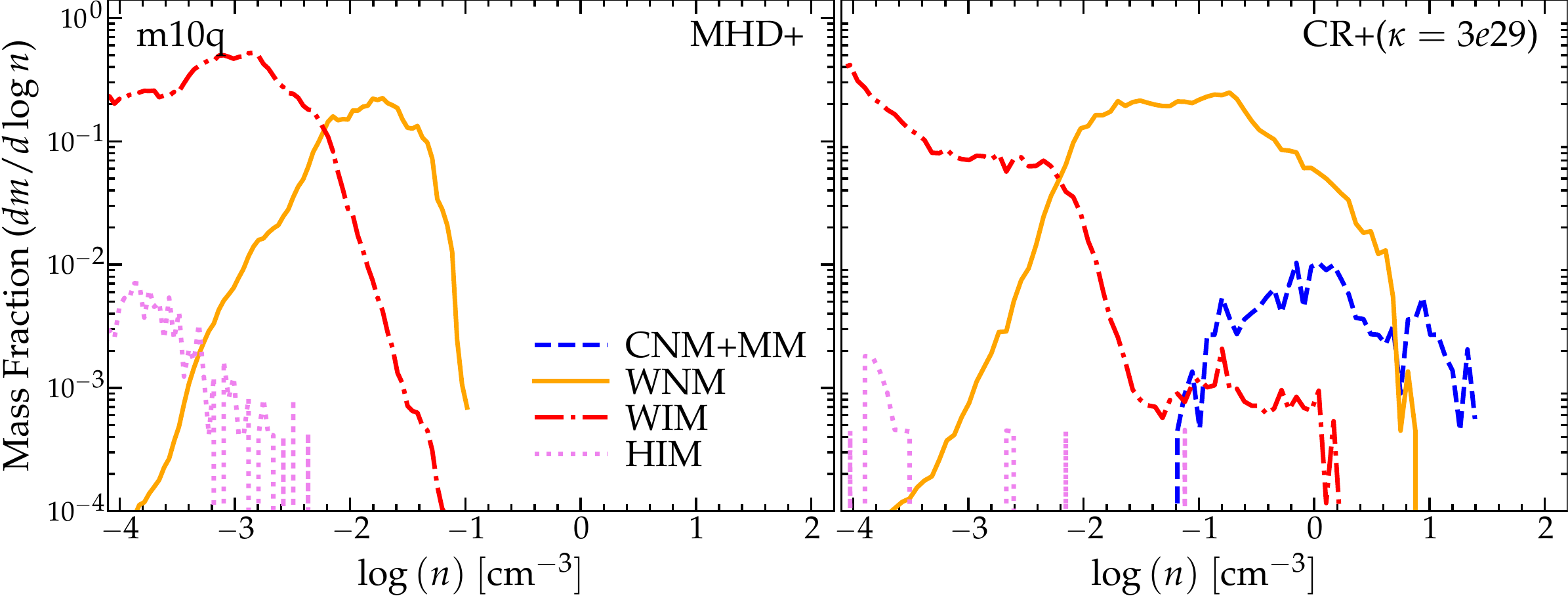}
\includegraphics[width={0.48\textwidth}]{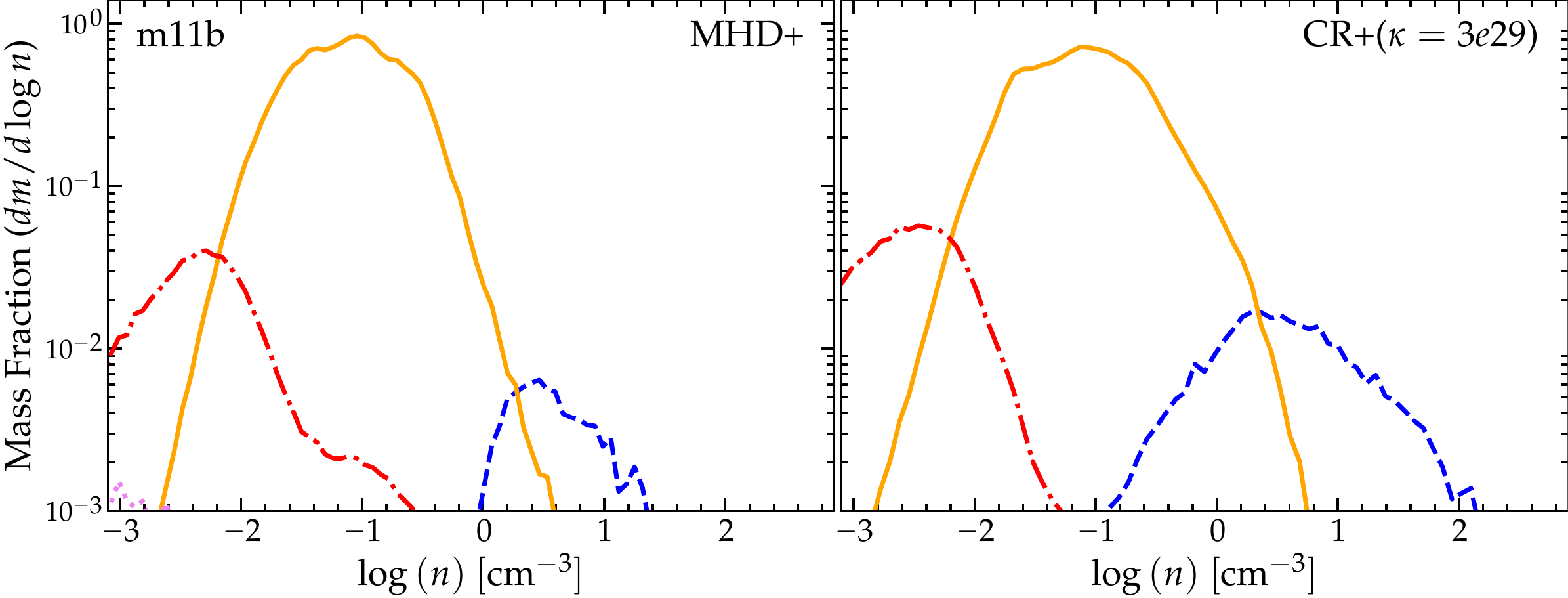}
\includegraphics[width={0.48\textwidth}]{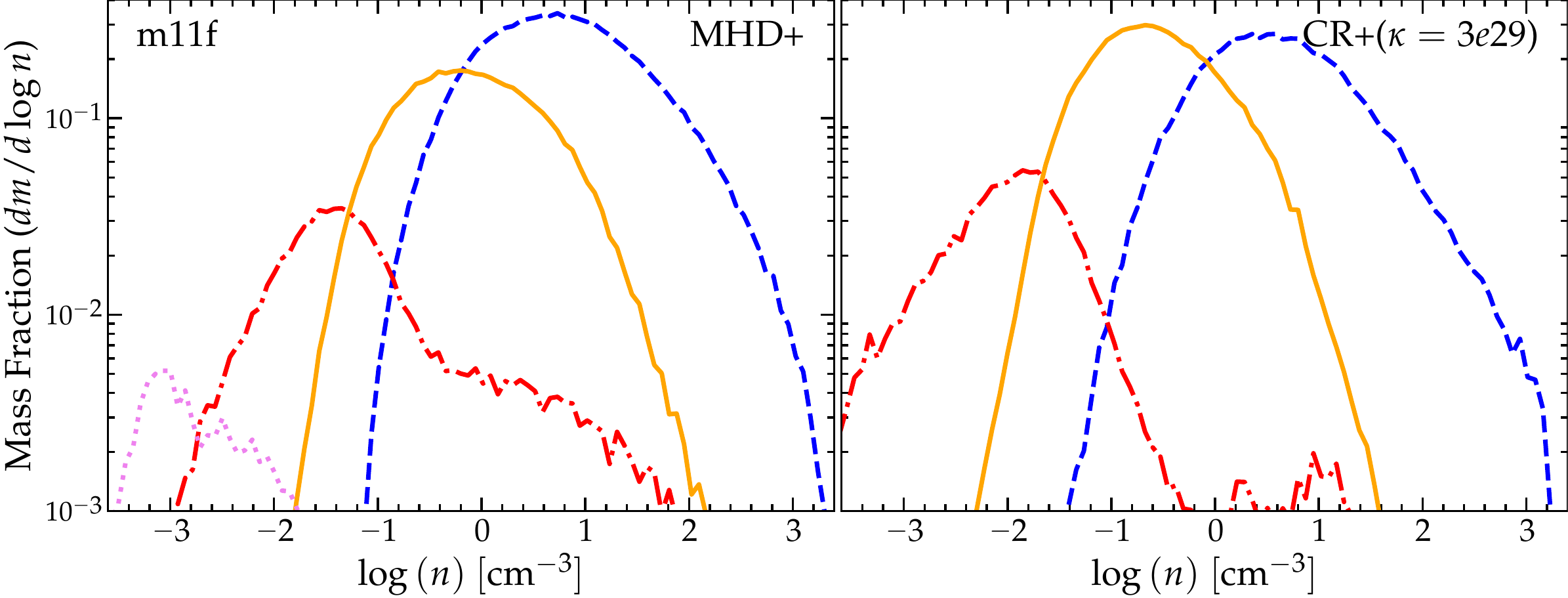}
\includegraphics[width={0.48\textwidth}]{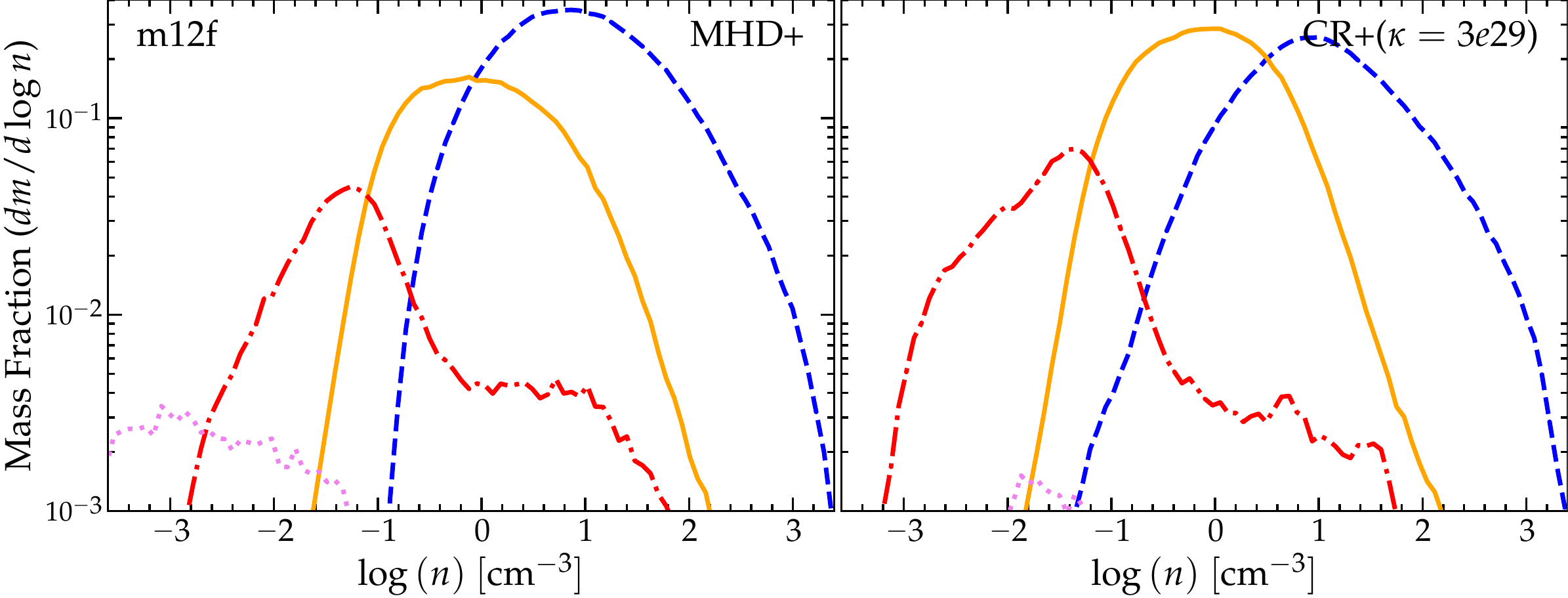}
\includegraphics[width={0.48\textwidth}]{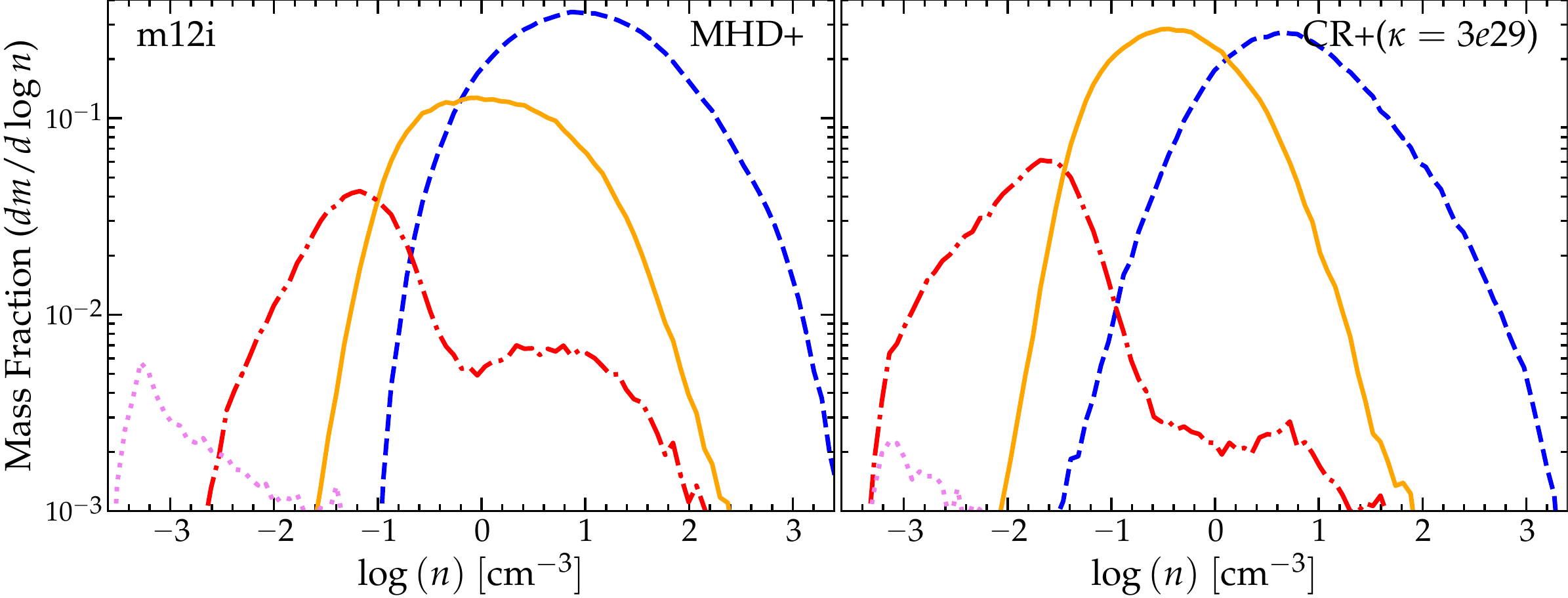}
    \end{centering}
    \vspace{-0.5cm}
        \caption{Phase distribution of gas in the ISM (within $<10\,$kpc of the galaxy center). We plot the {\em mass-weighted} distribution of gas as a function of density (per $\log_{10}(n)$), normalized so the integral over all curves equals unity, in representative galaxies ({\bf m10q}, {\bf m11b}, {\bf m11f}, {\bf m12f}, {\bf m12i}), comparing ``MHD+'' and ``CR+($\kappa=3e29$)'' runs (``Hydro+'' closely resembles ``MHD+''; see \citealt{su:2016.weak.mhd.cond.visc.turbdiff.fx}). Phases are defined as: cold neutral + molecular medium (CNM+MM; neutral gas with $T<1000\,$K), warm neutral medium (WNM; neutral and $T>1000\,$K), warm ionized medium (WIM; ionized and $T<10^{5}\,$K), hot ionized medium (HIM; ionized and $T>10^{5}$\,K). The qualitative features are similar in CR+ and Hydro+/MHD+ runs; however the CR+ runs (including ``CR+($\kappa=3e28$)'') support relatively more WNM (mostly shifting gas from CNM to WNM), making the two more comparable in MW-mass galaxies. The HIM is also somewhat suppressed in the CR+ runs: it is still created, but escapes the galaxy more easily. 
    \label{fig:phases}}
\end{figure}

\subsection{Magnetic and CR Energies, Pressures, and Heating Rates}
\label{sec:results:pressures}

Fig.~\ref{fig:profile.pressure} shows the gas thermal ($=n\,k_{B}\,T$), magnetic ($=|{\bf B}|^{2}/8\pi$), CR ($=(\gamma_{\rm cr}-1)\,e_{\rm cr}$), and kinetic ($=\rho\,|{\bf v}|^{2}/2$) pressures (equivalent to energy densities), as a function of radial distance $r$ from the galaxy center at $z=0$ (in our standard physics suite). We focus on a representative sub-set of our ensemble, but the qualitative results shown here are generic to the ensemble of simulations. Each property is computed for each resolution element, exactly as they are determined in-code (self-consistently), and we show the $5-95\%$ inclusion contour of these properties at each galacto-centric radius. Because of our Lagrangian numerical method, this is effectively a {\em gas-mass-weighted} distribution of these values. To contrast, we therefore also plot the {\em volume-averaged} values in spherical shells. Note that where most of the gas is in a thin disk (e.g.\ in {\bf m12i} at $r \lesssim 10\,$kpc), this means the spherically-volume-weighted average value of certain quantities (if they are concentrated in the thin disk) will be lower than its midplane value (closer to the mass-weighted average value) by a factor $\sim H/R$ (the ratio of the disk scale-height $H$ to radius $R$). We compare these to the ``gravitational'' or approximate {\em local} virial energy density, $\equiv \rho\,V_{c}^{2}/2 \equiv \rho\,G\,M_{\rm enc}(<r)/2\,r$. Using a different definition based on the total potential gives quite similar results for our purposes here. 

Note that the kinetic energy density here is defined as $\rho\,|{\bf v}|^{2}$, with ${\bf v}$ defined relative to the mean velocity of the whole galaxy, i.e.\ this {\em includes} rotation. This is done in part for simplicity because the dwarfs do not have strong rotation and separating rotation vs.\ dispersion (even in simulations) is in general quite challenging \citep[see][]{elbadry:fire.morph.momentum}. It also allows us to immediately see (whether primarily in rotation or dispersion) where the gas is primarily ``held up'' by kinetic energy. 

We see (discussed below) that magnetic pressure is almost always sub-dominant ($\beta \equiv P_{\rm thermal}/P_{\rm magnetic} \gg 1$), while CR pressure  can dominate and maintain virial equilibrium in massive halos, especially in our high-$\kappa$ runs, at radii outside the galaxy  from $\sim 30-200\,$kpc. 

Fig.~\ref{fig:profile.heating} considers a similar comparison of the radial profiles of  the gas cooling rate, computed in-code at $z=0$ in each  radial annulus using our full cooling function, to the heating rate from the CR ``streaming loss'' (gyro-resonant \Alf-wave heating) and ``collisional'' (thermalized hadronic+Compton) terms. These are always sub-dominant to the gas cooling rates (dominated by gas at $\sim 10^{4}-10^{6}\,$K, near the peak  of the cooling curve).

\subsection{Internal Galaxy Properties: Metallicities, Star Formation Rates, ISM Phase Structure, Morphologies, Angular Momentum}
\label{sec:results:internal}

\begin{figure}
\begin{centering}
\includegraphics[width={0.42\columnwidth}]{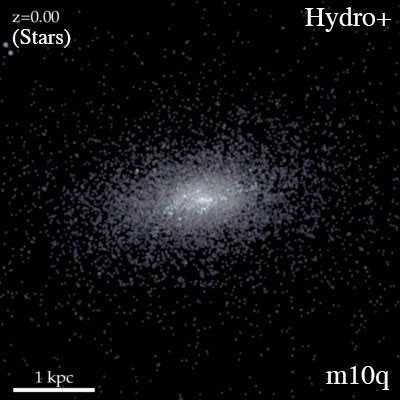}
\includegraphics[width={0.42\columnwidth}]{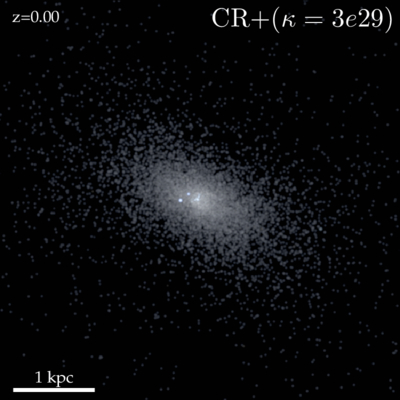} \\
\includegraphics[width={0.42\columnwidth}]{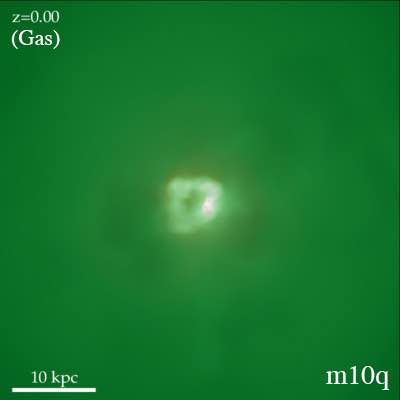}
\includegraphics[width={0.42\columnwidth}]{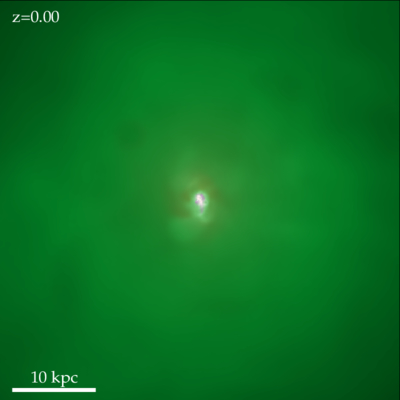} \\
\includegraphics[width={0.42\columnwidth}]{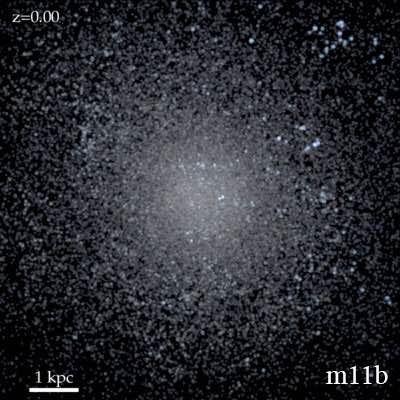}
\includegraphics[width={0.42\columnwidth}]{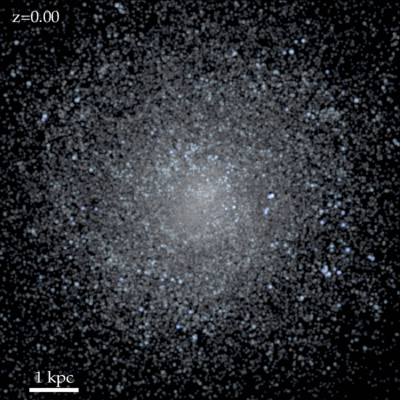} \\
\includegraphics[width={0.42\columnwidth}]{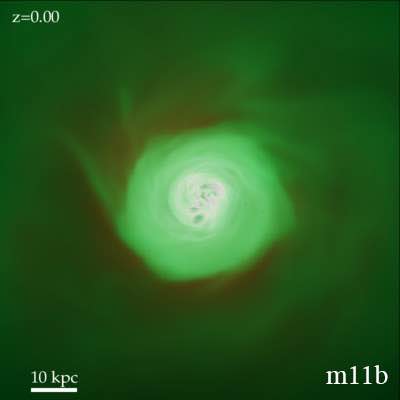}
\includegraphics[width={0.42\columnwidth}]{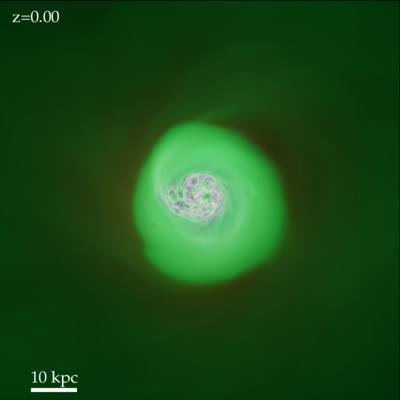} \\
\includegraphics[width={0.42\columnwidth}]{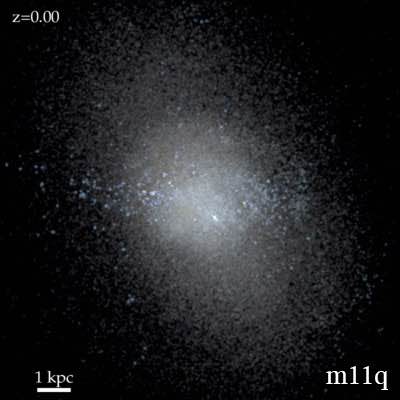}
\includegraphics[width={0.42\columnwidth}]{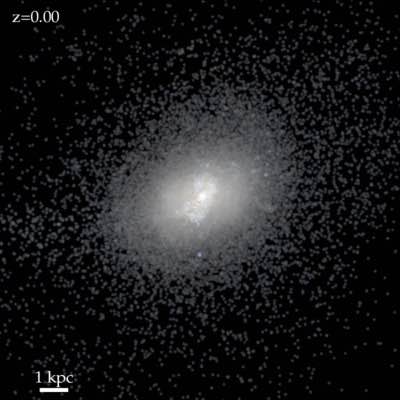} \\
\includegraphics[width={0.42\columnwidth}]{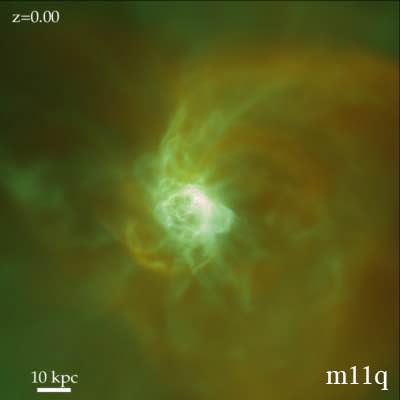}
\includegraphics[width={0.42\columnwidth}]{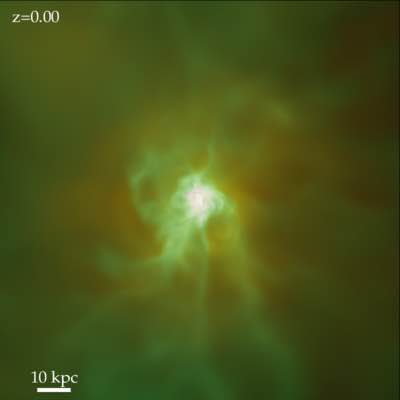} \\
    \end{centering}
    \vspace{-0.25cm}
        \caption{Morphologies of dwarfs ({\bf m10q}, {\bf m11b}, {\bf m11q}), at $z=0$. Visual is a $ugr$ composite, ray-tracing starlight (attenuated by dust in the simulation), with a log-stretch ($\sim 4\,$dex range). Gas is a 3-band volume render showing ``hot'' ($T\gg 10^{5}\,$K; {\em red}), ``warm/cool'' ($T\sim 10^{4}-10^{5}$\,K; {\em green}), and ``cold (neutral)'' ($T\ll 10^{4}\,$K; {\em magenta}) phases. We compare ``Hydro+'' ({\em left}) and ``CR+($\kappa=3e29$)'' ({\em right}); there is no large systematic difference.
    \label{fig:morph.dwarfs}}
\end{figure}

\begin{figure*}
\begin{centering}
\includegraphics[width={0.24\textwidth}]{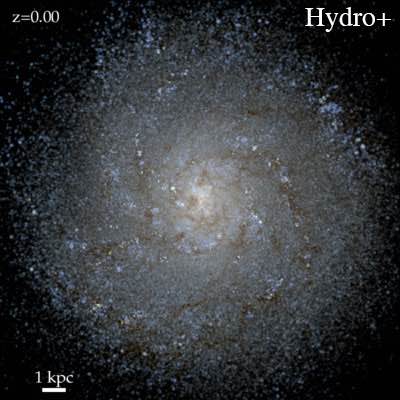}
\includegraphics[width={0.24\textwidth}]{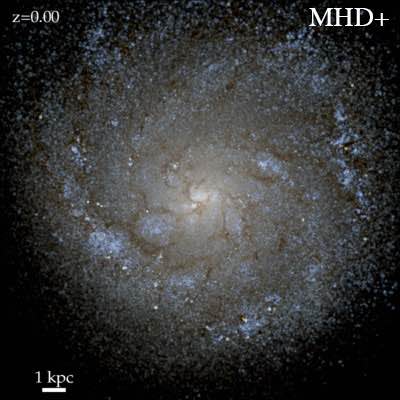}
\includegraphics[width={0.24\textwidth}]{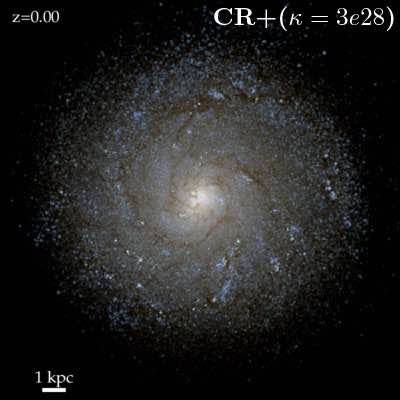}
\includegraphics[width={0.24\textwidth}]{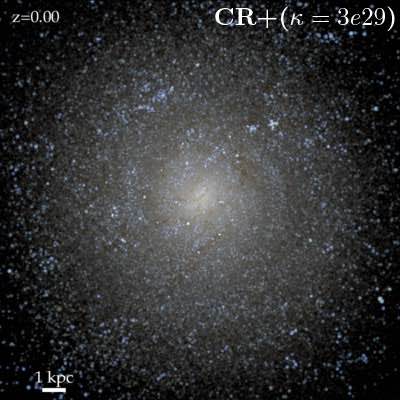} \\
\includegraphics[width={0.24\textwidth}]{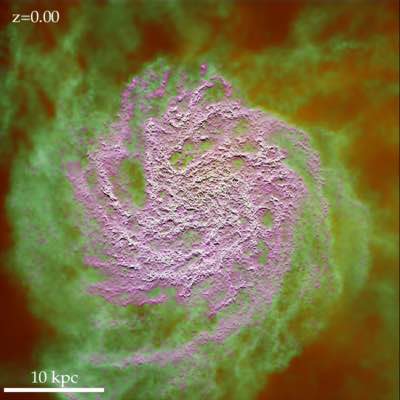}
\includegraphics[width={0.24\textwidth}]{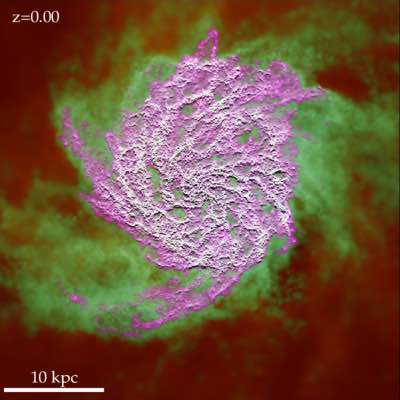}
\includegraphics[width={0.24\textwidth}]{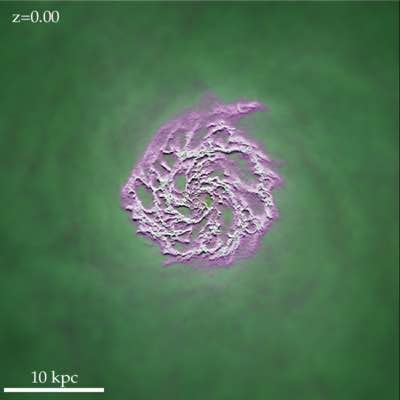}
\includegraphics[width={0.24\textwidth}]{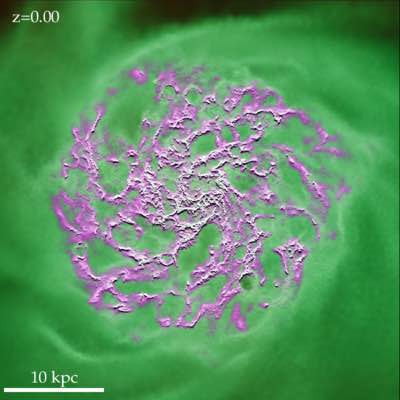} \\
    \end{centering}
    \vspace{-0.25cm}
        \caption{Visual \&\ gas morphologies of the intermediate-mass galaxy {\bf m11f} ($M_{\ast} \sim 1.5-3\times10^{10}\,M_{\odot}$, $M_{\rm halo}\sim 5\times10^{11}\,M_{\sun}$), as Fig.~\ref{fig:morph.dwarfs}. 
We compare ``Hydro+'', ``MHD+'', ``CR+($\kappa=3e28$)'', ``CR+($\kappa=3e29$)'' ({\em left-to-right}). 
        ``Hydro+'' and ``MHD+'' exhibit no differences. In stellar/visual and gas {\em disk} morphology, all runs are broadly similar. The ``CR+($\kappa=3e28$)'' run is somewhat more compact as shown in Fig.~\ref{fig:cr.demo.m11}, as a consequence of slightly more efficient star formation (slightly higher stellar mass). The ``CR+($\kappa=3e29$)'' run is slightly later type (less dusty with less well-defined arms), consistent with its factor $\sim 2$ lower stellar mass (essentially identical to the morphology of the ``Hydro+'' run at an earlier time, when it was similar mass). The gas disks evolve accordingly, although the volume-filling factor of warm gas is slightly higher in the CGM (with slightly less-sharp cold/neutral structures). The CGM differs dramatically in the CR runs, especially with low-$\kappa$ where it is warm/cool gas-dominated (this extends  beyond the region shown, but we defer CGM studies to future work). 
        Similar-mass runs (e.g.\ {\bf m11g}, {\bf m11h}, {\bf m11d}) show very similar systematic effects.
    \label{fig:morph.m11f}}
\end{figure*}

\begin{figure*}
\begin{centering}
\includegraphics[width={0.24\textwidth}]{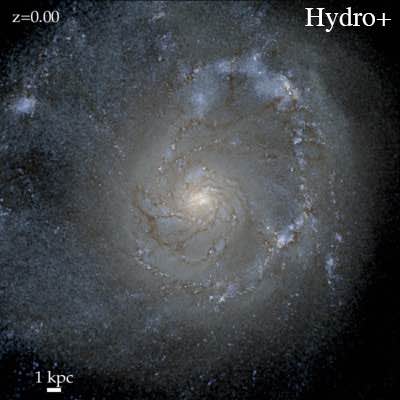}
\includegraphics[width={0.24\textwidth}]{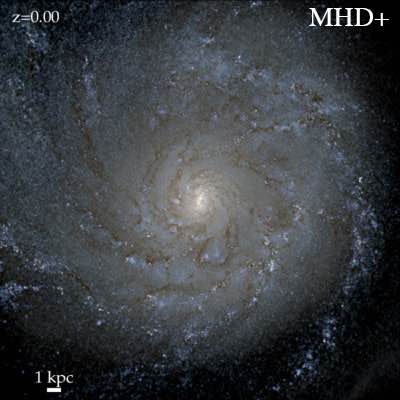}
\includegraphics[width={0.24\textwidth}]{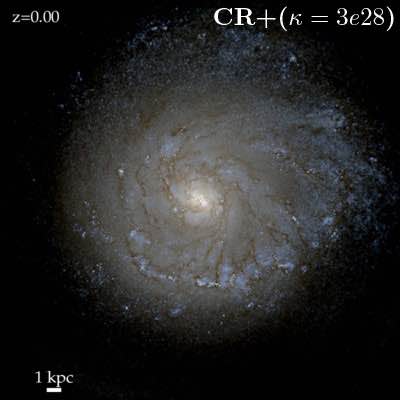}
\includegraphics[width={0.24\textwidth}]{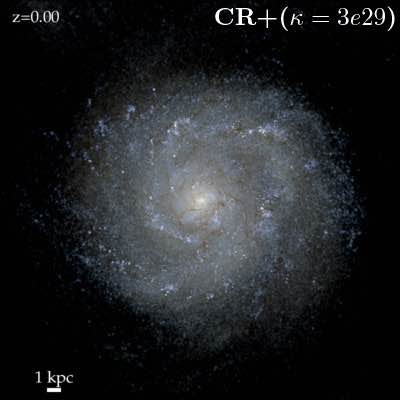} \\
\includegraphics[width={0.24\textwidth}]{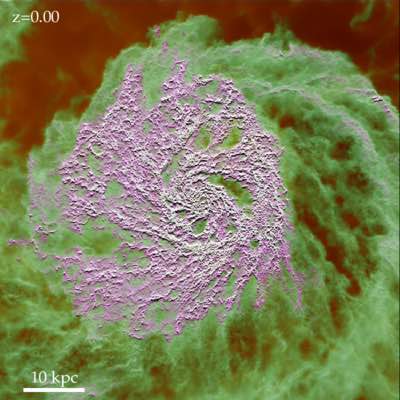}
\includegraphics[width={0.24\textwidth}]{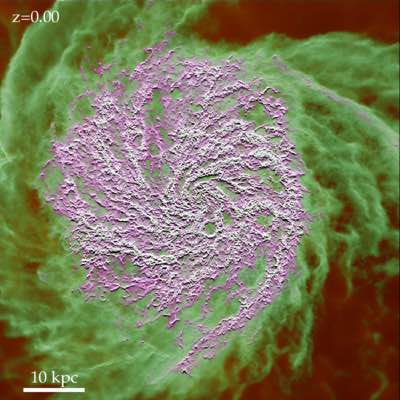}
\includegraphics[width={0.24\textwidth}]{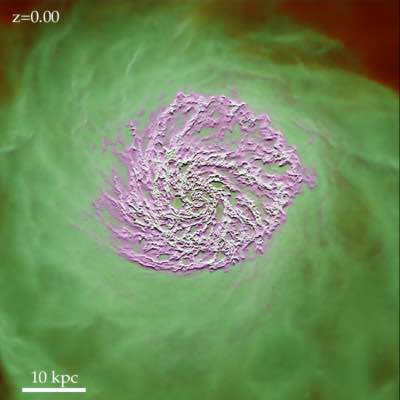}
\includegraphics[width={0.24\textwidth}]{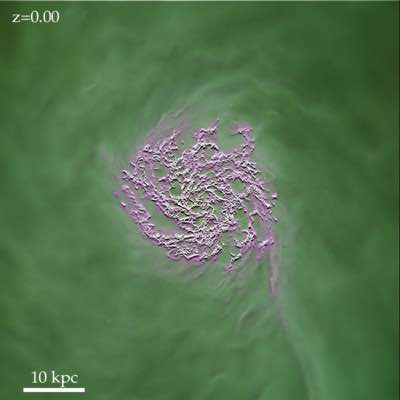} \\
    \end{centering}
    \vspace{-0.25cm}
        \caption{Visual \&\ gas morphologies of the MW-mass galaxy {\bf m12f} ($M_{\ast}\sim 3-8\ \times 10^{10}M_{\sun}$, $M_{\rm halo}\sim 10^{12}\,M_{\sun}$), as Fig.~\ref{fig:morph.dwarfs}. Similar trends appear as Fig.~\ref{fig:morph.m11f}. ``Hydro+'' and ``MHD+'' runs do not differ, while ``CR+($\kappa=3e28$)'' and ``CR+($\kappa=3e29$)'' runs have earlier/later-type morphologies and redder/bluer colors, respectively, but these simply follow from their higher/lower stellar masses (see Fig.~\ref{fig:cr.demo.m12}) -- again, the high-$\kappa$ run  resembles the ``Hydro+'' or ``MHD+'' run from an earlier time when the masses were more similar. All CR runs exhibit substantially enhanced warm/cool gas in the inner CGM. Here this is more dramatic at high-$\kappa$, because the gas is so heavily depleted in the low-$\kappa$ run that what remains is very tenuous. 
    \label{fig:morph.m12f}}
\end{figure*}

\begin{figure*}
\begin{centering}
\includegraphics[width={0.24\textwidth}]{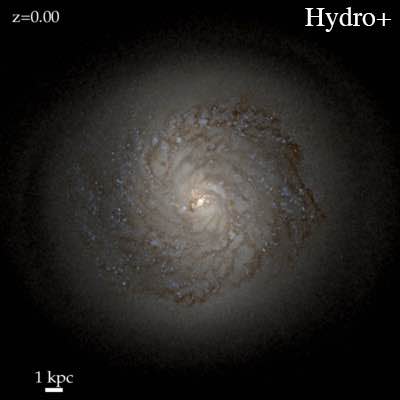}
\includegraphics[width={0.24\textwidth}]{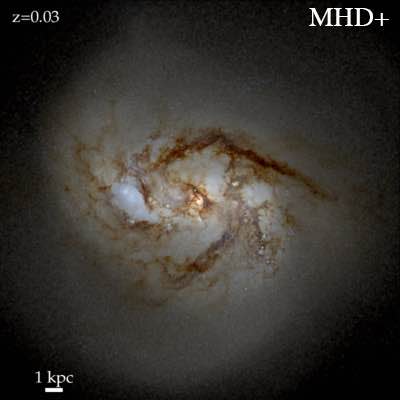}
\includegraphics[width={0.24\textwidth}]{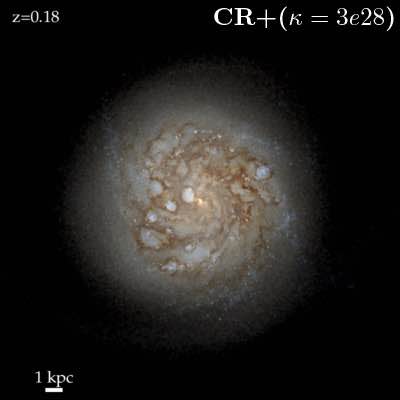}
\includegraphics[width={0.24\textwidth}]{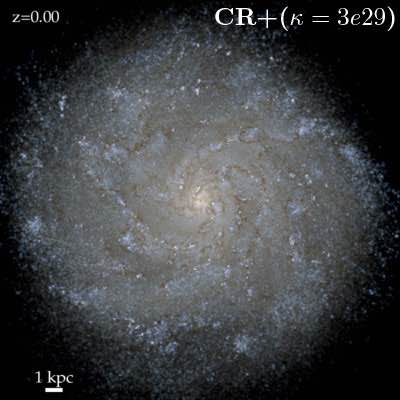} \\
\includegraphics[width={0.24\textwidth}]{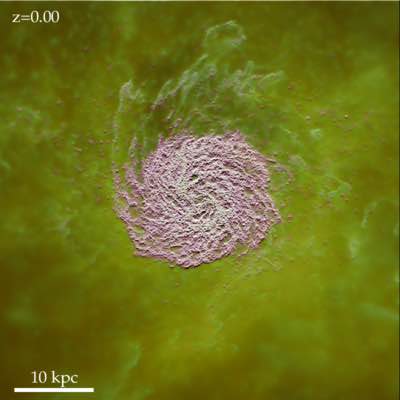}
\includegraphics[width={0.24\textwidth}]{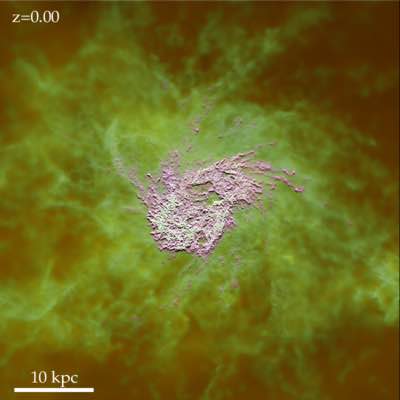}
\includegraphics[width={0.24\textwidth}]{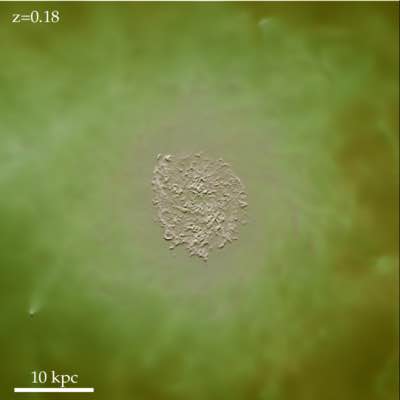}
\includegraphics[width={0.24\textwidth}]{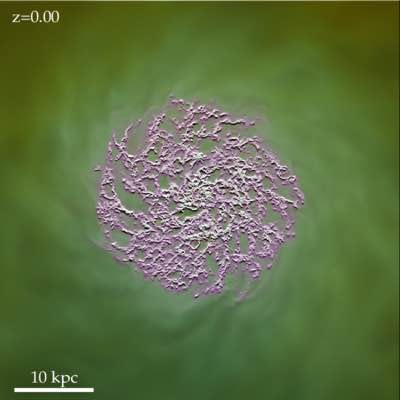} \\
    \end{centering}
    \vspace{-0.25cm}
        \caption{Visual \&\ gas morphologies of the MW-mass galaxy {\bf m12m}, as Fig.~\ref{fig:morph.dwarfs}. Similar trends appear as in {\bf m12f}. Note that as this halo is more massive, even the ``CR+'' runs are more dominated by hot gas in the CGM, and the Hydro+/MHD+/CR+($\kappa=3e28$) runs (with higher masses) are notably redder in color (with older, more metal-rich stellar populations formed at $z\sim 0.5-1$) and have prominent stellar bars (MHD+ is being perturbed by a minor merger passage whose timing is slightly different owing to the different mass). CR+($\kappa=3e29$) is less massive and later-type.
    \label{fig:morph.m12m}}
\end{figure*}

\begin{figure*}
\begin{centering}
\includegraphics[width={0.24\textwidth}]{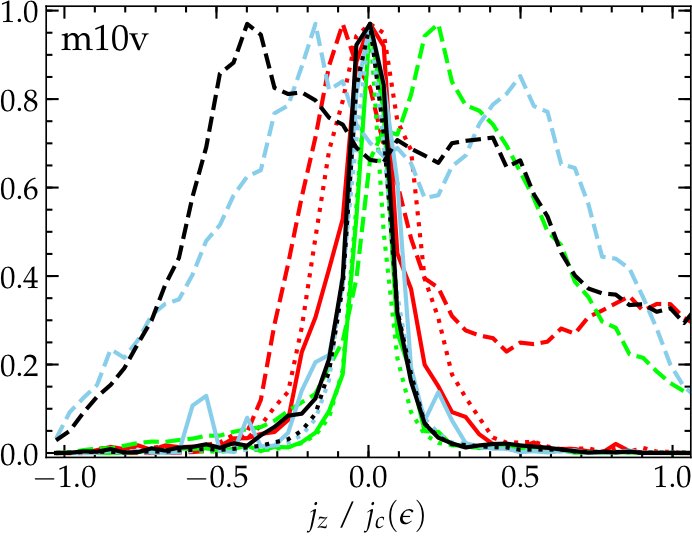}
\includegraphics[width={0.24\textwidth}]{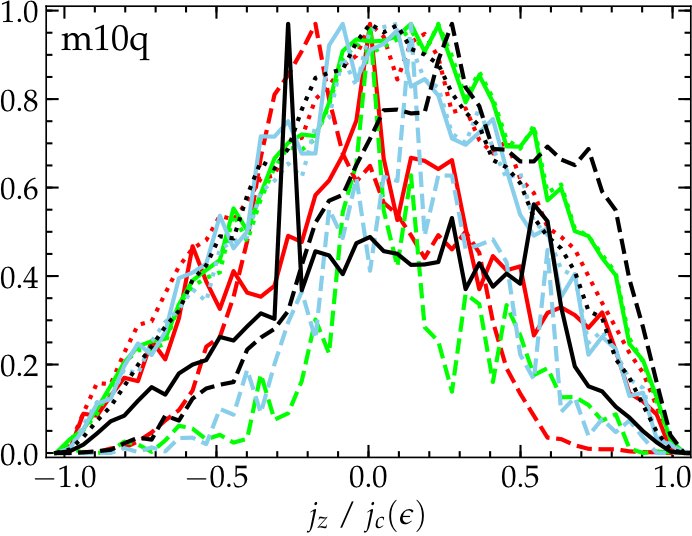}
\includegraphics[width={0.24\textwidth}]{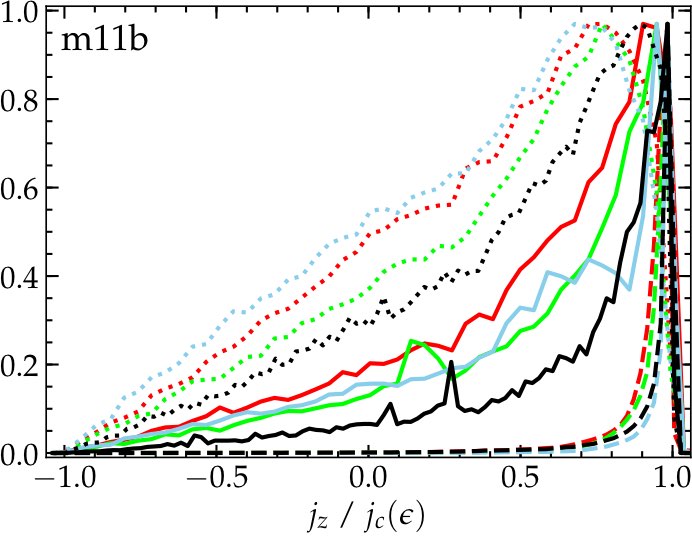}
\includegraphics[width={0.24\textwidth}]{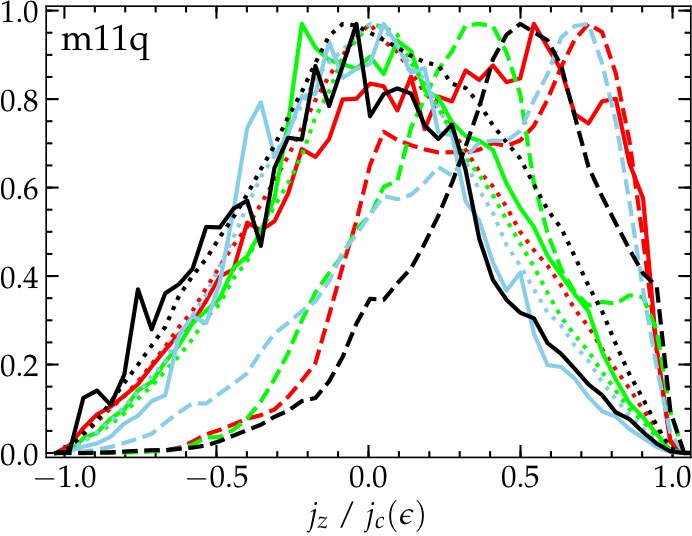} \\
\includegraphics[width={0.24\textwidth}]{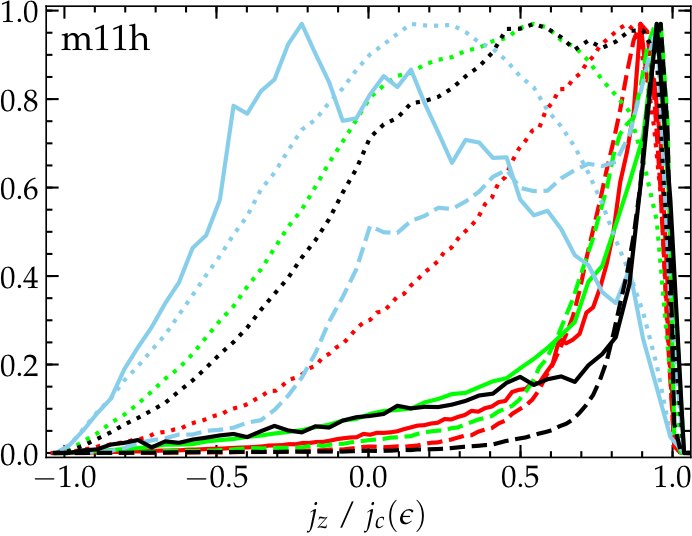}
\includegraphics[width={0.24\textwidth}]{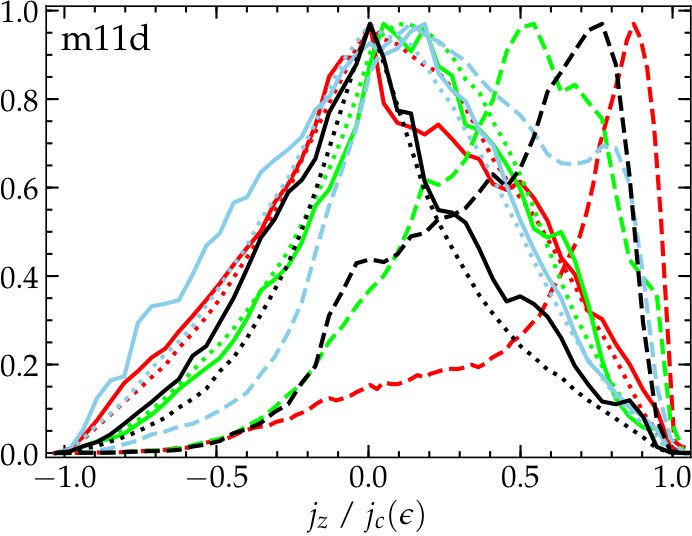}
\includegraphics[width={0.24\textwidth}]{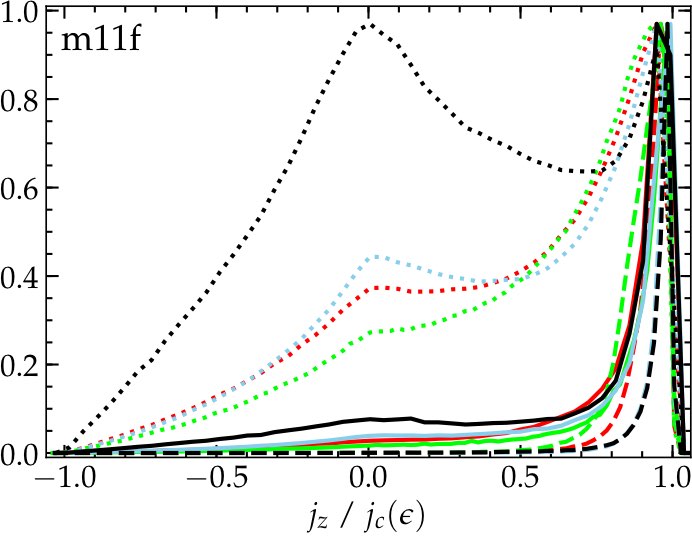}
\includegraphics[width={0.24\textwidth}]{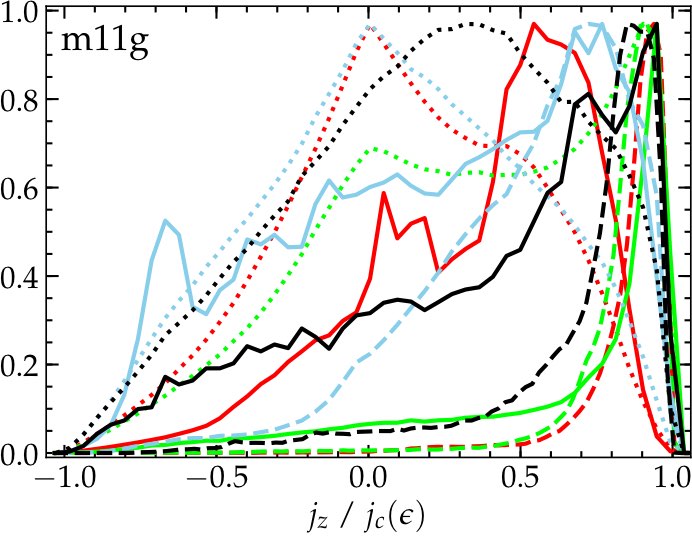} \\
\includegraphics[width={0.24\textwidth}]{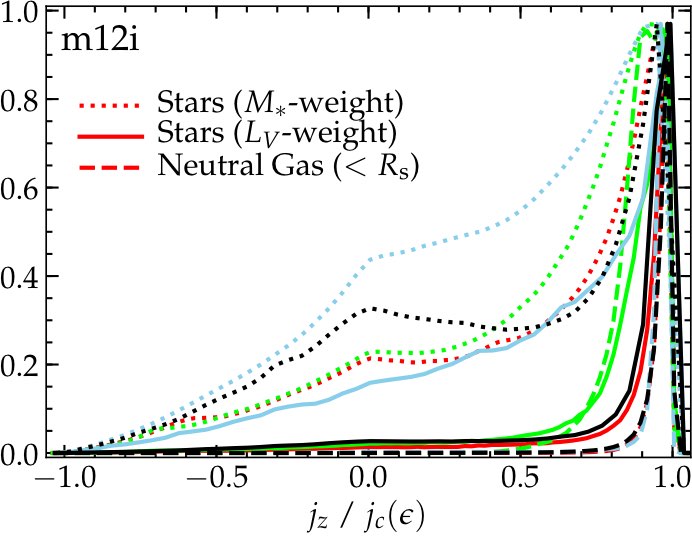}
\includegraphics[width={0.24\textwidth}]{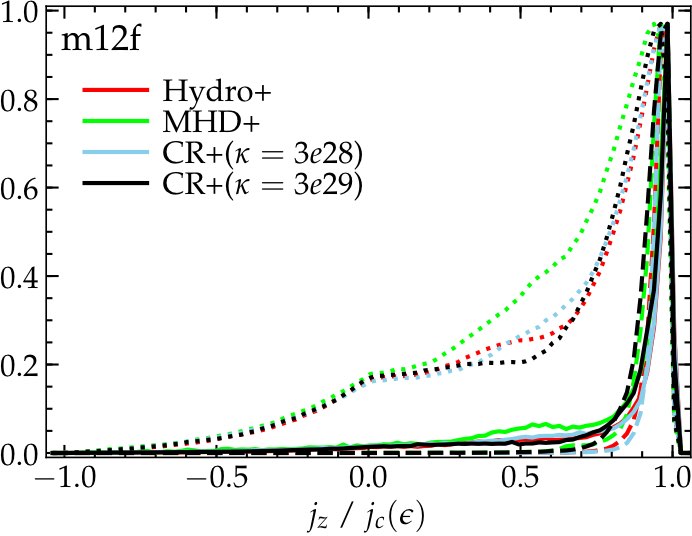}
\includegraphics[width={0.24\textwidth}]{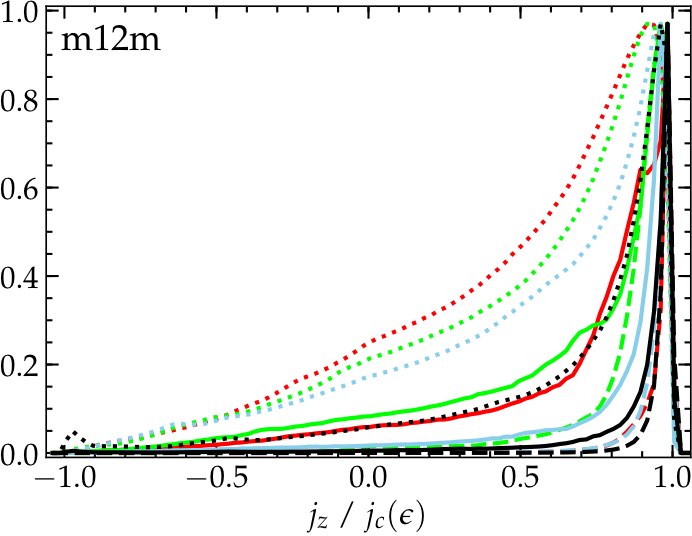} \\
    \end{centering}
    \vspace{-0.25cm}
        \caption{Kinematic morphology of our simulated galaxies. We plot the distribution of specific angular momentum (${\bf j}$, specifically the component $j_{z}$ along the total angular momentum axis) versus the specific angular momentum of a test particle on a perfectly-aligned circular orbit ($j_{c}[\epsilon]$) with the same specific energy ($\epsilon$) -- so $+1$ represents a perfectly-aligned circular orbit, $-1$ anti-aligned, and $0$ a perpendicular or radial orbit. We compare the distribution for stars weighted by stellar mass ($M_{\ast}$), $V$-band luminosity ($L_{V}$), and neutral gas inside the halo scale radius, for different galaxies. For each, we compare ``Hydro+'' ({\em red}), ``MHD+'' ({\em green}), ``CR+($\kappa=3e28$)'' ({\em blue}), ``CR+($\kappa=3e29$)'' ({\em black}). The results here largely mirror those from the visual morphologies in Figs.~\ref{fig:morph.dwarfs}-\ref{fig:morph.m12m}. Low-mass dwarfs are primarily dispersion-supported in all cases, higher-mass galaxies and younger stars (more prominent in $L_{V}$ vs.\ $M_{\ast}$ weighting) are systematically more disk-dominated. There are minor run-to-run variations (largely stochastic), and when the disk forms at late times, some of the ``CR+($\kappa=3e29$)'' runs which specifically suppress that late-time star formation (e.g.\ {\bf m11f}) have a smaller fraction of their stellar mass in the disk, as expected (although note the gas is still disky, and the $L_{V}$-weighted distribution, since it is dominated by the younger stars, is also strongly disk-dominated). In no case do any physics studied here change spheroidal galaxies to disky or vice versa. 
    \label{fig:j.distrib}}
\end{figure*}

We next survey a more detailed set of internal galaxy properties. 

Fig.~\ref{fig:mgal.mhalo} shows each of the ``primary'' galaxies in our suite (Table~\ref{tbl:sims}) on the stellar mass-halo mass  and stellar mass-stellar metallicity relations at $z=0$. These re-affirm the trends suggested in Figs.~\ref{fig:cr.demo.m10}-\ref{fig:cr.demo.m12}: our ``MHD+'' and ``CR+($\kappa=3e28$)'' runs do not deviate significantly from ``Hydro+,'' while the ``CR+($\kappa=3e29$)'' runs show systematically lower stellar mass in massive halos ($M_{\rm halo} \gtrsim 10^{10.5-11}\,M_{\odot}$) by a factor $\sim 2-3$, but all runs move along a single (relatively tight) mass-metallicity relation. In addition to the ``primary'' galaxies, we randomly select $\sim50$ additional non-satellite galaxies within the box\footnote{These galaxies are chosen within the high-resolution ``zoom-in'' region ($<1\%$ contamination by low-resolution dark matter particles inside their virial radii), but $>500$\,kpc away from the ``primary'' and outside the virial radius of any more massive halo. We select from the boxes {\bf m12i}, {\bf m12f}, {\bf m12m} (each with resolution $m_{i,\,1000}=7$) because these have relatively large zoom-in regions.} surrounding our {\bf m12} halos, to  compare with the ``primary'' dwarfs. Although the dwarfs in the larger {\bf m12} boxes are simulated at lower resolution ($m_{i,\,1000}=7$) compared to the smaller {\bf m09} or {\bf m10} boxes around a single dwarf (reaching $m_{i,\,1000}=0.25$), the results are consistent. 

In Fig.~\ref{fig:mgal.mhalo} we also include observational estimates for guidance,  but stress that we are not attempting to compare rigorously (e.g.\ we are not matching what  is actually measured, comparing scatter, etc.) -- such comparisons are presented (for non-CR runs) in \citet{ma:2015.fire.mass.metallicity,hopkins:fire2.methods,ma:fire2.reion.gal.lfs,garrisonkimmel:local.group.fire.tbtf.missing.satellites,wheeler:ultra.highres.dwarfs}. All our galaxies lie within the approximate $\sim 5-95\%$ scatter inferred around  the median $M_{\ast}-M_{\rm halo}$ relation  at all masses shown \citep{sgk:2016.mgal.mhalo.lowmass.scatter}, although interestingly at the highest masses ($M_{\rm halo}  \gtrsim  10^{11.5}\,M_{\odot}$) the low-$\kappa$ CR and Hydro/MHD runs appear to be  systematically high in  $M_{\ast}$ while  the high-$\kappa$ CR runs are systematically low (bracketing the observed median). In metallicity the simulations  agree extremely well with  observations  at $M_{\ast} \gtrsim 10^{8}\,M_{\odot}$ but appear to fall more steeply in very faint dwarfs compared to  the Local Group satellites in \citet{kirby:2013.mass.metallicity}: this is explored in detail  in \citet{wheeler:ultra.highres.dwarfs}.

Fig.~\ref{fig:kslaw} shows the locations of the example galaxies from Figs.~\ref{fig:cr.demo.m10}-\ref{fig:cr.demo.m12} on the observed Schmidt-Kennicutt relation, at $z=0$. We specifically follow \citet{kennicutt98} and measure the SFR (defined as the average over the last $<10\,$Myr) and total neutral (HI+H$_{2}$) gas mass inside a projected circular aperture enclosing $90\%$ of the starlight (approximately equivalent to the RC2 sizes used therein, for MW-mass systems, but this allows us to include low-surface-brightness dwarfs), to define $\Sigma_{\rm SFR}$ and $\Sigma_{\rm gas}$. We also show the $\sim 90\%$ inclusion contour of observed galaxies compiled from \citet{kennicutt:m51.resolved.sfr,bigiel:2008.mol.kennicutt.on.sub.kpc.scales,genzel:2010.ks.law}, although this is again intended only for reference, as we are not attempting a rigorous comparison with observations (such a comparison is presented in \citealt{orr:ks.law}, for those interested) but only to discern systematic effects between different runs. Briefly, we see no apparent offset in the relation between any of our physics  suites.

Fig.~\ref{fig:phases} compares the distribution of gas phases (densities and temperatures) in the interstellar medium. We plot the distribution of mass as a function of density, in the different traditional ISM phases (molecular, cold neutral, warm neutral, warm  ionized, hot ionized), where for simplicity  we define  ``ISM'' as gas within $r<10\,$kpc (the exact threshold makes little difference). For simplicity, we only compare ``MHD+'' and ``CR+($\kappa=3e29$)'', as ``Hydro+'' and ``MHD+'' are nearly-identical (see \citealt{su:2016.weak.mhd.cond.visc.turbdiff.fx} for more detailed comparison), and ``CR+($\kappa=3e28$)'' is as well. Even with  higher-$\kappa$, the differences are very subtle (tens of percent shift in  cold-to-warm-neutral  medium mass). In a temperature-density  diagram, this subtle shift is nearly undetectable. We see similar effects examining the phase structure of outflows, specifically, but defer a detailed study of this to future work.

Figs.~\ref{fig:morph.dwarfs}, \ref{fig:morph.m11f}, \ref{fig:morph.m12f}, \ref{fig:morph.m12m} shows the visual morphologies of the galaxies at $z=0$ in mock HST images of starlight, as well as the gas morphology in different phases and on different spatial scales. The stellar images are mock $u/g/r$ composite ray-tracing images determined using {\small STARBURST99} to compute the age-and metallicity-dependent spectrum of each star (the same assumptions used in-code) and adopting a constant dust-to-metals ratio to use the gas and metals distribution determined in-code to attenuate and extinct the light; the gas images are volume-renderings with iso-temperature contours centered on broad (log-normal) temperature bands with dispersion $\sim 0.5$\,dex around ``cold,'' ``cool'' or ``warm,'' and ``hot'' gas (see \citealt{hopkins:2013.fire} for details of the rendering). For dwarfs, we see no major difference with any physics change. For massive halos, the high-$\kappa$ CR runs are later-type, consistent with their lower mass, and we see more warm (as compared to hot) gas in the CGM (beyond the disk). The CGM properties will be studied in greater detail in future work. 

We have also examined the morphology of the magnetic fields, specifically, but  find they  are highly tangled on  all scales here, in both ``MHD+'' and ``CR+'' runs, consistent with our more detailed studies in \citet{su:2016.weak.mhd.cond.visc.turbdiff.fx} and \citet{su:fire.feedback.alters.magnetic.amplification.morphology}.

Fig.~\ref{fig:j.distrib} shows the ``quantitative kinematic morphology'' (angular momentum distribution) of stars (weighted by stellar mass or visual luminosity) and neutral gas. In all of these, we again compare our ensemble of galaxies from Table~\ref{tbl:sims} across our ``core physics variations'' from Table~\ref{tbl:physics}. Again, the variations with physics  are weak, and (where present) consistent  with the morphological changes described  above.

\begin{figure*}
    \plotsidesize{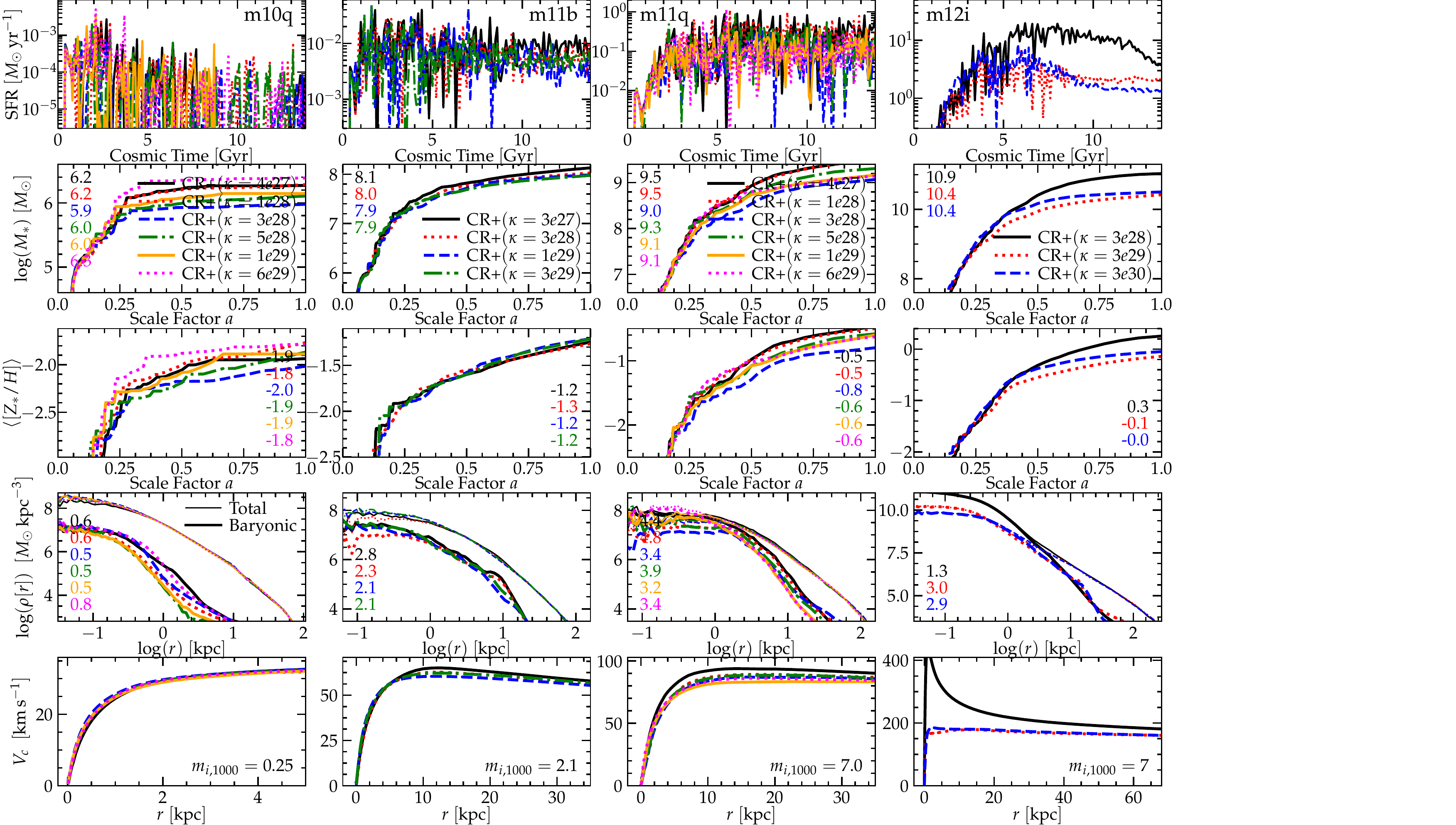}{0.99}
    \vspace{-0.25cm}
    \caption{As Figs.~\ref{fig:cr.demo.m10}-\ref{fig:cr.demo.m12}, varying the CR diffusion coefficient $\kappa$ more extensively, in a subset of our dwarf-through-MW mass runs. Very low $\kappa \lesssim 10^{28}\,{\rm cm^{2}\,s^{-1}}$ produce essentially no systematic effect at any mass scale. Intermediate $10^{28} \lesssim \kappa \lesssim 3\times 10^{29}$ may produce a {\em small} suppression of SF (factor of $\sim 0.1-0.3\,$dex, at most) in dwarfs. In massive, higher-density galaxies, these coefficients still produce no effect. Higher coefficients $\gg 10^{29}$, produce factor $\sim 2-3$ suppression of the SFR at $z\lesssim 1-2$ in massive halos (and corresponding suppression of their central densities).
    \label{fig:diffusioncoeff}}
\end{figure*}

\begin{figure}
    \plotonesize{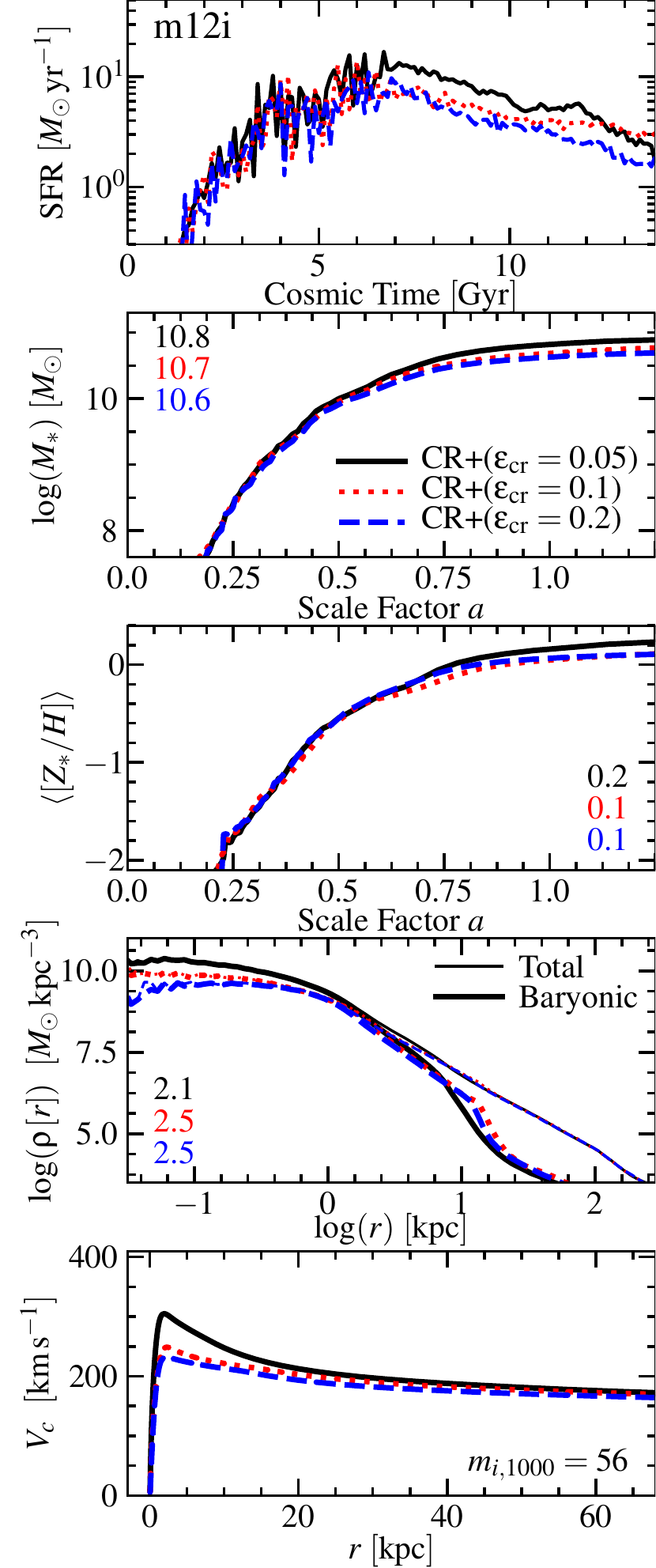}{0.85}
    \vspace{-0.25cm}
    \caption{As Figs.~\ref{fig:cr.demo.m10}-\ref{fig:cr.demo.m12}, comparing our low-resolution MW-mass {\bf m12i} run with ``full CR physics'' (CR+), and the {\em high} diffusion coefficient ($\kappa=3e29$), but varying the CR ``injection fraction'' $\epsilon_{\rm cr}$ (fraction of SNe ejecta kinetic energy assumed to go into CRs). The default value is $\epsilon_{\rm cr}=0.1$. Increasing/decreasing this produces systematically stronger/weaker suppression of SF, as expected, but the effect is relatively small compared to changes in the diffusion coefficient in Fig.~\ref{fig:diffusioncoeff}.
    \label{fig:epsilonCR}}
\end{figure}

\begin{figure*}
    \plotsidesize{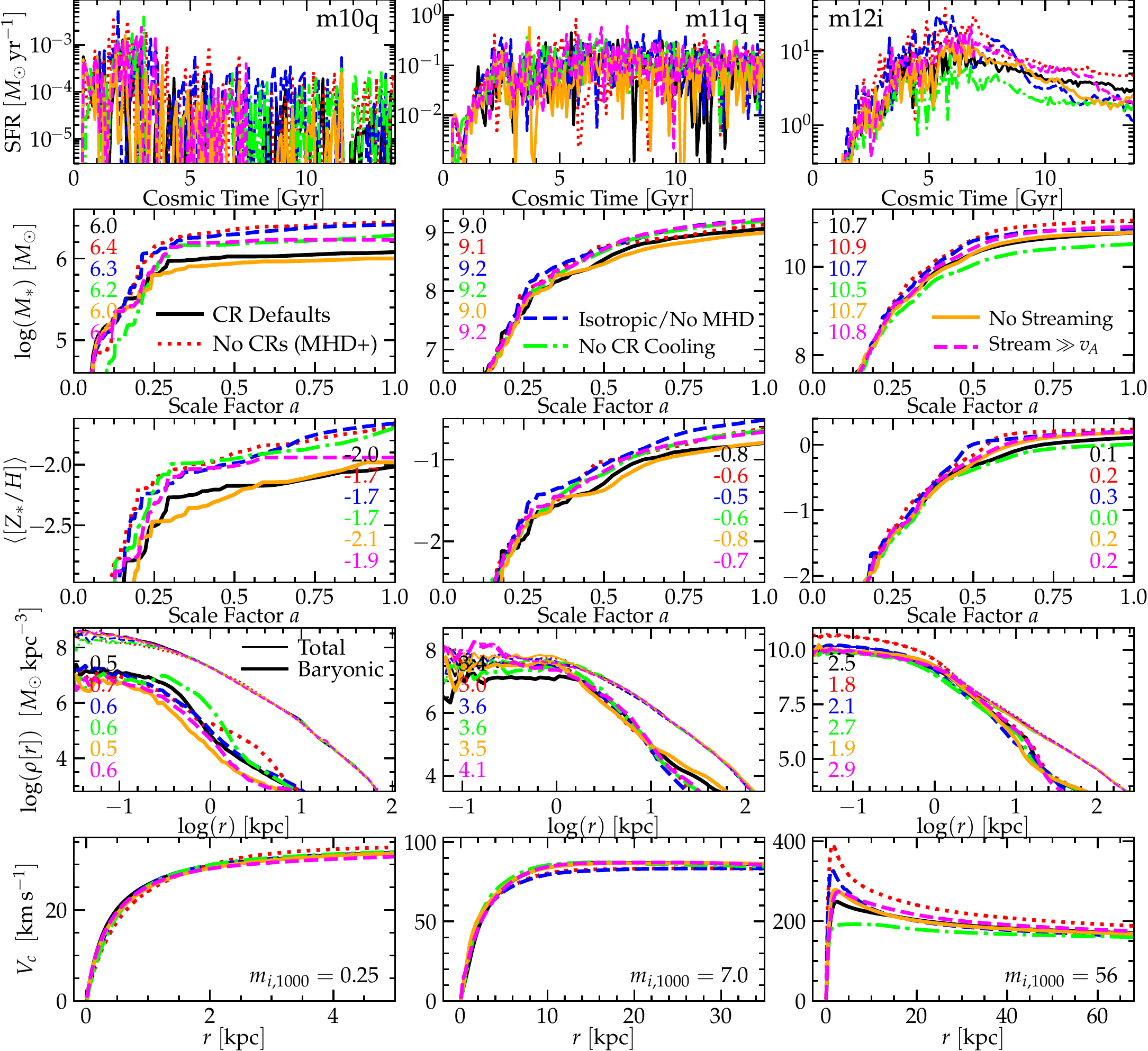}{0.99}
    \vspace{-0.25cm}
    \caption{As Fig.~\ref{fig:diffusioncoeff}, varying the CR physics in ``CR+'' runs with otherwise identical physics, at a few different mass scales. Here we choose $\kappa=3e28$ for the dwarfs ({\bf m10q} and {\bf m11q}) and $\kappa=3e29$ for the MW-mass halo, because these values give some of the strongest effects seen at each mass (making differences here more obvious). We compare:
    {\bf (1)} ``CR Default'': the default physics shown in all CR+ runs elsewhere. 
    {\bf (2)} ``No CRs (MHD+)'': the default MHD+ runs (without CR transport) for reference.
    {\bf (3)} ``Isotropic/No MHD'': CR runs without MHD, where (lacking a magnetic field) CR streaming and diffusion, as well as Spitzer-Braginskii conduction and viscosity, are assumed to be isotropic. 
    {\bf (4)} ``No CR Cooling'': Turning off hadronic \&\ Coulomb losses from CRs. 
    {\bf (5)} ``No Streaming'': Disabling CR streaming (setting the streaming velocity to zero).
    {\bf (6)} ``Stream $\gg v_{A}$'': Allowing super-\Alf{ic}/sonic streaming (streaming velocity $=3\,(c_{s}^{2}+v_{A}^{2})^{1/2}$, a multiple of the fastest MHD wavespeed). 
    In {\bf m10q} the stellar mass varies by a factor $\sim 2$ with these variations, but this is totally dominated by the amplitude of the high-$z$ burst around $z\sim 3$, so it is difficult to interpret. 
    These choices generally have small effects in {\bf m11q} (LMC-mass). 
    In {\bf m12i} (MW-mass), artificially removing CR losses/cooling leads to significantly stronger suppression of SF, as expected, while allowing highly super-\Alf{ic} streaming actually produces less suppression of SF (owing to enhanced CR streaming losses leaving a less-energetic CR halo), and allowing isotropic streaming (without MHD) produces the {\em least} effective suppression of SF, as the CRs {\em too efficiently} escape the galaxy and halo.
    \label{fig:cr.physics}}
\end{figure*}

\subsection{Variations in CR Transport Physics}
\label{sec:results:cr.transport}

We have widely-varied the CR transport coefficients $\kappa$ and $v_{\rm stream}$ in a subset of our simulations. The effect on $\gamma$-ray emission is shown  in Fig.~\ref{fig:Lgamma}. 

Fig.~\ref{fig:diffusioncoeff} shows the effects on {\em galaxy} properties (using the  same style as Fig.~\ref{fig:cr.demo.m10}) of varying $\kappa$ more widely and densely in our galaxies {\bf m10q}, {\bf m11b}, {\bf m11q}, {\bf m12i}. This confirms that CRs have weak effects for $\kappa \lesssim 10^{29}\,{\rm cm^{2}\,s^{-1}}$ and maximal effect in massive halos at $z\lesssim 1-2$ for $\kappa \sim 3-30\times10^{29}\,{\rm cm^{2}\,s^{-1}}$. 

Fig.~\ref{fig:epsilonCR} varies the fraction  $\epsilon_{\rm CR}$ of SNe energy which is injected into CRs, focusing on a low-resolution version of run {\bf m12i} with $\kappa=3\times10^{29}\,{\rm cm^{2}\,s^{-1}}$ (since MW-mass  halos with this diffusivity are where we see the most dramatic CR effects). As expected, increasing the CR energy input increases their effect, but the effect is quite weak (sub-linear).\footnote{Throughout, we have assumed CRs are injected at the sites of SNe: because the efficiency of CR acceleration scales with the shock velocity, models (and SN remnant observations) generally assign most of the acceleration to the ``fastest'' SNe shocks with $v\gtrsim 1000\,{\rm km\,s^{-1}}$ and ``swept up'' ISM mass $\sim M_{\rm ejecta} \sim 1-10\,M_{\odot}$, which is always un-resolved in our simulations. However one might imagine that if CRs are injected primarily outside the galaxy from e.g.\ structure-formation or superbubble-CGM shocks, then this would allow one to match the observations with lower $\kappa$ -- but only if we dramatically lowered the injection from SNe themselves (otherwise this would simply increase $L_{\gamma}$). So this is unlikely to change our conclusions here, but could be important for CRs from AGN (where jet termination shocks reach into the CGM).}

Fig.~\ref{fig:cr.physics} compares {\bf m10q}, {\bf m11q}, {\bf m12i} in a survey of basic CR physics: comparing our default ``CR+'' implementation (\S~\ref{sec:methods}) to runs (1) without MHD (where lacking ${\bf B}$ directions we assume CR diffusion/streaming, and conduction/viscosity, are isotropic, i.e.\ take the projection tensor $\hat{\bf B}\otimes \hat{\bf B} \rightarrow \mathbb{I}$); (2) without collisional (hadronic or Coulomb) CR losses; (3) without streaming (setting $v_{\rm stream}\rightarrow 0$); and (4) allowing super-sonic and super-\Alf{ic} streaming by setting the streaming speed to a few times the fastest MHD wavespeed $v_{\rm stream} \rightarrow 3\,(c_{s}^{2} + v_{A}^{2})^{1/2}$. These effects are discussed in detail below but their effects are generally small compared to including/excluding CRs at all.

Note that for Fig.~\ref{fig:Lgamma} we noted we have actually  run a large ensemble of simulations with lower $\kappa=3\times10^{28}\,{\rm cm^{2}\,s^{-1}}$ and the larger $v_{\rm stream} \rightarrow 3\,(c_{s}^{2} + v_{A}^{2})^{1/2}$: we have also compared the resulting galaxy properties but omit them for brevity as the difference owing to faster streaming is very small. We have also run a large suite with $\kappa=3\times10^{29}\,{\rm cm^{2}\,s^{-1}}$  varying $v_{\rm stream}  = (1-3)\,v_{A}$ more  modestly,  and find  (unsurprisingly) essentially no effect.

Additional purely-numerical tests are given in \citet{chan:2018.cosmicray.fire.gammaray} and Appendix~\ref{sec:additional.tests}. This includes modifications of the CR pressure tensor, form of the flux equation, and other detailed assumptions that stem from any two-moment expansion of the CR transport equations. We do not see large effects from these on our predictions, but of course can only survey a limited range of possibilities.

\subsection{High-Redshift Galaxies}
\label{sec:results:redshift}

Figs.~\ref{fig:HiZ}-\ref{fig:HiZ.profiles} repeats our earlier exercises comparing galaxy properties and magnetic+CR energy densities (from Figs.~\ref{fig:cr.demo.m10} and \ref{fig:profile.pressure}) for the suite of halos in Table~\ref{tbl:HiZ}, which reach large masses ($M_{\rm halo} > 10^{12}\,M_{\sun}$) at increasingly higher redshifts $z\sim 1-10$. These simulations are not run to $z=0$, hence we analyze them separately. 

We discuss the results in detail below, but briefly, we see no appreciable effects of CRs or MHD above  $z\gtrsim 1-2$.

\vspace{-0.5cm}
\section{Discussion}
\label{sec:discussion}

We now explore the implications of the results presented in \S~\ref{sec:results}. 

\vspace{-0.5cm}
\subsection{Magnetic Fields, Conduction, \&\ Viscosity}
\label{sec:discussion.mhd}

In \citet{su:2016.weak.mhd.cond.visc.turbdiff.fx}, we used similar FIRE-2 simulations to study the effects of magnetic fields and anisotropic Spitzer-Braginskii conduction and viscosity on galaxy properties. The study there included much more detailed measurements of properties like the gas phase distributions of the ISM and CGM, outflow properties, turbulence and energy balance in the ISM, magnetic field amplification, and more. There, we concluded that there was no appreciable systematic effect on any global galaxy properties from these physics. 

The ``MHD+'' simulations here improve on those studied in \citet{su:2016.weak.mhd.cond.visc.turbdiff.fx} in three significant ways. {\bf (1)}  Our mass resolution is an order-of-magnitude better, allowing us to much better-resolve the Field length and other small-scale effects. {\bf (2)} The simulation suite here is an order-of-magnitude larger, and all fully-cosmological, allowing us to assess and improve the statistics and avoid uncertainties owing to inevitable run-to-run stochastic variations in galaxy properties (discussed therein or in \citealt{keller:stochastic.gal.form.fx,genel:stochastic.gal.form.fx}). {\bf (3)} Our treatment of anisotropic conduction \&\ viscosity is more accurate. Specifically, a large body of recent work in the plasma physics literature (both theoretical and experimental) has shown that the parallel transport coefficients for heat and momentum (e.g.\ $\kappa_{\rm cond}$ and $\eta_{\rm visc}$) are strongly self-limited by micro-scale plasma instabilities (e.g.\ the Whistler, firehose, and mirror instabilities) in high-$\beta$ plasmas like the ISM and CGM \citep{kunz:firehose,komarov:conduction.vs.mirror.instab,komarov:heat.flux.suppression.ICM,riquelme:viscosity.limits,santos:2016.ion.temp.anisotropy.limits,roberg:2016.whistler.turb.conduction.suppression,roberg:2018.whistler.turb.conduction.suppression,tong:obs.solar.wind.electron.heat.flux.constraints,squire:2017.max.braginskii.scalings,squire:2017.kinetic.mhd.alfven,squire:2017.max.anisotropy.kinetic.mhd,komarov:whistler.instability.limiting.transport}. Our treatments in \S~\ref{sec:methods.mhd} include these effects, while our previous work did not. 

Despite this (or perhaps because of it), we confirm the conclusions of \citet{su:2016.weak.mhd.cond.visc.turbdiff.fx}: in every property we measure, the ``MHD+'' runs do not appear to systematically differ significantly from the ``Hydro+'' runs. This is perhaps not surprising: the only {\em physical} change of {\bf (1)}-{\bf (3)} above is {\bf (3)}, which has the effect of uniformly {\em decreasing} the magnitude of conduction and viscosity at high-$\beta$ (e.g.\ the ISM and CGM). Because of this, we will not discuss these variations further (for much more detailed discussion of {\em why} the results do not change, we refer to \citealt{su:2016.weak.mhd.cond.visc.turbdiff.fx}). 

Our goal here is to understand effects of magnetic fields on galaxies, not to consider a detailed comparison with observations. However, \citet{guszejnov:fire.gmc.props.vs.z} directly compared the values of ${\bf B}$ predicted in our MW-like {\bf m12i} simulation in atomic/molecular clouds to Zeeman observations from \citet{crutcher:cloud.b.fields} and found remarkably good agreement at all observed densities ($n \gtrsim 10\,{\rm cm^{-3}}$). In the ionized/warm ISM and thick disk/inner halo, observations are much more uncertain and model-dependent \citep[see][]{2017ARA&A..55..111H}, e.g.\ assuming gas density profiles and equipartition between CR and magnetic energy to infer $|{\bf B}|$ from rotation measures (RMs) or synchrotron emission ($I_{s}$). But preliminary estimates from our simulations show the combination of large ``clumping factors'' (e.g.\ $\langle |{\bf B}|^{4}\rangle / \langle |{\bf B}| \rangle^{4}$), violation of equipartition, and more extended gaseous halos can produce similar RM and $I_{s}$ to observations, often with much lower median ${\bf B}$ compared to simpler models. More detailed forward-modeling is clearly needed. In the extended CGM/halo, observations are lacking, but comparing our Fig.~\ref{fig:profile.pressure} to other cosmological simulations of MW and dwarf galaxies including stellar feedback shows that we predict similar, or somewhat higher, values of $|{\bf B}|$ as a function of either baryon density ($n/\langle n \rangle$) or galacto-centric distance $r/r_{\rm vir}$ \citep{2014MNRAS.445.3706V,2015MNRAS.453.3999M,2018MNRAS.479.3343M,2019MNRAS.484.2620K}. Older simulations not including stellar feedback show substantially lower $|{\bf B}|$ \citep{2008A&A...482L..13D,2008SSRv..134..311D,2018SSRv..214..122D}, consistent with recent studies of the role of turbulence and winds in field amplification \citep{su:fire.feedback.alters.magnetic.amplification.morphology,2018MNRAS.479.3343M}.

\vspace{-0.5cm}
\subsection{Cosmic Rays in Dwarf Galaxies}
\label{sec:discussion.dwarf}

In essentially every {\em galaxy} property we examine, the effects of CR physics on dwarf galaxies ($M_{\ast} \ll 10^{10}\,M_{\odot}$, $M_{\rm halo} \lesssim 10^{11}\,M_{\odot}$) are relatively small. Above in \S~\ref{sec:toy}, we argued that this is generically expected: in equilibrium for realistic star-forming galaxies on the observed ``main sequence'' of star formation, the ratio of CR pressure forces on gas in the halo, compared to gravitational forces, scales $\propto M_{\ast}/M_{\rm halo}$. Of course, other forms of feedback scale this way as well. But with other forms of stellar feedback in place (notably SNe), CR pressure is {\em relatively} inefficient at re-accelerating winds or stalling accretion on scales of order the galaxy and halo scale radius. One key point is that on small (ISM) scales, in dwarfs, with low metallicities and densities, SNe cool relatively inefficiently and can convert a large fraction of their ejecta energy into work and/or thermal energy which is not immediately radiated (see \citealt{hopkins:sne.methods}). Since the CR energy is only $\sim 10\%$ of the mechanical energy, by construction, this means mechanical energy will always dominate, if it is not efficiently radiated (as it would be in dense gas, in more massive galaxies). 

A second key point follows directly from the observations. In dwarf galaxies, almost all of the CRs escape efficiently from the ISM: but unlike massive galaxies where there is a large CGM ``hot halo'' extending to $>300\,$kpc which can ``confine'' the CRs and allow them to build up a substantial pressure profile, in dwarf galaxies there is no such hot halo (it cannot be built up from virial shocks, and outflows are much less confined as well so simply escape) and the size of the virial radius is much smaller ($\ll 100\,$kpc). So CRs in dwarfs simply escape from their CGM as well (and even if they stay, they have relatively little mass to support and prevent from accreting). 

We do see some effects of CRs, but these are small compared to the effects of different treatments of SNe and stellar radiation (see \citealt{hopkins:fire2.methods,hopkins:sne.methods,hopkins:radiation.methods}), and appear primarily only at lower $\kappa$ than allowed observationally. Table~\ref{tbl:sims} and Figs.~\ref{fig:cr.demo.m10}-\ref{fig:cr.demo.m12} (also \ref{fig:mgal.mhalo}, \ref{fig:diffusioncoeff}) show that at ultra-faint masses (e.g.\ {\bf m09} and {\bf m10v} or {\bf m10q} at very early times, with $M_{\ast} \lesssim 3\times10^{5}\,M_{\odot}$), runs with CRs have slightly {\em increased} stellar mass by $\sim 0.1-0.2$\,dex. At somewhat higher masses CRs can slightly suppress SF in dwarfs, for $\kappa$ within the range $\kappa \approx 10^{28}-10^{29}\,{\rm cm^{2}\,s^{-1}}$. This effect is maximal for {\bf m10q}, but even there is only a factor of $\sim 2$ (much smaller than many radiative or mechanical feedback effects in such small galaxies, which are only turning $\ll 1\%$ of their baryons into stars). By slightly higher masses ({\bf m11b}, {\bf m11q} at $M_{\ast} \sim 10^{7.5}-10^{8.5}\,M_{\sun}$) the effect weakens to $\sim 0.1$ dex or nothing at all ($M_{\ast}\gtrsim 10^{9}\,M_{\sun}$ in {\bf m11v}, {\bf m11c}, see Table~\ref{tbl:sims}). Figs.~ \ref{fig:kslaw}, \ref{fig:phases}, \ref{fig:morph.dwarfs} \&\ \ref{fig:j.distrib}, show that even where the effect on stellar masses is maximal it has no systematic qualitative effect on the visual/stellar or gas morphology, phase structure, or kinematics either within the galaxy itself or the inner CGM, nor on the star formation efficiency of the galaxy.

Figs.~\ref{fig:profile.pressure}-\ref{fig:profile.heating} illustrate why this is the case, in the context of the arguments above. In all cases, CR ``heating'' is vastly sub-dominant to gas cooling (as expected -- recall  the virial temperatures of these halos are near the peak of the cooling curve). Moreover, for higher-$\kappa$ ($\gtrsim 10^{29}\,{\rm cm^{2}\,s^{-1}}$), even with zero losses, the CR pressure in the galaxy and CGM is always sub-dominant to thermal pressure (SFRs are simply too low to support an energetically-dominant CR halo). By making $\kappa$ lower and trapping CRs one can build up more CR pressure close to the galaxy (keeping warm gas ``puffy'' and non-star-forming), but for $\kappa \lesssim 10^{28}\,{\rm cm^{2}\,s^{-1}}$ diffusion is too slow and much of the CR energy is lost collisionally. Even at the ``sweet spot'' where the CR pressure is maximized (about $\kappa\sim 3\times10^{28}\,{\rm cm^{2}\,s^{-1}}$), CR pressure only becomes comparable to thermal pressure (never strongly dominant) in the more massive dwarfs. 

Because these effects are somewhat fine-tuned, and CRs are never strongly dominant, they are also sensitive to other details of the physical treatment: Fig.~\ref{fig:cr.physics} shows that if we allow modestly super-\Alf{ic} streaming (increasing {\em both} the streaming speed and ``streaming loss'' term in these runs), the combination of enhanced losses and faster escape from the galaxy (extending but lowering the CR pressure) eliminates the (already small) effect of CRs within galaxies. Likewise, isotropic transport (effectively allowing slightly faster diffusion) allows the CRs to escape and weakens the effects. These conclusions are discussed and demonstrated in more detail in \citet{chan:2018.cosmicray.fire.gammaray} (and also consistent across our more extensive studies for e.g.\ Fig.~\ref{fig:Lgamma}). 

Finally, and perhaps most important, Fig.~\ref{fig:Lgamma} shows that the relatively low $\kappa$ required to produce an effect on dwarfs leads to a factor $\sim 10$ over-prediction of the observed $\gamma$-ray luminosity in low-surface-density galaxies. Recall, the observed points include the LMC, SMC, and M33, analogous to several of our simulated dwarfs. In order to match these, Fig.~\ref{fig:Lgamma} shows $\kappa > 10^{29}\,{\rm cm^{2}\,s^{-1}}$ is required, at which point the effect of CRs on most internal dwarf galaxy properties vanishes.

Briefly, we note that our conclusions here appear to contradict some claims in the literature that CRs can have a strong effect on SF in dwarf galaxies \citep[e.g.][]{jubelgas:2008.cosmic.ray.outflows,Boot13,Chen16}. However, to our knowledge, all such claims either (a) adopted low diffusion coefficients, $\kappa \ll 10^{29}\,{\rm cm^{2}\,s^{-1}}$, without comparing to the constraints from $\gamma$-ray fluxes which we argue prohibit such coefficients (in several such studies, collisional losses were not included at all, which allows CRs to artificially ``build up'' at low $\kappa$ when their energy should be lost); or (b) used idealized (non-cosmological) simulations, or simulations with very weak (or non-existent) stellar feedback from other sources (e.g.\ mechanical SNe and/or radiative feedback), such that the SFR and ratio $M_{\ast}/M_{\rm halo}$ was much higher than observed (which, according to the scalings in \S~\ref{sec:toy}, would allow CRs to have a large effect, but directly violates observations of galaxy stellar masses and SFRs). 

We do stress that CRs could still have an effect in the CGM, or ICM further away from the galaxy, at least at intermediate mass scales. This will be studied in detail in future work, but briefly, we note that even in the outer CGM or IGM out to $\sim 4\,R_{\rm vir}$, we see no obvious systematic effect of CRs in very low-mass halos (e.g.\ $M_{\rm halo} \lesssim 10^{10}\,M_{\sun}$, i.e.\ {\bf m10q} and smaller galaxies). However we do see some effects on the velocity field of gas even at surprisingly small halo mass scales (down to $M_{\rm halo} \sim 4\times10^{10}\,M_{\sun}$, corresponding to {\bf m11b}). In these intermediate-mass systems, Fig.~\ref{fig:profile.pressure} shows the CR pressure is not completely negligible around $\sim R_{\rm vir}$ (although it is not dominant), so this is plausible. But since our primary focus here is galaxy properties, we defer a more detailed investigation to future work.

In summary, for any observationally-allowed CR parameters explored here, the effects on any {\em galaxy} property studied are small (much smaller than effects of e.g.\ mechanical or radiative feedback).

\vspace{-0.5cm}
\subsection{Cosmic Rays in Intermediate and Milky Way-Mass Galaxies}
\label{sec:discussion.MW}

As we look at progressively more massive low-redshift galaxies, above $M_{\rm halo} \gtrsim 10^{11}\,M_{\odot}$ (at $z=0$), however, we see that CRs can have significant effects, as predicted in \S~\ref{sec:toy}. 

However, with a lower diffusion coefficient, $\kappa \lesssim 10^{29}\,{\rm cm^{2}\,s^{-1}}$, the effects on {\em galaxy} properties (in e.g.\ Figs.~\ref{fig:cr.demo.m11}-\ref{fig:cr.demo.m12}, \ref{fig:mgal.mhalo}-\ref{fig:phases},  \ref{fig:morph.m11f}-\ref{fig:j.distrib}, etc.) are very weak -- typically $\sim 0.1\,$ dex or so in SFR and stellar mass, at most. The effects at CRs at low-$\kappa$ are even weaker with super-\Alf{ic} streaming, as seen with dwarfs (see also  \citealt{chan:2018.cosmicray.fire.gammaray}). The reason is obvious in Fig.~\ref{fig:Lgamma}: for $\kappa \lesssim 10^{29}\,{\rm cm^{2}\,s^{-1}}$, the massive galaxies become proton calorimeters, i.e.\ lose most of their CR energy to collisions within the galaxy -- also as predicted in \S~\ref{sec:toy}. Note that the massive galaxies have higher central densities compared to the dwarfs (almost all have $\Sigma_{\rm central} \gtrsim 10^{-2}\,{\rm g\,cm^{-2}}$), so it requires larger $\kappa$ for CRs to escape without losing most of their energy. This is also obvious in Fig.~\ref{fig:profile.pressure} -- for these lower $\kappa$ values, the CR pressure/energy density outside the galaxy is order-of-magnitude below the predicted value if the CRs diffused without losses.

Fig.~\ref{fig:diffusioncoeff} shows that, as a result, the ``sweet spot'' where CRs have maximal effect occurs at $\kappa \sim 3-30\times10^{29}\,{\rm cm^{2}\,s^{-1}}$.\footnote{Note that although the low-resolution ($m_{i,\,1000}=56$) {\bf m12i} shown in Fig.~\ref{fig:diffusioncoeff} shows increasing effects of CRs going from $\kappa \sim 3\times10^{29}$ to $\kappa\sim 3\times10^{30}$, the low-resolution version of this simulation is a relatively high-density galaxy with $\Sigma_{\rm central} \sim 10^{-1.2}\,{\rm g\,cm^{-2}}$ in Fig.~\ref{fig:Lgamma}, so the higher $\kappa$ improves escape. We have run both higher resolution ($m_{i,\,1000}=7$) {\bf m12i} (where the galaxy is systematically less dense), and {\bf m11f} with $\kappa\sim 3\times10^{30}$, to limited redshift ($z\sim1$), and find in these less-dense ($\Sigma_{\rm central} \sim 10^{-(1.7-2.1)}\,{\rm g\,cm^{-2}}$) systems that the CR effects become weaker at this still-higher $\kappa$, as predicted if they escape ``too'' efficiently.} For these $\kappa$, Figs.~\ref{fig:cr.demo.m11}-\ref{fig:cr.demo.m12} show that as the the galaxies approach stellar masses $M_{\ast} \sim 10^{10}\,M_{\ast}$, or halo mass $\sim 10^{11}\,M_{\sun}$ (SFRs $\gtrsim 0.1-1\,{\rm M_{\sun}\,yr^{-1}}$) the CRs begin to have an effect suppressing SF. The suppressed SFRs in the ``CR+($\kappa=3e29$)'' runs tend to flatten once this mass scale is reached, while SFRs in the ``Hydro+'', ``MHD+'', and ``CR+($\kappa=3e28$)'' continue to rise, so at the peak of the ``Hydro+'' run SFRs ($z\sim 0$ for the lower-mass {\bf m11} runs, or $z\sim 1$ for the higher-mass {\bf m12} runs) the difference in SFR can be as large as factor $\sim 3-10$, although the integrated difference in stellar mass by $z\sim0$ is usually a more modest factor $\sim 2-3$. 

These effects do depend on redshift, as discussed below (\S~\ref{sec:discussion:redshift}) and shown in Figs.~\ref{fig:HiZ}-\ref{fig:HiZ.profiles}. Essentially all the effects we see from CRs are confined to relatively low redshifts $z\lesssim 1-2$.

Figs.~\ref{fig:epsilonCR}-\ref{fig:cr.physics} (see also Appendix~\ref{sec:additional.tests} \&\ the more detailed studies in \citealt{chan:2018.cosmicray.fire.gammaray}) show that the generic behaviors described above are not especially sensitive to other details of the CR transport physics, provided similar large $\kappa$ (discussed further in \S~\ref{sec:discussion.transport}). In Fig.~\ref{fig:epsilonCR} we systematically vary the fraction of SNe energy injected as CRs from $\epsilon_{\rm cr}\sim 0.05-0.2$: as expected, more efficient CR production produces stronger SFR suppression, but the effect is highly sub-linear (a factor of $\sim 4$ change in $\epsilon_{\rm cr}$ produces a factor $\sim 1.5$ change in stellar mass or SFR). Consistent with Fig.~\ref{fig:profile.pressure}, the effect is primarily a ``threshold'' effect: once {\em sufficient} CRs reach large radii to set up a halo that can pressure-support the cool gas, adding somewhat more produces little effect. Of course eventually if $\epsilon_{\rm cr}$ is too low, the CR halo cannot maintain pressure support, and the effect of CRs should rapidly vanish.

Fig.~\ref{fig:Lgamma} shows that for these MW-mass systems, the same $\kappa\sim 3\times10^{29}\,{\rm cm^{2}\,s^{-1}}$ which produces large effects on their SF histories appears to agree well with the spallation constraints in the MW and observed $\gamma$-ray luminosities in the MW and local galaxies, while the lower $\kappa \lesssim 10^{29}\,{\rm cm^{2}\,s^{-1}}$ (which produces weak effects on the galaxies) predicts excessive $\gamma$-ray flux. 

The effects of CRs {\em within} the galaxies, even at high-$\kappa$, appear quite weak, as expected (see \S~\ref{sec:toy}) -- properties like the stellar/visual and/or gas morphology and kinematics within the galactic disks, abundances/metallicities, baryonic and dark matter mass profiles, disk sizes and rotation curves, outflow rates, and star formation efficiencies (location on e.g.\ the Kennicutt-Schmidt relation) do not appear to be strongly altered at {\em any} $\kappa$ or mass scale we study, except insofar as the {\em total} galaxy mass and SFR shift up or down (changing e.g.\ the total gas supply, or total mass in metals produced, or total mass in baryons contributing to $V_{c}$). This is not surprising: in the disk midplane, we confirm the CR pressure gradients are sub-dominant to thermal and turbulent forces. But in the CGM around the galaxies, the CRs appear to have a direct and dramatic effect. 
We discuss this further in \S~\ref{sec:discussion.where.crs} below.

However Figs.~\ref{fig:profile.pressure} and \ref{fig:morph.m11f}-\ref{fig:morph.m12m} demonstrate that the effects of CRs on the CGM around these galaxies, at radii $\sim 10-100\,$kpc, are dramatic. In the cases where CRs suppress SF, they establish a high-pressure CR halo outside of the galactic disk (extending to or even past the virial radius), supporting a large reservoir of gas which is much cooler ($T \ll 10^{6}$\,K) than would be required to maintain thermal pressure equilibrium. In contrast in the ``MHD+'' or ``Hydro+'' cases denser and/or cooler halo gas cools rapidly, then falls onto the galaxy, leaving a virialized halo of only the ``leftover'' gas which is more tenuous. In the ``CR+'' runs around massive galaxies, Fig.~\ref{fig:profile.pressure} shows the CR pressure is dominant over thermal and magnetic pressure outside the disk, all the way to the virial radius (Fig.~\ref{fig:profile.heating} shows the direct CR heating is negligible). For low-$\kappa$ the CR pressure is suppressed owing to losses, but for the high-$\kappa$ runs the CR pressure profile agrees remarkably well with the analytic predictions in \S~\ref{sec:toy}, which also predict accurately the mass scale where this can support enough gas mass to suppress gas inflows and (ultimately) SF in the galaxies at a significant level. 
We discuss the dynamics of the CRs in the CGM in more detail below and in future work.

\vspace{-0.5cm}
\subsection{Dependence on Redshift (and Super-$L_{\ast}$ Massive Galaxies)}
\label{sec:discussion:redshift}

\begin{figure*}
    \plotsidesize{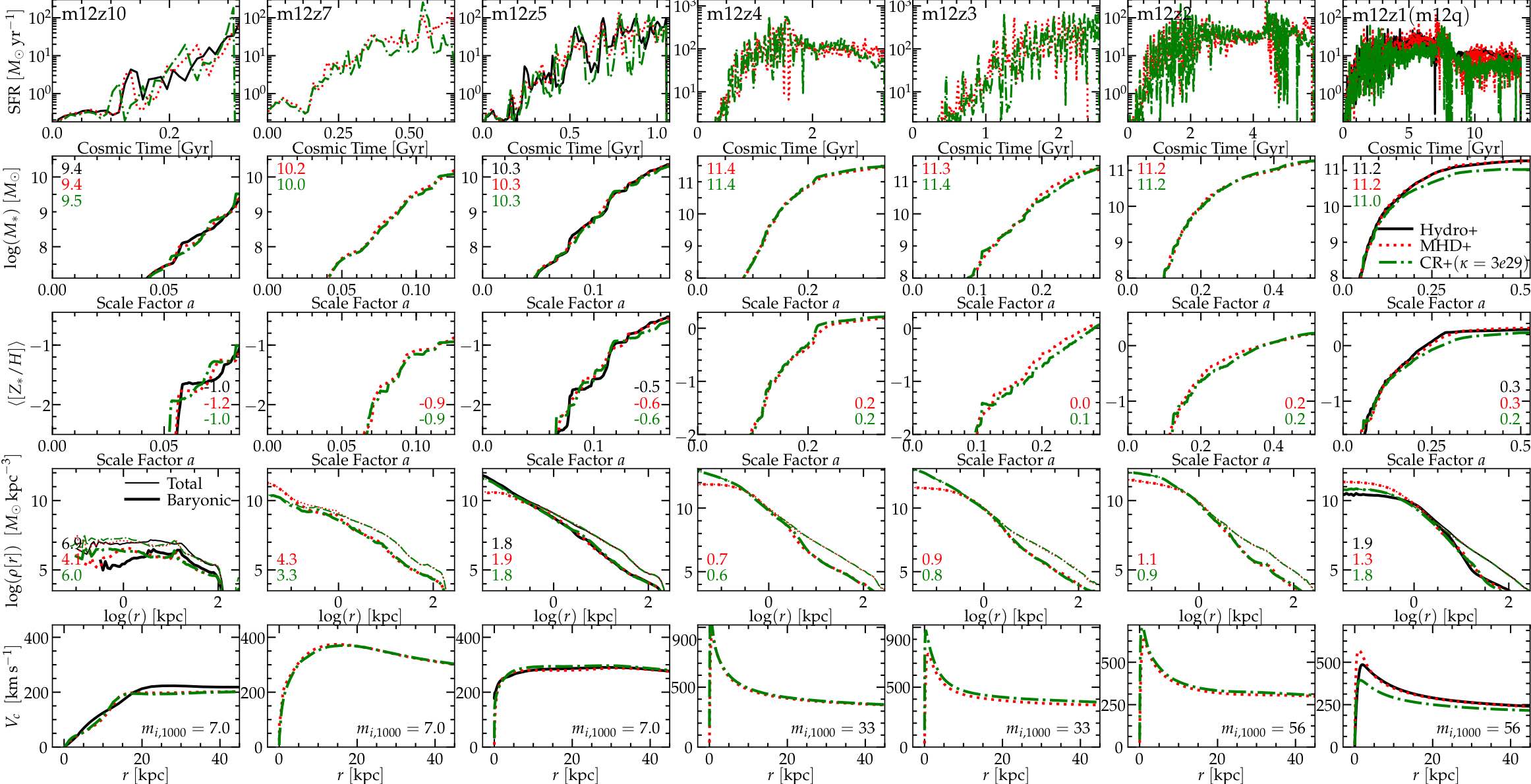}{0.99}
    \vspace{-0.25cm}
    \caption{As Fig.~\ref{fig:cr.demo.m10}, but comparing our high-redshift massive halos from Table~\ref{tbl:HiZ}. Each halo labeled {\bf m12zX} exceeds a halo mass $\gtrsim 10^{12}\,M_{\sun}$ at a redshift $z = z_{12} \sim${\bf X}. We run to at least this redshift, and in some cases somewhat further. Consistent with the lower-mass halos in Figs.~\ref{fig:cr.demo.m10}-Fig.~\ref{fig:cr.demo.m12} (where $z_{12}\lesssim 0$), in every case, CRs halo little or no effect at redshifts $z \gtrsim 1-2$. This is consistent with our analytic expectations (\S~\ref{sec:toy}): at high-$z$ the SFRs (and corresponding CR injection rates) are higher at a given mass (some reaching $\sim 1000\,M_{\sun}\,{\rm yr^{-1}}$ here), but the CGM densities/pressures are much higher so the CRs are not able to support the halo in virial equilibrium. Denser gas in galaxies also produces large CR losses during the peak starburst  epochs (all the systems with $\dot{M}_{\ast} \gtrsim 100\,M_{\sun}\,{\rm yr^{-1}}$ have $\gamma$-ray losses near calorimetric, and $\Sigma_{\rm central} \gtrsim 0.1\,{\rm g\,cm^{-2}}$ as in Fig.~\ref{fig:Lgamma}. As shown in \citet{su:2018.stellar.fb.fails.to.solve.cooling.flow}, lacking AGN (or some other) feedback, feedback from SNe alone cannot ``quench'' SF in massive halos and they over-cool, producing the extremely large central $V_{c}$ in several of the runs.
    \label{fig:HiZ}}
\end{figure*}

\begin{figure}
\begin{centering}
    \includegraphics[width={0.235\textwidth}]{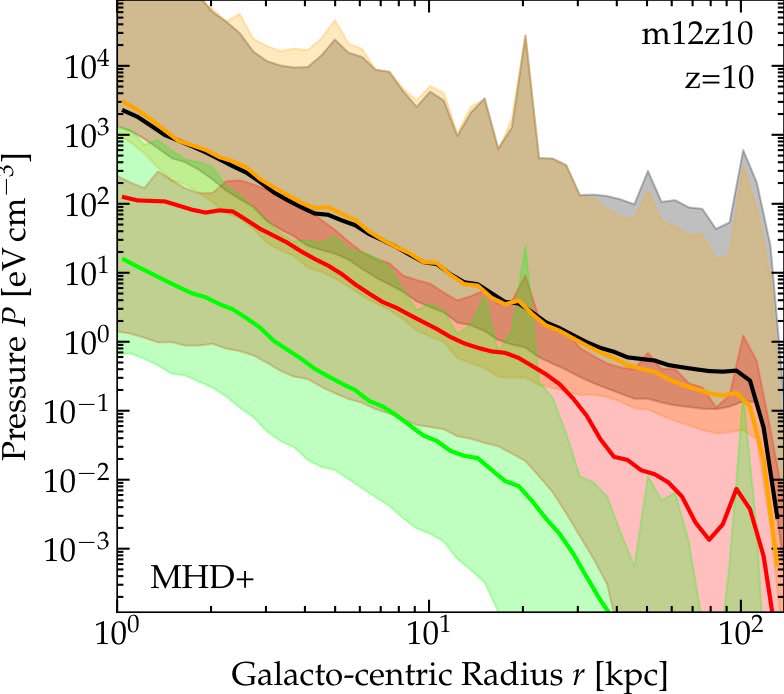}
    \includegraphics[width={0.235\textwidth}]{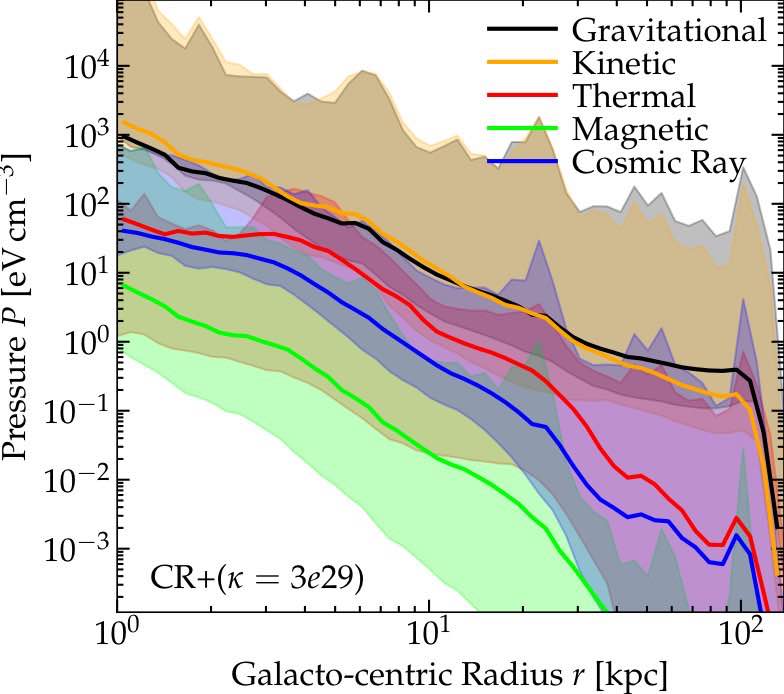} \\
    \includegraphics[width={0.235\textwidth}]{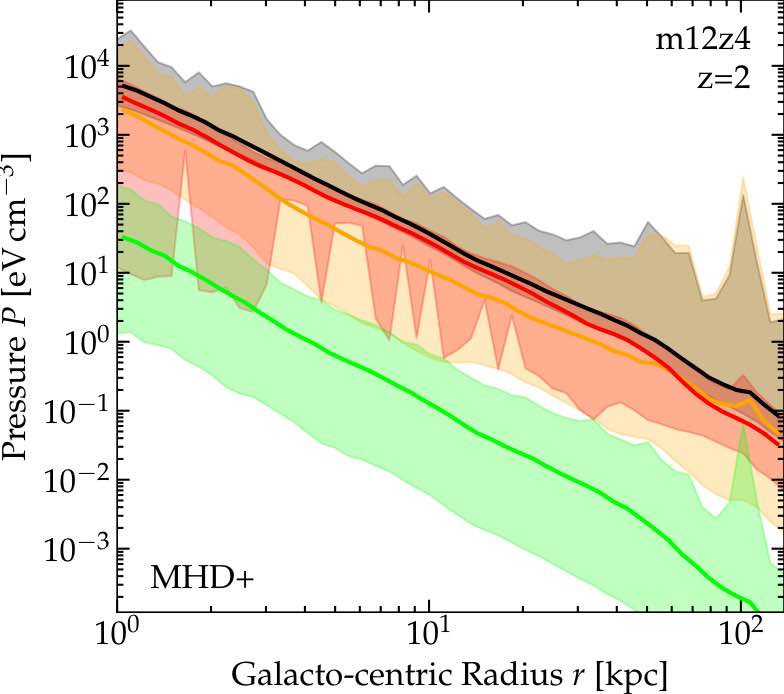}
    \includegraphics[width={0.235\textwidth}]{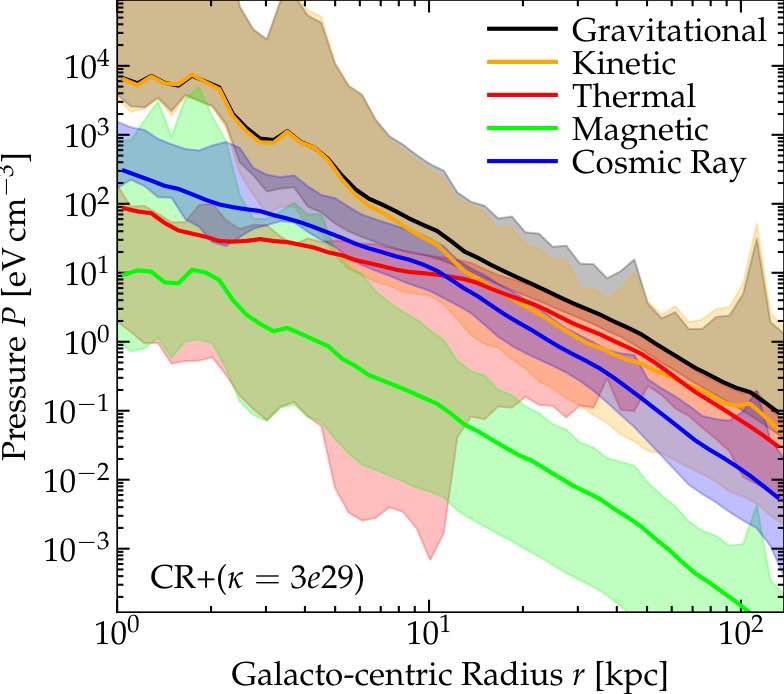} \\
    \includegraphics[width={0.235\textwidth}]{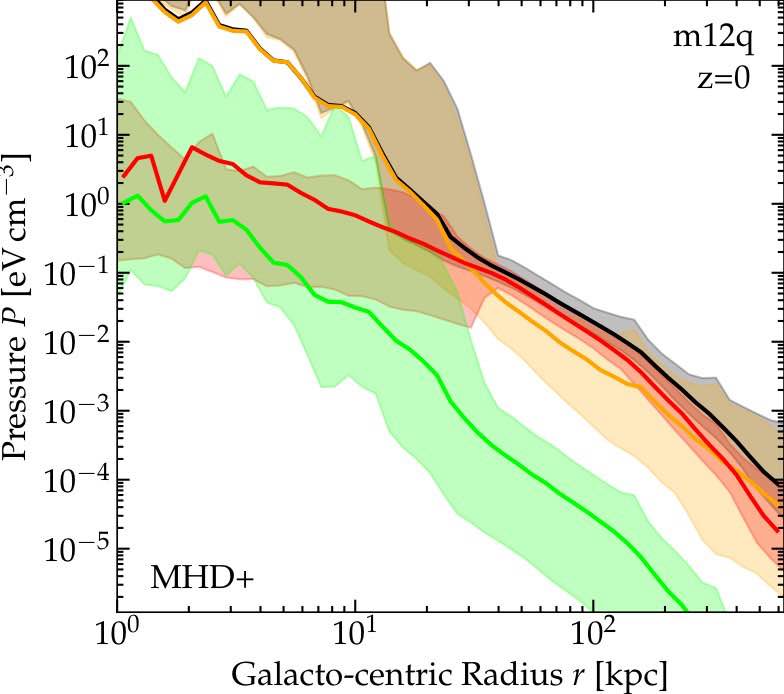} 
    \includegraphics[width={0.235\textwidth}]{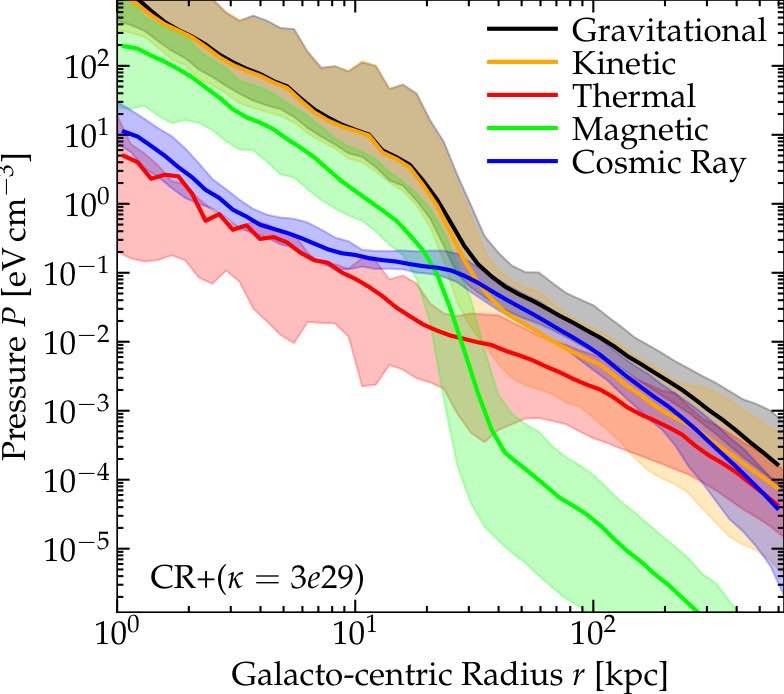} \\
\end{centering}
    \vspace{-0.25cm}
    \caption{Gas pressure profiles (as Fig.~\ref{fig:profile.pressure}) divided into thermal/magnetic/CR/kinetic energy and ``gravitational'' pressure needed for hydrostatic balance, for three representative examples of our high-redshift halos from Table~\ref{tbl:HiZ} \&\ Fig.~\ref{fig:HiZ}. In {\bf m12z10} at $z=10$ ({\em top}; {\bf m12z7}, {\bf m12z5}, and {\bf m12z4} at $z=7,5,4$, respectively, are similar), the CR pressure is slightly sub-dominant to thermal pressure, but {\em both} are much less (factors $\sim 30$) than the kinetic+virial pressure -- this arises because the halo gas is not in virial equilibrium, but is mostly free-falling onto the galaxy. Note the large scatter in CR pressure -- some gas (outflows at low density ``between'' inflowing filaments) has $P_{\rm cr} \sim P_{\rm gravity}$, but this is a small fraction of the mass. In {\bf m12z4} at its latest time run, $z=2$ ({\em middle}; {\bf m12z3} at $z=2.5$ is similar), CR and magnetic pressure are still sub-dominant to gravity, but less dramatically so (factor $\sim 3-10$). By {\bf m12q} at $z=0$ ({\em bottom}), CR pressure is close to virial in the halo ($r\gtrsim 30\,$kpc), allowing CRs to influence the star formation history.
    \label{fig:HiZ.profiles}}
\end{figure}

The galaxies of interest in \S~\ref{sec:discussion.MW}, where CRs have appreciable effects, have $M_{\rm  halo} \sim 10^{11-12}\,M_{\sun}$ at $z\sim 0$. It is natural to ask whether galaxies in this mass range at higher redshifts also experience similar effects of CR transport. 

First, note that high-redshift dwarfs are represented in our sample already, in Figs.~\ref{fig:cr.demo.m10}-\ref{fig:cr.demo.m12}, etc. These are simply the progenitors of the $z=0$ galaxies. Since we plot  their entire evolutionary history, it is straightforward to see in these figures that essentially {\em all} of the differences owing to CRs (at any mass scale) manifest only at relatively low redshifts $z\lesssim 1-2$. In most cases, it is quite notable: e.g.\ for galaxies {\bf m11i}, {\bf m11d}, {\bf m11h}, {\bf m11f}, {\bf m11g}, {\bf m12i}, {\bf m12f},  {\bf m12m}, the SFH, stellar mass, and metallicity (as well as all other properties,  such as morphology, outflow properties, etc.) in the CR+ runs closely track the MHD+ runs until $z\lesssim 1-2$, where they begin to diverge significantly. For the lower-mass galaxies,  this is less surprising: their progenitors at $z\gtrsim  1-2$ are small dwarfs (for which  the  effects of CRs are weak  at $z=0$). But note that some of the more massive galaxies (e.g.\ {\bf m12m} and {\bf m12f}) reach stellar masses $\gtrsim 10^{10}\,M_{\sun}$ and halo masses $\gtrsim 10^{11.5}\,M_{\sun}$ long before they begin to diverge -- well above the threshold where we see effects appear at $z=0$. This indicates that there is some additional redshift dependence here. 

To explore more massive halos at higher redshifts, i.e.\ those with $M_{\rm halo}\sim 10^{11-12}\,M_{\sun}$ at $z\sim 1-10$, Figs.~\ref{fig:HiZ}-\ref{fig:HiZ.profiles} consider the extended suite  of high-redshift,  high-mass galaxies from Table~\ref{tbl:HiZ}. These are halos which reach $\sim 10^{12}\,M_{\sun}$ (comparable to our most massive $z=0$ halos) already at various redshifts $z = z_{12} \sim 1-10$. The earliest of these represent progenitors of what will be massive galaxy clusters, by $z\sim 0$.  In Fig.~\ref{fig:HiZ} we can see essentially no detectable systematic effects of MHD or CRs on these halos at any redshift, {\em except} for the lowest-redshift example ({\bf m12z1}) which begins to show a modest suppression of its SFR and stellar mass only at $z\lesssim 1.5$. We have also examined the other properties in this paper (e.g.\ morphologies, gas phase distributions, outflow velocities) and similarly find no differences above $z\gtrsim 1-2$ (hence their not being shown here, for brevity).

This is actually predicted by the simple scalings in \S~\ref{sec:toy}. At high redshift, SFRs are higher at a given stellar mass (as obvious in Fig.~\ref{fig:HiZ}), so the CR production/injection rate is also higher (scaling $\propto t_{\rm Hubble}^{-1} \sim (1+z)^{3/2}$). However, the density of the CGM and IGM, and ram pressure of inflowing gas, is much higher (scaling $\sim (1+z)^{3}$). So CRs are unable to maintain a super-virial pressure which can suppress inflow/accretion in this dense gas (as in Fig.~\ref{fig:profile.pressure}). We have directly confirmed this, in fact, comparing the CR and virial pressure in Fig.~\ref{fig:HiZ.profiles} -- for the high-redshift massive halos, most of the inflow is under-pressurized relative to what would be needed to maintain virial equilibrium, with or without CRs.\footnote{In the high-redshift halos in Table~\ref{tbl:HiZ}, there is some low-density gas in the CGM for which the CR pressure exceeds or is comparable to that needed for virial equilibrium. This may imply that CRs have an effect in the CGM even where they have little effect on the bulk galaxy properties. However, this under-dense gas represents very little of the gas which accretes onto the galaxy (with or without CRs).} Moreover, owing to their high inflow rates, most of the star formation in the high-redshift systems occurs in starbursts with very high SFRs and, correspondingly, very high gas densities within  the galaxy (obeying the  extension of the observed Schmidt-Kennicutt law in Fig.~\ref{fig:kslaw} to  higher densities) -- we see  this directly in Fig.~\ref{fig:HiZ} where all the massive high-$z$ systems reach  $\dot{M}_{\ast} \gtrsim 100\,M_{\sun}\,{\rm yr^{-1}}$ and the couple most extreme reach $\dot{M}_{\ast} \gtrsim 1000\,M_{\sun}\,{\rm yr^{-1}}$. During these phases, the gas surface densities reach $\gg 0.1\,{\rm g\,cm^{-2}}$ (see Fig.~\ref{fig:kslaw}  and the more detailed studies in \citealt{hopkins:rad.pressure.sf.fb,orr:ks.law}), or $\gg 10^{3}\,M_{\sun}\,{\rm pc^{-2}}$. At these densities, we see in Fig.~\ref{fig:Lgamma} that essentially all observed galaxies,  and all  of our simulations (even with extremely high $\tilde{\kappa} > 10^{30}\,{\rm cm^{2}\,s^{-1}}$) approach the proton-calorimetric limit -- in other words, a substantial fraction of the CR energy is  lost to collisions in the dense ISM. We confirm this directly in the simulations in Fig.~\ref{fig:HiZ} in their ``peak starburst'' phases. Interestingly, there is some evidence that {\em away from the starburst} peaks, at the lowest SFRs, the CR+ runs exhibit slightly more-suppressed SFRs, but these phases of course contribute negligibly to the total SF and stellar mass (and therefore most other galaxy properties). 

Importantly, these simulations also highlight the critical need for some additional feedback beyond the stellar feedback mechanisms (SNe types Ia and II, stellar mass loss from O/B and AGB stars, radiation, cosmic rays accelerated  in SNe) and microphysical processes (magnetic fields, conduction, viscosity) studied here, in order to explain the suppression of SF (i.e.\ ``quenching'') in massive ($\gtrsim L_{\ast}$) galaxies. In \citet{su:2018.stellar.fb.fails.to.solve.cooling.flow}, we show this explicitly, in non-cosmological simulations of halos with $M_{\rm halo}\sim 10^{12-14}\,M_{\sun}$ at $z\sim 0$ -- demonstrating that all the physics studied here cannot solve or even substantially mitigate the ``cooling flow'' and quenching problems. Here in Figs.~\ref{fig:HiZ}-\ref{fig:HiZ.profiles} we are essentially showing the same for the high-redshift progenitors of these massive halos (at $z\sim 2-10$, when their halo masses were in the range $M_{\rm halo} \sim 10^{12-13}\,M_{\sun}$). Not only do these massive halos sustain high SFRs ($\dot{M}_{\ast} \gtrsim 10-100\,M_{\sun}\,{\rm yr^{-1}}$) as long as we run them (including to $z\sim 0$) -- i.e.\ clearly fail to quench -- but they also form extremely dense central bulges in their starbursts with (in the most extreme cases) central circular velocities approaching $\sim 1000\,{\rm km\,s^{-1}}$ (well in excess of the most massive galaxies observed). But these are precisely the systems ($M_{\rm halo} \gtrsim 10^{12-13}\,M_{\sun}$ at $z\sim 2$, and $\gtrsim 10^{14}\,M_{\sun}$ at $z\sim 0$, with bulge-dominated $M_{\ast} \gtrsim 2\times10^{11}\,M_{\sun}$) expected to host extremely massive super-massive black holes and quasars. And indeed, \citet{daa:BHs.on.FIRE} presented preliminary studies of the most extreme two  halos here ({\bf m12z4} and {\bf m12z3}) including models for black  hole growth (but not AGN feedback), where the black holes reached masses $\gtrsim 10^{8-9}\,M_{\sun}$. So it will clearly be of particular interest in the future to explore the effects of AGN feedback in these systems.

\vspace{-0.5cm}
\subsection{Where Do CRs Act Most Efficiently? (Effects Inside versus Outside Galaxies)}
\label{sec:discussion.where.crs}

We noted in \S~\ref{sec:discussion.MW} above that CRs appear to have very weak effects on {\em instantaneous} properties {\em within} galaxies, even in the regime (e.g.\ large-$\kappa$ and high-$M_{\ast}$) where they have a large cosmologically-integrated effect on galaxy masses and SFHs. 

\vspace{-0.5cm}
\subsubsection{Weak Effects Within Galaxies}

For example, in Figs.~\ref{fig:cr.demo.m10}-\ref{fig:cr.demo.m12}  \&\ \ref{fig:mgal.mhalo}, in the systems where CRs suppress SF, the metallicity and central circular velocity are also lower, accordingly, but these are essentially consistent with moving {\em along}, not off of, the observed mass-metallicity relation and Tully-Fisher relations, respectively \citep[for more detailed studies of those, see][]{ma:2015.fire.mass.metallicity,elbadry:HI.obs.gal.kinematics}. 
From these plots, we see that the galaxies with higher (lower) star formation rates have systematically higher (lower) baryonic masses in their centers, i.e.\ they appear to be shifting with the ``supply'' of gas. 
Fig.~\ref{fig:kslaw} shows this more rigorously, demonstrating that the different physics runs do not systematically differ in the normalization of the Schmidt-type relations -- i.e.\ they are not consuming gas faster/slower or more/less ``efficiently.'' Rather, the galaxies with suppressed star formation have moved {\em along} the relation. This is quite different from what occurs if we increase/decrease  the number or mechanical energy of supernovae, which systematically moves the relation down/up (for the same gas mass, fewer/more massive stars are required to regulate against gravitational collapse; see \citealt{hopkins:rad.pressure.sf.fb,ostriker.shetty:2011.turb.disk.selfreg.ks,cafg:sf.fb.reg.kslaw,agertz:sf.feedback.multiple.mechanisms,orr:ks.law}).
Furthermore, although we defer a detailed study of the effect of CRs on galactic winds to future work, we find that gas outflow rates and velocities {\em in the immediate vicinity of the galaxy} are not strongly influenced by CRs -- again unlike the case if we were to change the energetics or rate of SNe \citep[see][]{hopkins:fb.ism.prop,hopkins:stellar.fb.mergers,hopkins:2013.merger.sb.fb.winds,hopkins:sne.methods}. It is possible, of course, that CRs contribute to outflows via ``slow'' or ``gentle'' acceleration of material at sub-virial velocities, and this increases at larger and larger radii (discussed below) but they do not {\em directly} launch ``fast'' outflows ($V_{\rm launch} \gtrsim V_{\rm escape}$).

Similarly, Figs.~\ref{fig:morph.dwarfs}-\ref{fig:morph.m12m} show that the {\em galaxy-scale} stellar/visual and gas morphologies are only weakly modified even in any of our core suite. Where the stellar masses are strongly suppressed at high-$\kappa$ and high-$M_{\ast}$, the galaxies do tend to have slightly later-type visual morphologies, but this is entirely consistent with their lower masses -- they simply resemble the earlier-time versions of their ``Hydro+'' counterparts (i.e.\ they have simply evolved less along the mass-morphology sequence; for more details of that see \citealt{garrisonkimmel:fire.morphologies.vs.dm}). 
Conversely, some of the low-$\kappa$ runs which produce slightly {\em higher} stellar masses exhibit earlier-type morphologies.
In the gas within the disks there are some slight differences in the ``sharpness'' of features in the cool gas in Figs.~\ref{fig:morph.dwarfs}-\ref{fig:morph.m12m}, but nothing qualitatively distinct. Fig.~\ref{fig:phases} shows some non-trivial (but still quantitative, rather than qualitative) effects on the ISM distribution of phases: namely, CRs  can support more warm neutral gas within the galaxy. This is not surprising (they both heat and pressure-stabilize this phase of gas, without ionizing it significantly), but it is quantitatively non-negligible for detailed comparisons of ISM phase structure.
Also, in Fig.~\ref{fig:j.distrib} we see the CRs do not radically alter the kinematics of gas or stars (again, except insofar as they suppress the amount of low-redshift star formation; see \citealt{elbadry:fire.morph.momentum} for a detailed study of how this varies as a function of mass in our ``Hydro+'' runs). 

All of this is expected from \S~\ref{sec:toy} and Figs.~\ref{fig:profile.pressure}-\ref{fig:profile.heating}. Within the star-forming galactic disk, CRs can have roughly equipartition-level energy densities, but their large diffusivity means that (as observed) the CR scale-height/length is much larger than the cold star-forming gas scale height (let alone the size of structures like GMCs, cores, etc). This means that CR pressure gradients -- which actually determine the forces -- are usually order-of-magnitude smaller than the small-scale forces from gravity, magnetic fields, gas thermal pressure, and turbulent ram pressure, in the multi-phase ISM.

\vspace{-0.5cm}
\subsubsection{(Potentially) Strong Effects in the CGM}

On the other hand, we see strong CR effects in the CGM, even in many simulations (e.g.\ with lower $\kappa$) where the CRs do not have a large effect on galaxy masses. 

This is especially evident in Figs.~\ref{fig:morph.dwarfs}-\ref{fig:morph.m12m}: for dwarfs with halo masses $M_{\rm halo} \gtrsim 10^{11}\,M_{\sun}$, the ``CR+'' runs often feature much more prominent cool gas in the CGM. 
Figs.~\ref{fig:cr.demo.m10}-\ref{fig:cr.demo.m12} show that the {\em total} baryonic mass density on CGM scales ($\sim 10-100\,$kpc) is not drastically modified by CRs, although there are subtle (order-unity) changes evident in many cases (e.g.\ some of the intermediate-mass {\bf m11} halos with lower-$\kappa$ feature enhanced gas density on large scales). 
The runs where CRs dominate over thermal pressure in the CGM (see Fig.~\ref{fig:profile.pressure}) typically feature {\em modest} (factor $\sim 2-3$) overall enhancements of gas density at some intermediate range of radii of order the halo scale radius ($R_{s} \sim R_{\rm vir}/10$), with little effect on total gas density out to or beyond the virial radius $\sim R_{\rm vir}$. 

As noted above, Fig.~\ref{fig:profile.heating} shows this enhancement is not driven by CR ``heating'' via either collisional or streaming processes. Most of the CGM mass in these runs is in ``cool'' or ``warm'' CGM phases, where the cooling times are relatively short and  the temperature is maintained largely by photo-ionization equilibrium \citep{ji:fire.cr.cgm}, hence their low (sub-virial) thermal pressure. Rather, Fig.~\ref{fig:profile.pressure} shows that the ``maintenance'' of this gas  density profile owes to the CRs establishing a quasi-virial-equilibrium pressure profile on large scales. This is qualitatively similar to the findings of e.g.\ \citet{Sale14cos,Sale16,Chen16} in their cosmological simulations with CRs, despite their adopting different numerical methods and treatment of the CR and star formation/feedback ``microphysics,'' suggesting the conclusion is robust to these details.

These, plus the weak effects of CRs on essentially all star formation and outflow and internal galaxy properties discussed above, demonstrate that the CRs primarily operate as a ``preventive'' feedback mechanism, rather than an ``ejective'' or ``responsive'' feedback mechanism. Rather than launching strong outflows, or {\em removing} gas from the halo, or preventing gas which has already accreted into the galaxy from efficiently forming stars, the primary effect of CRs appears to be preventing gas in the halo  from actually accreting rapidly onto the galaxy.  These CR effects result in a more subtle re-arrangement of gas mass within the halo and between different phases.\footnote{Note that it is not possible here to completely dis-entangle the role of CRs preventing ``new'' gas from accreting onto the galaxy at all, versus ``lofting'' or ``gently/slowly accelerating'' cool gas in galactic fountains (at velocities $\ll V_{c}$) and preventing it from re-accreting. In an instantaneous sense these are the same thing: suppressing cool gas from falling into the galaxy. Future work following trajectories of individual gas elements will allow us to better disentangle these possibilities.} Even when  $\gtrsim L_{\ast}$ halos become dominated by a hot, virialized halo gas, cooling of that hot atmosphere or re-accretion of previously ejected gas can provide a substantial gas supply for late time star formation \citep{keres:hot.halos,muratov:2015.fire.winds,angles.alcazar:particle.tracking.fire.baryon.cycle.intergalactic.transfer}. Preventing such accretion (or re-accretion) has important consequences for the late time star formation and growth of galaxies.

Note that even a modest (factor $\sim 2$) effect on the gas density profile around the halo scale radius is significant -- this is the radius where most of the halo gas mass resides, and most of the total baryon supply is in the halo (especially in the lower-mass galaxies) rather than in stars, so this relatively modest effect can easily account for differences of factors $\sim 2-5$ in star formation rates and stellar masses {\em within} the galaxies.

\vspace{-0.5cm}
\subsection{Cosmic Ray Transport}
\label{sec:discussion.transport}

We now discuss the different detailed CR transport processes modeled here, their effects, and sensitivity to uncertain physics.

\vspace{-0.5cm}
\subsubsection{The Critical Role of the CR ``Effective Diffusivity'' (Transport Speed)}

In \S~\ref{sec:discussion.dwarf} \&\ \ref{sec:discussion.MW}, we noted the existence of a ``sweet spot'' in the CR diffusion coefficient $\kappa$, demonstrated throughout but especially in Fig.~\ref{fig:diffusioncoeff}. As predicted analytically in \S~\ref{sec:toy} and confirmed in the simulations in Fig.~\ref{fig:Lgamma}, at too-low $\kappa$, CRs take too long to escape dense gas, and lose their energy rapidly to collisional processes. Advection alone (e.g.\ CRs being ``carried'' out of galaxies in super-bubbles and cold outflows) is sub-dominant even at quite low $\kappa$. Even without collisional losses, it requires a relatively high $\kappa$ for CRs to build up in the CGM before trans-\Alf{ic} streaming (which is also ``lossy'') takes over. On the other hand, at arbitrarily high-$\kappa$, CRs would escape ``too efficiently'' even from the CGM/extended halo and the steady-state CR pressure ($\propto 1/\kappa$) would become too small to support gas or do any interesting work.

For dwarfs, this ``special'' value of $\kappa \sim 3\times10^{28}\,{\rm cm^{2}\,s^{-1}}$, but even there the effects of CRs are weak, and moreover the observations of $\gamma$-ray luminosities from dwarfs like the SMC, LMC, and M33 favor higher $\kappa \gtrsim 3\times10^{29}\,{\rm cm^{2}\,s^{-1}}$ (where they escape low-density dwarfs more efficiently and do little work). But for massive (MW-mass) systems, interestingly, the ``sweet spot'' in $\kappa \sim 3-30\times10^{29}\,{\rm cm^{2}\,s^{-1}}$ appears to neatly coincide with the observationally-favored values. 

It is also encouraging that the more detailed study of CR transport physics in isolated (non-cosmological) simulations in \citet{chan:2018.cosmicray.fire.gammaray} identified approximately the same critical $\kappa$ needed to reproduce the $\gamma$-ray observations. In detail we favor {\em slightly} higher $\kappa$ here,  owing to the larger gaseous halos present in cosmological simulations, which contribute to increasing the probability of CRs re-entering the disk (hence ``residence time'') before they escape, but this is expected. We also note that the value we quote here is the {\em parallel} diffusivity. If one assumed isotropic CR diffusion, the  corresponding isotropically-averaged $\tilde{\kappa}$ would be a factor $\sim 3$ lower. 

At both mass scales, we find this ``ideal'' range of $\kappa$ spans approximately an order-of-magnitude in range, and within this range, the predictions are not especially sensitive to $\kappa$ (i.e.\ changing $\kappa$ by a factor $\sim 2$ about this ``ideal'' value produces relatively small effects). So although the physical scalings of CR diffusivity remain poorly-understood, this does not necessarily require particular fine-tuning. 

This highlights, however, what is likely the most uncertain-but-important assumption in this work: that CR transport can be approximated with a constant diffusivity (or a fluid-like model at all). CR transport -- what gives rise to ``diffusive-like'' behavior at all -- remains deeply theoretically and observationally uncertain, as it is generally believed that diffusive behavior arises non-linearly from the interplay between different plasma instabilities and transport processes along field lines. For this reason we simplified and explored different, constant $\kappa$ models here. If, in nature, the ``effective'' $\kappa$ simply varies systematically from galaxy-to-galaxy (e.g.\ with cosmological environment or halo mass) this is not so problematic: it simply amounts to choosing a different ``effective $\kappa$'' as ``most appropriate'' for our different simulations. If the effective $\kappa$ varies on small spatial or time scales, this is also not such a concern, as the CRs diffuse sufficiently rapidly that these variations are averaged out on the large CGM scales ($\sim 10-100\,$kpc) on which they act (and indeed, because the transport is anisotropic, local turbulent magnetic field-line variations already mean that the actual local diffusivity is constantly varying by factors of several on small scales). 

The biggest cause for concern is if, in reality, there are large systematic variations in effectively diffusivity with some property that varies between galaxies and CGM (e.g.\ spatial scale, density, magnetic field strength). It is plausible to imagine models where $\kappa \sim 3\times10^{29}\,{\rm cm^{2}\,s^{-1}}$ in the ISM (producing the same $\gamma$-ray luminosity), but CRs ``de-couple'' or ``free-stream'' or escape much more rapidly in the CGM, dramatically reducing their effects on galaxy formation (see \S~\ref{sec:conclusions}).


\vspace{-0.5cm}
\subsubsection{(Trans-\Alf{ic}) CR Streaming}

We confirm the conclusion in \citet{chan:2018.cosmicray.fire.gammaray}, that streaming at trans-\Alf{ic} speeds (as parameterized here) appears to be sub-dominant in transport. For observationally-favored (high) $\kappa$ values, transport is sufficiently fast that streaming at speeds $\sim v_{A}$ can only take over as a dominant transport mechanism, and begin to dissipate a large fraction of the CR energy, outside a radius $\gtrsim 30-100\,$kpc (\S~\ref{sec:toy} and Fig.~\ref{fig:profile.heating}). Thermalized energy from streaming losses are never, in the simulations here, able to compete with gas cooling losses (Fig.~\ref{fig:profile.heating}). As a result, turning off ``streaming'' entirely (Fig.~\ref{fig:cr.physics}) produces only minor effects on galaxy properties, around the favored $\kappa$. 

Above, we argued that for CRs to have large effects, high ``effective'' $\kappa_{\ast}$ is needed in order for CRs to escape the dense star-forming regions of galaxies, and these high-$\kappa$ values are also favored by observations of dwarfs and MW-like galaxies (which appear to be well below calorimetric). Because the actual physics which regulates streaming speeds remains uncertain (and streaming losses will not appear in the $\gamma$-ray constraints), one might reasonably wonder whether the same effects might be accomplished by maintaining a relatively low diffusivity (e.g.\ $\kappa_{\|} = 3\times10^{28}\,{\rm cm^{2}\,s^{-1}}$) but increasing the streaming velocity by a modest factor. This is the experiment performed in Figs.~\ref{fig:Lgamma} and \ref{fig:cr.physics}. If the ``streaming losses'' scale with $v_{\rm stream}$ -- as assumed in this test for the sake of illustration -- then increasing $v_{\rm stream}$ is {\em not} equivalent to increasing $\kappa$. In fact, increasing $v_{\rm stream}$ and decreasing $\kappa$ in this manner leads to weaker, not stronger, effects of CRs on galaxy properties. And while increasing $v_{\rm stream}$ (with $\kappa$ constant) does slightly reduce the predicted $L_{\gamma}/L_{\rm SF}$ (Fig.~\ref{fig:Lgamma}), the effect is minor for the values considered here. There are two fundamental reasons for this. First, if the ``streaming loss'' term scales with $v_{\rm stream}$, then more of the CR energy is dissipated close to the galaxy (where cooling is efficient). Second, if streaming dominates then the effective diffusivity is $\kappa_{\ast} \sim v_{\rm stream}\,\ell_{\rm cr}$ (see \S~\ref{sec:methods:crplus}; where $\ell_{\rm cr} \equiv P_{\rm cr}/|\nabla_{\|} P_{\rm cr}|$), so for the large $\kappa_{\ast}$ favored we require $v_{\rm stream} \sim 1000\,{\rm km\,s^{-1}}(\kappa_{\ast}/3\times10^{29}\,{\rm cm^{2}\,s^{-1}})\,(\ell_{\rm cr}/{\rm kpc})^{-1}$), much larger than $v_{A}$ or $c_{s}$.

So, if we set $v_{\rm stream}$ to be {\em much} larger ($\gtrsim 500-1000\,{\rm km\,s^{-1}}$), and limit streaming losses to scale with the \Alf\ speed ($\sim v_{A}\,\nabla P_{\rm cr}$, as these come from the excited \Alf\ waves which cannot propagate faster than this), then streaming will become essentially degenerate with high-$\kappa$ diffusion. We see this indirectly via the fact that the maximum ``transport speed'' of CRs has little effect on the results once it is sufficiently large (Fig.~\ref{fig:cr.vmax}). How, micro-physically, these transport speeds arise remains uncertain.

\vspace{-0.5cm}
\subsubsection{Collisional and Streaming Losses}

As shown in Fig.~\ref{fig:profile.heating}, thermalized CR collisional or streaming losses are essentially never important as a source of heating the gas (they are always highly sub-dominant to cooling). However, they can be important as loss mechanism for the CRs themselves. 
Although it is sometimes (incorrectly) assumed that CRs are ``lossless'' (or ``don't cool''), the collisional loss timescale for CRs in dense gas on length scale $\sim \ell$ is shorter than the diffusion time if $n \gtrsim 10\,{\rm cm^{-3}}\,\tilde{\kappa}_{29}\,(\ell /{\rm kpc})^{-2}$. And indeed, essentially all observed starburst galaxies, which have very high nuclear gas densities (see Fig.~\ref{fig:Lgamma}) are consistent with being proton-calorimeters (i.e.\ all or most of the CR energy appears to be lost). 
For that reason, at either low-$\kappa$ ($\ll 10^{29}\,{\rm cm^{2}\,s^{-1}}$) or high gas surface densities, collisional losses play an important role in limiting CR energetics and effects on galaxies. Otherwise, in ``steady state'' very low-$\kappa$ would produce more trapping and allow one to artificially build up essentially arbitrarily high CR energy densities. 
On the other hand, once $\kappa$ is relatively high, in galaxies with central densities comparable to or less than the MW, then most of the CRs escape without decaying (e.g.\ $L_{\gamma}$ is well below calorimetric in Fig.~\ref{fig:Lgamma}). In these cases, which include most of the cases of greatest interest above, collisional losses necessarily play a minor role, and so de-activating these losses in Fig.~\ref{fig:cr.physics} produces relatively small effects under these conditions.

Streaming losses are similarly not dominant {\em if} $\kappa_{\ast}$ is sufficiently large, at least out to radii $r_{\rm stream} \sim \kappa/v_{\rm stream}$ where trans-\Alf{ic} streaming can begin to dominate the transport. Removing streaming losses with similar $v_{\rm stream}$ and $\kappa$ simply removes the losses at large $r$, making the equilibrium CR pressure profile fall somewhat less rapidly. But because, in this regime, it is already falling more rapidly than in the halo center (see Fig.~\ref{fig:profile.pressure}), this is a second-order effect. Of course, as discussed above, if we substantially increase the streaming velocity beyond the \Alf\ speed, and correspondingly increase the streaming loss rate (and shrink $r_{\rm stream}$ so the losses happen closer to the galaxy), then it can become important limiting the effect of CRs. Of course, as noted above, if we increase $v_{\rm stream}$ but remove or suppress the loss term so it does not scale accordingly, this is just equivalent to increasing $\kappa$.

\vspace{-0.5cm}
\subsubsection{Isotropic vs.\ Anisotropic Transport \&\ Magnetic Fields}

Fig.~\ref{fig:cr.physics} considers some experiments where we remove magnetic fields, and assume all transport processes are isotropic (instead of projecting them along field lines). There, and in the other galaxy and CGM properties studied here, we see no radical or qualitative change in behavior. In the dwarfs, the results are very similar; in the MW-mass experiment ({\bf m12i}), switching to isotropic transport and removing magnetic fields (Fig.~\ref{fig:cr.physics}) leads to a slightly weaker effect from CRs here, largely because the CRs escape more efficiently (but the difference is small compared to removing CRs entirely). 
In general, as shown in \citet{chan:2018.cosmicray.fire.gammaray}, the leading-order effect of anisotropic transport is to alter the effective (galaxy-averaged) $\kappa$, by a modest (order-unity) factor, with a net effect of somewhat lower diffusion coefficients required in isotropic diffusion case for similar effect.

Some of the reason for this is that magnetic fields are highly disordered on most scales, especially in the CGM. This is shown more rigorously in previous studies \citep[e.g.][]{su:fire.feedback.alters.magnetic.amplification.morphology}, and will be explored in subsequent work as well \citep{ji:fire.cr.cgm}. This is not surprising, as the CGM of $\sim L_{\ast}$ and dwarf galaxies is generically trans-sonically turbulent and has plasma $\beta \gg 1$ (so field lines are essentially passively advected into locally disordered configurations). So typical anisotropic ``suppression factors'' vary within order-unity values but strong, systematic suppression of CR transport is never really possible. This may differ in more massive halos (e.g.\ clusters), where the turbulent Mach numbers in the (much hotter) diffuse gas are expected to be much smaller.

Related to this, we see no evidence for CRs systematically ``opening up'' field lines on large scales. This is also not surprising, as we have already argued the primary role of CRs in the simulations here is not in violently ejecting material from small radii to large, but in quasi-statically maintaining the halo.

\vspace{-0.5cm}
\subsection{Resolution \&\ Numerical Effects}
\label{sec:numerics}

We have noted in several places above various numerical tests in our simulations. For example, we have run our standard physics set (``Hydro+'', ``MHD+'', ``CR+($\kappa=3e28$)'', ``CR+($\kappa=3e29$)'') at several different resolution levels in boxes {\bf m10q} (with resolution $m_{i,\,1000}=0.25,\,2.1,\,16$), {\bf m11q} ($m_{i,\,1000}=0.88,\,7,\,56$), {\bf m12i}, {\bf m12f}, and {\bf m12m} (each with $m_{i,\,1000}=7,\,56,\,450$), i.e.\ $60$ simulations. Some examples are shown in Appendix~\ref{sec:additional.tests}  (and in Fig.~\ref{fig:mgal.mhalo}). As shown exhaustively in \citet{hopkins:fire2.methods}, the dwarf galaxies {\bf m10q} and {\bf m11q} show very weak resolution-dependence in ``Hydro+'' runs (or ``MHD+'', in \citealt{su:2016.weak.mhd.cond.visc.turbdiff.fx}). We find the same for ``CR+'' runs, reinforcing our conclusions. As studied extensively in \citet{hopkins:fire2.methods}, our massive halos {\bf m12i}, {\bf m12f}, {\bf m12m} {\em do} show systematic resolution dependence in {\em all} the runs. This is of course important and the scalings and reasons for this are discussed in \citet{hopkins:fire2.methods}. What is important, for our purposes here, is that our qualitative conclusions about the {\em systematic} effects (or lack thereof) of MHD and CRs, are independent of resolution. Like the dwarf runs, at every resolution we find the MHD runs and low-$\kappa$ CR runs have little or no systematic effect on the massive halos, while the high-$\kappa$ CR runs suppress SF significantly, via the same physical mechanisms. The effect is slightly stronger in the higher-resolution runs because the galaxy is overall lower-density and lower-mass even in the ``Hydro+'' runs (this owes to better resolution of galactic winds ``venting'' and mixing in the CGM), which means (per Fig.~\ref{fig:Lgamma}) the collisional losses are somewhat less.

\citet{chan:2018.cosmicray.fire.gammaray} also considered extensive tests of the imposed maximum CR free-streaming speed $\tilde{c}$, and showed that as long as this is faster than typical bulk velocities of gas in the simulated systems, it has no effect  on  the results: Appendix~\ref{sec:additional.tests} shows this holds  in our simulations so long as $\tilde{c} \gtrsim 500\,{\rm  km\,s^{-1}}$. More detailed resolution tests and idealized validation tests of our numerical CR methods, and explicit comparison of the results of two-moment vs.\ one-moment approximations for the CR transport flux solver, are all presented in \citet{chan:2018.cosmicray.fire.gammaray}: none of these presents obvious numerical concerns here (see also Appendix~\ref{sec:additional.tests}).

We stress that extensive numerical tests of almost every other aspect of these simulations (resolution, force softening, hydrodynamic solvers, radiation-hydrodynamics solvers, etc.) are presented in \citet{hopkins:fire2.methods}, \citet{hopkins:sne.methods}, and \citet{hopkins:radiation.methods}, for our ``Hydro+'' models. For more detailed resolution and physics tests of the ``MHD+'' models, we refer to \citet{su:2016.weak.mhd.cond.visc.turbdiff.fx,su:fire.feedback.alters.magnetic.amplification.morphology}, and for extensive numerical validation tests of the MHD, conduction, viscosity, and anisotropic transport solvers we refer to \citet{hopkins:mhd.gizmo,hopkins:cg.mhd.gizmo,hopkins:gizmo.diffusion}.

\vspace{-0.5cm}
\section{Conclusions}
\label{sec:conclusions}

\subsection{Overview}

We present and study a suite of $> 150$ new high-resolution ($\sim 100-10000\,M_{\sun}$, $\sim 1-10\,$pc, $10-100\,$yr, $\sim 10^{3}-10^{4}\,{\rm cm^{-3}}$) FIRE-2 cosmological zoom-in simulations, with explicit treatment of stellar feedback (SNe Types Ia \&\ II, O/B \&\ AGB mass-loss, photo-heating and radiation pressure), magnetic fields, fully-anisotropic conduction and viscosity (accounting for saturation and their limitation by plasma instabilities at high-$\beta$), and cosmic rays. Our CR treatment accounts for injection in SNe shocks; advection, fully-anisotropic streaming and diffusion; and losses from collisional (hadronic+Coulomb), streaming, and adiabatic processes. We systematically survey different aspects of the CR physics and uncertain transport coefficients. We examine the effects of these physics on a range of galaxy properties including: stellar masses, star formation rates and histories, metallicities and abundances, stellar sizes and baryonic mass profiles, dark matter mass profiles, rotation curves, visual morphologies and kinematics of stars and gas, and gas phase distributions. We survey these properties across a suite of simulations spanning all redshifts, and masses at $z\sim 0-10$ ranging from ultra-faint dwarf ($M_{\ast}\sim 10^{4}\,M_{\sun}$, $M_{\rm halo}\sim 10^{9}\,M_{\sun}$) through Milky Way mass. We summarize our conclusions as follows:

\begin{itemize}

\item{We confirm the growing body of work showing that magnetic fields, physical conduction, and viscosity on resolved scales have little effect on any galaxy property studied (Figs.~\ref{fig:cr.demo.m10}-\ref{fig:cr.demo.m12} and \ref{fig:mgal.mhalo}-\ref{fig:kslaw}). These simulations reach higher resolution (sufficient to resolve the Field length in warm and hot gas with $T \gtrsim 2\times10^{5}\,{\rm K}\,(n/0.01\,{\rm cm^{-3}})^{0.4}$), and include more detailed treatment of the effect of plasma instabilities on transport coefficients, compared to our previous work \citep{su:2016.weak.mhd.cond.visc.turbdiff.fx}, but this only serves to reinforce that conclusion. It is of course possible there are important un-resolved effects which could be important via ``sub-grid'' effects (e.g.\ altering the effective cooling rates or stellar initial mass function).} 

\item{Magnetic fields are highly-tangled, at all mass scales we survey, with or without CR physics (see also \citealt{ji:fire.cr.cgm}). Per Fig.~\ref{fig:profile.pressure}, the plasma $\beta \equiv P_{\rm thermal}/P_{\rm magnetic}$ varies enormously in the ISM ($\beta \sim  0.1-100$) but in warm/hot phases is usually large (median $\sim 1-10$ in the diffuse ISM in MW-mass galaxies, and larger $\sim 10-30$ or $\sim 30-300$ in lower-mass $M_{\rm halo} \sim 10^{11}\,M_{\sun}$ and $\sim 10^{10}\,M_{\sun}$ dwarfs, respectively), consistent with many recent studies of amplification \citep[see e.g.][and references therein]{2018MNRAS.479.3343M}. It rises in the CGM with galacto-centric distance (to a median $\sim 100-1000$ at $\gtrsim R_{\rm vir}$, with local variations reaching $\sim 10^{4}$).}

\item{CRs have relatively weak effects on the {\em galaxy} properties studied, in dwarfs with $M_{\rm halo} \ll 10^{11}\,M_{\sun}$ ($M_{\ast} \ll 10^{10}\,M_{\sun}$), for essentially any physical CR parameters considered, once realistic mechanical and radiative feedback are already included (Figs.~\ref{fig:cr.demo.m10}, \ref{fig:mgal.mhalo}, \ref{fig:morph.dwarfs}). This is both predicted and easily understood from basic energetic considerations (\S~\ref{sec:toy}). Previous claims to the contrary have either failed to account for (a) the fact that realistic dwarfs have very low $M_{\ast}/M_{\rm halo}$ (and correspondingly low SFRs, hence SNe and CR energy injection rates), without much ``hot halo'' gas, (b) realistic supernova and radiative feedback which easily overwhelm the effects of CRs in dwarfs, and/or (c) observational constraints from $\gamma$-ray emission in dwarfs which place upper limits on the collisional loss rate and require that $>90-99\%$ of the CR energy escape without hadronic losses (prohibiting low transport speeds). It remains possible CRs modify the CGM in these dwarfs, or modify processes like ram-pressure stripping in dwarf satellites around massive galaxies (not studied in detail here).}

\item{CRs (from SNe) also have relatively weak effects on galaxy properties, at {\em any} mass scale, at high redshifts $z\gtrsim 1-2$ (Fig.~\ref{fig:HiZ}). Although SFRs (and therefore CR production rates) are higher (at a given mass) at high-$z$, the CGM density and ram pressure of dense, cold inflows (carrying most of the mass) is much higher, and so CR pressure is insufficient to strongly suppress those inflows (Fig.~\ref{fig:HiZ.profiles}). Moreover the periods of strongest inflow are often accompanied by dense starbursts within the galaxy where surface densities exceed $\gg 0.1\,{\rm g\,cm^{-3}}$ ($\gg 1000\,M_{\sun}\,{\rm pc^{-2}}$), during which CRs experience strong hadronic losses. Consistent with observed low-redshift starbursts at these densities, they become approximate proton calorimeters.}

\item{CRs {\em can} have significant effects in massive galaxies ($M_{\rm halo} \gtrsim 10^{11}\,M_{\sun}$, $M_{\ast} \gtrsim 10^{10}\,M_{\sun}$) at relatively low redshifts ($z\lesssim 1-2$), reducing their peak star formation rates by as much as factors $\sim 5$ and $z=0$ stellar masses by factors $\sim 2-3$ (Figs.~\ref{fig:cr.demo.m11}, \ref{fig:cr.demo.m12}, \ref{fig:mgal.mhalo}). This in turn significantly reduces the central peak in their rotation curves, and moves the galaxies along the mass-metallicity, Kennicutt-Schmidt, morphology-mass, and other mass-based scaling relations. This maximal effect requires effective diffusivities in the range $\kappa_{\ast} \sim 3\times10^{29}-3\times10^{30}\,{\rm cm^{2}\,s^{-1}}$ (which can arise from a combination of microphysical diffusion and/or streaming).}

\item{In these systems, the CRs primarily operate on the CGM, and have relatively little {\em direct} effect within the galaxies (e.g.\ Figs.~\ref{fig:kslaw}, \ref{fig:phases}, \ref{fig:j.distrib}). Essentially any model where CRs are sufficiently well-trapped in galaxies to, e.g.\ slow down SF directly in relative dense ($>1\,{\rm cm^{-3}}$) disk gas, or launch strong outflows (velocities $\gtrsim V_{c}$) is ruled out observationally and results in excessive CR losses (greatly limiting their net effect). We see essentially all galaxy properties move {\em along}, not off of, standard scaling relations. This is dramatically different from e.g.\ increasing the mechanical energy or rate of SNe. Instead, CR feedback is primarily ``preventive.'' In massive halos at low redshifts, with the appropriate $\kappa_{\ast}$, the CRs establish a quasi-virial-equilibrium pressure profile, which dominates over the gas thermal and magnetic pressure from $\sim 20-200\,$kpc, and supports warm+cool gas ($T\ll 10^{6}\,$K) which would otherwise cool and rain onto the galaxy (Fig.~\ref{fig:profile.pressure} and \citealt{ji:fire.cr.cgm}).}

\item{Given present uncertainties, the most important-yet-uncertain parameter determining the effects of CRs (especially in massive galaxies) is the effective diffusivity $\kappa_{\ast}$. If too low, CRs are trapped and (a) lose energy to collisional processes in dense gas (negating their ability to do work, and directly violating observational constraints from spallation and $\gamma$-ray emission), and (b) cannot propagate to large-enough radii to slow accretion in the outer halo. If $\kappa_{\ast}$ is too high, CRs should simply escape and their equilibrium pressure, even in the outer halo, would be too low to do interesting work. The ``sweet spot'' is approximately an order-of-magnitude in width, and agrees well in the simulations and simple analytic models.}

\item{We show that with $\kappa_{\ast}\sim 3-30\times10^{29}\,{\rm cm^{2}\,s^{-1}}$, cosmological simulations reproduce observed $\gamma$-ray emissivities (and similar constraints from spallation in the MW), at dwarf through Local Group through starburst-galaxy density/star formation rate scales (Fig.~\ref{fig:Lgamma}). This echoes the conclusion from recent non-cosmological simulations \citep{chan:2018.cosmicray.fire.gammaray}, for the first time in cosmological simulations.} 

\item{Cosmic ray ``heating'' of the gas (via energy transfer through collisional or streaming losses) is orders-of-magnitude smaller than gas cooling rates and never important for the gas (Fig.~\ref{fig:profile.heating}). It is however important as a loss mechanism for CRs, especially in dense starburst-type systems (which become proton calorimeters) or the outer halo (where trans-\Alf{ic} streaming can dominate).} 

\item{Around the effective diffusivities of interest, anisotropy and streaming at the \Alf\ speed play second-order roles in CR transport (Fig.~\ref{fig:cr.physics}). Anisotropy does not have radical obvious effects, but does lead to somewhat suppressed diffusion, especially in dense gas where losses can be large (necessitating larger-$\kappa_{\ast}$ to enable escape). Trans-\Alf{ic} streaming is too slow to account for the required $\kappa_{\ast}$ in the ISM and inner CGM, although it can dominate transport outside of some radius $\sim \kappa_{\ast} / v_{A}$ (in the outer halo).} 

\item{We present a simple analytic equilibrium model in \S~\ref{sec:toy}, which is able to at least qualitatively predict essentially all of the relevant CR effects here. Particularly in the mass and $\kappa_{\ast}$ range of greatest interest, this analytic model provides a remarkably accurate description of when CRs become important, the favored values of $\kappa_{\ast}$, and the resulting CR pressure profile and equilibrium gas density profile in the CGM of CR-dominated halos.}

\end{itemize}

\vspace{-0.5cm}
\subsection{Future Work \&\ More Massive Galaxies}

This study -- despite its length -- is far from comprehensive. In future work, we plan to explore a number of properties of these simulations in greater detail, including: their more detailed outflow and gas phase structure, properties of GMCs in the ISM and accretion onto the galactic disk, observable ions in the CGM, properties of resolved satellites around massive galaxies in the modified CR-dominated CGM. There may be more detailed tracers which can distinguish between different models in  greater detail, or areas where MHD/conduction/viscosity/CRs have large effects which we have failed to identify here. Moreover, even assuming our simple CR treatment is a reasonable representation, there are a number of basic physical processes meriting further investigation. For example it is unclear how the non-linear thermal instability operates in a stratified halo supported primarily by a CR pressure gradient (nor is it likely this can be fully resolved in the simulations here). It is not obvious how much of the additional cool gas in the halo in the high-$\kappa_{\ast}$, high-$M_{\ast}$ CR runs owes to ``pure suppression'' of new inflowing material vs.\ CRs ``gently'' (and slowly) re-accelerating or ``lofting'' cool gas which would otherwise recycle in galactic fountains. And it is worth exploring how the virial shock and transition ``out of'' the CR dominated-regime at large radii occur. These and many additional questions clearly merit exploration in idealized, high-resolution numerical experiments.

Future and parallel work will also explore the role of CRs in more massive ($> L_{\ast}$) galaxies. We predict and find in our simulations that the strength of the effects from CRs sourced via SNe scales as $\propto \dot{M}_{\ast} / M_{\rm halo}$, which is maximized around $\sim L_{\ast}$. In very massive halos (in nature), this declines both because (a) $M_{\ast}/M_{\rm halo}$ drops, and (b) star formation is ``quenched.'' As a result, our parallel study in \citet{su:2018.stellar.fb.fails.to.solve.cooling.flow} demonstrates that CRs sourced by SNe (including Ia's) cannot possibly solve the ``cooling flow problem'' and resist excessive cooling and star formation in very massive ($M_{\rm halo} \sim 10^{14}\,M_{\sun}$) halos. And here we show essentially the same at higher redshifts when these halos initially form most of their stars and dense, compact bulges (when their progenitor halo masses are $\lesssim 10^{12-13}\,M_{\sun}$). 

However, in those same massive galaxies, super-massive black holes (SMBHs) and AGN appear to dominate CR production (seen in e.g.\ jets and ``bubbles'') by orders-of-magnitude compared to SNe, and in this paper we {\em only} accounted for CRs produced in SNe. In a companion study \citep{su:2018.forms.of.feedback.vs.cooling.flow} we show that injection of $\sim 10^{43}\,{\rm erg\,s^{-1}}$ in CRs in a $\sim 10^{14}\,M_{\sun}$ halo -- modest for the SMBHs and AGN in those systems, but $\sim 100-1000$ times the production rate from SNe assuming $\sim 10\%$ of the SNe energy goes into CRs -- can have a dramatic effect on the halo gas and cooling flow in these massive  galaxies. Obviously it will be important to revisit these high-redshift halos with an explicit model for AGN feedback.

\vspace{-0.5cm}
\subsection{Fundamental Physical Uncertainties}

We wish to strongly emphasize that our conclusion is only that CRs {\em could} be important to massive galaxies at low redshifts, not that they {\em are} necessarily. We have parameterized our ignorance by adopting a simple two-moment model with fixed parallel diffusivity $\kappa_{\|}$, but the reality is that deep physical uncertainties remain. It is not clear what actual physics determines the ``effective diffusivity'' or transport parameters of CRs in all the regimes here, let alone how these parameters should scale as a function of local plasma properties (e.g.\ magnetic field strength, density, mean-free-path, strength of turbulence, etc.). Although constraints from detailed modeling of the Milky Way CR population and $\gamma$-ray observations of nearby galaxies favor an effective diffusivity $\kappa_{\ast} \gtrsim 10^{29}\,{\rm cm^{2}\,s^{-1}}$, this constraint is (a) only measured in a few $z=0$ galaxies at present, and (b) more importantly, only constrains the effective diffusivity or ``residence time'' in the relatively dense, star-forming galactic disk gas (where the hadronic collision rate, which scales as $n_{\rm cr}\,n_{\rm nucleon}$, is maximized). It is completely plausible to imagine models where $\kappa_{\ast}$ is $\sim 3\times10^{29}\,{\rm cm^{2}\,s^{-1}}$ in this gas, but once CRs reach the CGM at $r\gg 10\,$kpc (where the gas has low-density, very weak magnetic fields $\beta \sim 100-10^{4}$, and  long mean-free paths $\sim 1-100\,$pc), the transport speed rises dramatically and CRs simply ``escape'' rather than forming a high-pressure halo. This is, essentially, the classic ``leaky box'' model. Even if the CRs are confined in the halo on $\sim 100\,$kpc scales, it is not clear whether the local tight-coupling approximation is valid in the tenuous CGM -- i.e.\ can CR ``pressure'' actually be simply ``added'' to to the local gas stress tensor in the MHD equations? Or can CRs ``slip'' or couple with non-negligible lag? Fundamentally, our treatment of CRs as a fluid depends on assuming something about their ability to reach a micro-scale gyrotropic equilibrium distribution function, which may not be valid when scattering rates are low, i.e.\ transport speeds are very large (see e.g.\ \citealt{holcolmb.spitkovsky:saturation.gri.sims}). These are critical questions which could completely alter our conclusions (in the examples above, they would make the effects of CRs much weaker), and may fundamentally require PIC-type simulations that can follow explicit kinetic plasma processes to answer. 

Observationally, direct constraints on CRs in the regimes of interest are unfortunately very limited. Essentially the only such measure outside the MW is the $\gamma$-ray luminosity discussed extensively here. Recall, most of the CR energy/pressure is in $\sim$\,GeV protons. So constraints on the CR electrons (e.g.\ synchrotron), which tend to dissipate their energy far more efficiently and closer to sources, are not particularly illuminating for this specific question, nor are constraints on the high-energy CR population (which has very different diffusivity and contains negligible CR pressure). However, given that the regime of greatest interest is precisely where we predict CRs should have a large effect on the temperature and density distribution of CGM gas around $\sim L_{\ast}$ galaxies -- a topic of tremendous observational progress at present -- it is likely that indirect observational constraints from the CGM will represent, in the near future, the best path towards constraining the CR physics of greatest interest here.

\vspace{-0.7cm}
\acknowledgments 
We thank the anonymous referee for a number of insightful comments and suggestions. Support for PFH and co-authors was provided by an Alfred P. Sloan Research Fellowship, NSF Collaborative Research Grant \#1715847 and CAREER grant \#1455342, and NASA grants NNX15AT06G, JPL 1589742, 17-ATP17-0214. DK
was supported by NSF grant AST-1715101 and the Cottrell
Scholar Award from the Research Corporation for Science
Advancement. CAFG was supported by NSF through grants AST-1517491, AST-1715216, and CAREER award AST-1652522, by NASA through grant 17-ATP17-0067, and by a Cottrell Scholar Award from the Research Corporation for Science Advancement.
Numerical calculations were run on the Caltech compute cluster ``Wheeler,'' allocations from XSEDE TG-AST130039 and TG-AST120025 and PRAC NSF.1713353 supported by the NSF, and NASA HEC SMD-16-7592.\\

\vspace{-0.2cm}
\bibliography{ms_extracted}

\begin{thebibliography}{221}
\expandafter\ifx\csname natexlab\endcsname\relax\def\natexlab#1{#1}\fi

\bibitem[{{Achterberg}(1981)}]{achterberg:1981.cr.streaming}
{Achterberg}, A. 1981, \aap, 98, 161

\bibitem[{{Agertz} \&
  {Kravtsov}(2015)}]{agertz:sf.feedback.multiple.mechanisms}
{Agertz}, O., \& {Kravtsov}, A.~V. 2015, \apj, 804, 18

\bibitem[{{Agertz} {et~al.}(2013){Agertz}, {Kravtsov}, {Leitner}, \&
  {Gnedin}}]{agertz:2013.new.stellar.fb.model}
{Agertz}, O., {Kravtsov}, A.~V., {Leitner}, S.~N., \& {Gnedin}, N.~Y. 2013,
  \apj, 770, 25

\bibitem[{{Amato} \& {Blasi}(2018)}]{2018AdSpR..62.2731A}
{Amato}, E., \& {Blasi}, P. 2018, Advances in Space Research, 62, 2731

\bibitem[{{Angl{\'e}s-Alc{\'a}zar}
  {et~al.}(2017{\natexlab{a}}){Angl{\'e}s-Alc{\'a}zar}, {Faucher-Gigu{\`e}re},
  {Kere{\v s}}, {Hopkins}, {Quataert}, \&
  {Murray}}]{angles.alcazar:particle.tracking.fire.baryon.cycle.intergalactic.transfer}
{Angl{\'e}s-Alc{\'a}zar}, D., {Faucher-Gigu{\`e}re}, C.-A., {Kere{\v s}}, D.,
  {Hopkins}, P.~F., {Quataert}, E., \& {Murray}, N. 2017{\natexlab{a}}, \mnras,
  470, 4698

\bibitem[{{Angl{\'e}s-Alc{\'a}zar}
  {et~al.}(2017{\natexlab{b}}){Angl{\'e}s-Alc{\'a}zar}, {Faucher-Gigu{\`e}re},
  {Quataert}, {Hopkins}, {Feldmann}, {Torrey}, {Wetzel}, \& {Kere{\v
  s}}}]{daa:BHs.on.FIRE}
{Angl{\'e}s-Alc{\'a}zar}, D., {Faucher-Gigu{\`e}re}, C.-A., {Quataert}, E.,
  {Hopkins}, P.~F., {Feldmann}, R., {Torrey}, P., {Wetzel}, A., \& {Kere{\v
  s}}, D. 2017{\natexlab{b}}, \mnras, 472, L109

\bibitem[{{Armillotta} {et~al.}(2017){Armillotta}, {Fraternali}, {Werk},
  {Prochaska}, \& {Marinacci}}]{2017MNRAS.470..114A}
{Armillotta}, L., {Fraternali}, F., {Werk}, J.~K., {Prochaska}, J.~X., \&
  {Marinacci}, F. 2017, \mnras, 470, 114

\bibitem[{{Banda-Barrag{\'a}n} {et~al.}(2018){Banda-Barrag{\'a}n}, {Federrath},
  {Crocker}, \& {Bicknell}}]{2018MNRAS.473.3454B}
{Banda-Barrag{\'a}n}, W.~E., {Federrath}, C., {Crocker}, R.~M., \& {Bicknell},
  G.~V. 2018, \mnras, 473, 3454

\bibitem[{{Beck} {et~al.}(2012){Beck}, {Lesch}, {Dolag}, {Kotarba}, {Geng}, \&
  {Stasyszyn}}]{2012MNRAS.422.2152B}
{Beck}, A.~M., {Lesch}, H., {Dolag}, K., {Kotarba}, H., {Geng}, A., \&
  {Stasyszyn}, F.~A. 2012, \mnras, 422, 2152

\bibitem[{{Beck}(2009)}]{2009ASTRA...5...43B}
{Beck}, R. 2009, Astrophysics and Space Sciences Transactions, 5, 43

\bibitem[{{Beck} {et~al.}(1996){Beck}, {Brandenburg}, {Moss}, {Shukurov}, \&
  {Sokoloff}}]{1996ARA&A..34..155B}
{Beck}, R., {Brandenburg}, A., {Moss}, D., {Shukurov}, A., \& {Sokoloff}, D.
  1996, \araa, 34, 155

\bibitem[{Bell(2004)}]{Bell.cosmic.rays}
Bell, A.~R. 2004, \mnras, 353, 550

\bibitem[{{Bigiel} {et~al.}(2008){Bigiel}, {Leroy}, {Walter}, {Brinks}, {de
  Blok}, {Madore}, \& {Thornley}}]{bigiel:2008.mol.kennicutt.on.sub.kpc.scales}
{Bigiel}, F., {Leroy}, A., {Walter}, F., {Brinks}, E., {de Blok}, W.~J.~G.,
  {Madore}, B., \& {Thornley}, M.~D. 2008, \aj, 136, 2846

\bibitem[{{Blasi} \& {Amato}(2012)}]{blasi:cr.propagation.constraints}
{Blasi}, P., \& {Amato}, E. 2012, Journal of Cosmology and Astroparticle
  Physics, 1, 010

\bibitem[{{Bonaca} {et~al.}(2017){Bonaca}, {Conroy}, {Wetzel}, {Hopkins}, \&
  {Kere{\v s}}}]{bonaca:gaia.structure.vs.fire}
{Bonaca}, A., {Conroy}, C., {Wetzel}, A., {Hopkins}, P.~F., \& {Kere{\v s}}, D.
  2017, \apj, 845, 101

\bibitem[{{Booth} {et~al.}(2013){Booth}, {Agertz}, {Kravtsov}, \&
  {Gnedin}}]{Boot13}
{Booth}, C.~M., {Agertz}, O., {Kravtsov}, A.~V., \& {Gnedin}, N.~Y. 2013,
  \apjl, 777, L16

\bibitem[{{Boulares} \& {Cox}(1990)}]{Boul90}
{Boulares}, A., \& {Cox}, D.~P. 1990, \apj, 365, 544

\bibitem[{{Braginskii}(1965)}]{braginskii:viscosity}
{Braginskii}, S.~I. 1965, Reviews of Plasma Physics, 1, 205

\bibitem[{{Breitschwerdt} {et~al.}(1991){Breitschwerdt}, {McKenzie}, \&
  {Voelk}}]{Brei91}
{Breitschwerdt}, D., {McKenzie}, J.~F., \& {Voelk}, H.~J. 1991, \aap, 245, 79

\bibitem[{{Breitschwerdt} {et~al.}(1993){Breitschwerdt}, {McKenzie}, \&
  {Voelk}}]{Brei93}
---. 1993, \aap, 269, 54

\bibitem[{{Brook} {et~al.}(2014){Brook}, {Di Cintio}, {Knebe}, {Gottl{\"o}ber},
  {Hoffman}, {Yepes}, \& {Garrison-Kimmel}}]{brook:2014.mgal.mhalo.local.group}
{Brook}, C.~B., {Di Cintio}, A., {Knebe}, A., {Gottl{\"o}ber}, S., {Hoffman},
  Y., {Yepes}, G., \& {Garrison-Kimmel}, S. 2014, \apjl, 784, L14

\bibitem[{{Br{\"u}ggen} \& {Scannapieco}(2016)}]{2016ApJ...822...31B}
{Br{\"u}ggen}, M., \& {Scannapieco}, E. 2016, \apj, 822, 31

\bibitem[{{Bryan} \& {Norman}(1998)}]{bryan.norman:1998.mvir.definition}
{Bryan}, G.~L., \& {Norman}, M.~L. 1998, \apj, 495, 80

\bibitem[{{Butsky} \& {Quinn}(2018)}]{Buts18}
{Butsky}, I., \& {Quinn}, T.~R. 2018, ArXiv e-prints

\bibitem[{{Chan} {et~al.}(2018){Chan}, {Keres}, {Hopkins}, {Quataert}, {Su},
  {Hayward}, \& {Faucher-Giguere}}]{chan:2018.cosmicray.fire.gammaray}
{Chan}, T.~K., {Keres}, D., {Hopkins}, P.~F., {Quataert}, E., {Su}, K.~Y.,
  {Hayward}, C.~C., \& {Faucher-Giguere}, C.~A. 2018, arXiv e-prints,
  arXiv:1812.10496

\bibitem[{{Chan} {et~al.}(2015){Chan}, {Kere{\v s}}, {O{\~n}orbe}, {Hopkins},
  {Muratov}, {Faucher-Gigu{\`e}re}, \& {Quataert}}]{chan:fire.dwarf.cusps}
{Chan}, T.~K., {Kere{\v s}}, D., {O{\~n}orbe}, J., {Hopkins}, P.~F., {Muratov},
  A.~L., {Faucher-Gigu{\`e}re}, C.-A., \& {Quataert}, E. 2015, \mnras, 454,
  2981

\bibitem[{{Chen} {et~al.}(2016){Chen}, {Bryan}, \& {Salem}}]{Chen16}
{Chen}, J., {Bryan}, G.~L., \& {Salem}, M. 2016, \mnras, 460, 3335

\bibitem[{{Choi} \& {Stone}(2012)}]{2012ApJ...747...86C}
{Choi}, E., \& {Stone}, J.~M. 2012, \apj, 747, 86

\bibitem[{{Colbrook} {et~al.}(2017){Colbrook}, {Ma}, {Hopkins}, \&
  {Squire}}]{colbrook:passive.scalar.scalings}
{Colbrook}, M.~J., {Ma}, X., {Hopkins}, P.~F., \& {Squire}, J. 2017, \mnras,
  467, 2421

\bibitem[{Cole {et~al.}(2000)Cole, Lacey, Baugh, \&
  Frenk}]{cole:durham.sam.initial}
Cole, S., Lacey, C.~G., Baugh, C.~M., \& Frenk, C.~S. 2000, \mnras, 319, 168

\bibitem[{{Courteau} {et~al.}(2007){Courteau}, {Dutton}, {van den Bosch},
  {MacArthur}, {Dekel}, {McIntosh}, \& {Dale}}]{courteau:disk.scalings}
{Courteau}, S., {Dutton}, A.~A., {van den Bosch}, F.~C., {MacArthur}, L.~A.,
  {Dekel}, A., {McIntosh}, D.~H., \& {Dale}, D.~A. 2007, \apj, 671, 203

\bibitem[{{Cowie} \& {McKee}(1977)}]{cowie:1977.evaporation}
{Cowie}, L.~L., \& {McKee}, C.~F. 1977, \apj, 211, 135

\bibitem[{{Crutcher} {et~al.}(2010){Crutcher}, {Wandelt}, {Heiles},
  {Falgarone}, \& {Troland}}]{crutcher:cloud.b.fields}
{Crutcher}, R.~M., {Wandelt}, B., {Heiles}, C., {Falgarone}, E., \& {Troland},
  T.~H. 2010, \apj, 725, 466

\bibitem[{{Cummings} {et~al.}(2016){Cummings}, {Stone}, {Heikkila}, {Lal},
  {Webber}, {J{\'o}hannesson}, {Moskalenko}, {Orlando}, \&
  {Porter}}]{2016ApJ...831...18C}
{Cummings}, A.~C., {et~al.} 2016, \apj, 831, 18

\bibitem[{{Dobbs} {et~al.}(2011){Dobbs}, {Burkert}, \&
  {Pringle}}]{dobbs:2011.why.gmcs.unbound}
{Dobbs}, C.~L., {Burkert}, A., \& {Pringle}, J.~E. 2011, \mnras, 413, 528

\bibitem[{{Dolag} {et~al.}(2008){Dolag}, {Bykov}, \&
  {Diaferio}}]{2008SSRv..134..311D}
{Dolag}, K., {Bykov}, A.~M., \& {Diaferio}, A. 2008, \ssr, 134, 311

\bibitem[{{Donnert} {et~al.}(2018){Donnert}, {Vazza}, {Br{\"u}ggen}, \&
  {ZuHone}}]{2018SSRv..214..122D}
{Donnert}, J., {Vazza}, F., {Br{\"u}ggen}, M., \& {ZuHone}, J. 2018, \ssr, 214,
  122

\bibitem[{{Dorfi} \& {Breitschwerdt}(2012)}]{Dorf12}
{Dorfi}, E.~A., \& {Breitschwerdt}, D. 2012, \aap, 540, A77

\bibitem[{{Dubois} \& {Teyssier}(2008)}]{2008A&A...482L..13D}
{Dubois}, Y., \& {Teyssier}, R. 2008, \aap, 482, L13

\bibitem[{{El-Badry} {et~al.}(2018{\natexlab{a}}){El-Badry}, {Bradford},
  {Quataert}, {Geha}, {Boylan-Kolchin}, {Weisz}, {Wetzel}, {Hopkins}, {Chan},
  {Fitts}, {Kere{\v s}}, \&
  {Faucher-Gigu{\`e}re}}]{elbadry:HI.obs.gal.kinematics}
{El-Badry}, K., {et~al.} 2018{\natexlab{a}}, \mnras, 477, 1536

\bibitem[{{El-Badry} {et~al.}(2018{\natexlab{b}}){El-Badry}, {Quataert},
  {Wetzel}, {Hopkins}, {Weisz}, {Chan}, {Fitts}, {Boylan-Kolchin}, {Kere{\v
  s}}, {Faucher-Gigu{\`e}re}, \&
  {Garrison-Kimmel}}]{elbadry:fire.morph.momentum}
---. 2018{\natexlab{b}}, \mnras, 473, 1930

\bibitem[{{El-Badry} {et~al.}(2018{\natexlab{c}}){El-Badry}, {Bland-Hawthorn},
  {Wetzel}, {Quataert}, {Weisz}, {Boylan-Kolchin}, {Hopkins},
  {Faucher-Gigu{\`e}re}, {Kere{\v s}}, \&
  {Garrison-Kimmel}}]{elbadry:most.ancient.mw.halo.stars.kicked}
---. 2018{\natexlab{c}}, \mnras, 480, 652

\bibitem[{{En{\ss}lin} {et~al.}(2011){En{\ss}lin}, {Pfrommer}, {Miniati}, \&
  {Subramanian}}]{ensslin:2011.cr.transport.clusters}
{En{\ss}lin}, T., {Pfrommer}, C., {Miniati}, F., \& {Subramanian}, K. 2011,
  \aap, 527, A99

\bibitem[{{En{\ss}lin} {et~al.}(2007){En{\ss}lin}, {Pfrommer}, {Springel}, \&
  {Jubelgas}}]{Enss07}
{En{\ss}lin}, T.~A., {Pfrommer}, C., {Springel}, V., \& {Jubelgas}, M. 2007,
  \aap, 473, 41

\bibitem[{{Escala} {et~al.}(2018){Escala}, {Wetzel}, {Kirby}, {Hopkins}, {Ma},
  {Wheeler}, {Kere{\v s}}, {Faucher-Gigu{\`e}re}, \&
  {Quataert}}]{escala:turbulent.metal.diffusion.fire}
{Escala}, I., {et~al.} 2018, \mnras, 474, 2194

\bibitem[{{Everett} {et~al.}(2008){Everett}, {Zweibel}, {Benjamin}, {McCammon},
  {Rocks}, \& {Gallagher}}]{Ever08}
{Everett}, J.~E., {Zweibel}, E.~G., {Benjamin}, R.~A., {McCammon}, D., {Rocks},
  L., \& {Gallagher}, III, J.~S. 2008, \apj, 674, 258

\bibitem[{{Evoli} {et~al.}(2017){Evoli}, {Gaggero}, {Vittino}, {Di Bernardo},
  {Di Mauro}, {Ligorini}, {Ullio}, \& {Grasso}}]{evoli:dragon2.cr.prop}
{Evoli}, C., {Gaggero}, D., {Vittino}, A., {Di Bernardo}, G., {Di Mauro}, M.,
  {Ligorini}, A., {Ullio}, P., \& {Grasso}, D. 2017, Journal of Cosmology and
  Astroparticle Physics, 2, 015

\bibitem[{{Farber} {et~al.}(2018{\natexlab{a}}){Farber}, {Ruszkowski}, {Yang},
  \& {Zweibel}}]{Farb18}
{Farber}, R., {Ruszkowski}, M., {Yang}, H.-Y.~K., \& {Zweibel}, E.~G.
  2018{\natexlab{a}}, \apj, 856, 112

\bibitem[{{Farber} {et~al.}(2018{\natexlab{b}}){Farber}, {Ruszkowski}, {Yang},
  \& {Zweibel}}]{farber:decoupled.crs.in.neutral.gas}
---. 2018{\natexlab{b}}, \apj, 856, 112

\bibitem[{{Faucher-Giguere} {et~al.}(2015){Faucher-Giguere}, {Hopkins},
  {Keres}, {Muratov}, {Quataert}, \&
  {Murray}}]{faucher-giguere:2014.fire.neutral.hydrogen.absorption}
{Faucher-Giguere}, C.-A., {Hopkins}, P.~F., {Keres}, D., {Muratov}, A.~L.,
  {Quataert}, E., \& {Murray}, N. 2015, \mnras, 449, 987

\bibitem[{{Faucher-Gigu{\`e}re} {et~al.}(2013){Faucher-Gigu{\`e}re},
  {Quataert}, \& {Hopkins}}]{cafg:sf.fb.reg.kslaw}
{Faucher-Gigu{\`e}re}, C.-A., {Quataert}, E., \& {Hopkins}, P.~F. 2013, \mnras,
  433, 1970

\bibitem[{{Federrath} {et~al.}(2014){Federrath}, {Schober}, {Bovino}, \&
  {Schleicher}}]{federrath:supersonic.turb.dynamo}
{Federrath}, C., {Schober}, J., {Bovino}, S., \& {Schleicher}, D.~R.~G. 2014,
  \apjl, 797, L19

\bibitem[{{Feldmann} {et~al.}(2016){Feldmann}, {Hopkins}, {Quataert},
  {Faucher-Gigu{\`e}re}, \& {Kere{\v
  s}}}]{feldmann.2016:quiescent.massive.highz.galaxies.fire}
{Feldmann}, R., {Hopkins}, P.~F., {Quataert}, E., {Faucher-Gigu{\`e}re}, C.-A.,
  \& {Kere{\v s}}, D. 2016, \mnras, 458, L14

\bibitem[{{Field}(1965)}]{field:length}
{Field}, G.~B. 1965, \apj, 142, 531

\bibitem[{{Fu} {et~al.}(2017){Fu}, {Xia}, \&
  {Shen}}]{fu:2017.m33.revised.cr.upper.limit}
{Fu}, L., {Xia}, Z.~Q., \& {Shen}, Z.~Q. 2017, \mnras, 471, 1737

\bibitem[{{Gaggero} {et~al.}(2015){Gaggero}, {Urbano}, {Valli}, \&
  {Ullio}}]{gaggero:2015.cr.diffusion.coefficient}
{Gaggero}, D., {Urbano}, A., {Valli}, M., \& {Ullio}, P. 2015, \prd, 91, 083012

\bibitem[{{Gallazzi} {et~al.}(2005){Gallazzi}, {Charlot}, {Brinchmann},
  {White}, \& {Tremonti}}]{gallazzi:ssps}
{Gallazzi}, A., {Charlot}, S., {Brinchmann}, J., {White}, S.~D.~M., \&
  {Tremonti}, C.~A. 2005, \mnras, 362, 41

\bibitem[{{Garrison-Kimmel} {et~al.}(2017{\natexlab{a}}){Garrison-Kimmel},
  {Bullock}, {Boylan-Kolchin}, \&
  {Bardwell}}]{sgk:2016.mgal.mhalo.lowmass.scatter}
{Garrison-Kimmel}, S., {Bullock}, J.~S., {Boylan-Kolchin}, M., \& {Bardwell},
  E. 2017{\natexlab{a}}, \mnras, 464, 3108

\bibitem[{{Garrison-Kimmel} {et~al.}(2017{\natexlab{b}}){Garrison-Kimmel},
  {Hopkins}, {Wetzel}, {El-Badry}, {Sanderson}, {Bullock}, {Ma}, {van de
  Voort}, {Hafen}, {Faucher-Gigu{\`e}re}, {Hayward}, {Quataert}, {Keres}, \&
  {Boylan-Kolchin}}]{garrisonkimmel:fire.morphologies.vs.dm}
{Garrison-Kimmel}, S., {et~al.} 2017{\natexlab{b}}, \mnras, in press
  arXiv:1712.03966

\bibitem[{{Garrison-Kimmel} {et~al.}(2018){Garrison-Kimmel}, {Hopkins},
  {Wetzel}, {Bullock}, {Boylan-Kolchin}, {Keres}, {Faucher-Giguere},
  {El-Badry}, {Lamberts}, {Quataert}, \&
  {Sanderson}}]{garrisonkimmel:local.group.fire.tbtf.missing.satellites}
---. 2018, \mnras, submitted, arXiv:1806.04143

\bibitem[{{Genel} {et~al.}(2018){Genel}, {Bryan}, {Springel}, {Hernquist},
  {Nelson}, {Pillepich}, {Weinberger}, {Pakmor}, {Marinacci}, \&
  {Vogelsberger}}]{genel:stochastic.gal.form.fx}
{Genel}, S., {et~al.} 2018, \apj, in press, arXiv:1807.07084

\bibitem[{{Genzel} {et~al.}(2010)}]{genzel:2010.ks.law}
{Genzel}, R., {et~al.} 2010, \mnras, 407, 2091

\bibitem[{{Ginzburg} \& {Ptuskin}(1985)}]{Ginz85}
{Ginzburg}, V.~L., \& {Ptuskin}, V.~S. 1985, Astrophysics and Space Physics
  Reviews, 4, 161

\bibitem[{{Girichidis} {et~al.}(2018){Girichidis}, {Naab}, {Hanasz}, \&
  {Walch}}]{Giri18}
{Girichidis}, P., {Naab}, T., {Hanasz}, M., \& {Walch}, S. 2018, ArXiv e-prints

\bibitem[{{Girichidis} {et~al.}(2016){Girichidis}, {Naab}, {Walch}, {Hanasz},
  {Mac Low}, {Ostriker}, {Gatto}, {Peters}, {W{\"u}nsch}, {Glover}, {Klessen},
  {Clark}, \& {Baczynski}}]{Giri16}
{Girichidis}, P., {et~al.} 2016, \apjl, 816, L19

\bibitem[{{Grenier} {et~al.}(2015){Grenier}, {Black}, \& {Strong}}]{Gren15}
{Grenier}, I.~A., {Black}, J.~H., \& {Strong}, A.~W. 2015, \araa, 53, 199

\bibitem[{{Griffin} {et~al.}(2016){Griffin}, {Dai}, \&
  {Thompson}}]{griffin:2016.arp220.detection.gammarays}
{Griffin}, R.~D., {Dai}, X., \& {Thompson}, T.~A. 2016, \apjl, 823, L17

\bibitem[{{Guo} \& {Oh}(2008)}]{guo.oh:cosmic.rays}
{Guo}, F., \& {Oh}, S.~P. 2008, \mnras, 384, 251

\bibitem[{{Guo} {et~al.}(2016){Guo}, {Tian}, \& {Jin}}]{2016ApJ...819...54G}
{Guo}, Y.-Q., {Tian}, Z., \& {Jin}, C. 2016, \apj, 819, 54

\bibitem[{{Guszejnov} {et~al.}(2019){Guszejnov}, {Grudi{\'c}}, {Offner},
  {Boylan-Kolchin}, {Faucher-Gig{\`e}re}, {Wetzel}, {Benincasa}, \&
  {Loebman}}]{guszejnov:fire.gmc.props.vs.z}
{Guszejnov}, D., {Grudi{\'c}}, M.~Y., {Offner}, S. S.~R., {Boylan-Kolchin}, M.,
  {Faucher-Gig{\`e}re}, C.-A., {Wetzel}, A., {Benincasa}, S.~M., \& {Loebman},
  S. 2019, arXiv e-prints, arXiv:1910.01163

\bibitem[{{Hahn} \& {Abel}(2011)}]{hahn:2011.music.code.paper}
{Hahn}, O., \& {Abel}, T. 2011, \mnras, 415, 2101

\bibitem[{{Han}(2017)}]{2017ARA&A..55..111H}
{Han}, J.~L. 2017, \araa, 55, 111

\bibitem[{{Hanasz} {et~al.}(2013){Hanasz}, {Lesch}, {Naab}, {Gawryszczak},
  {Kowalik}, \& {W{\'o}lta{\'n}ski}}]{Hana13}
{Hanasz}, M., {Lesch}, H., {Naab}, T., {Gawryszczak}, A., {Kowalik}, K., \&
  {W{\'o}lta{\'n}ski}, D. 2013, \apjl, 777, L38

\bibitem[{{Hanasz} {et~al.}(2009){Hanasz}, {W{\'o}lta{\'n}ski}, \&
  {Kowalik}}]{Hana09}
{Hanasz}, M., {W{\'o}lta{\'n}ski}, D., \& {Kowalik}, K. 2009, \apjl, 706, L155

\bibitem[{{Harper-Clark} \& {Murray}(2011)}]{harper-clark:2011.gmc.sims}
{Harper-Clark}, E., \& {Murray}, N. 2011, in Computational Star Formation;
  Cambridge University Press, ed. {J.~Alves, B.~G.~Elmegreen, J.~M.~Girart \&
  V.~Trimble}, Vol. 270 (Cambridge, UK: Cambridge University Press), 235--238

\bibitem[{{Haverkorn}(2015)}]{2015ASSL..407..483H}
{Haverkorn}, M. 2015, in Astrophysics and Space Science Library, Vol. 407,
  Magnetic Fields in Diffuse Media, ed. A.~{Lazarian}, E.~M. {de Gouveia Dal
  Pino}, \& C.~{Melioli}, 483

\bibitem[{{Heckman} {et~al.}(2000){Heckman}, {Lehnert}, {Strickland}, \&
  {Armus}}]{heckman:superwind.abs.kinematics}
{Heckman}, T.~M., {Lehnert}, M.~D., {Strickland}, D.~K., \& {Armus}, L. 2000,
  \apjs, 129, 493

\bibitem[{{Holcomb} \&
  {Spitkovsky}(2019)}]{holcolmb.spitkovsky:saturation.gri.sims}
{Holcomb}, C., \& {Spitkovsky}, A. 2019, \apj, 882, 3

\bibitem[{{Holman} {et~al.}(1979){Holman}, {Ionson}, \&
  {Scott}}]{holman:1979.cr.streaming.speed}
{Holman}, G.~D., {Ionson}, J.~A., \& {Scott}, J.~S. 1979, \apj, 228, 576

\bibitem[{{Hopkins}(2015)}]{hopkins:gizmo}
{Hopkins}, P.~F. 2015, \mnras, 450, 53

\bibitem[{{Hopkins}(2016)}]{hopkins:cg.mhd.gizmo}
---. 2016, \mnras, 462, 576

\bibitem[{{Hopkins}(2017)}]{hopkins:gizmo.diffusion}
---. 2017, \mnras, 466, 3387

\bibitem[{{Hopkins} \&
  {Conroy}(2017)}]{hopkins.conroy.2015:metal.poor.star.abundances.dust}
{Hopkins}, P.~F., \& {Conroy}, C. 2017, \apj, 835, 154

\bibitem[{{Hopkins} {et~al.}(2013{\natexlab{a}}){Hopkins}, {Cox}, {Hernquist},
  {Narayanan}, {Hayward}, \& {Murray}}]{hopkins:stellar.fb.mergers}
{Hopkins}, P.~F., {Cox}, T.~J., {Hernquist}, L., {Narayanan}, D., {Hayward},
  C.~C., \& {Murray}, N. 2013{\natexlab{a}}, \mnras, 430, 1901

\bibitem[{{Hopkins} {et~al.}(2018{\natexlab{a}}){Hopkins}, {Grudic}, {Wetzel},
  {Keres}, {Gaucher-Giguere}, {Ma}, {Murray}, \&
  {Butcher}}]{hopkins:radiation.methods}
{Hopkins}, P.~F., {Grudic}, M.~Y., {Wetzel}, A.~R., {Keres}, D.,
  {Gaucher-Giguere}, C.-A., {Ma}, X., {Murray}, N., \& {Butcher}, N.
  2018{\natexlab{a}}, \mnras, submitted, arXiv:1811.12462

\bibitem[{{Hopkins} {et~al.}(2014){Hopkins}, {Keres}, {Onorbe},
  {Faucher-Giguere}, {Quataert}, {Murray}, \& {Bullock}}]{hopkins:2013.fire}
{Hopkins}, P.~F., {Keres}, D., {Onorbe}, J., {Faucher-Giguere}, C.-A.,
  {Quataert}, E., {Murray}, N., \& {Bullock}, J.~S. 2014, \mnras, 445, 581

\bibitem[{{Hopkins} {et~al.}(2013{\natexlab{b}}){Hopkins}, {Kere{\v s}},
  {Murray}, {Hernquist}, {Narayanan}, \&
  {Hayward}}]{hopkins:2013.merger.sb.fb.winds}
{Hopkins}, P.~F., {Kere{\v s}}, D., {Murray}, N., {Hernquist}, L., {Narayanan},
  D., \& {Hayward}, C.~C. 2013{\natexlab{b}}, \mnras, 433, 78

\bibitem[{{Hopkins} {et~al.}(2013{\natexlab{c}}){Hopkins}, {Narayanan}, \&
  {Murray}}]{hopkins:virial.sf}
{Hopkins}, P.~F., {Narayanan}, D., \& {Murray}, N. 2013{\natexlab{c}}, \mnras,
  432, 2647

\bibitem[{{Hopkins} {et~al.}(2011){Hopkins}, {Quataert}, \&
  {Murray}}]{hopkins:rad.pressure.sf.fb}
{Hopkins}, P.~F., {Quataert}, E., \& {Murray}, N. 2011, \mnras, 417, 950

\bibitem[{{Hopkins} {et~al.}(2012){Hopkins}, {Quataert}, \&
  {Murray}}]{hopkins:fb.ism.prop}
---. 2012, \mnras, 421, 3488

\bibitem[{{Hopkins} \& {Raives}(2016)}]{hopkins:mhd.gizmo}
{Hopkins}, P.~F., \& {Raives}, M.~J. 2016, \mnras, 455, 51

\bibitem[{{Hopkins} {et~al.}(2018{\natexlab{b}}){Hopkins}, {Wetzel}, {Kere{\v
  s}}, {Faucher-Gigu{\`e}re}, {Quataert}, {Boylan-Kolchin}, {Murray},
  {Hayward}, {Garrison-Kimmel}, {Hummels}, {Feldmann}, {Torrey}, {Ma},
  {Angl{\'e}s-Alc{\'a}zar}, {Su}, {Orr}, {Schmitz}, {Escala}, {Sanderson},
  {Grudi{\'c}}, {Hafen}, {Kim}, {Fitts}, {Bullock}, {Wheeler}, {Chan},
  {Elbert}, \& {Narayanan}}]{hopkins:fire2.methods}
{Hopkins}, P.~F., {et~al.} 2018{\natexlab{b}}, \mnras, 480, 800

\bibitem[{{Hopkins} {et~al.}(2018{\natexlab{c}}){Hopkins}, {Wetzel}, {Kere{\v
  s}}, {Faucher-Gigu{\`e}re}, {Quataert}, {Boylan-Kolchin}, {Murray},
  {Hayward}, \& {El-Badry}}]{hopkins:sne.methods}
---. 2018{\natexlab{c}}, \mnras, 477, 1578

\bibitem[{{Ipavich}(1975)}]{Ipav75}
{Ipavich}, F.~M. 1975, \apj, 196, 107

\bibitem[{{Jacob} {et~al.}(2018){Jacob}, {Pakmor}, {Simpson}, {Springel}, \&
  {Pfrommer}}]{Jaco18}
{Jacob}, S., {Pakmor}, R., {Simpson}, C.~M., {Springel}, V., \& {Pfrommer}, C.
  2018, \mnras, 475, 570

\bibitem[{{Ji} {et~al.}(2019){Ji}, {Chan}, {Hummels}, {Hopkins}, {Stern},
  {Kere{\v{s}}}, {Quataert}, {Faucher-Gigu{\`e}re}, \&
  {Murray}}]{ji:fire.cr.cgm}
{Ji}, S., {et~al.} 2019, \mnras, in press, arXiv:1909.00003, arXiv:1909.00003

\bibitem[{{Jiang} \& {Oh}(2018)}]{jiang.oh:2018.cr.transport.m1.scheme}
{Jiang}, Y.-F., \& {Oh}, S.~P. 2018, \apj, 854, 5

\bibitem[{{J{\'o}hannesson} {et~al.}(2016){J{\'o}hannesson}, {Ruiz de Austri},
  {Vincent}, {Moskalenko}, {Orlando}, {Porter}, {Strong}, {Trotta}, {Feroz},
  {Graff}, \& {Hobson}}]{2016ApJ...824...16J}
{J{\'o}hannesson}, G., {et~al.} 2016, \apj, 824, 16

\bibitem[{{Jubelgas} {et~al.}(2008){Jubelgas}, {Springel}, {En{\ss}lin}, \&
  {Pfrommer}}]{jubelgas:2008.cosmic.ray.outflows}
{Jubelgas}, M., {Springel}, V., {En{\ss}lin}, T., \& {Pfrommer}, C. 2008, \aap,
  481, 33

\bibitem[{{Jun} \& {Jones}(1999)}]{1999ApJ...511..774J}
{Jun}, B.-I., \& {Jones}, T.~W. 1999, \apj, 511, 774

\bibitem[{{Jun} \& {Norman}(1996{\natexlab{a}})}]{1996ApJ...472..245J}
{Jun}, B.-I., \& {Norman}, M.~L. 1996{\natexlab{a}}, \apj, 472, 245

\bibitem[{{Jun} \& {Norman}(1996{\natexlab{b}})}]{1996ApJ...465..800J}
---. 1996{\natexlab{b}}, \apj, 465, 800

\bibitem[{{Jun} {et~al.}(1995){Jun}, {Norman}, \&
  {Stone}}]{1995ApJ...453..332J}
{Jun}, B.-I., {Norman}, M.~L., \& {Stone}, J.~M. 1995, \apj, 453, 332

\bibitem[{{Kannan} {et~al.}(2014){Kannan}, {Stinson}, {Macci{\`o}}, {Brook},
  {Weinmann}, {Wadsley}, \&
  {Couchman}}]{kannan:2013.early.fb.gives.good.highz.mgal.mhalo}
{Kannan}, R., {Stinson}, G.~S., {Macci{\`o}}, A.~V., {Brook}, C., {Weinmann},
  S.~M., {Wadsley}, J., \& {Couchman}, H.~M.~P. 2014, \mnras, 437, 3529

\bibitem[{{Katz} {et~al.}(2019){Katz}, {Martin-Alvarez}, {Devriendt}, {Slyz},
  \& {Kimm}}]{2019MNRAS.484.2620K}
{Katz}, H., {Martin-Alvarez}, S., {Devriendt}, J., {Slyz}, A., \& {Kimm}, T.
  2019, \mnras, 484, 2620

\bibitem[{{Katz} {et~al.}(1996){Katz}, {Weinberg}, \&
  {Hernquist}}]{katz:treesph}
{Katz}, N., {Weinberg}, D.~H., \& {Hernquist}, L. 1996, \apjs, 105, 19

\bibitem[{{Keller} {et~al.}(2018){Keller}, {Wadsley}, {Wang}, \&
  {Kruijssen}}]{keller:stochastic.gal.form.fx}
{Keller}, B.~W., {Wadsley}, J.~W., {Wang}, L., \& {Kruijssen}, J.~M.~D. 2018,
  \mnras, in press, arXiv:1803.05445

\bibitem[{{Kennicutt}(1998)}]{kennicutt98}
{Kennicutt}, Jr., R.~C. 1998, \apj, 498, 541

\bibitem[{{Kennicutt} {et~al.}(2007)}]{kennicutt:m51.resolved.sfr}
{Kennicutt}, Jr., R.~C., {et~al.} 2007, \apj, 671, 333

\bibitem[{{Kere{\v s}} {et~al.}(2009){Kere{\v s}}, {Katz}, {Dav{\'e}},
  {Fardal}, \& {Weinberg}}]{keres:fb.constraints.from.cosmo.sims}
{Kere{\v s}}, D., {Katz}, N., {Dav{\'e}}, R., {Fardal}, M., \& {Weinberg},
  D.~H. 2009, \mnras, 396, 2332

\bibitem[{{Kere{\v s}} {et~al.}(2005){Kere{\v s}}, {Katz}, {Weinberg}, \&
  {Dav{\'e}}}]{keres:hot.halos}
{Kere{\v s}}, D., {Katz}, N., {Weinberg}, D.~H., \& {Dav{\'e}}, R. 2005,
  \mnras, 363, 2

\bibitem[{{Kim} \& {Ostriker}(2015)}]{2015ApJ...815...67K}
{Kim}, C.-G., \& {Ostriker}, E.~C. 2015, \apj, 815, 67

\bibitem[{{Kirby} {et~al.}(2013){Kirby}, {Cohen}, {Guhathakurta}, {Cheng},
  {Bullock}, \& {Gallazzi}}]{kirby:2013.mass.metallicity}
{Kirby}, E.~N., {Cohen}, J.~G., {Guhathakurta}, P., {Cheng}, L., {Bullock},
  J.~S., \& {Gallazzi}, A. 2013, \apj, 779, 102

\bibitem[{{Komarov} {et~al.}(2018){Komarov}, {Schekochihin}, {Churazov}, \&
  {Spitkovsky}}]{komarov:whistler.instability.limiting.transport}
{Komarov}, S., {Schekochihin}, A.~A., {Churazov}, E., \& {Spitkovsky}, A. 2018,
  Journal of Plasma Physics, 84, 905840305

\bibitem[{{Komarov} {et~al.}(2016){Komarov}, {Churazov}, {Kunz}, \&
  {Schekochihin}}]{komarov:conduction.vs.mirror.instab}
{Komarov}, S.~V., {Churazov}, E.~M., {Kunz}, M.~W., \& {Schekochihin}, A.~A.
  2016, \mnras, 460, 467

\bibitem[{{Komarov} {et~al.}(2014){Komarov}, {Churazov}, {Schekochihin}, \&
  {ZuHone}}]{komarov:heat.flux.suppression.ICM}
{Komarov}, S.~V., {Churazov}, E.~M., {Schekochihin}, A.~A., \& {ZuHone}, J.~A.
  2014, \mnras, 440, 1153

\bibitem[{{Korsmeier} \& {Cuoco}(2016)}]{2016PhRvD..94l3019K}
{Korsmeier}, M., \& {Cuoco}, A. 2016, \prd, 94, 123019

\bibitem[{{Kroupa}(2001)}]{kroupa:2001.imf.var}
{Kroupa}, P. 2001, \mnras, 322, 231

\bibitem[{{Krumholz} \& {Gnedin}(2011)}]{krumholz:2011.molecular.prescription}
{Krumholz}, M.~R., \& {Gnedin}, N.~Y. 2011, \apj, 729, 36

\bibitem[{{Kulpa-Dybe{\l}} {et~al.}(2015){Kulpa-Dybe{\l}}, {Nowak},
  {Otmianowska-Mazur}, {Hanasz}, {Siejkowski}, \&
  {Kulesza-{\.Z}ydzik}}]{Kulp15}
{Kulpa-Dybe{\l}}, K., {Nowak}, N., {Otmianowska-Mazur}, K., {Hanasz}, M.,
  {Siejkowski}, H., \& {Kulesza-{\.Z}ydzik}, B. 2015, \aap, 575, A93

\bibitem[{{Kulpa-Dybe{\l}} {et~al.}(2011){Kulpa-Dybe{\l}}, {Otmianowska-Mazur},
  {Kulesza-{\.Z}ydzik}, {Hanasz}, {Kowal}, {W{\'o}lta{\'n}ski}, \&
  {Kowalik}}]{Kulp11}
{Kulpa-Dybe{\l}}, K., {Otmianowska-Mazur}, K., {Kulesza-{\.Z}ydzik}, B.,
  {Hanasz}, M., {Kowal}, G., {W{\'o}lta{\'n}ski}, D., \& {Kowalik}, K. 2011,
  \apjl, 733, L18

\bibitem[{{Kulsrud} \& {Pearce}(1969)}]{kulsrud.1969:streaming.instability}
{Kulsrud}, R., \& {Pearce}, W.~P. 1969, \apj, 156, 445

\bibitem[{{Kulsrud}(2005)}]{kulsrud:plasma.astro.book}
{Kulsrud}, R.~M. 2005, {Plasma physics for astrophysics} (Princeton, N.J. :
  Princeton University Press)

\bibitem[{{Kunz} {et~al.}(2014){Kunz}, {Schekochihin}, \&
  {Stone}}]{kunz:firehose}
{Kunz}, M.~W., {Schekochihin}, A.~A., \& {Stone}, J.~M. 2014, Physical Review
  Letters, 112, 205003

\bibitem[{{Lacki} {et~al.}(2011){Lacki}, {Thompson}, {Quataert}, {Loeb}, \&
  {Waxman}}]{lacki:2011.cosmic.ray.sub.calorimetric}
{Lacki}, B.~C., {Thompson}, T.~A., {Quataert}, E., {Loeb}, A., \& {Waxman}, E.
  2011, \apj, 734, 107

\bibitem[{{Lazarian}(2016)}]{lazarian:2016.cr.wave.damping}
{Lazarian}, A. 2016, \apj, 833, 131

\bibitem[{{Lee} {et~al.}(2017){Lee}, {Hopkins}, \&
  {Squire}}]{lee:dynamics.charged.dust.gmcs}
{Lee}, H., {Hopkins}, P.~F., \& {Squire}, J. 2017, \mnras, 469, 3532

\bibitem[{{Leitherer} {et~al.}(1999)}]{starburst99}
{Leitherer}, C., {et~al.} 1999, \apjs, 123, 3

\bibitem[{{Lopez} {et~al.}(2018){Lopez}, {Auchettl}, {Linden}, {Bolatto},
  {Thompson}, \& {Ramirez-Ruiz}}]{lopez:2018.smc.below.calorimetric.crs}
{Lopez}, L.~A., {Auchettl}, K., {Linden}, T., {Bolatto}, A.~D., {Thompson},
  T.~A., \& {Ramirez-Ruiz}, E. 2018, \apj, in press, arXiv:1807.06595

\bibitem[{{Ma} {et~al.}(2016){Ma}, {Hopkins}, {Faucher-Gigu{\`e}re}, {Zolman},
  {Muratov}, {Kere{\v s}}, \& {Quataert}}]{ma:2015.fire.mass.metallicity}
{Ma}, X., {Hopkins}, P.~F., {Faucher-Gigu{\`e}re}, C.-A., {Zolman}, N.,
  {Muratov}, A.~L., {Kere{\v s}}, D., \& {Quataert}, E. 2016, \mnras, 456, 2140

\bibitem[{{Ma} {et~al.}(2017{\natexlab{a}}){Ma}, {Hopkins}, {Feldmann},
  {Torrey}, {Faucher-Gigu{\`e}re}, \& {Kere{\v s}}}]{ma:radial.gradients}
{Ma}, X., {Hopkins}, P.~F., {Feldmann}, R., {Torrey}, P.,
  {Faucher-Gigu{\`e}re}, C.-A., \& {Kere{\v s}}, D. 2017{\natexlab{a}}, \mnras,
  466, 4780

\bibitem[{{Ma} {et~al.}(2017{\natexlab{b}}){Ma}, {Hopkins}, {Wetzel}, {Kirby},
  {Angl{\'e}s-Alc{\'a}zar}, {Faucher-Gigu{\`e}re}, {Kere{\v s}}, \&
  {Quataert}}]{ma:2016.disk.structure}
{Ma}, X., {Hopkins}, P.~F., {Wetzel}, A.~R., {Kirby}, E.~N.,
  {Angl{\'e}s-Alc{\'a}zar}, D., {Faucher-Gigu{\`e}re}, C.-A., {Kere{\v s}}, D.,
  \& {Quataert}, E. 2017{\natexlab{b}}, \mnras, 467, 2430

\bibitem[{{Ma} {et~al.}(2018{\natexlab{a}}){Ma}, {Hopkins}, {Garrison-Kimmel},
  {Faucher-Gigu{\`e}re}, {Quataert}, {Boylan-Kolchin}, {Hayward}, {Feldmann},
  \& {Kere{\v s}}}]{ma:fire2.reion.gal.lfs}
{Ma}, X., {et~al.} 2018{\natexlab{a}}, \mnras, 478, 1694

\bibitem[{{Ma} {et~al.}(2018{\natexlab{b}}){Ma}, {Hopkins}, {Boylan-Kolchin},
  {Faucher-Gigu{\`e}re}, {Quataert}, {Feldmann}, {Garrison-Kimmel}, {Hayward},
  {Kere{\v s}}, \&
  {Wetzel}}]{ma:fire.reionization.epoch.galaxies.clumpiness.luminosities}
---. 2018{\natexlab{b}}, \mnras, 477, 219

\bibitem[{{Mannheim} \& {Schlickeiser}(1994)}]{Mann94}
{Mannheim}, K., \& {Schlickeiser}, R. 1994, \aap, 286, 983

\bibitem[{{Mao} \& {Ostriker}(2018)}]{Mao18}
{Mao}, S.~A., \& {Ostriker}, E.~C. 2018, \apj, 854, 89

\bibitem[{{Marinacci} {et~al.}(2015){Marinacci}, {Vogelsberger}, {Mocz}, \&
  {Pakmor}}]{2015MNRAS.453.3999M}
{Marinacci}, F., {Vogelsberger}, M., {Mocz}, P., \& {Pakmor}, R. 2015, \mnras,
  453, 3999

\bibitem[{{Markevitch} \& {Vikhlinin}(2007)}]{2007PhR...443....1M}
{Markevitch}, M., \& {Vikhlinin}, A. 2007, \physrep, 443, 1

\bibitem[{{Martin}(1999)}]{martin99:outflow.vs.m}
{Martin}, C.~L. 1999, \apj, 513, 156

\bibitem[{{Martin} {et~al.}(2010){Martin}, {Scannapieco}, {Ellison}, {Hennawi},
  {Djorgovski}, \& {Fournier}}]{martin:2010.metal.enriched.regions}
{Martin}, C.~L., {Scannapieco}, E., {Ellison}, S.~L., {Hennawi}, J.~F.,
  {Djorgovski}, S.~G., \& {Fournier}, A.~P. 2010, \apj, 721, 174

\bibitem[{{Martin-Alvarez} {et~al.}(2018){Martin-Alvarez}, {Devriendt}, {Slyz},
  \& {Teyssier}}]{2018MNRAS.479.3343M}
{Martin-Alvarez}, S., {Devriendt}, J., {Slyz}, A., \& {Teyssier}, R. 2018,
  \mnras, 479, 3343

\bibitem[{{McCourt} {et~al.}(2015){McCourt}, {O'Leary}, {Madigan}, \&
  {Quataert}}]{2015MNRAS.449....2M}
{McCourt}, M., {O'Leary}, R.~M., {Madigan}, A.-M., \& {Quataert}, E. 2015,
  \mnras, 449, 2

\bibitem[{{McKenzie} \& {Voelk}(1982)}]{mckenzie.voelk:1982.cr.equations}
{McKenzie}, J.~F., \& {Voelk}, H.~J. 1982, \aap, 116, 191

\bibitem[{{Mitra} {et~al.}(2017){Mitra}, {Dav{\'e}}, {Simha}, \&
  {Finlator}}]{2017MNRAS.464.2766M}
{Mitra}, S., {Dav{\'e}}, R., {Simha}, V., \& {Finlator}, K. 2017, \mnras, 464,
  2766

\bibitem[{{Moster} {et~al.}(2013){Moster}, {Naab}, \&
  {White}}]{moster:2013.abundance.matching.sfhs}
{Moster}, B.~P., {Naab}, T., \& {White}, S.~D.~M. 2013, \mnras, 428, 3121

\bibitem[{{Muratov} {et~al.}(2015){Muratov}, {Kere{\v s}},
  {Faucher-Gigu{\`e}re}, {Hopkins}, {Quataert}, \&
  {Murray}}]{muratov:2015.fire.winds}
{Muratov}, A.~L., {Kere{\v s}}, D., {Faucher-Gigu{\`e}re}, C.-A., {Hopkins},
  P.~F., {Quataert}, E., \& {Murray}, N. 2015, \mnras, 454, 2691

\bibitem[{{Muratov} {et~al.}(2017){Muratov}, {Kere{\v s}},
  {Faucher-Gigu{\`e}re}, {Hopkins}, {Ma}, {Angl{\'e}s-Alc{\'a}zar}, {Chan},
  {Torrey}, {Hafen}, {Quataert}, \&
  {Murray}}]{muratov:2016.fire.metal.outflow.loading}
{Muratov}, A.~L., {et~al.} 2017, \mnras, 468, 4170

\bibitem[{{O{\~n}orbe} {et~al.}(2015){O{\~n}orbe}, {Boylan-Kolchin}, {Bullock},
  {Hopkins}, {Kere{\v s}}, {Faucher-Gigu{\`e}re}, {Quataert}, \&
  {Murray}}]{onorbe:2015.fire.cores}
{O{\~n}orbe}, J., {Boylan-Kolchin}, M., {Bullock}, J.~S., {Hopkins}, P.~F.,
  {Kere{\v s}}, D., {Faucher-Gigu{\`e}re}, C.-A., {Quataert}, E., \& {Murray},
  N. 2015, \mnras, 454, 2092

\bibitem[{{O{\~n}orbe} {et~al.}(2014){O{\~n}orbe}, {Garrison-Kimmel}, {Maller},
  {Bullock}, {Rocha}, \& {Hahn}}]{onorbe:2013.zoom.methods}
{O{\~n}orbe}, J., {Garrison-Kimmel}, S., {Maller}, A.~H., {Bullock}, J.~S.,
  {Rocha}, M., \& {Hahn}, O. 2014, \mnras, 437, 1894

\bibitem[{{Orr} {et~al.}(2018){Orr}, {Hayward}, {Hopkins}, {Chan},
  {Faucher-Gigu{\`e}re}, {Feldmann}, {Kere{\v s}}, {Murray}, \&
  {Quataert}}]{orr:ks.law}
{Orr}, M.~E., {et~al.} 2018, \mnras, 478, 3653

\bibitem[{{Ostriker} \&
  {Shetty}(2011)}]{ostriker.shetty:2011.turb.disk.selfreg.ks}
{Ostriker}, E.~C., \& {Shetty}, R. 2011, \apj, 731, 41

\bibitem[{{Pakmor} {et~al.}(2016){Pakmor}, {Pfrommer}, {Simpson}, \&
  {Springel}}]{Pakm16}
{Pakmor}, R., {Pfrommer}, C., {Simpson}, C.~M., \& {Springel}, V. 2016, \apjl,
  824, L30

\bibitem[{{Pakmor} \& {Springel}(2013)}]{2013MNRAS.432..176P}
{Pakmor}, R., \& {Springel}, V. 2013, \mnras, 432, 176

\bibitem[{{Parker}(1992)}]{Park92}
{Parker}, E.~N. 1992, \apj, 401, 137

\bibitem[{{Parrish} {et~al.}(2012){Parrish}, {McCourt}, {Quataert}, \&
  {Sharma}}]{2012MNRAS.422..704P}
{Parrish}, I.~J., {McCourt}, M., {Quataert}, E., \& {Sharma}, P. 2012, \mnras,
  422, 704

\bibitem[{{Pettini} {et~al.}(2003){Pettini}, {Madau}, {Bolte}, {Prochaska},
  {Ellison}, \& {Fan}}]{pettini:2003.igm.metal.evol}
{Pettini}, M., {Madau}, P., {Bolte}, M., {Prochaska}, J.~X., {Ellison}, S.~L.,
  \& {Fan}, X. 2003, \apj, 594, 695

\bibitem[{{Piontek} \& {Ostriker}(2005)}]{2005ApJ...629..849P}
{Piontek}, R.~A., \& {Ostriker}, E.~C. 2005, \apj, 629, 849

\bibitem[{{Piontek} \& {Ostriker}(2007)}]{2007ApJ...663..183P}
---. 2007, \apj, 663, 183

\bibitem[{{Reynolds} {et~al.}(2005){Reynolds}, {McKernan}, {Fabian}, {Stone},
  \& {Vernaleo}}]{2005MNRAS.357..242R}
{Reynolds}, C.~S., {McKernan}, B., {Fabian}, A.~C., {Stone}, J.~M., \&
  {Vernaleo}, J.~C. 2005, \mnras, 357, 242

\bibitem[{{Rieder} \& {Teyssier}(2017)}]{2017MNRAS.471.2674R}
{Rieder}, M., \& {Teyssier}, R. 2017, \mnras, 471, 2674

\bibitem[{{Riquelme} {et~al.}(2016){Riquelme}, {Quataert}, \&
  {Verscharen}}]{riquelme:viscosity.limits}
{Riquelme}, M.~A., {Quataert}, E., \& {Verscharen}, D. 2016, \apj, 824, 123

\bibitem[{{Roberg-Clark} {et~al.}(2016){Roberg-Clark}, {Drake}, {Reynolds}, \&
  {Swisdak}}]{roberg:2016.whistler.turb.conduction.suppression}
{Roberg-Clark}, G.~T., {Drake}, J.~F., {Reynolds}, C.~S., \& {Swisdak}, M.
  2016, \apjl, 830, L9

\bibitem[{{Roberg-Clark} {et~al.}(2018){Roberg-Clark}, {Drake}, {Reynolds}, \&
  {Swisdak}}]{roberg:2018.whistler.turb.conduction.suppression}
---. 2018, Physical Review Letters, 120, 035101

\bibitem[{{Ro{\v s}kar} {et~al.}(2014){Ro{\v s}kar}, {Teyssier}, {Agertz},
  {Wetzstein}, \& {Moore}}]{roskar:2014.stellar.rad.fx.approx.model}
{Ro{\v s}kar}, R., {Teyssier}, R., {Agertz}, O., {Wetzstein}, M., \& {Moore},
  B. 2014, \mnras, 444, 2837

\bibitem[{{Ruszkowski} {et~al.}(2017){Ruszkowski}, {Yang}, \&
  {Zweibel}}]{Rusz17}
{Ruszkowski}, M., {Yang}, H.-Y.~K., \& {Zweibel}, E. 2017, \apj, 834, 208

\bibitem[{{Salem} \& {Bryan}(2014)}]{Sale14}
{Salem}, M., \& {Bryan}, G.~L. 2014, \mnras, 437, 3312

\bibitem[{{Salem} {et~al.}(2016){Salem}, {Bryan}, \& {Corlies}}]{Sale16}
{Salem}, M., {Bryan}, G.~L., \& {Corlies}, L. 2016, \mnras, 456, 582

\bibitem[{{Salem} {et~al.}(2014){Salem}, {Bryan}, \& {Hummels}}]{Sale14cos}
{Salem}, M., {Bryan}, G.~L., \& {Hummels}, C. 2014, \apjl, 797, L18

\bibitem[{{Sanderson} {et~al.}(2017){Sanderson}, {Garrison-Kimmel}, {Wetzel},
  {Keung Chan}, {Hopkins}, {Kere{\v s}}, {Escala}, {Faucher-Gigu{\`e}re}, \&
  {Ma}}]{sanderson:stellar.halo.mass.vs.fire.comparison}
{Sanderson}, R.~E., {et~al.} 2017, \apj, in press, arXiv:1712.05808

\bibitem[{{Santos-Lima} {et~al.}(2016){Santos-Lima}, {Yan}, {de Gouveia Dal
  Pino}, \& {Lazarian}}]{santos:2016.ion.temp.anisotropy.limits}
{Santos-Lima}, R., {Yan}, H., {de Gouveia Dal Pino}, E.~M., \& {Lazarian}, A.
  2016, \mnras, 460, 2492

\bibitem[{{Sarazin}(1988)}]{sarazin:coulomb.log}
{Sarazin}, C.~L. 1988, {X-ray emission from clusters of galaxies} (Cambridge:
  Cambridge University Press)

\bibitem[{{Sato} {et~al.}(2009){Sato}, {Martin}, {Noeske}, {Koo}, \&
  {Lotz}}]{sato:2009.ulirg.outflows}
{Sato}, T., {Martin}, C.~L., {Noeske}, K.~G., {Koo}, D.~C., \& {Lotz}, J.~M.
  2009, \apj, 696, 214

\bibitem[{{Seligman} {et~al.}(2018){Seligman}, {Hopkins}, \&
  {Squire}}]{seligman:2018.mhd.rdi.sims}
{Seligman}, D., {Hopkins}, P.~F., \& {Squire}, J. 2018, \mnras, submitted,
  arXiv:1810.09491

\bibitem[{{Sharma} {et~al.}(2009){Sharma}, {Chandran}, {Quataert}, \&
  {Parrish}}]{2009ApJ...699..348S}
{Sharma}, P., {Chandran}, B.~D.~G., {Quataert}, E., \& {Parrish}, I.~J. 2009,
  \apj, 699, 348

\bibitem[{{Sharma} {et~al.}(2010){Sharma}, {Parrish}, \&
  {Quataert}}]{2010ApJ...720..652S}
{Sharma}, P., {Parrish}, I.~J., \& {Quataert}, E. 2010, \apj, 720, 652

\bibitem[{{Sijacki} \& {Springel}(2006)}]{2006MNRAS.371.1025S}
{Sijacki}, D., \& {Springel}, V. 2006, \mnras, 371, 1025

\bibitem[{{Simpson} {et~al.}(2016){Simpson}, {Pakmor}, {Marinacci}, {Pfrommer},
  {Springel}, {Glover}, {Clark}, \& {Smith}}]{Simp16}
{Simpson}, C.~M., {Pakmor}, R., {Marinacci}, F., {Pfrommer}, C., {Springel},
  V., {Glover}, S.~C.~O., {Clark}, P.~C., \& {Smith}, R.~J. 2016, \apjl, 827,
  L29

\bibitem[{{Skilling}(1971)}]{skilling:1971.cr.diffusion}
{Skilling}, J. 1971, \apj, 170, 265

\bibitem[{{Skilling}(1975)}]{1975MNRAS.172..557S}
---. 1975, \mnras, 172, 557

\bibitem[{{Socrates} {et~al.}(2008){Socrates}, {Davis}, \&
  {Ramirez-Ruiz}}]{Socr08}
{Socrates}, A., {Davis}, S.~W., \& {Ramirez-Ruiz}, E. 2008, \apj, 687, 202

\bibitem[{{Somerville} \& {Primack}(1999)}]{somerville99:sam}
{Somerville}, R.~S., \& {Primack}, J.~R. 1999, \mnras, 310, 1087

\bibitem[{{Songaila}(2005)}]{songaila:2005.igm.metal.evol}
{Songaila}, A. 2005, \aj, 130, 1996

\bibitem[{{Sparre} {et~al.}(2017){Sparre}, {Hayward}, {Feldmann},
  {Faucher-Gigu{\`e}re}, {Muratov}, {Kere{\v s}}, \&
  {Hopkins}}]{sparre.2015:bursty.star.formation.main.sequence.fire}
{Sparre}, M., {Hayward}, C.~C., {Feldmann}, R., {Faucher-Gigu{\`e}re}, C.-A.,
  {Muratov}, A.~L., {Kere{\v s}}, D., \& {Hopkins}, P.~F. 2017, \mnras, 466, 88

\bibitem[{Spitzer \& H\"arm(1953)}]{spitzer:conductivity}
Spitzer, L., \& H\"arm, R. 1953, Phys. Rev., 89, 977

\bibitem[{{Springel} \& {Hernquist}(2003)}]{springel:lcdm.sfh}
{Springel}, V., \& {Hernquist}, L. 2003, \mnras, 339, 312

\bibitem[{{Squire} \& {Hopkins}(2017)}]{squire.hopkins:turb.density.pdf}
{Squire}, J., \& {Hopkins}, P.~F. 2017, \mnras, 471, 3753

\bibitem[{{Squire} {et~al.}(2017{\natexlab{a}}){Squire}, {Kunz}, {Quataert}, \&
  {Schekochihin}}]{squire:2017.kinetic.mhd.alfven}
{Squire}, J., {Kunz}, M.~W., {Quataert}, E., \& {Schekochihin}, A.~A.
  2017{\natexlab{a}}, Physical Review Letters, 119, 155101

\bibitem[{{Squire} {et~al.}(2017{\natexlab{b}}){Squire}, {Quataert}, \&
  {Kunz}}]{squire:2017.max.anisotropy.kinetic.mhd}
{Squire}, J., {Quataert}, E., \& {Kunz}, M.~W. 2017{\natexlab{b}}, Journal of
  Plasma Physics, 83, 905830613

\bibitem[{{Squire} {et~al.}(2017{\natexlab{c}}){Squire}, {Schekochihin}, \&
  {Quataert}}]{squire:2017.max.braginskii.scalings}
{Squire}, J., {Schekochihin}, A.~A., \& {Quataert}, E. 2017{\natexlab{c}}, New
  Journal of Physics, 19, 055005

\bibitem[{{Steidel} {et~al.}(2010){Steidel}, {Erb}, {Shapley}, {Pettini},
  {Reddy}, {Bogosavljevi{\'c}}, {Rudie}, \&
  {Rakic}}]{steidel:2010.outflow.kinematics}
{Steidel}, C.~C., {Erb}, D.~K., {Shapley}, A.~E., {Pettini}, M., {Reddy}, N.,
  {Bogosavljevi{\'c}}, M., {Rudie}, G.~C., \& {Rakic}, O. 2010, \apj, 717, 289

\bibitem[{{Strong} {et~al.}(2007){Strong}, {Moskalenko}, \& {Ptuskin}}]{Stro07}
{Strong}, A.~W., {Moskalenko}, I.~V., \& {Ptuskin}, V.~S. 2007, Annual Review
  of Nuclear and Particle Science, 57, 285

\bibitem[{{Su} {et~al.}(2018{\natexlab{a}}){Su}, {Hayward}, {Hopkins},
  {Quataert}, {Faucher-Gigu{\`e}re}, \& {Kere{\v
  s}}}]{su:fire.feedback.alters.magnetic.amplification.morphology}
{Su}, K.-Y., {Hayward}, C.~C., {Hopkins}, P.~F., {Quataert}, E.,
  {Faucher-Gigu{\`e}re}, C.-A., \& {Kere{\v s}}, D. 2018{\natexlab{a}}, \mnras,
  473, L111

\bibitem[{{Su} {et~al.}(2017){Su}, {Hopkins}, {Hayward}, {Faucher-Gigu{\`e}re},
  {Kere{\v s}}, {Ma}, \& {Robles}}]{su:2016.weak.mhd.cond.visc.turbdiff.fx}
{Su}, K.-Y., {Hopkins}, P.~F., {Hayward}, C.~C., {Faucher-Gigu{\`e}re}, C.-A.,
  {Kere{\v s}}, D., {Ma}, X., \& {Robles}, V.~H. 2017, \mnras, 471, 144

\bibitem[{{Su} {et~al.}(2018{\natexlab{b}}){Su}, {Hopkins}, {Hayward}, {Ma},
  {Faucher-Gigu{\`e}re}, {Kere{\v s}}, {Orr}, \&
  {Robles}}]{su:2018.stellar.fb.fails.to.solve.cooling.flow}
{Su}, K.-Y., {Hopkins}, P.~F., {Hayward}, C.~C., {Ma}, X.,
  {Faucher-Gigu{\`e}re}, C.-A., {Kere{\v s}}, D., {Orr}, M.~E., \& {Robles},
  V.~H. 2018{\natexlab{b}}, \mnras, in press, arXiv:1809.09120

\bibitem[{{Su} {et~al.}(2018{\natexlab{c}}){Su}, {Hopkins}, {Hayward}, {Ma},
  {Faucher-Gigu{\`e}re}, {Kere{\v s}}, {Orr}, \&
  {Robles}}]{su:2018.forms.of.feedback.vs.cooling.flow}
---. 2018{\natexlab{c}}, \mnras, submitted, arXiv:1809.09120

\bibitem[{{Su} {et~al.}(2018{\natexlab{d}}){Su}, {Hopkins}, {Hayward},
  {Faucher-Gigu{\`e}re}, {Kere{\v{s}}}, {Ma}, {Orr}, {Chan}, \&
  {Robles}}]{su:turb.crs.quench}
{Su}, K.-Y., {et~al.} 2018{\natexlab{d}}, \mnras, submitted, arXiv:1812.03997,
  arXiv:1812.03997

\bibitem[{{Tang} {et~al.}(2014){Tang}, {Wang}, \&
  {Tam}}]{tang:2014.ngc.2146.proton.calorimeter}
{Tang}, Q.-W., {Wang}, X.-Y., \& {Tam}, P.-H.~T. 2014, \apj, 794, 26

\bibitem[{{Tasker}(2011)}]{tasker:2011.photoion.heating.gmc.evol}
{Tasker}, E.~J. 2011, \apj, 730, 11

\bibitem[{{Thomas} \&
  {Pfrommer}(2018)}]{thomas.pfrommer.18:alfven.reg.cr.transport}
{Thomas}, T., \& {Pfrommer}, C. 2018, \mnras, submitted, arXiv:1805.11092,
  arXiv:1805.11092

\bibitem[{{Thompson}(2000)}]{2000ApJ...534..915T}
{Thompson}, C. 2000, \apj, 534, 915

\bibitem[{{Tong} {et~al.}(2018){Tong}, {Bale}, {Salem}, \&
  {Pulupa}}]{tong:obs.solar.wind.electron.heat.flux.constraints}
{Tong}, Y., {Bale}, S.~D., {Salem}, C., \& {Pulupa}, M. 2018, ArXiv e-prints

\bibitem[{{Uhlig} {et~al.}(2012){Uhlig}, {Pfrommer}, {Sharma}, {Nath},
  {En{\ss}lin}, \& {Springel}}]{uhlig:2012.cosmic.ray.streaming.winds}
{Uhlig}, M., {Pfrommer}, C., {Sharma}, M., {Nath}, B.~B., {En{\ss}lin}, T.~A.,
  \& {Springel}, V. 2012, \mnras, 423, 2374

\bibitem[{{Vazza} {et~al.}(2014){Vazza}, {Br{\"u}ggen}, {Gheller}, \&
  {Wang}}]{2014MNRAS.445.3706V}
{Vazza}, F., {Br{\"u}ggen}, M., {Gheller}, C., \& {Wang}, P. 2014, \mnras, 445,
  3706

\bibitem[{{Vladimirov} {et~al.}(2012){Vladimirov}, {J{\'o}hannesson},
  {Moskalenko}, \& {Porter}}]{vladimirov:cr.highegy.diff}
{Vladimirov}, A.~E., {J{\'o}hannesson}, G., {Moskalenko}, I.~V., \& {Porter},
  T.~A. 2012, \apj, 752, 68

\bibitem[{{Wang} \& {Abel}(2009)}]{2009ApJ...696...96W}
{Wang}, P., \& {Abel}, T. 2009, \apj, 696, 96

\bibitem[{{Wang} \&
  {Fields}(2018)}]{wang:2018.starbursts.are.proton.calorimeters}
{Wang}, X., \& {Fields}, B.~D. 2018, \mnras, 474, 4073

\bibitem[{{Weiner} {et~al.}(2009)}]{weiner:z1.outflows}
{Weiner}, B.~J., {et~al.} 2009, \apj, 692, 187

\bibitem[{{Wentzel}(1968)}]{wentzel:1968.mhd.wave.cr.coupling}
{Wentzel}, D.~G. 1968, \apj, 152, 987

\bibitem[{{Wetzel} {et~al.}(2016){Wetzel}, {Hopkins}, {Kim},
  {Faucher-Gigu{\`e}re}, {Kere{\v s}}, \& {Quataert}}]{wetzel.2016:latte}
{Wetzel}, A.~R., {Hopkins}, P.~F., {Kim}, J.-h., {Faucher-Gigu{\`e}re}, C.-A.,
  {Kere{\v s}}, D., \& {Quataert}, E. 2016, \apjl, 827, L23

\bibitem[{{Wheeler} {et~al.}(2015){Wheeler}, {O{\~n}orbe}, {Bullock},
  {Boylan-Kolchin}, {Elbert}, {Garrison-Kimmel}, {Hopkins}, \& {Kere{\v
  s}}}]{wheeler:dwarf.satellites}
{Wheeler}, C., {O{\~n}orbe}, J., {Bullock}, J.~S., {Boylan-Kolchin}, M.,
  {Elbert}, O.~D., {Garrison-Kimmel}, S., {Hopkins}, P.~F., \& {Kere{\v s}}, D.
  2015, \mnras, 453, 1305

\bibitem[{{Wheeler} {et~al.}(2017){Wheeler}, {Pace}, {Bullock},
  {Boylan-Kolchin}, {O{\~n}orbe}, {Elbert}, {Fitts}, {Hopkins}, \& {Kere{\v
  s}}}]{wheeler.2015:dwarfs.isolated.not.rotating}
{Wheeler}, C., {et~al.} 2017, \mnras, 465, 2420

\bibitem[{{Wheeler} {et~al.}(2018){Wheeler}, {Hopkins}, {Pace},
  {Garrison-Kimmel}, {Boylan-Kolchin}, {Wetzel}, {Bullock}, {Keres},
  {Faucher-Giguere}, \& {Quataert}}]{wheeler:ultra.highres.dwarfs}
---. 2018, arXiv e-prints, arXiv:1812.02749

\bibitem[{{Wiener} {et~al.}(2013{\natexlab{a}}){Wiener}, {Oh}, \&
  {Guo}}]{wiener:cr.supersonic.streaming.deriv}
{Wiener}, J., {Oh}, S.~P., \& {Guo}, F. 2013{\natexlab{a}}, \mnras, 434, 2209

\bibitem[{{Wiener} {et~al.}(2017){Wiener}, {Pfrommer}, \& {Oh}}]{Wien17}
{Wiener}, J., {Pfrommer}, C., \& {Oh}, S.~P. 2017, \mnras, 467, 906

\bibitem[{{Wiener} {et~al.}(2013{\natexlab{b}}){Wiener}, {Zweibel}, \&
  {Oh}}]{Wien13}
{Wiener}, J., {Zweibel}, E.~G., \& {Oh}, S.~P. 2013{\natexlab{b}}, \apj, 767,
  87

\bibitem[{{Wise} {et~al.}(2012){Wise}, {Abel}, {Turk}, {Norman}, \&
  {Smith}}]{wise:2012.rad.pressure.effects}
{Wise}, J.~H., {Abel}, T., {Turk}, M.~J., {Norman}, M.~L., \& {Smith}, B.~D.
  2012, \mnras, 427, 311

\bibitem[{{Wojaczy{\'n}ski} \&
  {Nied{\'z}wiecki}(2017)}]{wjac:2017.4945.gamma.rays}
{Wojaczy{\'n}ski}, R., \& {Nied{\'z}wiecki}, A. 2017, \apj, 849, 97

\bibitem[{{Yan} \&
  {Lazarian}(2008)}]{yan.lazarian.2008:cr.propagation.with.streaming}
{Yan}, H., \& {Lazarian}, A. 2008, \apj, 673, 942

\bibitem[{{Zhu} {et~al.}(2014){Zhu}, {Stone}, \&
  {Bai}}]{zhu:2014.dust.power.spectra.mri.turbulent.disks}
{Zhu}, Z., {Stone}, J.~M., \& {Bai}, X.-N. 2014, \apj, in press,
  arXiv:1405.2778

\bibitem[{{Zirakashvili} {et~al.}(1996){Zirakashvili}, {Breitschwerdt},
  {Ptuskin}, \& {Voelk}}]{Zira96}
{Zirakashvili}, V.~N., {Breitschwerdt}, D., {Ptuskin}, V.~S., \& {Voelk}, H.~J.
  1996, \aap, 311, 113

\bibitem[{{Zweibel}(2013)}]{Zwei13}
{Zweibel}, E.~G. 2013, Physics of Plasmas, 20, 055501

\end{thebibliography}

\begin{appendix}

\vspace{-0.5cm}
\section{Additional Numerical Tests}
\label{sec:additional.tests}

\begin{figure}
    \includegraphics[width=0.99\columnwidth]{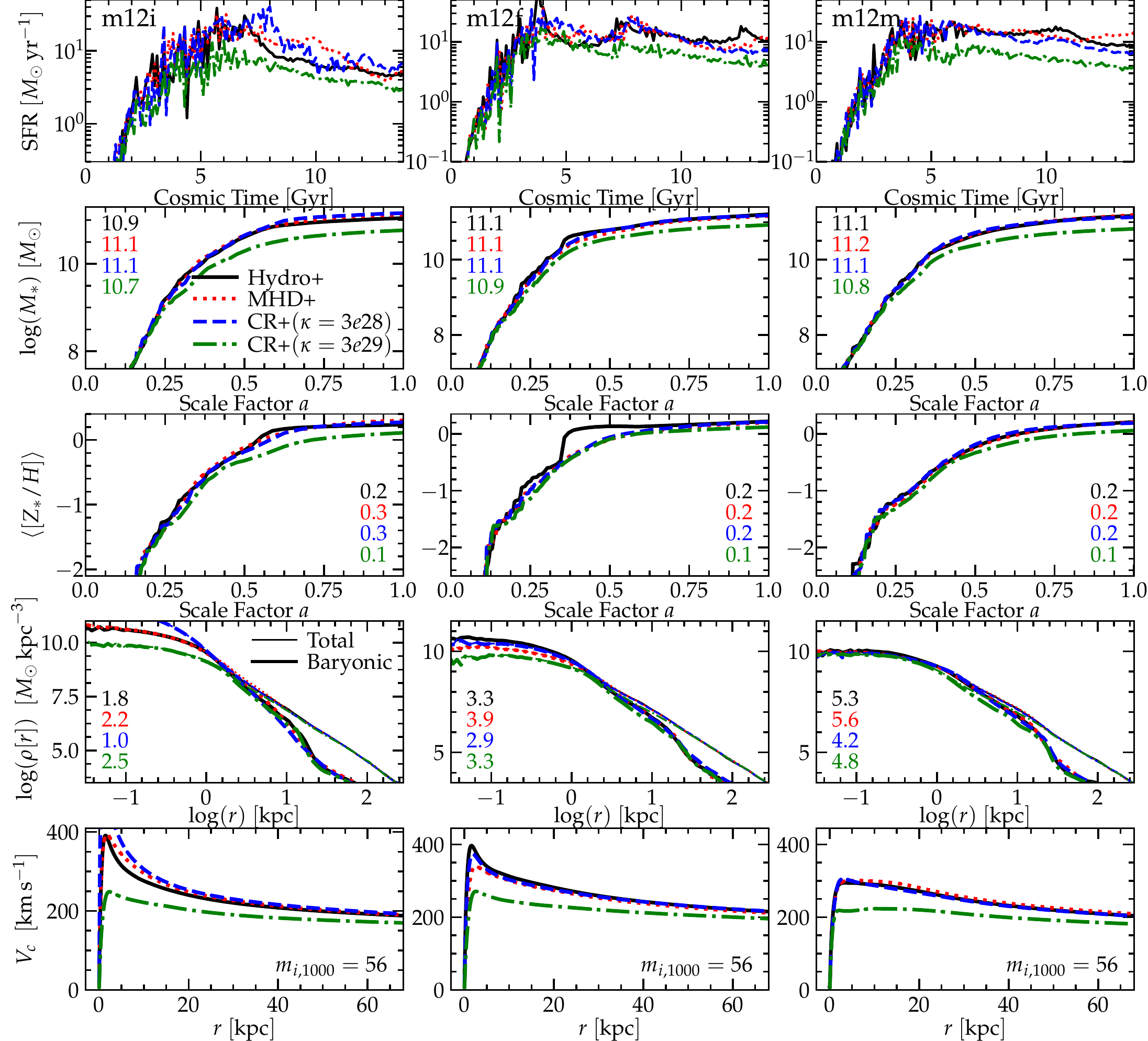}
    \vspace{-0.25cm}
    \caption{As Fig.~\ref{fig:cr.demo.m12}, at order-of-magnitude lower resolution. At lower resolution for our MW-mass systems, all the galaxies exhibit higher SFRs and stellar masses, as studied extensively in \citet{hopkins:fire2.methods}. However, the systematic effects (or lack thereof) of MHD and CRs are robust across resolution. Lower-resolution runs of our dwarfs ({\bf m10v}, {\bf m10q}, {\bf m11q}) show no significant resolution dependence with or without CRs (as shown in \citealt{hopkins:fire2.methods}).
    \label{fig:cr.demo.m12.LR}}
\end{figure}

\begin{figure}
    \plotonesize{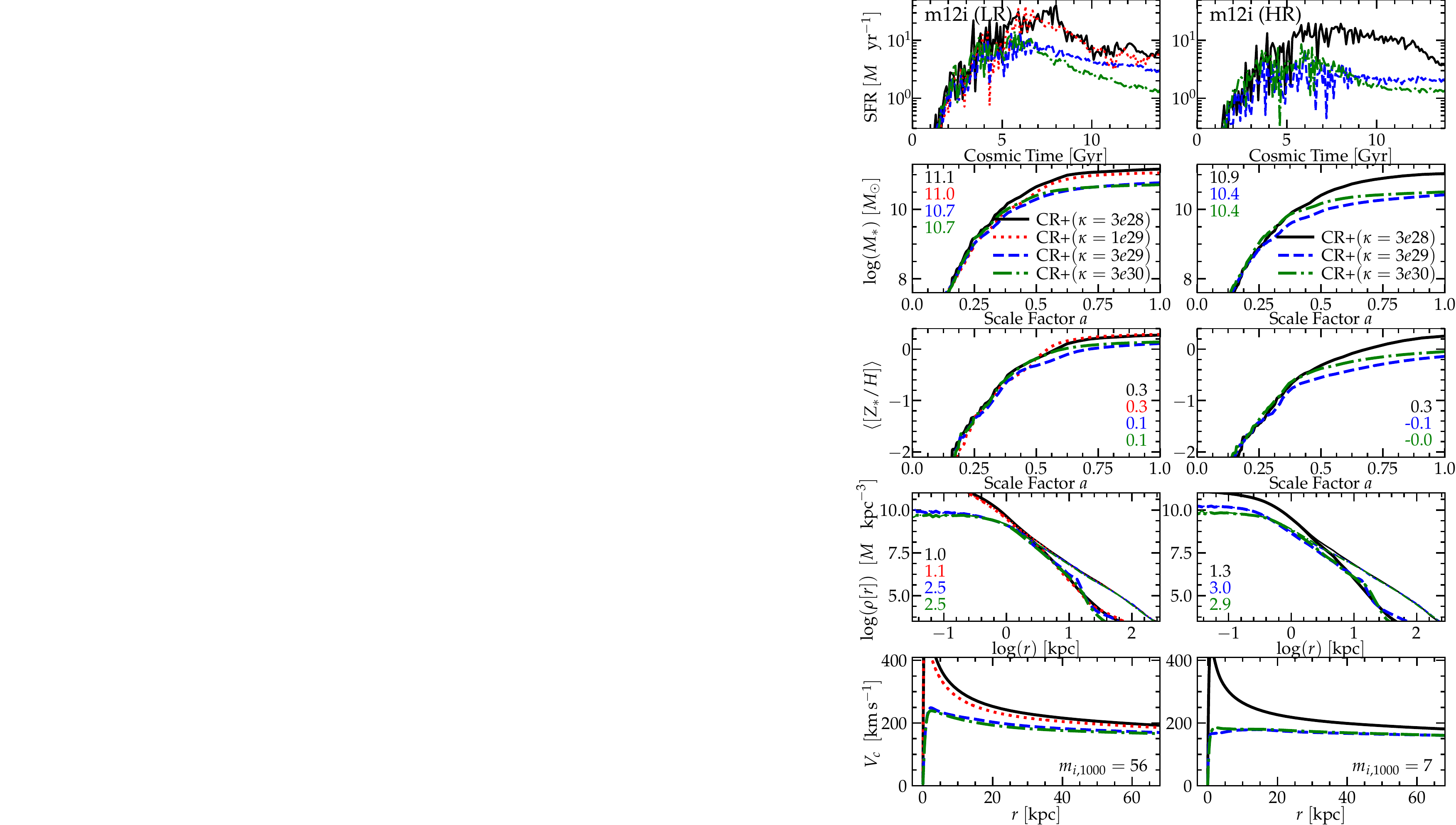}{0.99}
    \vspace{-0.25cm}
    \caption{As Fig.~\ref{fig:cr.demo.m12.LR}, varying the CR diffusion coefficient $\kappa$ more extensively, in both our default (high-resolution ``HR''; {\em right}) and low-resolution (``LR'' from Fig.~\ref{fig:cr.demo.m12.LR}; {\em left}) versions of {\bf m12i}. Although the stellar masses and central densities do depend on resolution, the systematic effect of CRs is similar at all resolution levels, for all values of $\kappa$ explored.
    \label{fig:diffusioncoeff.res}}
\end{figure}

\begin{figure}
    \plotonesize{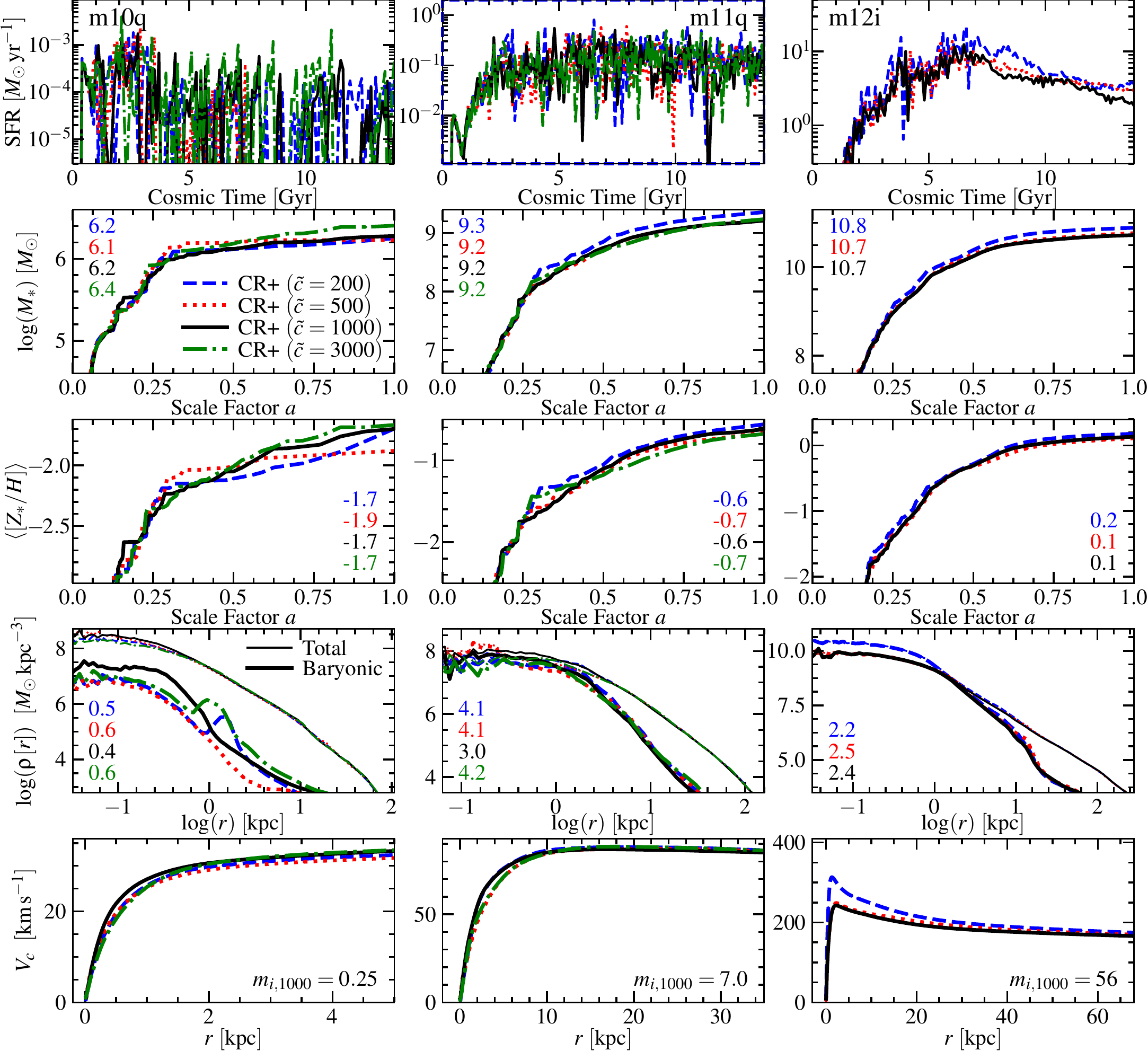}{0.99}
    \vspace{-0.25cm}
        \caption{As Fig.~\ref{fig:cr.demo.m12.LR}, comparing our ``CR+'' runs ($\kappa=3e28$ for the dwarfs {\bf m10q} \&\ {\bf m11q}, and $\kappa=3e29$ for {\bf m12i}), varying the numerical ``maximum CR free-streaming speed'' $\tilde{c}$ (values in ${\rm km\,s^{-1}}$). For values $\tilde{c}\gtrsim 200\,{\rm km\,s^{-1}}$ (fast enough that CRs can ``outpace'' most bulk rotation and outflow motion), we see excellent agreement, so our results the main text  should be insensitive to this.
    \label{fig:cr.vmax}}
\end{figure}

\begin{figure}
    \plotonesize{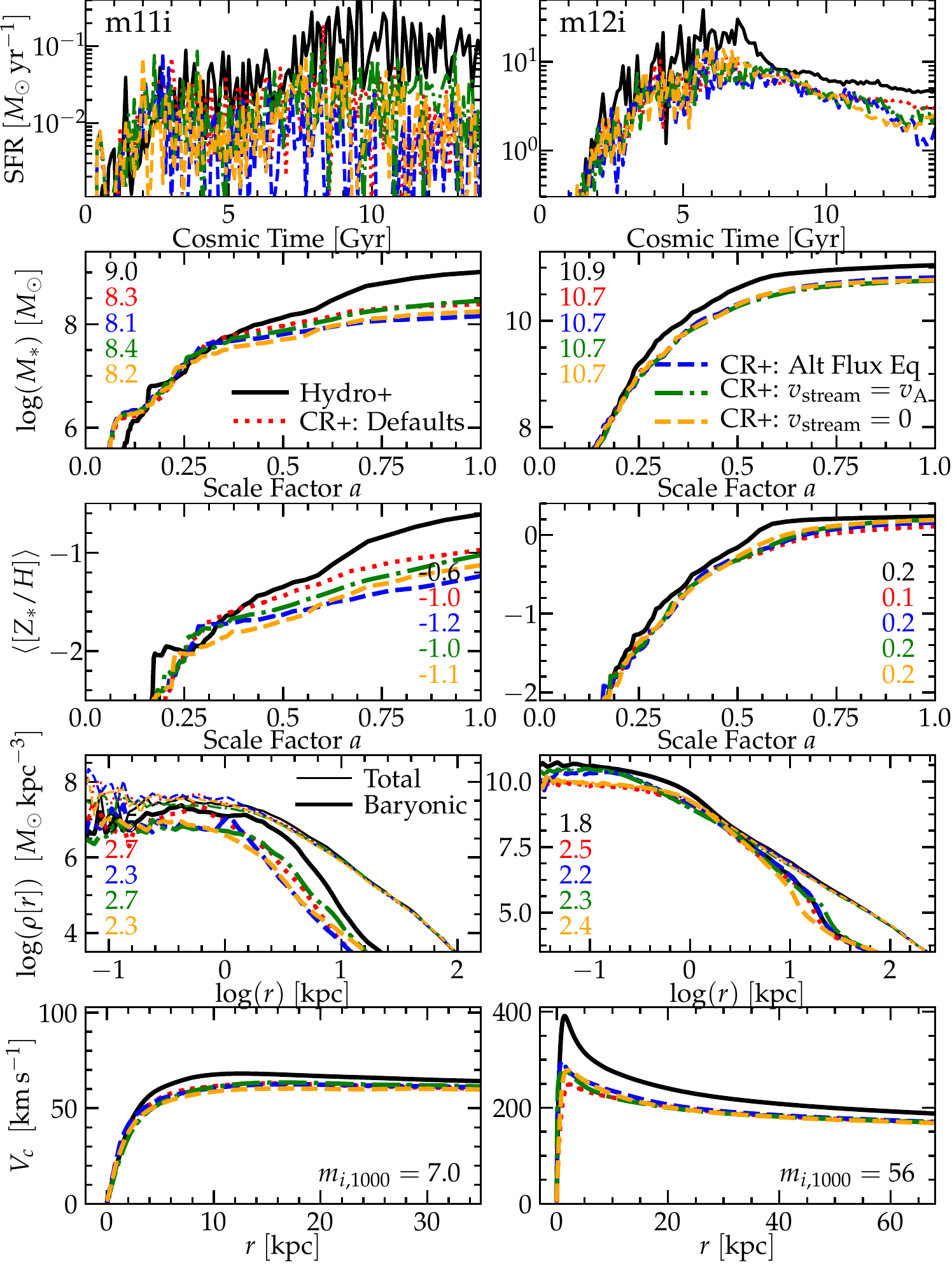}{0.99}
    \vspace{-0.25cm}
        \caption{As Fig.~\ref{fig:cr.demo.m12.LR}, comparing our default ``CR+($\kappa=3e29$)'' runs (with a ``Hydro+'' run for comparison) in {\bf m11i} and {\bf m12i}, with {\bf (1)} a different form for the CR flux equation (Eq.~\ref{eqn:cr.flux.tp}), {\bf (2)} streaming speed $=v_{A}$ (instead of our  default $=3\,v_{A}$), and {\bf (3)} streaming speed $=0$ but retaining the ``streaming losses'' ($\sim v_{A}\,\nabla P_{\rm cr}$). None of these variations have appreciable  effects on our conclusions.
    \label{fig:cr.fluxeqn}}
\end{figure}

In this Appendix we present some examples of additional  numerical tests, discussed in \S~\ref{sec:numerics} in the main text. 

Figs.~\ref{fig:cr.demo.m12.LR}-\ref{fig:diffusioncoeff.res} show resolution tests, for MW-mass halos and varied diffusivity $\kappa$. As noted in \S~\ref{sec:numerics} we have also extensively varied the resolution in our dwarf halos {\bf m10q} and {\bf m11q}, but since (a) those halos exhibit no large effect from either MHD or CRs (at any $\kappa$) and (b) our extensive resolution studies in \citet{hopkins:fire2.methods} (as well as \citealt{su:2016.weak.mhd.cond.visc.turbdiff.fx} and \citealt{chan:2018.cosmicray.fire.gammaray}) have shown the dwarf properties in ``Hydro+'' and ``MHD+'' runs are extremely insensitive to resolution, it is not surprising that we find there is no change at any resolution level for these halos (this is also demonstrated in Fig.~\ref{fig:mgal.mhalo}). Therefore we focus on MW mass  where both  CR physics and resolution have much larger effects. 

As discussed in \S~\ref{sec:numerics}, Figs.~\ref{fig:cr.demo.m12.LR}-\ref{fig:diffusioncoeff.res} do show a systematic resolution dependence in our massive {\bf m12} halos (compare also Fig.~\ref{fig:cr.demo.m12.LR} to \ref{fig:cr.demo.m12}). In massive halos, lower-resolution runs produce larger stellar masses and, correspondingly, higher central baryonic densities  and enhanced circular velocities. This is all studied in extensive detail in \citet{hopkins:fire2.methods}. Critically, for our study here, we see the same {\em systematic} effects (or lack thereof) of MHD and CRs at {\em all} resolution levels.  

In Fig.~\ref{fig:cr.vmax}, we systematically compare halos {\bf m10q}, {\bf m11q}, {\bf m12i}, chosen with the values of $\kappa$ where effects on the galaxy are maximal, varying the numerically-imposed maximum CR free-streaming speed $\tilde{c}$. As expected, we see that this has little effect on our conclusions (variations are smaller than stochastic run-to-run variations) for essentially all values $\tilde{c} \gg 200\,{\rm km\,s^{-1}}$, fast enough that CRs can (when they {\em should}, according to the diffusivity $\kappa$ and streaming speed $v_{\rm stream}$) diffuse or stream faster than local gas rotation/outflow velocities. Otherwise, CR effects are somewhat weaker, as the CRs can spend more time artificially trapped in dense gas (where collisions sap energy). This is consistent with our more detailed numerical study in \citet{chan:2018.cosmicray.fire.gammaray}.

\citet{chan:2018.cosmicray.fire.gammaray} also describe and test in detail the CR flux equation (Eq.~\ref{eqn:cr.flux}), i.e.\ the equation for the diffusive CR flux $\tilde{\bf F}_{\rm cr}$. They showed that using alternative forms of this equation, or simply omitting it entirely and solving directly a single 0th-moment (pure diffusion) CR energy equation (setting $\tilde{\bf F}_{\rm cr} \equiv \kappa_{\ast}\,\nabla_{\|} e_{\rm cr}$) give essentially identical results. We confirm this in our cosmological simulations in Fig.~\ref{fig:cr.fluxeqn}. We re-run a low and high-mass halo where CRs have a large effect ({\bf m11i} and {\bf m12i}), adopting the alternative formulation of the flux equation from \citet{thomas.pfrommer.18:alfven.reg.cr.transport}:
\begin{align}
\label{eqn:cr.flux.tp} \frac{\hat{\tilde{\bf F}}_{\rm cr}}{\tilde{c}^{2}}\,{\Bigl[} \frac{\partial |\tilde{\bf F}_{\rm cr}|}{\partial t} &+ \nabla\cdot\left( {\bf v}_{\rm gas} \,| \tilde{\bf F}_{\rm cr}| \right) + \tilde{\bf F}_{\rm cr}\cdot\{ (\hat{\tilde{\bf F}}_{\rm cr}\cdot\nabla)\,{\bf v}_{\rm gas} \} {\Bigr]} \\
\nonumber & + \nabla_{\|} P_{\rm cr} = -\frac{(\gamma_{\rm cr}-1)}{\kappa_{\ast}}\,\tilde{\bf F}_{\rm cr}
\end{align}
with $\hat{\tilde{\bf F}}_{\rm cr} \equiv \hat{\bf B}$ by definition. As discussed in \citet{chan:2018.cosmicray.fire.gammaray}, this is essentially the same as our formulation, up  to whether $\hat{\tilde{\bf F}}_{\rm cr}$ appears inside or outside of the derivative terms. The \citet{thomas.pfrommer.18:alfven.reg.cr.transport} formulation  arises if we  assume the perpendicular fluxes that arise  when  magnetic fields rotate are instantaneously converted into gyrotropic motion by micro-scale instabilities; this and the assumption of frame in which the motion is gyrotropic produce the small ``pseudo-force'' correction $\tilde{\bf F}_{\rm cr}\cdot\{ (\hat{\tilde{\bf F}}_{\rm cr}\cdot\nabla)\,{\bf v}_{\rm gas} \}$. In practice, the choice of Eq.~\ref{eqn:cr.flux.tp} instead of our default Eq.~\ref{eqn:cr.flux} only produces differences below the CR mean free path, and has no appreciable effect here. Unsurprisingly, the flux equation  from  \citet{jiang.oh:2018.cr.transport.m1.scheme}, which gives behavior ``in-between'' our Eq.~\ref{eqn:cr.flux}  and Eq.~\ref{eqn:cr.flux.tp} above (as shown in \citealt{chan:2018.cosmicray.fire.gammaray}) is even  more similar  to  our default. 

We have also experimented with a modified CR pressure term in the hydrodynamic equation motivated by \citet{jiang.oh:2018.cr.transport.m1.scheme}: replacing $\nabla P_{\rm cr} \rightarrow \nabla_{\bot}\,P_{\rm cr} - (\gamma_{\rm cr}-1)\,({\bf F}-{\bf v}_{\rm st}\,[e_{\rm cr} + P_{\rm cr}])/\kappa_{\ast} \approx  \nabla P_{\rm cr} + \tilde{c}^{-2}\,[\partial\tilde{\bf F}/\partial t + \nabla \cdot ({\bf v}_{\rm gas}\tilde{\bf F}_{\rm cr})]$, i.e.\ keeping the perpendicular CR pressure but only the parallel component contributing to CR scattering. Because this correction only enters in the $\sim 1/\tilde{c}^{2}$ terms, and because field lines are tangled, this has only weak effects.

Finally, Fig.~\ref{fig:cr.fluxeqn} also considers two additional variants of the streaming speed: taking $v_{\rm stream}=v_{A}$ (as compared to our default $=3\,v_{A}$), or setting $v_{\rm stream}=0$ but {\em retaining} the ``streaming loss'' term: in all of these the loss term scales as $v_{A}\,\nabla P_{\rm cr}$. Given the much larger variations to the streaming speed considered in the  main text (which produce small effects), it is expected that the effect of these variations is very small.

\end{appendix}

\end{document}